\documentclass{article}
\usepackage{authblk}
\usepackage{xcolor}
\usepackage{tcolorbox}
\usepackage{booktabs}
\usepackage{tikz}
\usepackage{amsmath}
\usepackage{colortbl}
\usepackage{enumerate}
\usepackage{graphicx} 
\usepackage{comment}
\usepackage{multirow}
\usepackage[a4paper, margin=2cm]{geometry}
\usepackage{amssymb}
\usepackage{mathtools}
\usepackage{subcaption}
\usepackage{array}
\usepackage{makecell} 
\usepackage{placeins} 

\renewcommand{\arraystretch}{1.2}

\newcommand{\singlecell}[2]{%
\begin{tikzpicture}[baseline=(current bounding box.center)]
\fill[#1!30] (0,0) rectangle (1,1);
\node at (0.5,0.5) {\Large #2};
\end{tikzpicture}%
}

\newcommand{\splitH}[4]{%
\begin{tikzpicture}[baseline=(current bounding box.center)]
\fill[#1!30] (0,0.5) rectangle (1,1);
\fill[#3!30] (0,0)   rectangle (1,0.5);
\node at (0.5,0.75) {\Large #2};
\node at (0.5,0.25) {\Large #4};
\end{tikzpicture}%
}

\newcommand{\splitV}[4]{%
\begin{tikzpicture}[baseline=(current bounding box.center)]
\fill[#1!30] (0,0)   rectangle (0.5,1);
\fill[#3!30] (0.5,0) rectangle (1,1);
\node at (0.25,0.5) {\Large #2};
\node at (0.75,0.5) {\Large #4};
\end{tikzpicture}%
}

\newcommand{\new}{\textcolor{black}}
\newcommand{\newnew}{\textcolor{black}}

\begin{document}

\title{\new{A Comprehensive Analysis of the Relation between Intermittency and Dissipation in Non-Homogeneous Turbulent Flows}}

\date{September 2026}

\author{F.H. Schmitt$^{\,\mathrm{a,b}}$, J. Peinke$^{\,\mathrm{b}}$ and M. Obligado$^{\,\mathrm{c,d}}$}

\affil{\textsuperscript{a} Univ. Grenoble Alpes, CNRS, Grenoble INP, LEGI, 38000 Grenoble, France}
\affil{\textsuperscript{b} Institut für Physik und ForWind, Universität Oldenburg, Küpkersweg 70, 26129 Oldenburg, Germany}
\affil{\textsuperscript{c} Univ. Lille, CNRS, ONERA, Arts et Métiers ParisTech, Centrale Lille, FRE 2017 - LMFL - Laboratoire de Mécanique des fluides de Lille - Kampé de Feriet, F-59000 Lille, France}
\affil{\textsuperscript{d} Institut universitaire de France (IUF), Paris, France}

\maketitle

\section*{Abstract} 

\new{A recent study, investigating a variety of turbulent flows, has presented evidence that intermittency and dissipation exhibit consistent scaling behavior. Departing from homogeneous isotropic turbulence (HIT), it has been shown for turbulent wakes, grid-turbulence and a turbulent jet, that the dissipation parameter $C_\varepsilon$ and the intermittency parameter $\mu$ can take various values while their product remains constant, independent of the Reynolds number $Re_\lambda$. While this previous study was limited to hot-wire anemometry for centerline data, we present here a follow-up study showing that this relation also holds for non-centerline data, including cases with shear. Under similar constraints, we identify two further relations: the product of $\gamma - 1$ (with $\gamma$ the absolute value of the slope of the energy spectrum in the inertial range) and $C_\varepsilon$ also remains constant. In addition, $\gamma$ varies linearly with the Kolmogorov parameter $C_k$. The primary constraint is that only data are considered for which the large-scale increment probability density function exhibits a Gaussian distribution, {indicating random large scale driving}. The set of varying $C_\varepsilon$, $\mu$, $\gamma$, and $C_k$ is reduced by their relations to three constants, which  enables the definition of a turbulence state that can be uniquely characterized by any of these parameters, for instance $C_\varepsilon$, independent of shear, flow type, or Reynolds number. We demonstrate that $C_\varepsilon$ depends directly on $Re_\lambda$ and the turbulence intensity $TI$ for a given flow case defined by a set of boundary conditions. As the values of $C_\varepsilon$, $\mu$, $\gamma$, and $C_k$ are not constant for the turbulent flows considered here, but instead  cover a broad range of values around their nominal values for HIT, we interpret this as a generalization of HIT to non-homogeneous turbulent flows.  We provide a detailed account of the post-processing and validation procedures, demonstrating for each considered turbulence quantity that the results are not dependent on the specific methodology employed. Moreover, we present a comprehensive analysis of the streamwise evolution of the turbulence quantities and their spatial patterns within the flow. Finally, we perform a forward analysis showing that, given a measured value of $C_\varepsilon$, the corresponding values of $\mu$, $\gamma$, and $C_k$ can be predicted within a defined level of uncertainty. These findings enable a fundamental new approach to turbulence, in which key leading-order features in turbulence are interconnected. The generalized framework presented here recovers the established values and trends reported for HIT, and expands them to cover inhomogeneous situations.} 

\FloatBarrier
\clearpage


\begin{table}[h!]
\centering

\begin{minipage}{0.45\textwidth}
\section*{\new{Nomenclature}}
\centering
\begin{tabular}{lcc}
\toprule
Symbol & Quantity & Unit \\
\midrule
$A^\prime$ & general const. &  \\
$C_k$ & Kolmogorov param. &  \\
$C_n$ & $n^{\mathrm{th}}$ ord. struct. funct. prefact. &  \\
$C_z$ & zero-crossing parameter &  \\
$C_\varepsilon$ & diss. param. &  \\
$\mathrm{D}$ & dimension &  \\
$d$ & diameter & m \\
$d^*$ & general diameter & m \\
$e_r(x)$ & energy transf. rate at scale $r$ & $\mathrm{m}^2/\mathrm{s}^3$ \\
$E(k)$ & energy spectr. density & $\mathrm{m}^3/\mathrm{s}^2$ \\
${e_c}$ & Castaing error at one scale &  \\
$e_s$ & subsampling error &  \\
$F$ & flatness&  \\
$F_u$ & flatn. of velocity time series &  \\
$f$ & sampling frequency & 1/s \\
$H_0$ & null hypothesis &  \\
$H_{1}$ & alternative hypothesis &  \\
$k$ & ang. wavenum. & 1/m \\
$k_{i}$ & ang. wavenum. (initial) & 1/m \\
$k_{s}$ & ang. wavenum. (small) & 1/m \\
$k_{l}$ & ang. wavenum. (large) & 1/m \\
$k_{void}$ & ang. wavenum. (void) & 1/m \\
$L$ & integral length scale & m \\
$L_e$ & integral length scale & m \\
$L_s$ & integral length scale & m \\
$L_z$ & integral length scale & m \\
$l_c$ & cutoff length scale & m \\
$M$ & mesh size & m \\
$m_1$ & expo. &  \\
$m_2$ & expo. &  \\
$N_s$ & num. of subsampling segments  &  \\
$N_v$ & num. of velocity time series &  \\
$n$ & ord. of moments &  \\
$n_L$ & num. of integral length scales &  \\
$n_z$ & zero-crossing density & 1/m \\
$p(\cdot)$ & probability density funct. &  \\
$\hat{p}(\cdot)$ & probability density funct. &  \\
$Re$ & Reynolds num. &  \\
$Re_L$ & loc. Reynolds num. based on $L$ &  \\
$Re_\lambda$ & loc. Reynolds num. based on $\lambda$ &  \\
$Re_G$ & glob. Reynolds num. &  \\
$R^2$ & coeff. of determination &  \\
$R^2_\mu$ & coeff. of determination for $\mu$ &  \\
$R_{uu}$ & auto-correlation funct. &  \\
$r$ & scale/spatial lag & m \\
$r_{1/\mathrm{e}}$ & scale for auto-correlation funct.  & m \\
\bottomrule
\end{tabular}
\end{minipage}
\hfill
\begin{minipage}{0.45\textwidth}
\centering
\begin{tabular}{lcc}
\toprule
Symbol & Quantity & Unit \\
\midrule
$r_{l}$ & scale (large) & m \\
$r_{s}$ & scale (small) & m \\
$S_n(r)$ & $n^{\mathrm{th}}$ ord. struct. funct. & m$^n$/s$^n$ \\
$S_u$ & skewn. of velocity time series &  \\
$s$ & solidity &  \\
$T$ & sampling duration & s \\
$TI$ & turbulence intensity &  \\
$t$ & time & s \\
$u$ & velocity fluctuations & m/s \\
$\hat{u}$ & velocity & m/s \\
$\overline{u}$ & mean velocity & m/s \\
$u^\prime$ & stand. deviation of fluct. & m/s \\
$\overline{u}_\infty$ & inflow velocity & m/s \\
$u_r(x)$ & increments, dep. on scale $r$ & m/s \\
$x$ & $x$-component & m \\
$y$ & $y$-component & m \\
$Z_i$ & zero crossings & m \\
\midrule
$\alpha$ & turbulence const. &  \\
$\beta$ & turbulence const. &  \\
$\gamma$ & abs. slope of spectr. law &  \\
$\gamma^*$ & mod. abs. slope of spectr. law &  \\
$\Delta(\cdot)$ & difference &  \\
$\Delta Z$ & diff. of zero crossings & m \\
$\varepsilon$ & mean diss. rate & $\mathrm{m}^2/\mathrm{s}^3$ \\
$\varepsilon_r(x)$ & diss. rate, dep. on scale $r$ & $\mathrm{m}^2/\mathrm{s}^3$ \\
$\varepsilon_d$ & mean diss. rate & $\mathrm{m}^2/\mathrm{s}^3$ \\
$\varepsilon_g$ & mean diss. rate  & $\mathrm{m}^2/\mathrm{s}^3$ \\
$\varepsilon_z$ & mean diss. rate  & $\mathrm{m}^2/\mathrm{s}^3$ \\
$\eta$ & length scale & m \\
$\theta$ & turbulence const. &  \\
$\lambda$ & length scale & m \\
$\lambda_d$ & length scale & m \\
$\lambda_z$ & length scale & m \\
$\Lambda^2(r)$ & shape parameter &  \\
$\Lambda_0^2$ & shape parameter at large scales &  \\
$\mu$ & intermittency parameter &  \\
$\mu^*$ & mod. intermittency parameter &  \\
$\mu_{\Lambda^2}$ & intermittency parameter &  \\
$\mu_z$ & intermittency parameter &  \\
$\mu_{\varepsilon}^{\prime}$ & intermittency parameter &  \\
$\mu_{\varepsilon}^{\prime\prime}$ & intermittency parameter &  \\
$\nu$ & kinematic viscosity & $\mathrm{m}^2/\mathrm{s}$ \\
$\phi$ & turbulence const. &  \\
$\Pi_i$ & theorem of Vaschy/Buckingham &  \\
$\sigma^2$ & var. &  \\
$\sigma_0^2$ & parameter &  \\
\bottomrule
\end{tabular}
\end{minipage}

\end{table}

\clearpage  

\begin{table}[h!]
\centering

\begin{minipage}{0.45\textwidth}
\centering
\begin{tabular}{lcc}
\toprule
Symbol & Quantity & Unit \\
\midrule
$\sigma_r^2$ & var. of $u_r(x)$ &$\mathrm{m}^2/\mathrm{s}^2$ \\
$\sigma_{r,\mathrm{ln}(\varepsilon/\overline{\varepsilon})}^2$ & var. of ln($\varepsilon_r/\overline{\varepsilon_r}$) & \\
$\sigma_{r,u_r|\varepsilon_r}^2$ & var. of $u_r|\varepsilon_r$ & $\mathrm{m}^2/\mathrm{s}^2$\\
$\sigma_s^2$ & var. of subsampl. quantity &  \\
$\tau$ & time lag & s \\
$\chi^2$ & prob. distribution & - \\
$\zeta_n$ & $n^{\mathrm{th}}$ ord. struct. funct. expo. &  \\
\midrule
$\overline{(\cdot)}$ & average &  \\
$\widetilde{(\cdot)}$ & median &  \\
$x|y$ & $x$ conditioned on $y$ &  \\
$(\cdot)_{cl}$ & detrended &  \\
\midrule
HIT & homo. isotropic turbulence & \\
\bottomrule
\end{tabular}
\end{minipage}
\hfill
\begin{minipage}{0.45\textwidth}
\centering
\begin{tabular}{lcc}
\toprule
Symbol & Quantity & Unit \\
\midrule
HWA & hot-wire anemometry & \\
ICM & individual check method & \\
KHMH & Kármán-Howarth-Monin-Hill & \\
K41 & theory of Kolmogorov 1941 & \\
K62 & theory of Kolmogorov 1962 & \\
KCB & Kolmogorov-Castaing-Beck & \\
NSE & Navier--Stokes equation & \\
PDF & probability density funct. &  \\
PTM & plateau threshold method & \\
SSM & subsampling method & \\
SST & scaling-structured turbulence &  \\
VTS & velocity time series & \\
ZCM & zero crossing method & \\
\bottomrule
\end{tabular}
\end{minipage}

\end{table}

\FloatBarrier

\tableofcontents  

\FloatBarrier

\section{\new{Introduction}} 

\new{Since the first half of the nineteenth century, the Navier--Stokes equations (NSE) have been known (\emph{cf.}~\cite{tamburrino2024navier}), providing a complete momentum balance for viscous Newtonian fluids and inherently encompassing the phenomenon of turbulence. The description and prediction of turbulent flows are crucial for many applications, ranging from biological problems like blood flow to engineering problems like wind energy~(\emph{cf.}~\cite{stevens2017flow, neunaber2020distinct}). Despite their apparent completeness, the NSE remain far from being solved in a general sense, as it is still unknown whether smooth solutions exist in three dimensions~\cite{fefferman2006existence, sreenivasan2025turbulence}. Furthermore, even the determination of averaged velocity fields for arbitrary boundary conditions remains an open problem~\cite{lesieur1987turbulence}.}

\new{To circumvent these difficulties, turbulent flows are commonly described using reduced models involving phenomenological laws and parameters that are not known \emph{a priori}. These parameters are typically estimated through dimensional analysis and/or experimental observations. Homogeneous isotropic turbulence (HIT) is arguably the most widely investigated reduced model. Homogeneity implies that the statistical description of the flow is unaffected by a spatial shift of the coordinate system. Isotropy imposes an additional constraint by requiring these statistics to be invariant with respect to rotations of the coordinate axes~\cite{pope2001turbulent}. In practical turbulent flows, however, both conditions are seldom fulfilled exactly over the complete spatial extent of the flow. They are therefore commonly treated as local approximations, assumed to be valid only within a sufficiently small three-dimensional subregion~\cite{pope2001turbulent}. For HIT in a statistically steady state, it was long assumed that such parameters become universal constants, at least at sufficiently high Reynolds numbers. Considerable effort has therefore been devoted to determining their values with increasing accuracy~\cite{praskovsky1994measurements, sreenivasan1995universality, arneodo1996structure, praskovsky1997comprehensive, yeung1997universality}.}

\new{Over the past two decades, however, growing evidence has challenged this assumption, particularly concerning dissipation in turbulent flows. Following the work of Vassilicos~\cite{vassilicos2015dissipation}, several studies have shown that in many flows the dissipation parameter (also known as the dissipation constant), defined as,}

\begin{equation} \new{
    C_{\varepsilon} = \frac {\varepsilon \: L} {u^{\prime 3}},}
    \label{dissipation_constant}
\end{equation}

\noindent \new{where $u^{\prime}$ denotes the root-mean-square value of the streamwise fluctuating velocity, $L$ the integral length scale and $\varepsilon$ the mean turbulent kinetic energy dissipation rate, depends on the Reynolds number, even under nominally HIT conditions. How $C_\varepsilon$ and other turbulence parameters generally evolve as the flow departs from HIT remains an open question.}

\new{In a recent study, Schmitt \emph{et al.}~\cite{schmitt2024universal} addressed these issues by focusing on dissipation and intermittency, see also~\cite{schmitt2026small}. Intermittency manifests itself through the occurrence of extreme events, which become increasingly frequent toward smaller scales. The intermittency parameter $\mu$ quantifies how strongly the prevalence of these extreme events increases across the scales~\cite{frisch1995turbulence}. It was shown in the recent study, that for inhomogeneous turbulence, the intermittency parameter not only departs significantly from its HIT value, but is also inversely proportional to the dissipation parameter $C_{\varepsilon}$. Moreover, this relation was extended to include the slope of the energy spectrum in the inertial range $\gamma$, and the prefactor of the spectral law $C_k$, commonly referred to as the Kolmogorov parameter~\cite{sreenivasan1995universality}.} 

\new{However, their analysis relied exclusively on centerline measurements in turbulent wakes, grid-generated turbulence and turbulent jets, where shear effects remain relatively weak. More generally, many experimental and numerical studies of turbulence focus on restricted flow configurations, limited datasets, and specific assumptions or diagnostics, thereby providing only partial insight into the inherently multi-dimensional nature of turbulence.}

\new{Building on the work of Schmitt \emph{et al.}~\cite{schmitt2024universal} and motivated by the limitations imposed by strict assumptions, particularly those associated with HIT, the present work seeks to extend the underlying concept toward inhomogeneous turbulence. This extension does not challenge the validity or relevance of HIT. Rather, it builds upon its framework in order to include a broader range of turbulent flows that retain comparable statistical characteristics. This broader class of turbulent flows is referred to as scaling-structured turbulence (SST).}

{Consequently, the present work aims to extend this analysis to non-centerline measurements in turbulent wakes, grid-generated turbulence and turbulent jets, spanning different levels of mean shear. In total, more than 5000 velocity time series (VTS) are analyzed. The aim is to develop a first general understanding of SST. However, particular caution is required when exploring a regime that has not yet been systematically characterized. The extensive scope of the present study reflects this caution, as sufficient attention is devoted to carefully examining the assumptions, methods, and implications associated with this broader turbulence framework. We also have to point out that in this paper we analyze only the velocity component in the flow direction and we use several definitions which are based on HIT. It is clear that, if the original definitions are retained purely formally, their well-established physical interpretation within HIT can no longer be taken for granted. Nevertheless, applying these definitions in their mathematical and statistical form yields consistent and informative results, which may in turn be regarded as justification for the present approach.}

\new{Evidence is presented that, within SST, the relation between dissipation and intermittency remains valid even in the presence of strong shear and is independent of the particular methods used to estimate $\mu$ and $C_\varepsilon$. Furthermore, two additional relations are demonstrated, one between $\gamma$ and $C_\varepsilon$, and another between $\gamma$ and $C_k$.}

\new{The paper is structured as follows. In chapter~\ref{beyond}, the motivation for developing a new approach to generalize HIT is presented in detail, together with the data selection that forms the basis of this approach. In addition, relevant quantities are identified and the underlying assumptions are outlined, leading to the formulation of scientific hypotheses that further specify the proposed framework and provide the foundation of this work. Chapter~\ref{new_laws} presents the results, including three novel relations for SST. In chapter~\ref{methods}, the methods to determine all relevant quantities are systematically examined and validated to ensure that the results for all data are not influenced by the employed methods. This is additionally demonstrated using eight representative VTS. Chapters~\ref{control} and~\ref{spatial} examine the dependence on conventional control parameters, such as $TI$ and $Re_\lambda$, and on the spatial position within the flow, respectively. Chapter~\ref{discussion} provides a broader discussion of the results, the data-selection procedure, and the underlying assumptions, together with interpretations of the findings and a further justification for extending the analysis beyond HIT. Finally, Chapter~\ref{conclusion} summarizes the main conclusions of the work.}

\section{\new{Beyond Kolmogorov}} 
\label{beyond}

\new{Statistically stationary turbulence is typically characterized by distinct phenomena, such as energy transfer, dissipation, and intermittency, whose respective magnitudes are quantified by dimensionless parameters. As turbulence exists in many different flows with distinct boundary conditions and many degrees of freedom due to its dependence on 3~D space, the main focus was often placed on the quantification of  dimensionless parameters in HIT. The reduced model of HIT is often viewed in the context of Kolmogorov, who was among the pioneers in establishing this idealized state as a foundation for turbulence research~\cite{frisch1995turbulence}. The use of HIT in turbulence research can be viewed as an important simplification, which may be set in analogy to the use of point masses in classical mechanics, enabling direct access to the underlying dynamics. Within the last 85 years, considerable effort has been devoted to understanding HIT and to improve Kolmogorov's models, leading to a well-established theoretical framework of this particular state of turbulence, which is generally considered to emerge asymptotically, as the Reynolds number approaches infinity~(\emph{cf.} for a recent review~\cite{sreenivasan2025turbulence}). Note that various definitions of the Reynolds number exist, all representing the ratio between inertial and viscous forces. For turbulence research away from boundaries, the Taylor-length based Reynolds number, $Re_\lambda = u^{\prime} \, \lambda / \nu$, has proven to be particularly useful. Here, $\lambda = (15 \: \nu \: u^{\prime 2}/\varepsilon)^{1/2}$ denotes the Taylor length scale which provides an estimate of the limit of the inertial range toward small scales, and $\nu$ the kinematic viscosity of the fluid~\cite{davidson2015turbulence}. Unless otherwise stated, this definition is used throughout and referred to simply as the Reynolds number.}

\subsection{\new{Inherent Challenges of Turbulence Research and a Way Forward}} 

\new{However, there are several problems that arise with the focus on HIT. On the experimental side, it is challenging to quantify the degree of homogeneity and isotropy of a turbulent flow. Although mathematically exact definitions exist, they are generally only applicable in some limits in experiments, as they require, at minimum, knowledge of the full Reynolds stress tensor at every point in 3~D space~\cite{pope2001turbulent, davidson2015turbulence}. It is therefore standard practice for HIT 
studies to retain only data exhibiting special properties, \textit{e.g.} an inertial-range energy spectrum with a slope close to $-5/3$, with respect to Kolmogorov's turbulence theory of 1941~\cite{kolmogorov1941local} (hereafter denoted as K41 and further explained below). On the post-processing side, quantifying deviations from the $-5/3$ scaling is equally challenging, as estimating such a slope requires a clearly defined fitting range, which itself is ambiguous to determine.} 

\new{For turbulence research on dimensionless parameters within HIT, this makes it hard to distinguish between scatter arising from the thresholds applied during post-processing, variability due to the inherently high-dimensional nature of turbulence, and deviations of physical origin. This applies not only to the slope of the energy spectrum but also to all other turbulence quantities.} 

\new{Even if one were to attempt to quantify the regions of the flow in which HIT can be reasonably assumed, and those in which it cannot, this idealized state represents only a small subset of real turbulent flows. In most practical applications, turbulence is inherently inhomogeneous and occurs not in the limit of infinite Reynolds numbers.}

\new{In this contribution, we will not focus on the important corrections to the Kolmogorov picture of turbulence, which usually concern higher-order statistical moments. Instead, we will explore how the basic HIT concept can be applied to turbulent flows that deviate from HIT. Our  concern is closely reflected in the perspective of Sreenivasan and Schumacher~\cite{sreenivasan2025turbulence}, who caution against reducing the turbulence problem to an overly restricted set of questions: \textit{``We believe that the greater danger is to define the problem so narrowly that its solution produces neither intellectual excitement nor practical impact''}. This work therefore follows a line of research that retains HIT as an important reference framework while deliberately extending the analysis beyond the HIT paradigm, as pursued in several recent studies by Vassilicos~\cite{chen2022scalings}, Dubrulle~\cite{dubrulle2019beyond}, Schmitt~\cite{schmitt2024universal} and others.}

\new{ We propose a way forward by conducting an experiment with the following characteristic,}

\new{
\begin{itemize} 
    \item measuring fluctuations with high temporal resolution using simple and robust techniques, such as hot wire,
    \item investigation of many different flow configurations, such as wakes, grid-turbulence, and jets,
    \item measurements at all locations where turbulence is assumed, yielding a large amount of high-quality data,
    \item application of simple post-processing methods to each phenomenon, enabling overall consistency checks,
    \item while relaxing both a lower bound on the local Reynolds number and the strict selection criteria associated with HIT, yet retaining key HIT-like features such as general scaling behavior in the energy spectrum.
\end{itemize} }

\noindent \new{Since measurements taken at different locations in various turbulent flows may yield VTS in which turbulence characteristics are too weak to be clearly identified, new selection criteria are required.}

\subsection{\new{Data and Data Selection as the Foundation of a New Approach}}
\label{data selection}

\new{The experimental data used in this work originate from 1$\,$D hot-wire anemometry (HWA) measurements. HWA is commonly employed to obtain a characteristic statistical description of HIT. The measurements were obtained during four independent campaigns (including grid-generated turbulence~\cite{ferran2023characterising, mora2019energy}, an axisymmetric turbulent jet~\cite{renner2001experimental} and planar and axisymmetric wakes~\cite{schmitt2024universal}), comprising a total of $5172$ VTS. While the free-jet and grid-turbulence datasets are identical to those used in the aforementioned work (where only centerline data were used), the wake dataset has been extended to include non-centerline measurements. For details about the different cases, see table~\ref{tab:PhD measurements in LEGI 2023} in the appendix~\cite{SM}, where the case names, \textit{e.g.} C4 are listed and the characteristic of the corresponding setup is described. Exemplary results will be shown for eight datasets, which will be explained later.}

\new{To define selection criteria for discarding VTS, we first define their general characteristics. Suitable selection criteria should be based on clear and measurable thresholds that can be evaluated using 1~D hot-wire measurements, have a well-defined physical meaning, and be kept to the minimum number necessary.}

\new{In the previous work on this topic, Schmitt~\emph{et al.}~\cite{schmitt2024universal} employed two sets of selection criteria to demonstrate the differences arising from applying two levels of relaxed restrictions compared to HIT. With the complete dataset now available (also non-centerline data), we are able to establish the following set of selection criteria, that more closely fulfills the aforementioned rules of quality.}

\new{The first selection criterion concerns the requirement of statistical stationarity implied by HIT. As there is no reason to relax this criterion within the present approach, all selected VTS must contain a sufficient number of data points to ensure statistical convergence. As this can be easily achieved by choosing an appropriate sampling time, all VTS in our dataset satisfy this criterion with $n_{L} > 1000$ (the smallest value of $n_L$ within our complete dataset is $4 \times 10^3$). $n_L=T \: \overline{u}/L$ represents the amount of integral length scales captured by the VTS, with $\overline{u}$ as the local mean velocity and $T$ as the sampling time. Stationarity is verified additionally by analyzing subsamples of the given datasets. }

\new{The second selection criterion concerns the temporal resolution of the VTS. As dissipation, one of the key phenomena of turbulence, occurs at the smallest scales, it is essential to assess the extent to which these scales are resolved. The smallest relevant scales are typically characterized by the Kolmogorov length scale $\eta := (\nu^3 / \varepsilon)^{1/4}$. In addition, we define $l_c$ as the smallest length scale for which the corresponding structures can be resolved without any visually discernible noise contamination in the energy spectrum obtained from hot-wire measurements. All selected VTS must satisfy $l_c/\eta < 15$. Since the maximum value of $l_c/\eta$ within our complete dataset is $12.8$, all VTS in our dataset fulfil this criterion. Notably, this criterion is also independent from any assumption regarding homogeneity and isotropy. As a reference, using $k = 2\pi/l_c$, the maximum value $l_c/\eta = 12.8$ corresponds to $k \,\eta \approx 0.5$. Melina~\textit{et al.}~\cite{melina2016vortex} reported that, at $k \, \eta = 0.5$, the dissipation rate $\varepsilon$ is underestimated by approximately $7\,\%$.}

\begin{figure}[htbp]
    \centering
    \begin{subfigure}[t]{0.49\textwidth}
        \centering        
        \includegraphics[width=\linewidth]{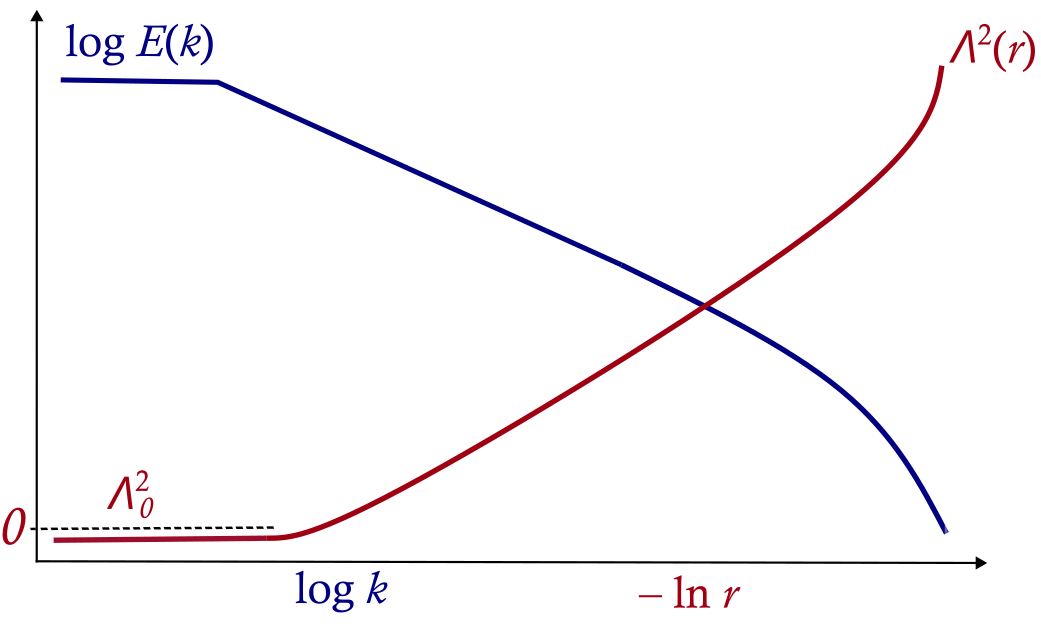}
        \caption{}
    \end{subfigure}
    \hfill
    \begin{subfigure}[t]{0.49\textwidth}
        \centering        
        \includegraphics[width=\linewidth]{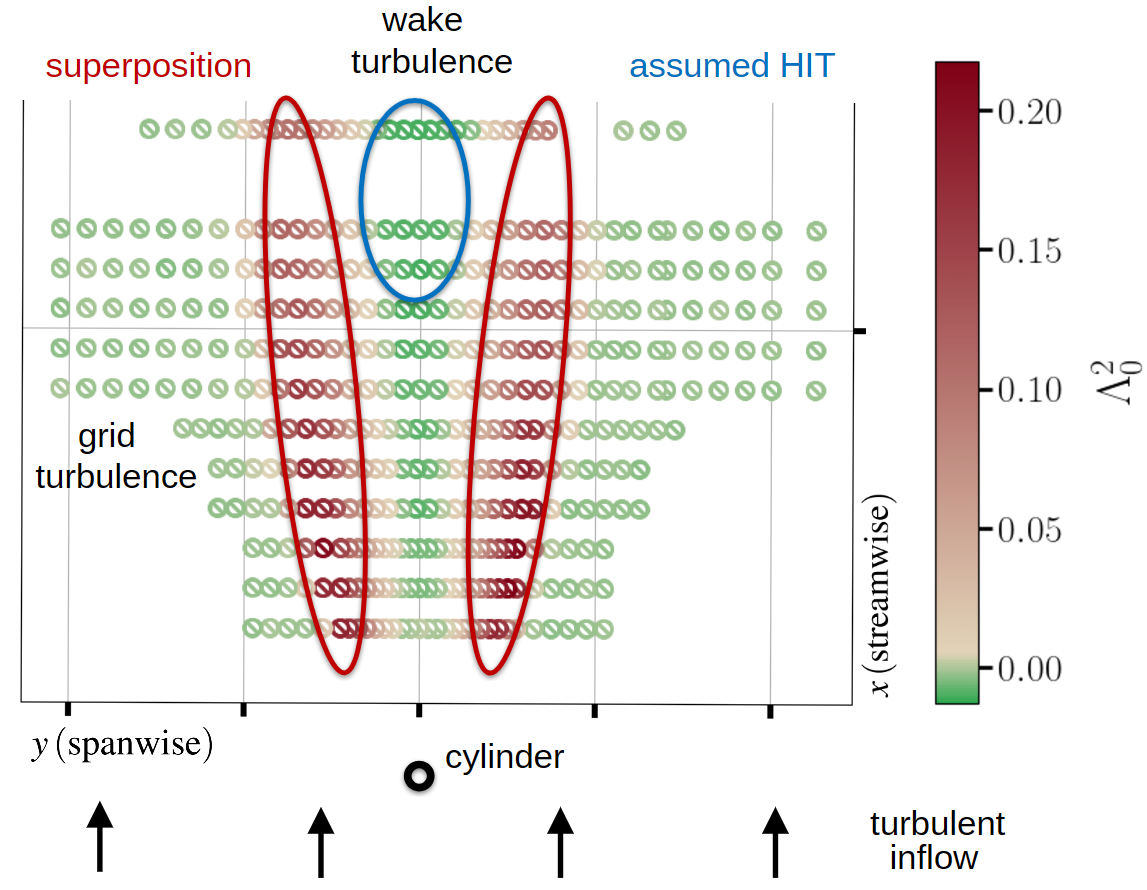}
        \caption{}
    \end{subfigure}
    \caption{\new{a) Schematic representation of the turbulence cascade. The blue curve shows the energy spectral density $E(k)$. The red curve represents the evolution of the shape parameter $\Lambda^2(r)$. Note that in this sketch, $\Lambda_0^2$ is approximately zero. b) Included and excluded data based on their value of $\Lambda_0^2$ for one case (C4) where a turbulent wake is created by cylinder, accompanied by a turbulent inflow generated by a passive grid. Note that the transition from red to green occurs exactly at $\Lambda_0^2 = 0.005$. The $x$ and $y$ axis represent the streamwise and spanwise coordinates, respectively, while the color displays the $\Lambda_0^2$ value for each measurement point. The black arrows indicate the turbulent inflow and the black circle schematically represents the cylinder. All green points exhibit scaling behavior, whereas the points circled in red display large-scale non-Gaussianity and are therefore excluded from used data. The red points correspond precisely to the region where the background turbulence from the turbulent inflow and the wake turbulence superimpose. The blue circle schematically shows, which data are usually taken when considering HIT only.}}
    \label{data_selection}
\end{figure}

\new{All further selection criteria are based on two-point statistics. Figure~\ref{data_selection} a) illustrates the concept of the following criteria. The inherent correlations in turbulence imply that it cannot be completely characterized by one-point statistics alone. Following Morales~\emph{et al.}~\cite{morales2012characterization}, we define quantities as one-point statistics if they are invariant under any reordering of the VTS. To illustrate this, consider a VTS of velocity fluctuations in which the data points are rearranged according to their magnitude. If the estimation of a given quantity yields significantly different results for the ordered and unordered VTS, the corresponding method necessarily involves at least two-point statistics. Notably this distinction does not depend on the temporal resolution of the VTS, and for composite quantities, the highest order involved determines the overall order.}

\new{The third criterion is based on K41, which predicts a scaling law for the energy spectrum within the inertial range for $\eta \ll r \ll L$ of HIT~\cite{kolmogorov1941local}, such that}

\begin{equation} \new{
     E(k)=C_k \:\: \varepsilon^{2/3} \: k^{-5/3},}
     \label{HIT_equation_energy_spectrum}
\end{equation} 

\noindent \new{where $E(k)$ denotes the energy spectral density, $k$ the angular wavenumber and $C_k$ the Kolmogorov parameter (also known as the Kolmogorov constant). Within this framework, the spectral exponent is fixed at -5/3, based on dimensional arguments and therefore also known as the five-thirds law~\cite{davidson2015turbulence}. At the same time, the non-dimensional prefactor is assumed to be constant, although no theoretical value has been derived. Consequently, an empirical value of $C_k \approx 0.5$ has been reported~\cite{sreenivasan1995universality}. Although Kolmogorov introduced refinements in 1962~\cite{kolmogorov1962refinement} (hereafter K62), these corrections to the $-5/3$ scaling exponent remain comparatively small within the framework of HIT. Nevertheless, several studies have reported measurable deviations from the K41-exponent of $-5/3$ within general turbulence~\cite{neunaber2020distinct, rodriguez2023not}. Accordingly, our third criterion only requires the presence of power-law scaling within the inertial range of $E(k)$ without any constraint on the value of the scaling exponent. In particular, we retain only data if $E(k)$ follows the relation}

\begin{equation} \new{
     E(k)=C_k \:\: \varepsilon^{2/3} \: k^{-5/3} \: (k \: \eta / 2 \: \pi)^{-\gamma + 5/3},}
     \label{equation_energy_spectrum}
\end{equation} 

\noindent \new{within the inertial range, where $\gamma$ denotes the absolute value of the slope of the spectrum. Note that $k$ is normalized using the $\eta$, the Kolmogorov length scale, following Mydlarski \& Warhaft~\cite{mydlarski1996onset}. This ensures that variations in $\gamma$ do not violate dimensional consistency, and for $\gamma = 5/3$, eq.~(\ref{HIT_equation_energy_spectrum}) is recovered. Within our approach, we explicitly allow $\gamma$ to take any value above a lower bound of 1.25. The lower bound is imposed because spectral fits become invalid as the slope approaches a value of 1. Furthermore we allow $C_k$ to take any positive value.}

\new{Due to the Wiener-Khintchine theorem~\cite{wiener1930generalized,khintchine1934korrelationstheorie}, the energy spectral density ($k$-space) is directly related to the autocorrelation function ($r$-space) and thus to the second order structure function ($r$-space). In general, structure functions are defined as $S_n(r) := \overline{{u_r(x)}^n}$ of a given order $n$, where $u_r(x) := u(x+r) - u(x)$ defines the velocity increments with the spatial lag $r$ and the overline denotes a spatial average. The measured VTS $\hat{u}(t)$ is decomposed using Reynolds decomposition, such that the velocity fluctuations are given by $u(t) = \hat{u}(t) - \overline{u}$. $u(x)$ is built using Taylor's hypothesis of frozen turbulence, where time $t = x/\overline{u}$ with $x$ denoting the streamwise distance~\cite{davidson2015turbulence} (a closer look at Taylor's hypothesis of frozen turbulence is provided below within the general assumptions). Thus, equivalent to eq.~(\ref{HIT_equation_energy_spectrum}) and thus also based on K41, the structure functions should scale as~\cite{davidson2015turbulence},}

\begin{equation} \new{
    S_n(r) = C_n \: (\varepsilon \, r)^{\zeta_n},}
    \label{HIT_n_order_SF}
\end{equation}

\noindent \new{within the inertial range, where $C_n$ is assumed to be an $n$-dependent universal prefactor and $\zeta_n$ should scale as $n/3$ for HIT~\cite{davidson2015turbulence}. In particular, within the inertial range, the scaling exponent of the second order structure function is given for K41 by $\zeta_2 = \gamma -1$, yielding $\zeta_2 = 2/3$ for $\gamma = 5/3$, commonly referred to as the \emph{two-thirds law}~\cite{davidson2015turbulence}.  For the third order structure function, $\zeta_3=1$ should hold, based on the Kármán-Howarth equation, which provides an exact energy balance in $r$-space for HIT, directly derived from the NSE. This result is associated with the well-known \emph{fourth-fifths} law, which specifies the corresponding exact prefactor $C_3 = -4/5$~\cite{davidson2015turbulence}. Since the present study also considers turbulence with significant shear beyond the regime of HIT, we relax, among others, the constraints on $\zeta_2, \zeta_3$ and $C_3$. In this context, the classical Kármán-Howarth equation is more appropriately replaced by its generalized form, the Kármán-Howarth-Monin-Hill (KHMH) equation~\cite{hill2002exact, vassilicos2015dissipation}. In contrast to the homogeneous case, the KHMH equation contains additional terms that account for spatial inhomogeneity, allowing for energy transport not only across scales but also between points with distinct $x$-coordinates.} 

\FloatBarrier

\new{The fourth criterion follows a similar rationale but focuses on the phenomenon of intermittency and its associated scaling laws. Intermittency can be interpreted as the deviation from self-similarity (self-similarity in the sense of K41) as one probes turbulence at smaller and smaller scales~\cite{frisch1995turbulence}. K41~\cite{kolmogorov1941local} neglects intermittency as it does not take fluctuations of the scale-dependent dissipation rates $\varepsilon_r(x)$ into account, assuming $\varepsilon_r(x)= \mathrm{const.}$, for HIT $\varepsilon_r(x)$ can be approximated as~\cite{naert1998conditional},}

\begin{equation} \new{
    \varepsilon_r(x) = \frac{15\, \nu}{r} \int_{x-r/2}^{x+r/2} \bigg(\frac{\mathrm{d}u(x)}{\mathrm{d}x}\bigg)^2 \, \mathrm{d}x.} 
    \label{equation_epsilon_r}
\end{equation}

\new{This assumption was criticized by Landau, since in reality, a stochastic energy cascade from large to small scales automatically leads to a fluctuating  $\varepsilon_r(x) \neq \mathrm{const.}$~\cite{kolmogorov1962refinement}. Subsequently, Kolmogorov introduced an intermittency correction within his K62 theory to account for fluctuations in the scale-dependent dissipation rate $\varepsilon_r(x)$~\cite{kolmogorov1962refinement}. While this view is now widely accepted among the turbulence community, the precise form of the resulting distribution $p(\varepsilon_r(x))$ remains an open question.}

\new{The theory of K62 is based on a log-normal distribution of $\varepsilon_r(x)$, which arises from the assumption of a multiplicative cascade. Commonly intermittency was defined through the parameter $\mu = 2-\zeta_6$, where $\zeta_6$ denotes the slope of the sixth order structure function within the inertial range~\cite{kolmogorov1962refinement}. Building on this foundation of K62, Castaing and co-workers redefined $\mu$ such that the quantification of intermittency was based on the evolution of the normalized spatial velocity increment PDFs across scales $r$ within the inertial range, rather than on an \emph{a priori} structure-function model~\cite{castaing1990velocity, arneodo1996structure}. The defining characteristic for the increment PDFs is the so-called shape parameter, which is defined as~\cite{chilla1996multiplicative}}

\begin{equation} \new{
    \Lambda^2(r) = \frac {\ln \left(F(u_r(x))/3 \right)} {4},}
    \label{equation_lambda_one}
\end{equation}

\noindent \new{with ln denoting the natural logarithm and $F$ denoting the flatness ($=3$ for Gaussian distribution), such that $\Lambda^2=0$ indicates a Gaussian distribution of the increments at a certain scale (at least regarding the fourth-order central moment), while positive values indicate a heavy tail distribution. {A brief note on the notation: in recent works, the symbols $\lambda^2$, $\Lambda^2$ and $\Lambda_0^2$ have been used with definitions that differ from those adopted here. To avoid confusion with the Taylor length scale $\lambda$, we consistently use the capital $\Lambda$ denoting intermittency-related quantities.} Note that this approach neglects the skewness of the distribution. Following Kolmogorov~\cite{kolmogorov1962refinement} and Castaing~\cite{castaing1990velocity}, the evolution of the shape parameter $\Lambda^2(r)$ should follow the relation}

\begin{equation} \new{
    \Lambda^2(r) = \Lambda_0^2 + \mu / 9 \:\: \mathrm{ln}(L/r),}
    \label{equation_lambda_two}
\end{equation}

\noindent {within the inertial range, where $\mu$ is the intermittency parameter, $L$ is the integral length scale and $\Lambda_0^2$ is a constant that depends on the boundary conditions of the flow (see figure~\ref{data_selection} a). Both Kolmogorov and Castaing assumed $\mu$ to be a universal constant in HIT and Arneodo~\emph{et al.} reported $\mu = 0.26$~\cite{arneodo1996structure} within the framework of HIT. Although we explicitly allow $\mu$ to take any positive value within our approach, our fourth criterion requires the presence of such scaling behavior, \textit{e.g.} we require that equation~(\ref{equation_lambda_two}) holds, see also figure~\ref{data_selection} a). To ensure the validity of this scaling, two selection conditions are applied. All VTS are discarded if the fit of $\Lambda^2(r)$ used to determine $\mu$ yields a coefficient of determination $R^2_\mu$ below 0.99. Second, all VTS are discarded if the Castaing error $\overline{e_c}$, defined below, exceeds 0.1. This indicates that the shape parameters $\Lambda^2(r=\lambda)$ and $\Lambda^2(r=L)$ are not well determined (details are provided in subsection~\ref{Intermittency parameter}). Note that, in this work, we exclusively adopt the definition of intermittency introduced by Kolmogorov~\cite{kolmogorov1962refinement} and Castaing~\cite{castaing1990velocity} in which intermittency is characterized by the scale-dependent evolution of the normalized increment PDF. For a distinction from other definitions of intermittency in turbulence, the reader is referred to~\cite{schmitt2024universal}. {The use of $\Lambda^2(r)$ to characterize intermittency has also the advantage that with $\Lambda^2(r)$, the PDFs of the velocity increment $p(u_r(x))$ are known, as will be discussed later and shown in figure~\ref{figure_eight_examples_PDF}.}

\new{Finally, the fifth selection criterion is directly related to the previous one and requires that the shape of the increment PDF at large scales is nearly Gaussian. This is physically meaningful as the value of $\mu$ is only properly defined as a departure from a Gaussian distribution at large scales. In consequence, only VTS with $\Lambda_0^2$ values below 0.005 are retained. This requirement ensures that the cascade starts from a quasi-Gaussian large-scale state, so that the measured value of $\mu$ reflects the scale-dependent build-up of non-Gaussianity (see figure~\ref{data_selection}). The application of the selection procedure, especially the fifth criterion is illustrated in figure~\ref{data_selection} b) where $\Lambda_0^2$ is shown as a colormap over the streamwise and spanwise coordinates for a cylinder-wake case with a wooden-grid inflow. The red-colored measurements exhibit non-Gaussianity at large scales, consequently, all red-circled points were discarded. This region corresponds to the spatial overlap of two distinct turbulence types. In contrast, for all other measurements, $\Lambda_0^2$ approximately vanishes and since the other three criteria are also satisfied for the vast majority, these VTS are retained. Note that the non-Gaussianity of large scales alone is also sometimes referred to as intermittency, for instance in~\cite{zhou2023appearance}. A general remark should also be made regarding higher-order moments, \textit{e.g.}, for $n>4$, which constitute an important component of turbulence characterization. The present study uses explicitly only moments of order $n\leq4$, to determine $ \Lambda^2(r)$, as explained below. As $\Lambda^2(r)$ also determines the form of the PDFs of the velocity increment $p(u_r(x))$ and their tails, higher order moments are also implicitly grasped. However, we must admit that our approach here is based on K62, including problems for moments of $n > 10$. A more detailed investigation of higher-order statistics is left for future work.}

\new{Altogether, this study presents only main results obtained from VTS that satisfy}

\begin{enumerate}
    \setcounter{enumi}{0}
    \item \new{Statistical stationarity of the VTS} \color{blue} $\rightarrow \: n_{L} > 1000$\color{black},
    \item {Temporal resolution of the VTS} \color{blue} $\rightarrow \: l_c/\eta < 15$\color{black},
    \item \new{Inertial range scaling of $E(k)$ in eq. (\ref{equation_energy_spectrum})} \color{blue} $\: \rightarrow \: \gamma > 1.25$\color{black}, 
    \item \new{Inertial range scaling of $\Lambda^2(r)$ in eq. (\ref{equation_lambda_two})} \color{blue} $\: \rightarrow \:  R_{\mu}^2 > 0.99, \:\:\: \overline{e_c} < 0.1$\color{black}, 
    \item \new{Gaussianity at large-scale spatial velocity increment PDFs in eq.(\ref{equation_lambda_two})} \color{blue} $ \: \rightarrow \: \Lambda_0^2 < 0.005$\color{black},
\end{enumerate}

\noindent \new{and thus correspond to scaling-structured turbulence (SST). After this conditioning, 1819 VTS remain. Of all discarded VTS, 62$\,\%$ were excluded based on their value of $\Lambda_0^2$. A further 22$\,\%$ were rejected because they did not exhibit sufficiently clear signatures of turbulence. We provide further details on why VTS were mainly discarded in chapter~\ref{spatial}).}

\FloatBarrier

\subsection{\new{Identification of Relevant Scalar Quantities for Statistical Description}}

\new{In the previous work on this topic by Schmitt \emph{et al.}, the focus was placed on, in addition to the Reynolds number, the scalar quantities $C_\varepsilon$, $\mu$, $\Lambda_0^2$, $\gamma$, and $C_k$, as they govern the equations describing the most extensively studied phenomena in turbulence: dissipation, intermittency, and energy transfer. In the present work, we adopt a different perspective by addressing the following questions.} 

\new{
\begin{itemize}
    \item Which independent scalar quantities are required for the statistical description of SST, stemming from a 1$\,$D VTS,
    \item and which are the corresponding dimensionless constants ($\Pi$-parameters) that allow a statistical description of SST following Vaschy~\cite{vaschy1892lois} and Buckingham~\cite{buckingham1914physically}?
\end{itemize}}

\noindent \new{To address the first question, table~\ref{table_which_quantities_are_important_all} provides an overview which scalar quantities are necessary to fully describe the lower-order ($n \leq 4)$ one-point and two-point statistics, associated with SST, stemming from a 1$\,$D VTS. The quantities listed in the first column are estimated from the measured spatial signal $\hat{u}(x)$. These quantities are related to the large scale velocity increments $u_r(x)$ ($r \geq L$). The second column contains quantities estimated by using $u_r(x)$ ($L \geq r \geq \lambda$), which covers the inertial-range velocity increments. Finally, the spatial derivative of the fluctuations $\partial u/ \partial x$ covers the two-point statistics of the smallest turbulence scales ($r \approx \eta$) and its quantities are summarized in the third column.} 

\new{The mean value of $\hat{u}(x)$ is described by $\overline{u}$, whereas the mean values of the increments and the derivative vanish by definition. The variance of $\hat{u}(x)$ is fully characterized by $u^\prime$. For the increments within the inertial range, the $r$-dependent variance is described by $\gamma$ and $C_k$. The variance of the derivative is proportional to the ratio between $\varepsilon$ and $\nu$. As mentioned in the data selection, the skewness is generally not considered in the present approach. This is due to the fact that no relation was found between the skewness and any other quantity. The flatness of the large-scale increments $\Lambda_0^2$, and thus the flatness $F_u$ of $\hat{u}(x)$, is fixed when restricting to SST. $\Lambda_0^2$ (and $F_u$) can be interpreted as control parameters determining whether the flow resides within the SST regime. The scale-dependent flatness of the increments within the inertial range is fully characterized by $\mu$. Since the focus is on the cascade, and thus mostly on the inertial range, the flatness of the derivative is not considered.}

\begin{table}[h]
    \centering
    \renewcommand{\arraystretch}{2} 
    \begin{tabular}{lccc}
        \hline
        & signal $\hat{u}(x)$ & increments $u_r(x)$ ($L \geq r \geq \lambda$) & derivative $\partial u/ \partial x$ ($r \approx \eta$) \\
        \hline
        mean &  $\overline{u}$ & 0 &  0 \\
        \hline
        variance & $u^{\prime 2}$ & $\gamma, C_k$ & $\varepsilon/ \nu$ \\
        \hline
        skewness & not considered & not considered & not considered\\
        \hline
        flatness & $F_u \approx 3$ ($\Lambda_0^2 \approx 0$) &  $\mu$ & not considered\\
        \hline
    \end{tabular}
    \caption{\new{Scalar quantities derived from a $1\,$D VTS obtained in SST. $\hat{u}(x)$ denotes the velocity spatial signal, $u_r(x)$ ($L \geq r \geq \lambda$) the velocity increments within the inertial range, and $\partial u/ \partial x$ the derivative of $u(x)$ at the smallest scales $r \approx \eta$. Moreover, $\overline{u}$ denotes the mean velocity at the acquisition point, $u^{\prime}$ the standard deviation of the fluctuations, $\gamma$ the absolute slope of the energy spectrum within the inertial range, $C_k$ the Kolmogorov parameter, $\varepsilon$ the mean dissipation rate, $\nu$ the kinematic viscosity, $F_u$ the flatness of the VTS, $\Lambda_0^2$ the shape parameter at the large scales, and $\mu$ the intermittency parameter.}}
    \label{table_which_quantities_are_important_all}
\end{table}

\FloatBarrier

{To characterize SST, based on a $1\,$D VTS, we need to know the following 12 scalar quantities: $\nu$ (m²/s), $\overline{u}$ (m/s), $u^\prime$ (m/s), $L$ (m), $\varepsilon$ (m²/s³), $\lambda$ (m), $\eta$ (m), $\gamma$, $C_k$, $\mu$, $\Lambda_0^2$, and $F_u$. To address the second question regarding non-dimensional quantities, $\gamma$, $C_k$, $\mu$, $\Lambda_0^2$, and $F_u$ are already dimensionless. } 

{Next, we seek corresponding dimensionless quantities for the five remaining dimensional quantities {($\lambda=(15 \: \nu \: u^{\prime 2}/\varepsilon)^{1/2}$ and $\eta=(\nu^3 / \varepsilon)^{1/4}$ can be neglected). According to the $\Pi$-theorem of Vaschy~\cite{vaschy1892lois} and Buckingham~\cite{buckingham1914physically}, we define three additional dimensionless quantities as follows:}

\begin{itemize} 
    \item $\overline{u}$ $\rightarrow$ \color{black} $u^\prime / \overline{u} = TI$ \color{black} as the turbulence intensity,
    \item $\varepsilon$ $\rightarrow$ \color{black} $\varepsilon \: L/u^{\prime3} = C_{\varepsilon}$ \color{black} as the dissipation parameter,
    \item $\nu$ $\rightarrow$ \color{black} $L \: u^\prime / \nu = Re_L = Re_{\lambda}^2 \: C_{\varepsilon}/15$, \color{black} with $Re_{\lambda}$ as a local Reynolds number.
\end{itemize} 

\FloatBarrier

\noindent \new{This means that the non-dimensional quantities in table~\ref{table_which_quantities_are_important} are enough to fully describe SST. Thus all dimensional quantities in table~\ref{table_which_quantities_are_important_all} are captured in table~\ref{table_which_quantities_are_important}. In essence, the relevant non-dimensional quantities are $TI, F_u, Re_\lambda, C_\varepsilon, \gamma, C_k, \mu, \Lambda_0^2$.} 

\FloatBarrier

\begin{table}[h]
    \centering
    \renewcommand{\arraystretch}{2} 
    \begin{tabular}{cc}
        \hline
        one-point & two-point (order matters) \\
        \hline
         $TI, F_u$ & $Re_\lambda, C_\varepsilon, \gamma, C_k, \mu, \Lambda_0^2$\\
        \hline
    \end{tabular}
    \caption{\new{Non-dimensional scalar quantities derived from a $1\,$D VTS obtained in SST: $TI$ the turbulence intensity, $F_u$ the flatness of the VTS, $Re_{\lambda}$ the Reynolds number based on the Taylor length scale, $C_\varepsilon$ the dissipation parameter, $\gamma$ the absolute slope of the energy spectrum within the inertial range and $C_k$ the Kolmogorov parameter, $\mu$ the intermittency parameter, and $\Lambda_0^2$ the large-scale non-Gaussianity. The column indicates whether the quantity corresponds to a one-point or a two-point statistic.}}
    \label{table_which_quantities_are_important}
\end{table}

\new{As mentioned above, we have selected eight VTS to show exemplary results. Table~\ref{table_eight_examples} lists all just defined characteristic quantities for these  eight VTS. a), b), c), e), f) and g) display VTS that satisfy the restriction criteria and taken together, span almost the full range of $\mu$-values. d) also fulfills the restriction criteria and exhibits the same value of $\mu$ as c) but with a $Re_\lambda$ more than three times smaller. h) however does not meet the restriction criteria since $\Lambda_0^2$ shows clear signatures of non-Gaussianity at large scales. The symbol $f$ denotes the sampling frequency, $x/d^*$ the streamwise distance $x$ normalized by $d^*$, the characteristic width of the turbulence generator, $S_u$ the skewness of the VTS, $\alpha$ the product of $\mu$ and $C_\varepsilon$, $\beta$ the product of $(\gamma - 1)$ and $C_\varepsilon$ and $\phi = ({\theta - \mathrm{log}(C_k)})/{\gamma}$, fixing the fitting constant $\theta$ to $\theta = 2.99$. Note that these eight representative VTS, are unchanged through the paper and are always indicated in the same way.}

\begin{table} [h!]
	\centering
       \setlength{\tabcolsep}{2pt} 
        \scriptsize
	\begin{tabular}{lcccccccccccccccccccccc}
		\toprule
		   & $T\:$(s) & $f\:$(kHz) &$x/d^*$ & $\overline{u}\:$(m/s) & $TI\:$($\%$) & $Re_{\lambda}$ & $S_u$ & $F_u$ & $L\:$(cm)& $\lambda\:$(mm)& $\eta\:$(mm) & $\varepsilon\:$(m²/s³) & ${\Lambda}^2_0$ & $C_\varepsilon$ & $\mu$ & $\gamma$ & $C_k$ & \cellcolor{blue!30} $\alpha$ & \cellcolor{blue!30} $\beta$ & \cellcolor{blue!30} $\phi$\\
		\midrule
		   a) & 240 & 50 & 45.2 & 7.52 & 2.55 & 91 & 0.02 & \cellcolor{green!50} 2.93 & 2.77 & 7.25 & 0.39 & 0.2 & \cellcolor{green!50} -0.003 & 0.63 & 0.15 & 1.31 & 2.71 & 0.09 & 0.19 & 1.95\\
           \midrule
		   b) & 300 & 50 & 17 & 17.12 & 3.13 & 123 & -0.05 & \cellcolor{green!50} 2.81 & 1.34 & 3.5 & 0.16 & 5.3 & \cellcolor{green!50} -0.008 & 0.46 & 0.19 & 1.44 & 1.71 & 0.09 & 0.2 & 1.92\\
           \midrule
		   c) & 120 & 50 & 7 & 7.38 & 21.89 & 760 & 0.11 & \cellcolor{green!50} 2.94 & 13.2 & 7.21 & 0.13 & 11.53 & \cellcolor{green!50} -0.01 & 0.36 & 0.26 & 1.6 & 0.84 & 0.1 & 0.22 & 1.92\\
           \midrule
		   d) & 120 & 50 & 11 & 10.03 & 7.75 & 182 & -0.11 & \cellcolor{green!50} 2.93 & 1.53 & 3.67 & 0.14 & 9.83 & \cellcolor{green!50} -0.003 & 0.34 & 0.26 & 1.53 & 1.3 & 0.09 & 0.18 & 1.88\\
           \midrule
		   e) & 120 & 50 & 25 & 10.31 & 9.98 & 227 & -0.21 & \cellcolor{green!50} 2.84 & 1.38 & 3.4 & 0.11 & 21.4 & \cellcolor{green!50} -0.007 & 0.27 & 0.39 & 1.74 & 0.46 & 0.11 & 0.2 & 1.91\\
           \midrule
		   f) & 120 & 50 & 14 & 9.62 & 16.13 & 182 & -0.08 & \cellcolor{green!50} 3.02 & 0.45 & 1.8 & 0.07 & 172.4 & \cellcolor{green!50} 0.001 & 0.21 & 0.5 & 1.93 & 0.36 & 0.11  & 0.19 & 1.88\\
           \midrule
		   g) & 120 & 50 & 8 & 9.64 & 21.37 & 232 & -0.17 & \cellcolor{green!50} 2.95 & 0.42 & 1.72 & 0.06 & 329.22 & \cellcolor{green!50} -0.002 & 0.16 & 0.7 & 2.13 & 0.07 & 0.11 & 0.18 & 1.92\\
           \midrule
		   h) & 300 & 50 & 2 & 9.08 & 6.61 & 109 & -0.94 & \cellcolor{red!50} 7.81 & 0.71 & 2.79 & 0.14 & 10.65 & \cellcolor{red!50} 0.147 & 0.35 & 0.77 & 1.8 & 0.3 & 0.27 & 0.28 & 1.95\\
		\bottomrule
	\end{tabular}
	\caption{\new{The characteristics are presented for the eight example VTSs that are used consistently throughout this work. $T$ as the sampling time, $f$ as the sampling frequency, $x/d^*$ as the streamwise distance $x$ normalized by $d^*$, the characteristic width of the turbulence generator, $\overline{u}$ as the mean velocity at the acquisition point, $u^{\prime}/\overline{u}$ as the turbulence intensity, $Re_{\lambda}$ as the Reynolds number based on the Taylor length scale, $S_u$ as the skewness of the VTS, $F_u$ as the flatness of the VTS, $L$ as the integral length scale, $\lambda$ as the Taylor length scale, $\eta$ as the Kolmogorov length scale, $\varepsilon$ as the mean dissipation rate, $C_\varepsilon$ as the dissipation parameter, $\mu$ as the intermittency parameter, $\alpha$ (highlighted blue) as the product of $\mu$ and $C_\varepsilon$, ${\Lambda_0}^2$ as the shape parameter on the large scales, $\gamma$ as the negative slope of the energy spectrum within the inertial range, $C_k$ as the Kolmogorov parameter, $\beta$ (highlighted blue) as the product of $(\gamma - 1)$ and $C_\varepsilon$ and $\phi = ({\theta - \mathrm{log}(C_k)})/{\gamma}$ (highlighted blue), keeping $\theta = 2.99$ constant. The origin of $\theta$ is explained below. The data stem from the cases G20, G24, C8, D11, C6, C1, C1, G23, respectively in order of appearance. a), b), c), e), f) and g) display VTS that satisfy the restriction criteria and taken together, span almost the full range of $\mu$-values. d) also fulfills the restriction criteria and exhibits the same value of $\mu$ as c) but with a $Re_\lambda$ more than three times smaller. h) however does not meet the restriction criteria since $\Lambda_0^2$ shows clear signatures of non-Gaussianity at large scales. Values of $F_u$ and $\Lambda_0^2$ highlighted in green indicate Gaussianity at the large scales and are therefore retained in the analysis. In contrast, values of $F_u$ and $\Lambda_0^2$ highlighted in red denote non-Gaussianity at the large scales and are consequently discarded. For details about the cases see table~\ref{tab:PhD measurements in LEGI 2023} in the appendix~\cite{SM}.}}
	\label{table_eight_examples}
\end{table}

\FloatBarrier

\subsection{\new{Assumptions and Scientific Hypothesis}}
\label{assumptions}

\new{Next, we will explain the following two assumptions of our approach:} \\

\noindent \new{\underline{Taylor assumption}: The Taylor's hypothesis of frozen turbulence can still be used even outside homogeneous isotropic turbulence, however, regarded purely formally as a coordinate transformation} \\
    
\noindent \new{\underline{Dissipation assumption}: The methods of estimating the mean dissipation rate developed for HIT can be extended to inhomogeneous turbulence by calculating a 1$\,$D-surrogate of the mean dissipation rate} \\

\noindent \new{When analyzing turbulence data stemming from HWA, it is unavoidable to use Taylor's hypothesis of frozen turbulence. Since we measure a VTS at one point in space, we face the problem that most theories and methods are based on fluctuations in physical space. To overcome this, the local mean velocity $\overline{u}$ is used for transformation. The underlying idea is that the turbulence is seen as ``frozen'' and thus advected with the mean velocity of the flow. While this is a good approximation for HIT, the question arises if physical interpretation can be applied to inhomogeneous flows, which is often denied~\cite{roy2021deviations, jacobitz2024revisiting}. However, our Taylor assumption is that we can use it purely formally as a coordinate transformation from \textit{time} to \textit{space} in order to have the right units. The application of Taylor's hypothesis outside the regime of HIT is well established in experimental studies of turbulent shear flows, for example in the works of Anselmet~\emph{et al.}~\cite{anselmet1984high} and Saddoughi \& Veeravalli~\cite{saddoughi1994local}.}

\new{Another question concerns the estimation of the mean dissipation rate $\varepsilon$ in inhomogeneous flows. For HIT, it is common practice to estimate $\varepsilon$ by integrating the dissipation spectrum $k^2 \, E(k)$~\cite{davidson2015turbulence}. The dissipation assumption states that the same methods can be applied for inhomogeneous flows, acknowledging that only a 1$\,$D-surrogate of the mean dissipation rate is obtained. The application of the 1$\,$D-surrogate of the mean dissipation rate outside the regime of HIT is, for example, used in turbulent boundary layer flows~\cite{nedic2017dissipation} and fractal-grid turbulence~\cite{vassilicos2015dissipation}. This approach is further justified by our focus on the relative behavior of turbulence quantities rather than their absolute values. All quantities are respectively derived from the same VTS, and consequently from the same surrogate representation. We therefore accept that the resulting estimate of $\varepsilon$ does not correspond to the true three-dimensional mean dissipation rate, but rather serves, as well as all other quantities, as a consistent surrogate measure within the present framework. \new{In chapter~\ref{discussion}, this assumption is further justified based on the behavior of $p(\varepsilon_r(x))$.}}

\new{Finally, the scientific hypotheses underlying this work are introduced. The central question concerns the behavior of the turbulence quantities associated with SST, as summarized in table~\ref{table_which_quantities_are_important}. In this framework, $\Lambda_0^2 \approx 0$ and $F_u \approx 3$ are fixed due to the selection of data. Regarding the remaining parameters, several scenarios are conceivable for their mutual behavior within SST but outside the HIT regime. One possibility is that some or all of the quantities $C_\varepsilon$, $\mu$, $\gamma$, and $C_k$ depart from their HIT behavior (in the easiest case remain constant) without exhibiting any systematic relation either to the control parameters $Re_\lambda$ and $TI$ or to one another. This scenario is formulated as the null hypothesis:}\\

\noindent \new{\underline{$\boldsymbol{H_0}$}: When departing HIT within the framework of SST, there is no order regarding the values of $TI$, $Re_\lambda$, $C_\varepsilon$, $\mu$, $\gamma$, $C_k$.}\\

\noindent \new{In contrast, a more general underlying structure may persist beyond the HIT regime. In this case, some or all of the parameters $C_\varepsilon$, $\mu$, $\gamma$, and $C_k$ may either retain their HIT behavior or depart from it in a systematic manner, exhibiting relationships with the control parameters $Re_\lambda$ and $TI$, with one another, or with both. This scenario is formulated as the alternative hypothesis:}\\

\noindent \new{\underline{$\boldsymbol{H_{1}}$}: When departing HIT within the framework of SST, there is a systematic structure regarding the values of $TI$, $Re_\lambda$, $C_\varepsilon$, $\mu$, $\gamma$, $C_k$.}\\

\noindent \new{The following chapter presents the results obtained from testing these hypotheses.}

\section{\new{New Empirical Relations between Parameters}}
\label{new_laws}

\noindent \new{Here, selected results of the study are presented, considering all VTS that satisfy the restrictions imposed for SST. This chapter is divided into three subsections, each examining the relation between two of the four parameters $C_\varepsilon$, $\mu$, $\gamma$, and $C_k$. $C_\varepsilon$ was determined according to eq.~(\ref{dissipation_constant}), where $\varepsilon$ was obtained by integrating the dissipation spectrum, while $L$ was calculated from the autocorrelation function by integrating up to its first crossing of 1/e, where e denotes Euler's number. The parameters $\mu$, $\gamma$, and $C_k$ were estimated using eqs.~(\ref{equation_lambda_two}) and~(\ref{equation_energy_spectrum}), with $\mu$ obtained from the former and $\gamma$ and $C_k$ from the latter. Further details on the thresholds, constraints, and additional criteria associated with these estimation procedures are provided in chapter~\ref{methods}.}

\subsection{\new{Relation between $\mu$ and $C_\varepsilon$}}
\label{law_1}

{Investigating the relationship between intermittency and dissipation, specifically between $\mu$ and $C_{\varepsilon}$, we found that, for data taken from the centerline of various flows}, their product remains constant $\mu \: C_{\varepsilon} = \alpha$ (\textit{cf.}~\cite{schmitt2024universal}). Extending this analysis to all VTS that fulfill the criteria introduced in section~\ref{data selection} and thus all VTS that fulfill SST conditions, we obtain the result shown in figure~\ref{figure_law_1} a). It can be seen that the data points, irrespective of the corresponding case, collapse on a reciprocal function suggesting the relation,}

\begin{equation} \new{
    \mu \, C_\varepsilon = \alpha \qquad\qquad \alpha = \mathrm{const.}>0, \qquad \mu, C_\varepsilon > 0,}
    \label{equation_law_1}
\end{equation}

\noindent \new{to be valid for general SST. Since many data points overlap in figure~\ref{figure_law_1} a), subplot b) provides a version of figure a) in which data clusters become visible through uniformly transparent markers. The highest marker density lies close to the reciprocal function, which is further highlighted by superimposing a window-averaged curve (orange) in the same figure, lying close to the reciprocal function. Other fitting curves, such as linear and quadratic fits, were also applied, however, the reciprocal function yields the best combination of a low number of degrees of freedom and a good coefficient of determination. Moreover, the PDF of $\alpha$ values $p(\alpha)$ is presented in subfigure c) revealing that the $\alpha$ values are well centered and slightly skewed toward larger values, suggesting a log-normal distribution. Other low-parameter fits, such as the Rayleigh distribution, were tested, but the log-normal fit best represents the empirical PDF. The evolution of the product $\mu \, C_\varepsilon$ for the centerline data (dashed orange curve) and for all data (dashed black curve) suggests that the variances are approximately similar, whereas the PDF of the centerline data is nearly Gaussian. The origin of this behavior is examined in section~\ref{discussion}. Note that the centerline definition is adopted from~\cite{schmitt2024universal}, where all measurements within $\pm10\,$mm of the centerline are classified as centerline data. Finally, figure d) shows the same relation as in a) for three representative cases, with the corresponding uncertainties indicated by error bars in both directions. The representative cases C1, C4, and D13 were selected because they span nearly the full range of observed $\mu$ and $C_\varepsilon$ values. The uncertainties are estimated based on the methods used to determine $\mu$ and $C_{\varepsilon}$, together with the corresponding robustness and consistency checks. The estimation of the uncertainties is described in detail in chapter~\ref{methods}. All uncertainties were treated symmetrically when constructing the error bars. Although the chosen uncertainty estimates appear reasonable, they do not call the observed trends into question, however they can account for part of the scatter of the data points around the reciprocal function. Taken together, the proposed relation between $\mu$ and $C_{\varepsilon}$ from \cite{schmitt2024universal} continues to hold even when non-centerline data, which naturally involve stronger shear, are included, and remains robust under closer examination.}

\begin{figure}[htbp]
    \centering
    \begin{subfigure}[t]{0.49\textwidth}
        \centering
        \includegraphics[width=\linewidth]{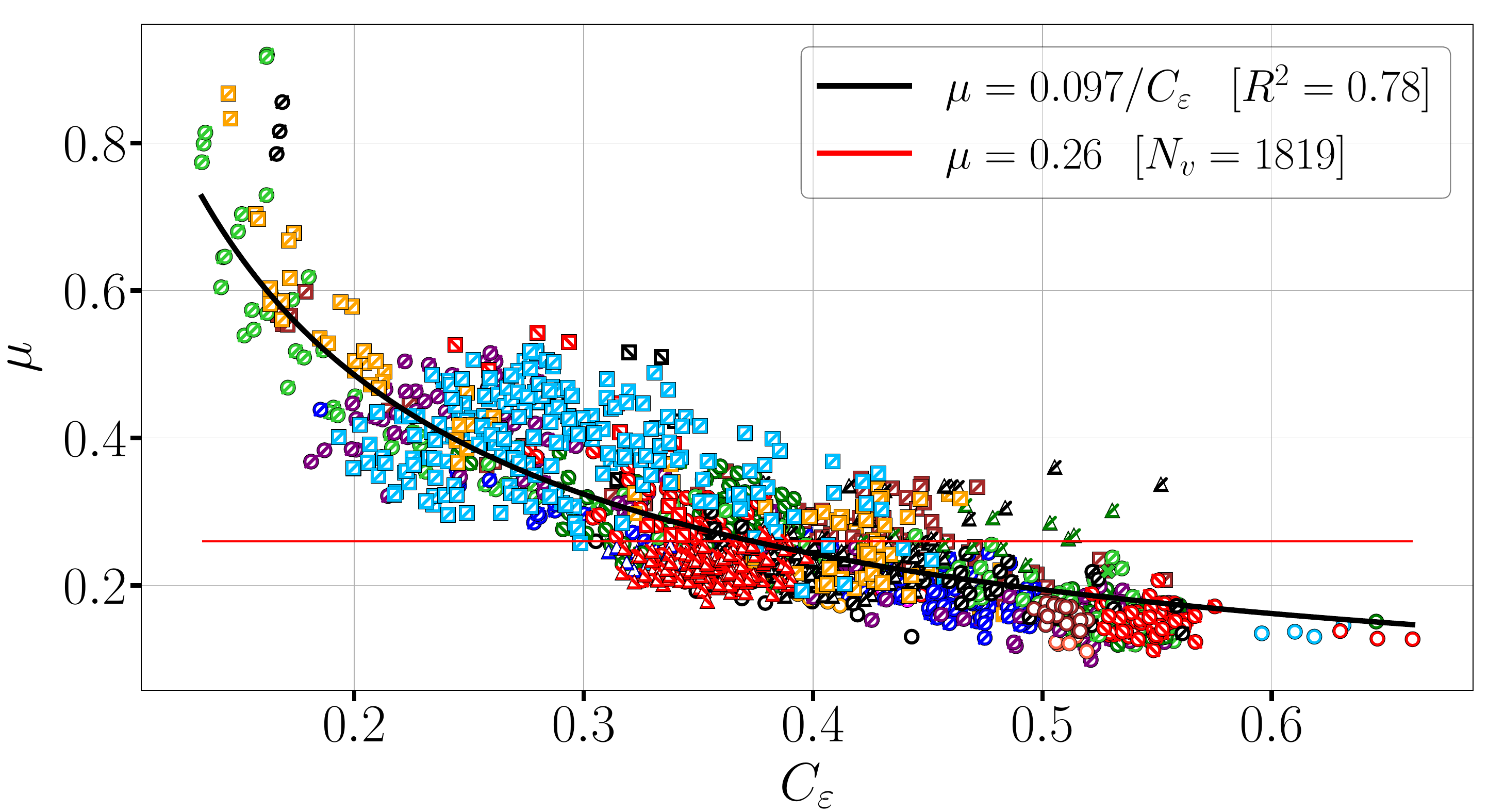}
        \caption{}
    \end{subfigure}
    \hfill
    \begin{subfigure}[t]{0.49\textwidth}
        \centering
        \includegraphics[width=\linewidth]{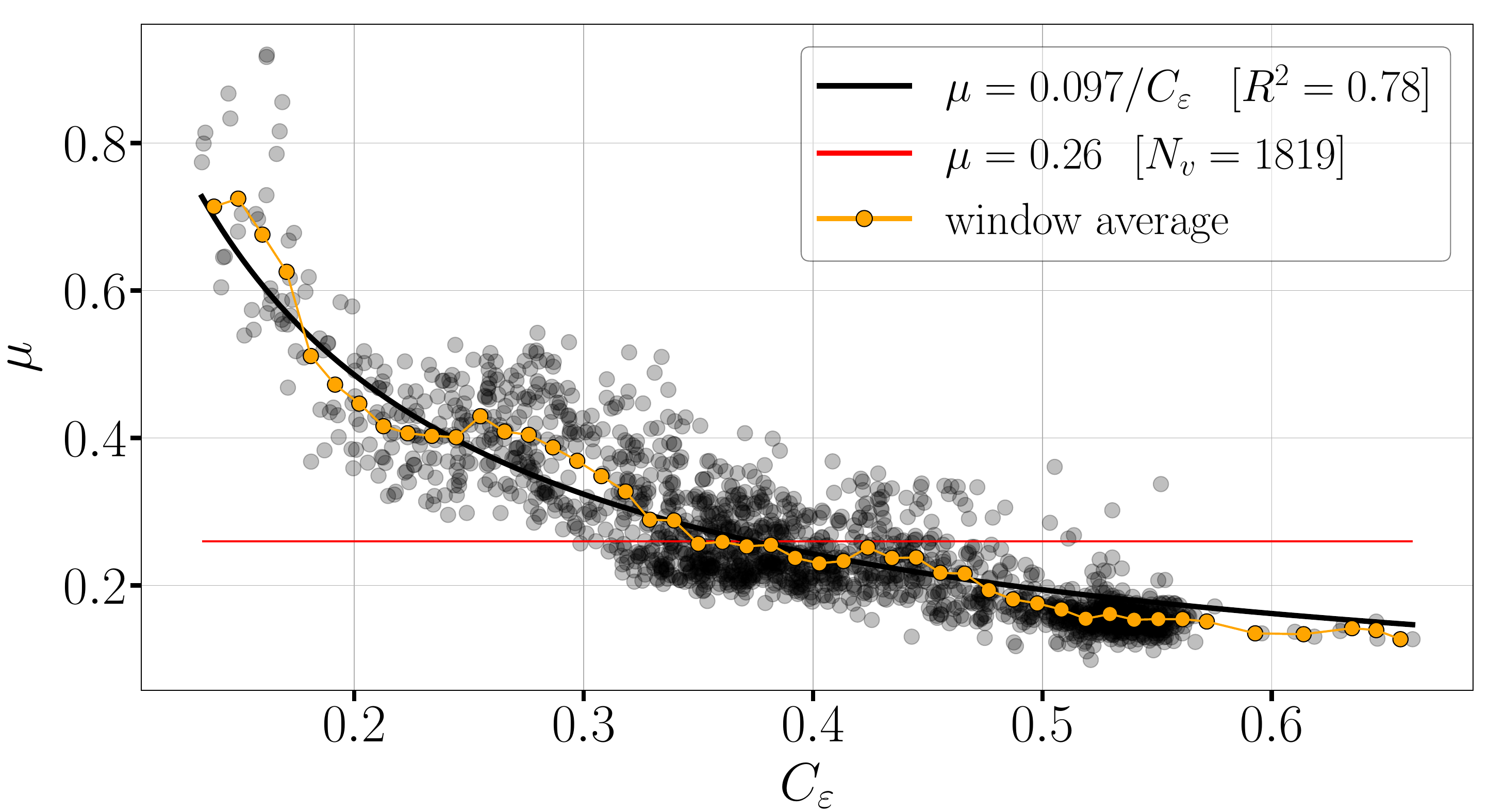}
        \caption{}
    \end{subfigure}
    \begin{subfigure}[t]{0.49\textwidth}
        \centering
        \includegraphics[width=\linewidth]{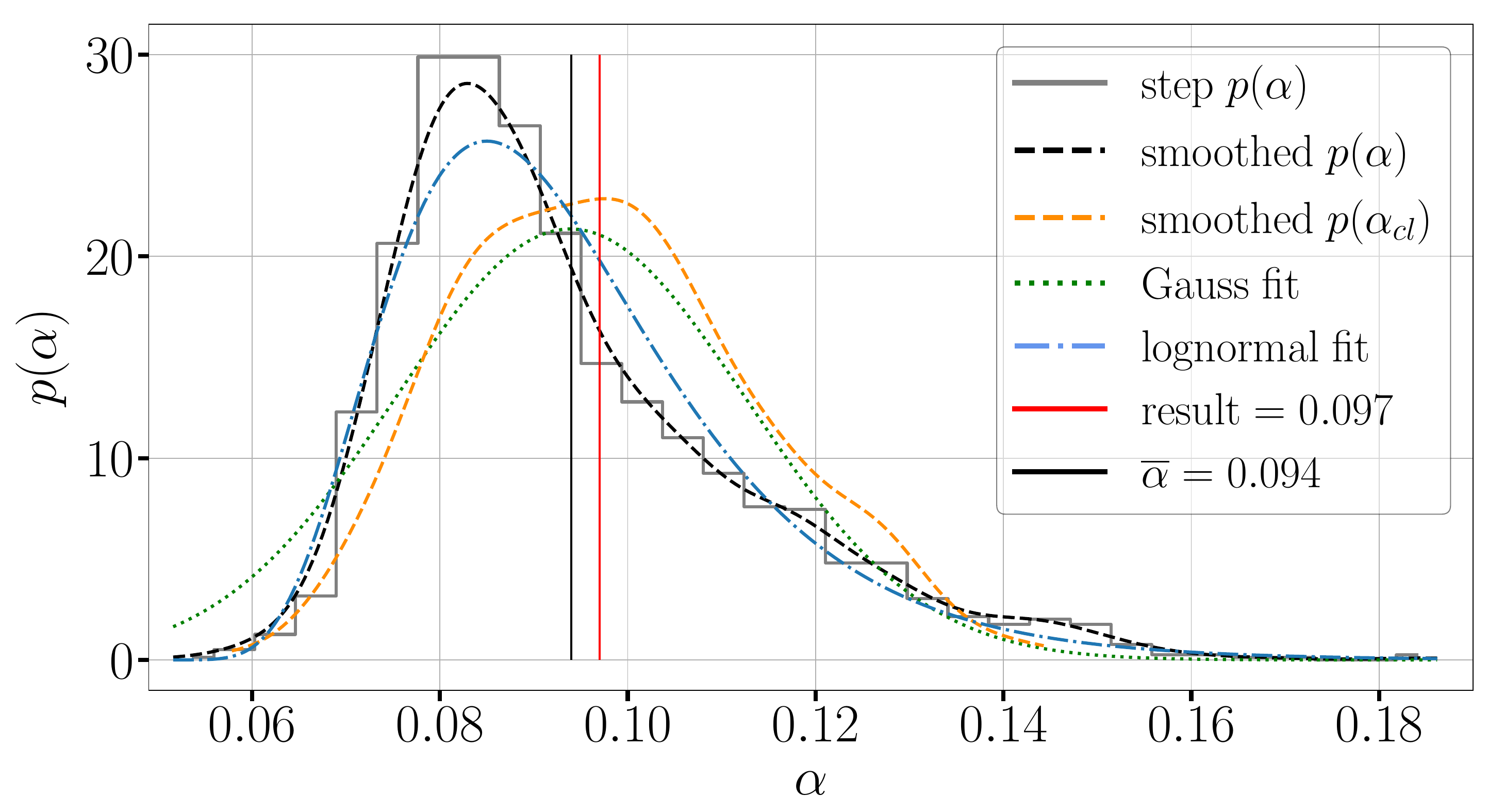}
        \caption{}
    \end{subfigure}
    \hfill
     \begin{subfigure}[t]{0.49\textwidth}
        \centering        \includegraphics[width=\linewidth]{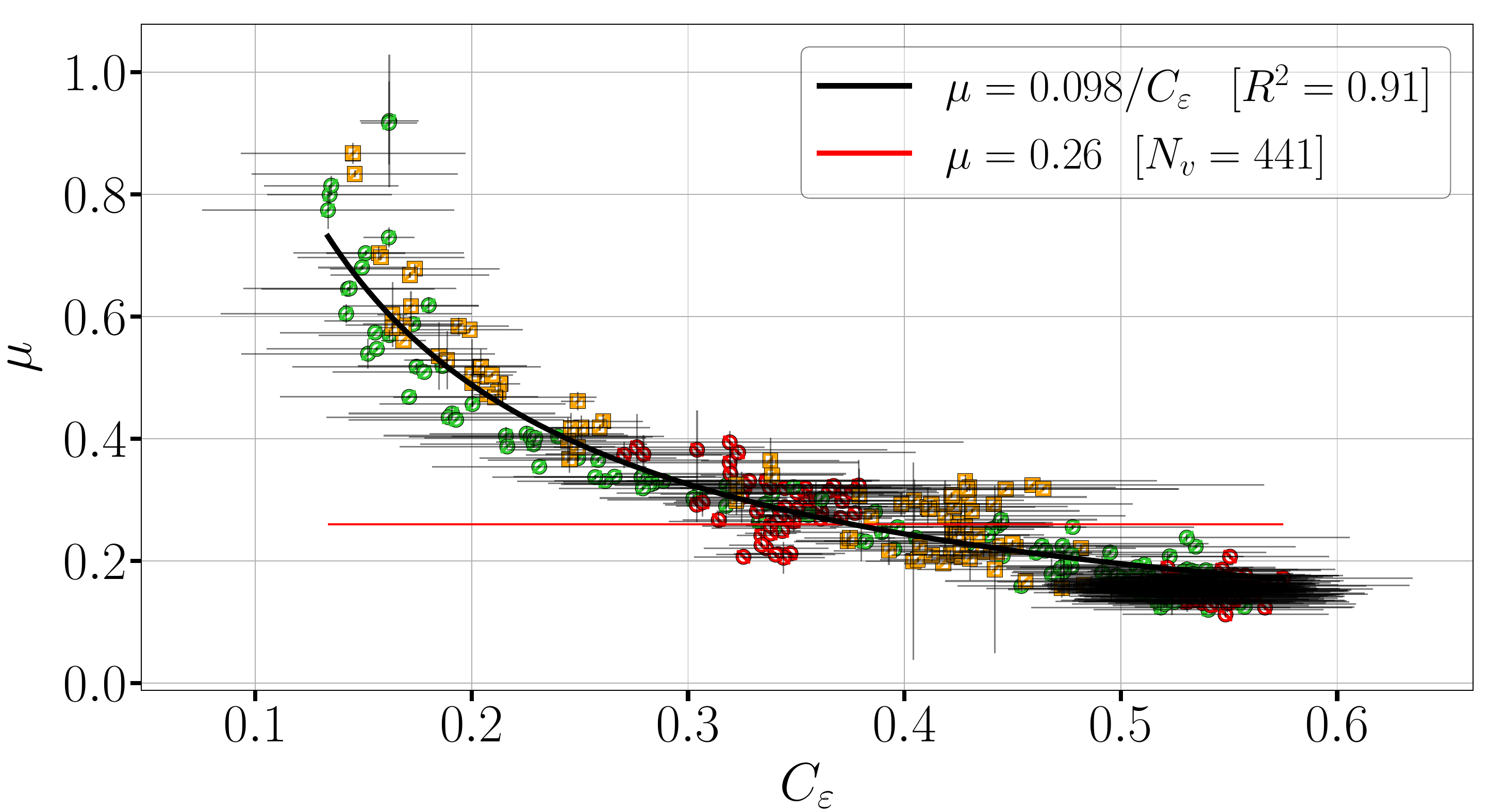}
        \caption{}
    \end{subfigure}
    \caption{\new{a), b), and d) present $\mu$ as a function of $C_\varepsilon$. The red line indicates a commonly accepted value for $\mu$ for HIT~\cite{arneodo1996structure}, and the black curve corresponds to a least-squares fit with $R^2$ being the coefficient of determination. a) and b) show all VTS used, whereas d) shows only the used VTS from cases C1, C4, and D13. Note that for panel b), all markers have the same shape and the same amount of transparency. Differences in color are solely due to overlapping markers. The orange curve in b) presents a window average out of 50 windows. Panel d) shows the errors in both $\mu$ and $C_\varepsilon$ based on the uncertainties of the used methods. Furthermore, the PDF of $\alpha$ values $p(\alpha=\mu \, C_\varepsilon)$ for all used data is presented in subfigure c) with a solid gray line (histogram), a dashed black line (smoothed histogram), a dotted green line (Gaussian fit), a dashdotted blue line (log-normal fit) and a dashed orange line (smoothed histogram for centerline data only). The solid red line indicates the result for $\alpha$ from the fit shown in a) while the black solid line represents the actual mean value of the ensemble of $\alpha$ values. In general, $N_v$ indicates the number of VTS shown in the plot. The symbols and corresponding configurations are shown and explained in table~\ref{tab:PhD measurements in LEGI 2023} in the appendix~\cite{SM}.}}
    \label{figure_law_1}
\end{figure} 

\FloatBarrier

\subsection{\new{Relation between $\gamma$ and $C_\varepsilon$}}
\label{law_2}

\new{Figure~\ref{figure_law_2} a) shows $\gamma - 1$ as a function of $C_\varepsilon$ for all VTS used. It can be seen that the data points, irrespective of the corresponding case collapse on a reciprocal function suggesting the relation,}

\begin{equation} \new{
    (\gamma - 1) \, C_\varepsilon = \beta\qquad\qquad\beta = \mathrm{const.}>0,\qquad(\gamma - 1), C_\varepsilon > 0,}
    \label{equation_law_2}
\end{equation}

\noindent \new{to be valid for general SST. Since many data points overlap in figure~\ref{figure_law_2} a), subplot b) provides a version of figure a) in which data clusters become visible through uniformly transparent markers. The highest marker density lies close to the reciprocal function, which is further highlighted by superimposing a window-averaged curve (orange) in the same figure, lying close to the reciprocal function. Other fitting curves, such as linear and quadratic fits, were also applied; however, the reciprocal function yields the best combination of a low number of degrees of freedom and a good coefficient of determination. Moreover, the PDF of $\beta$ values $p(\beta)$ is presented in subfigure c) revealing that the $\beta$ values are well centered, suggesting a normal distribution. The evolution of the product $(\gamma - 1) \, C_\varepsilon$ for the centerline data (dashed orange curve) and for all data (dashed black curve) shows that the PDFs are similar. Finally, figure d) shows the same relation as in a) for three representative cases, with the corresponding uncertainties indicated by error bars in both directions. The representative cases C1, C4, and D13 were selected because they span nearly the full range of observed $\gamma$ and $C_\varepsilon$ values. The uncertainties are estimated based on the methods used to determine $\gamma$ and $C_{\varepsilon}$, together with the corresponding robustness and consistency checks. The estimation of the uncertainties is described in detail in chapter~\ref{methods}. All uncertainties were treated symmetrically when constructing the error bars. Although the chosen uncertainty estimates appear reasonable, they can account for part of the scatter without calling the observed trends into question. Taken together, the proposed relation between $\gamma$ and $C_{\varepsilon}$ remains robust under closer examination.}

\begin{figure}[htbp]
    \centering
    \begin{subfigure}[t]{0.49\textwidth}
        \centering
        \includegraphics[width=\linewidth]{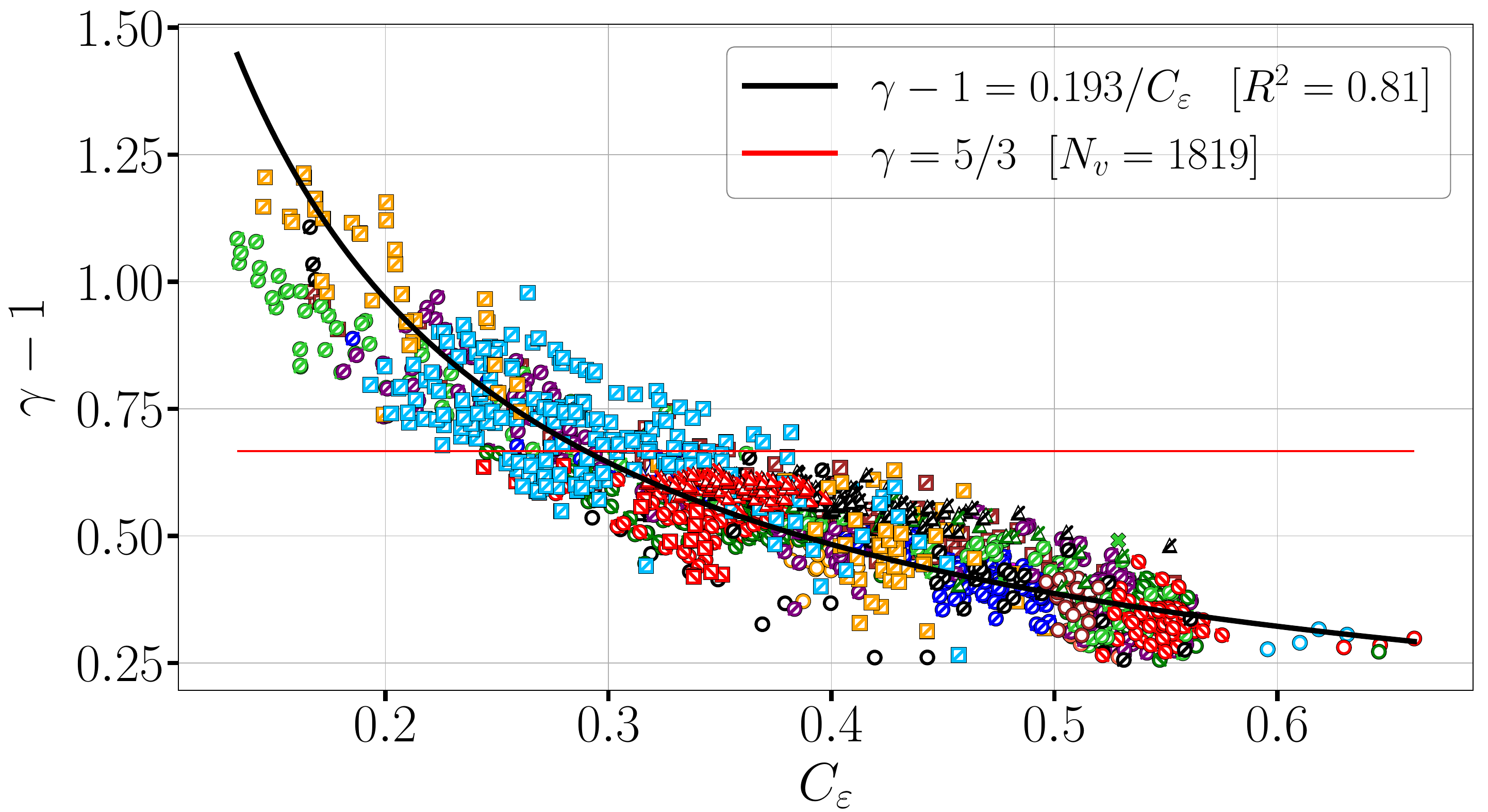}
        \caption{}
    \end{subfigure}
    \hfill
    \begin{subfigure}[t]{0.49\textwidth}
        \centering
        \includegraphics[width=\linewidth]{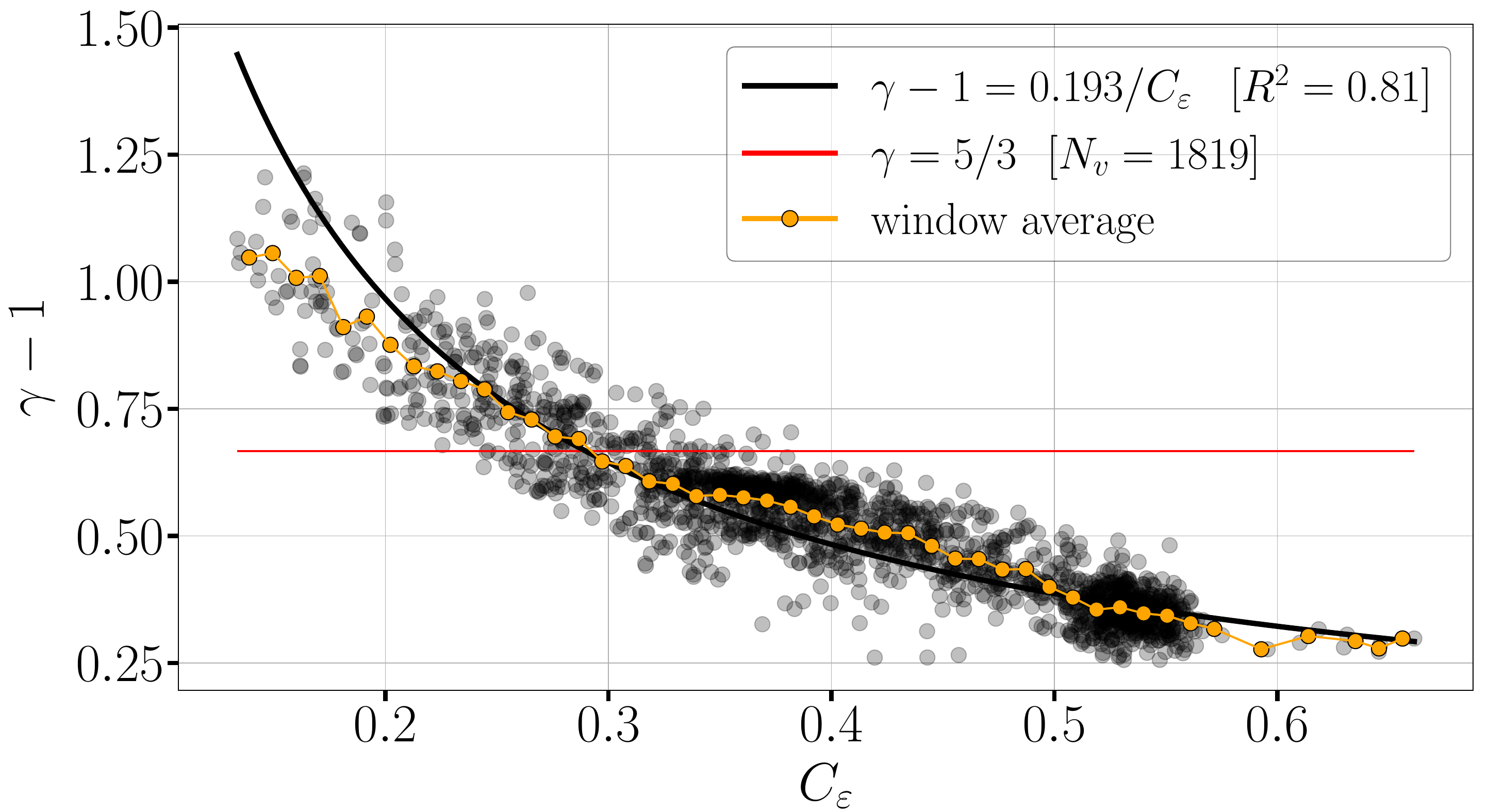}
        \caption{}
    \end{subfigure}
    \begin{subfigure}[t]{0.49\textwidth}
        \centering
        \includegraphics[width=\linewidth]{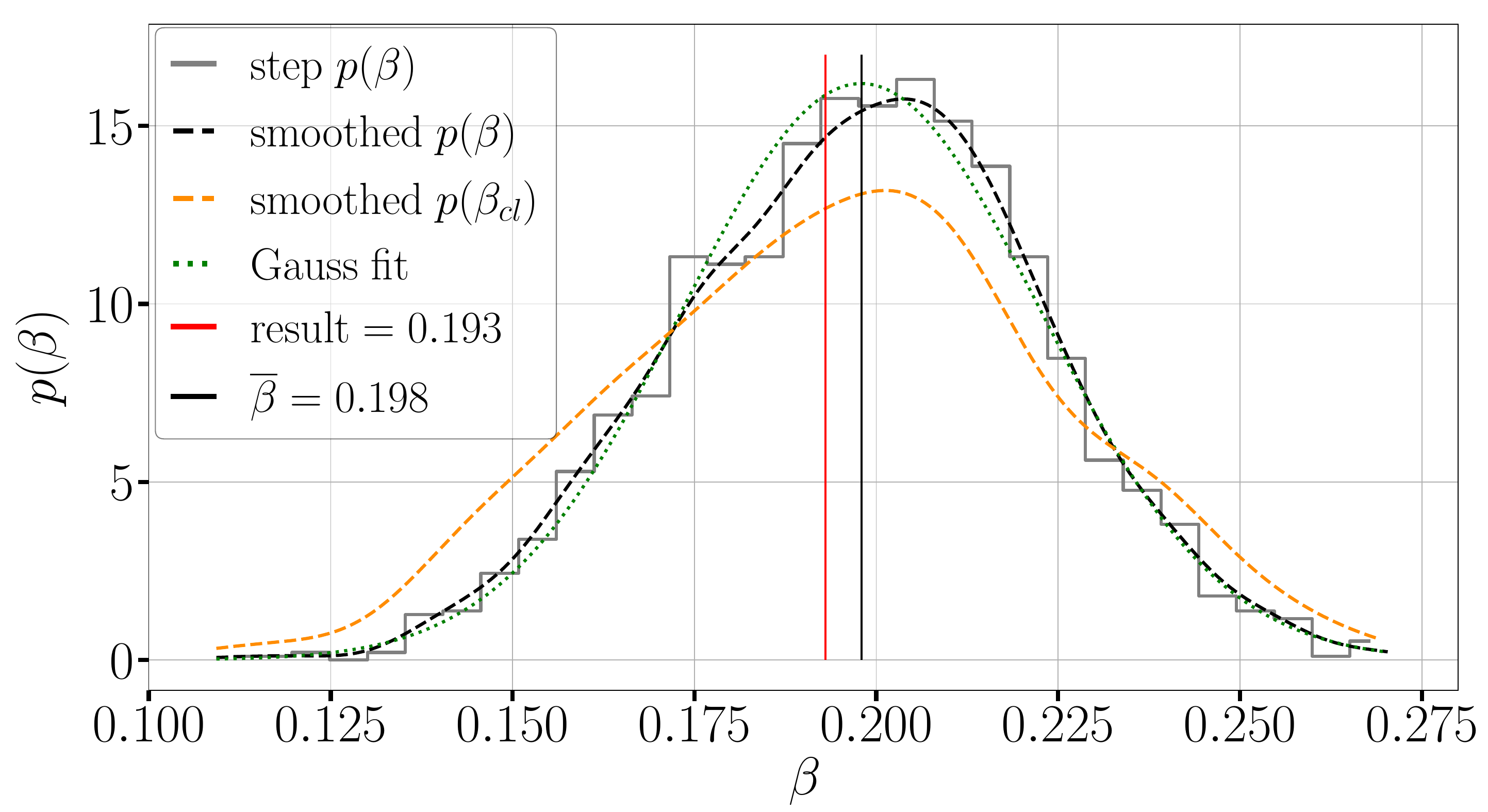}
        \caption{}
    \end{subfigure}
    \hfill
    \begin{subfigure}[t]{0.49\textwidth}
        \centering        \includegraphics[width=\linewidth]{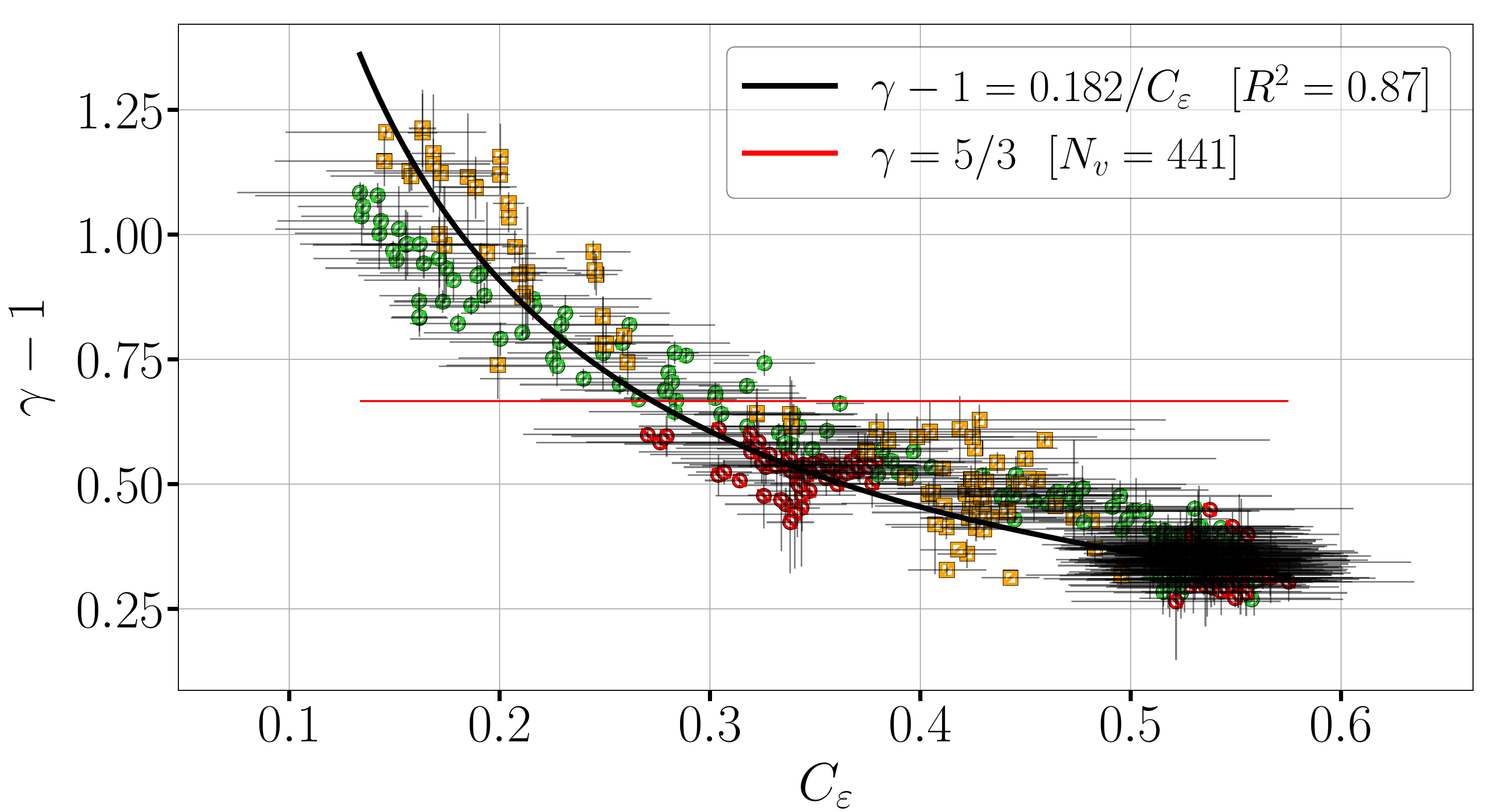}
        \caption{}
    \end{subfigure}
    \caption{\new{a), b), and d) present $\gamma - 1$ as a function of $C_\varepsilon$. The red line indicates a commonly accepted value for $\gamma$ for HIT~\cite{sreenivasan1995universality}, and the black curve corresponds to a least-squares fit with $R^2$ being the coefficient of determination. a) and b) show all VTS used, whereas d) shows only the used VTS from cases C1, C4, and D13. Note that for panel b), all markers have the same shape and the same amount of transparency. Differences in color are solely due to overlapping markers. The orange curve in b) presents a window average out of 50 windows. Panel d) shows the errors in both $\gamma - 1$ and $C_\varepsilon$ based on the uncertainties of the used methods. Furthermore, the PDF of $\beta$ values $p(\beta=(\gamma - 1) \, C_\varepsilon)$ for all used data is presented in subfigure c) with a solid gray line (histogram), a dashed black line (smoothed histogram), a dotted green line (Gaussian fit), and a dashed orange line (smoothed histogram for centerline data only). The solid red line indicates the result for $\beta$ from the fit shown in a) while the black solid line represents the actual mean value of the ensemble of $\beta$ values. In general, $N_v$ indicates the number of VTS shown in the plot. The symbols and corresponding configurations are shown and explained in table~\ref{tab:PhD measurements in LEGI 2023} in the appendix~\cite{SM}.}}
    \label{figure_law_2}
\end{figure}

\FloatBarrier

\subsection{\new{Relation between $C_k$ and $\gamma$}}
\label{law_3}

\new{The evidence supporting the third equation is presented next. Figure~\ref{figure_law_3} a) presents $\mathrm{log}(C_k)$ as a function of $\gamma$ for all VTS used. It can be seen that the data points, irrespective of the corresponding case collapse on a linear function suggesting the relation,}

\begin{equation} \new{
    \frac{\theta - \mathrm{log}(C_k)}{\gamma} = \phi \qquad\qquad \phi, \theta = \mathrm{const.}>0, \qquad (\theta - \mathrm{log}(C_k)), \gamma, C_k > 0,
    \label{equation_law_3}}
\end{equation} 

\noindent \new{to be valid for general SST, keeping $\theta = 2.99$ constant, which is consistent with preliminary findings in \cite{schmitt2024universal} on an indicative relation of $\gamma$ and $C_k$. Relation (\ref{equation_law_3}) is not straightforward since on one side, $\gamma$ or $\gamma - 1$ can be used, and on the other side, $C_k$ can be derived using different ways of normalizing the spectral law (\ref{equation_energy_spectrum}). Furthermore applying the logarithm, the choice of the base needs to be answered. Figure~\ref{figure_law_3} a) results from the parameter choices that yield the least amount of scatter. Since many data points overlap in figure~\ref{figure_law_3} a), subplot b) provides a version of figure a) in which data clusters become visible through uniformly transparent markers. The highest marker density lies close to the linear function, which is further highlighted by superimposing a window-averaged curve (orange) in the same figure, lying very close to the linear function. Moreover, the PDF of $\phi$ values $p(\phi)$ is presented in subfigure c) revealing that the $\phi$ values are well centered, suggesting a normal distribution. The evolution of $\phi$ for the centerline data (dashed orange curve) and for all data (dashed black curve) shows that the PDFs are similar. Finally, figure d) shows the same relation as in a) for three representative cases, with the corresponding uncertainties indicated by error bars in both directions. The representative cases C1, C4, and D13 were selected because they span nearly the full range of observed $\gamma$ and $C_k$ values. The uncertainties are estimated based on the methods used to determine $\gamma$ and $C_k$, together with the corresponding robustness and consistency checks. The estimation of the uncertainties is described in detail in chapter~\ref{methods}. All uncertainties were treated symmetrically when constructing the error bars. Although the chosen uncertainty estimates appear reasonable, they do not call the observed trends into question, however they can account for part of the scatter of the data points around the reciprocal function. Taken together, the proposed relation between $\gamma$ and $C_k$ remains robust under closer examination.} \\

\begin{figure}[htbp]
    \centering
    \begin{subfigure}[t]{0.49\textwidth}
        \centering
        \includegraphics[width=\linewidth]{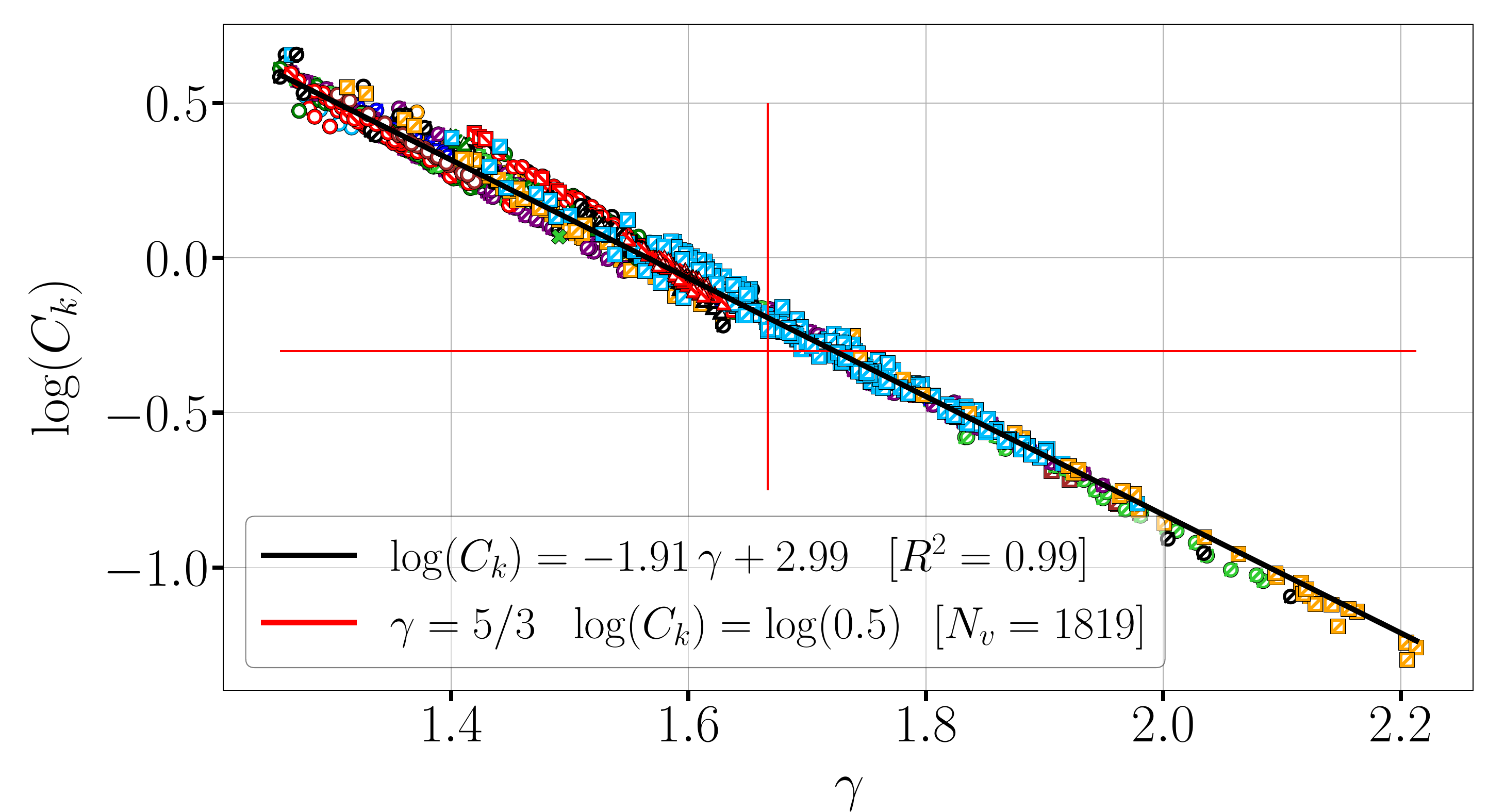}
        \caption{}
    \end{subfigure}
    \hfill
    \begin{subfigure}[t]{0.49\textwidth}
        \centering
        \includegraphics[width=\linewidth]{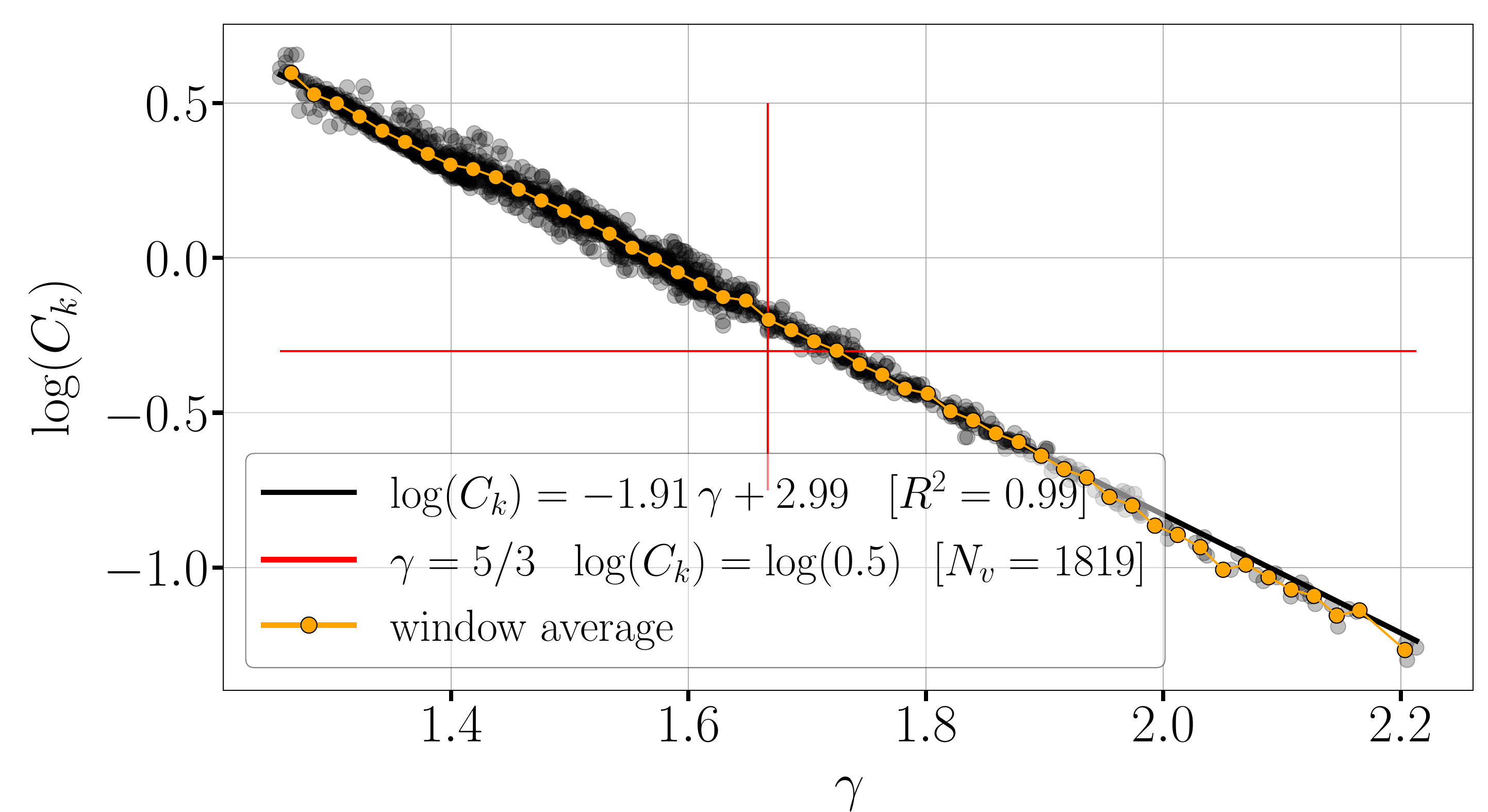}
        \caption{}
    \end{subfigure}
    \begin{subfigure}[t]{0.49\textwidth}
        \centering
        \includegraphics[width=\linewidth]{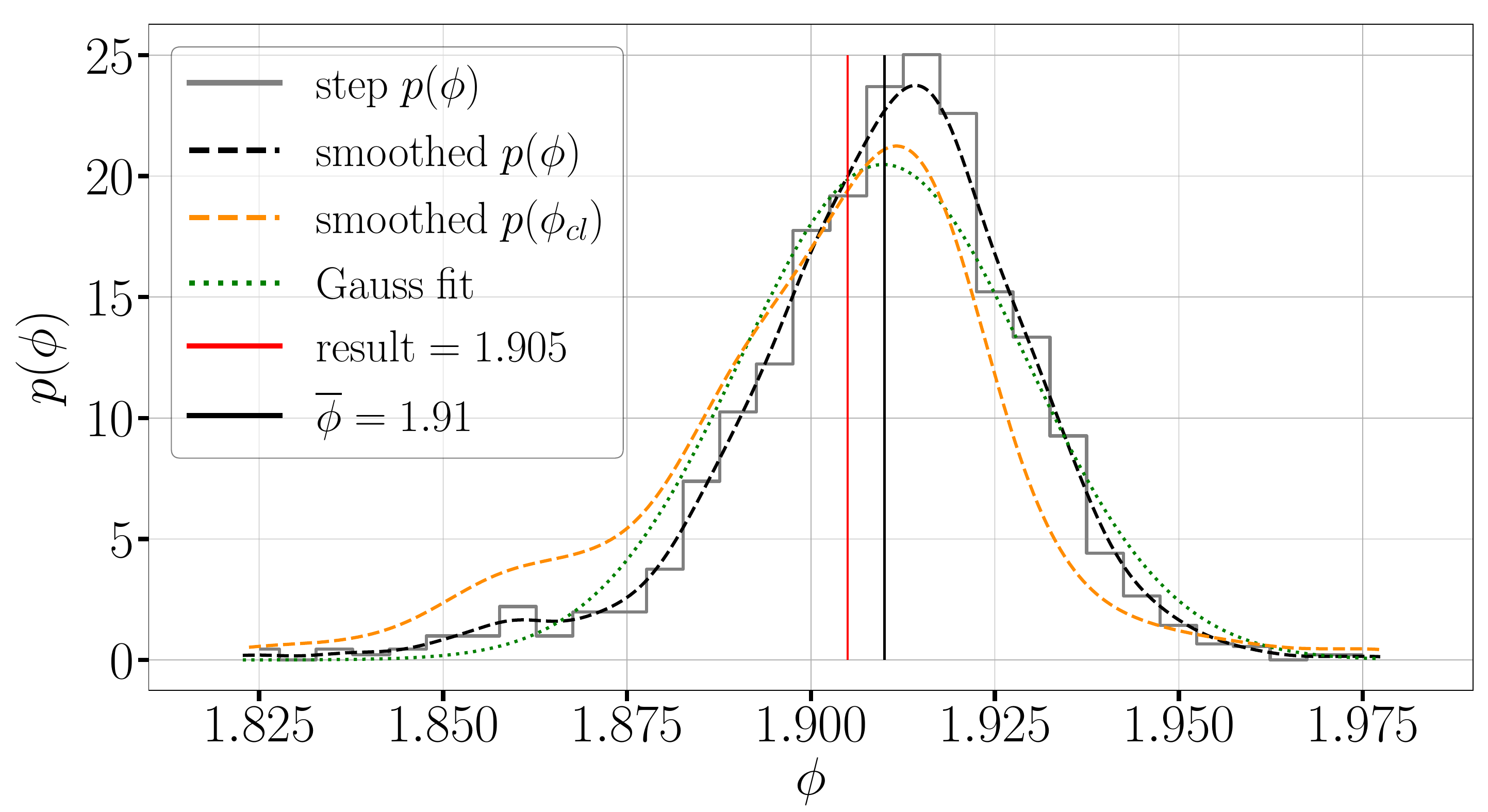}
        \caption{}
    \end{subfigure}
    \hfill
    \begin{subfigure}[t]{0.49\textwidth}
        \centering        \includegraphics[width=\linewidth]{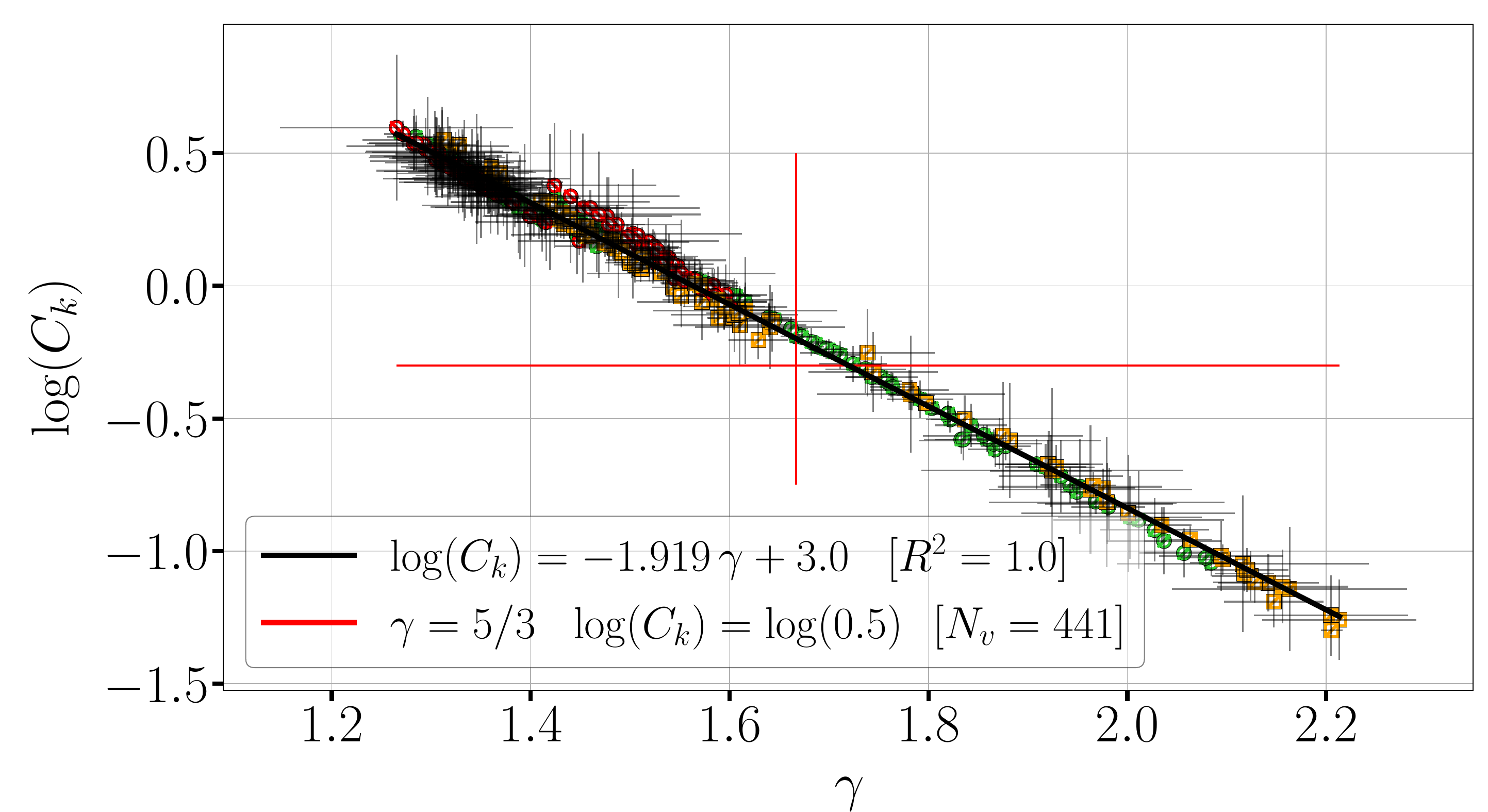}
        \caption{}
    \end{subfigure}
    \caption{a), b), and d) present $\mathrm{log}(C_k)$ as a function of $\gamma$. The red lines indicate commonly accepted values for $C_k$ and $\gamma$ for HIT~\cite{sreenivasan1995universality}, and the black curve corresponds to a least-squares fit with $R^2$ being the coefficient of determination. a) and b) show all VTS used, whereas d) shows only the used VTS from cases C1, C4, and D13. Note that for panel b), all markers have the same shape and the same amount of transparency. Differences in color are solely due to overlapping markers. The orange curve in b) presents a window average out of 50 windows. Panel d) shows the errors in both $\mathrm{log}(C_k)$ and $\gamma$ based on the uncertainties of the used methods. Furthermore, the PDF of $\phi$ values $p(\phi= ({\theta - \mathrm{log}(C_k)})/{\gamma})$ for all used data is presented with a solid gray line (histogram), a dashed black line (smoothed histogram), a dotted green line (Gaussian fit), and a dashed orange line (smoothed histogram for centerline data only). The solid red line indicates the result for $\phi$ from the fit shown in a) while the black solid line represents the actual mean value of the ensemble of $\phi$ values. In general, $N_v$ indicates the number of VTS shown in the plot. {The symbols and corresponding configurations are shown and explained in table~\ref{tab:PhD measurements in LEGI 2023} in the appendix~\cite{SM}.}}
    \label{figure_law_3}
\end{figure}

{To conclude this chapter, we see as a first point that the relation between $\mu$ and $C_\varepsilon$ is not limited to centerline data, but remains valid for off-centerline measurements. Second, two additional general relations are established, one linking $\gamma$ to $C_\varepsilon$ and another linking $\gamma$ to $C_k$. All together, the results indicate that the four quantities $C_\varepsilon$, $\mu$, $\gamma$, $C_k$ are related to each other. The consequences of these results will be discussed later. In the following chapter, we examine how the estimated parameter values depend on the specific methods used for their determination.}

\FloatBarrier

\section{\new{Methodological Robustness and Consistency}}
\label{methods}

\new{For most of the quantities listed in table~\ref{table_which_quantities_are_important}, as well as for the underlying variables from which they are derived, their estimation is not straightforward. Instead, several methods are often available for determining the same quantity, and these may yield different results. Such discrepancies can arise, for instance, from measurement noise or from assumptions inherent to a particular turbulence model on which the estimation method is based. Consequently, the relations presented in chapter~\ref{new_laws} can only be regarded as general laws if they remain independent of the specific methods used to estimate the individual quantities. In this chapter, we describe in detail which methods are used to estimate the 10 relevant quantities $\nu$, $\overline{u}$, $u^\prime$, $L$, $\varepsilon$, $F_u$, $\gamma$, $C_k$, $\mu$, and $\Lambda_0^2$ that are the base for a full statistical description of SST (according to table~\ref{table_which_quantities_are_important}) and how we make sure that the results are not biased due to the method itself. First, an overview is provided of the methods that were applied throughout the analysis or used for the estimation of several quantities.}

\subsection{\new{Overview of the Methodology}} 

We begin with several general remarks on the pre- and post-processing applied to all VTS. Three procedures are then introduced that are used repeatedly for different purposes throughout the analysis. First, the individual check method (ICM) is described, which is used to assess the general quality of the overall scaling behaviour of a VTS. This is followed by the zero-crossing method (ZCM), which is employed in the determination of several quantities. Finally, the subsampling method (SSM) is presented as an empirical approach for estimating uncertainties.

As a part of the pre- and post-processing, each VTS was first analyzed using the main processing routine, which generated an individual diagnostic figure containing $E(k)$ and $\Lambda^2(r)$ together with the corresponding fitted scaling ranges, similar to figure~\ref{data selection} a). Subsequently, an individual quality check was performed for every VTS by visually inspecting the respective diagnostic figure. This step was implemented in a separate code with an interactive interface, through which each VTS was classified as accepted, dismissed, or recalled if one of the fitted scaling ranges was judged to be misplaced.

For recalled VTS, the limits associated with the visually identified inadequate fit were discarded and new fitting bounds were selected manually. The corresponding VTS were then reprocessed using the main algorithm with the updated limits and, based on the resulting diagnostic figure, were finally either accepted or dismissed. The assessment criteria were based on the clear presence of an inertial range in both $E(k)$ and $\Lambda^2(r)$. Diagnostic figures were accepted when well-defined inertial ranges were visible and the corresponding fits of $E(k)$ and $\Lambda^2(r)$ were appropriately positioned. Figures were recalled when an inertial range appeared to be present but the fitted range was misplaced, for example when it extended predominantly into the dissipation range. Only a very small number of VTS were dismissed because no clear inertial range could be identified in either $E(k)$ or $\Lambda^2(r)$.

This visually guided procedure is justified by the fact that the subsequent selection criteria and consistency checks independently test whether the expected scaling behavior is genuinely present. In addition, the main post-processing routines implemented in Python were validated for consistency against other established analysis codes.

\new{The so-called zero-crossing method (ZCM) analyzes the statistical distribution of zero crossings of the fluctuating velocity $u(x)$, where the VTS $u(t)$ is transformed using Taylor hypothesis. Several studies have demonstrated that the zero crossings of $u(x)$ are closely related to key characteristics of turbulent flows. In particular, the zero-crossing density $n_z$ can be used to estimate the Taylor length scale $\lambda$~\cite{liepmann1953counting, sreenivasan1983zero, mazellier2008turbulence, mora2019experimental, ferran2023characterising}. Liepmann \& Robinson~\cite{liepmann1953counting} first applied Rice's theorem~\cite{rice1945mathematical} to turbulent flows and showed that $\lambda$ is proportional to the inverse of the zero-crossing density, namely}

\begin{equation} \new{
    n_z^{-1} =  \overline{\Delta Z} = C_z \, \pi \, \lambda,}
    \label{equation_rice_theorem}
\end{equation}

\noindent \new{where $\Delta Z$ denotes the distance between successive zero crossings $Z_{i+1}$ and $Z_i$, and $C_z$ is a parameter of order unity that accounts for the non-Gaussianity of the velocity derivative (with $C_z=1$ for a Gaussian process with a Gaussian derivative).}

\new{The subsampling method (SSM), a kind of bootstrap method, was performed to estimate the intrinsic error for $\overline{u}$, $u^\prime$, $L$, $\varepsilon$, $\lambda$, $\eta$, $F_u$, $S_u$, $TI$, $Re_\lambda$, $C_\varepsilon$, $\gamma$, and $C_k$~\cite{hardle2003bootstrap, lahiri2013resampling}. The intrinsic error $e_s$ is defined as}

\begin{equation} \new{
    e_s = \frac{\sigma_s}{\sqrt{N_s}},}
    \label{equation_intrinsic}
\end{equation}

\noindent {where the VTS is divided in $N_s$ equidistant non-overlapping segments, each containing one million points. $\sigma_s$ is the standard deviation of the ensemble of individual values of each segment, obtained with the same method used for the entire VTS. $\sigma_s$ is divided by the square root of the number of segments to obtain the uncertainty associated with the complete dataset. Since the subsampling method is a measure of the random error of our measurement, the question regarding the systematic error remains. A systematic error, such as $\pm1\,\%$ or $\pm0.1\,\mathrm{m/s}$ of the estimated hot-wire velocity cannot be reliably propagated given the applied processing methods. However, significant systematic errors are unlikely, given the diversity of independent measurement campaigns: the data stem from different measurement campaigns, which were conducted independently by four different researchers in three different labs using three different HWA-acquisition setups.}  

\new{Table~\ref{table_consistency_check} presents an overview of all used methods depending on the quantity, which are then described in detail below. Hereby, PTM indicates the plateau threshold method, which serves to assess the consistency of the inertial fitting range and is explained in detail below.  In the following, each quantity is discussed individually:}

\begin{table}[h]
    \centering
    \renewcommand{\arraystretch}{2} 
    \begin{tabular}{lcc}
        \hline
        & used method & consistency check \\
        \hline
        \hline
        \cellcolor{darkgray!30} $\nu$ & fluid parameter & - \\
        \hline
        \cellcolor{darkgray!30} $\overline{u}$ & mean & SSM \\
        \hline
        \cellcolor{darkgray!30} $u^\prime$ & $2^{nd}$ cent. moment & SSM \\
        \hline
        \cellcolor{darkgray!30} $L$ & auto-correlation & trend in $x$-direction + zero crossing of $\Lambda^2$ + ZCM + SSM + ICM\\
        \hline
        \cellcolor{darkgray!30} $\varepsilon$ & diss. spectrum & extension increase + deriv. of VTS + ZCM + SSM + ICM\\
        \hline
        \cellcolor{teal!30} $F_u$ & $4^{th}$ cent. moment & SSM \\
        \hline
        \cellcolor{teal!30} $\gamma$ & fit $E(k)$ & deriv. plat. (length, slope, $R^2$) + PTM + lower limit + $\zeta_2$ + SSM + ICM \\
        \hline
        \cellcolor{teal!30} $C_k$ & fit $E(k)$, norm. $2 \, \pi/\eta$ & deriv. plat. (length, slope, $R^2$) + PTM + $C_2$ + SSM + ICM\\
        \hline
        \cellcolor{teal!30} $\mu$ & fit $\Lambda^2(r)$ & deriv. plat. (length, slope, $R^2$) + PTM + Castaing fit + ZCM + $p(\varepsilon_r)$ + ICM\\
        \hline
        \cellcolor{teal!30} $\Lambda_0^2$ & fit $\Lambda^2(r>L)$ & signal flatness $F_u$ + ICM \\
        \hline
    \end{tabular}
    \caption{\new{1$\,$D quantities of turbulence are listed along with both the applied method and the procedures used to verify its consistency. Dimensional quantities are highlighted in gray while the non-dimensional are highlighted in turquoise. PTM stands for plateau threshold method, ZCM stands for zero-crossing method, SSM denotes the subsampling method and ICM stands for an individual check method, meaning that each main check figure is examined individually and based on this, the results of the corresponding VTS are retained, discarded or recalled.}}
    \label{table_consistency_check}
\end{table}

\new{As the fluid in all datasets consists of air, the kinematic viscosity $\nu$ is taken with the constant value of $15.32 \; 10^{-6} \, \mathrm{m^2/s}$ is taken. Regarding $\overline{u}$, $u^\prime$ and $F_u$ all three quantities are determined by their definition without any assumptions. Since there are no alternative methods, and all VTS fulfill $n_{L,min} = 1000$, no further checks of consistency for both quantities are carried out.} 

\new{All remaining quantities listed in table~\ref{table_consistency_check} are discussed individually in the following sections according to a common structure. First, the respective quantity is introduced together with the main challenges associated with its estimation. This is followed by an overview of the different versions of the quantity estimations considered in this work. The version ultimately used in the subsequent analyses, \textit{e.g.} in chapter~\ref{new_laws}, is then described in detail, including the complete estimation procedure and the quantity-specific checks used to assess the validity of the resulting values. Afterwards, the alternative estimation approaches are introduced. Then, the results obtained from these alternative approaches are compared with those of the method adopted for the main analysis. Finally, for most quantities, an uncertainty associated with the method ultimately adopted for the main analysis is derived from one of the corresponding consistency checks.}

\FloatBarrier

\subsection{\new{Integral Length Scale}}

\new{The integral length scale $L$ gives an estimate of the largest length scales within the cascade. A common approach to estimate $L$ is to integrate the normalized autocorrelation function $R_{uu}(\tau \, \overline{u})$ until it first crosses 0. $\tau$ denotes the time lag at which the VTS loses self-correlation. Applying the Taylor hypothesis, the time lag is transformed into a spatial lag $r = \tau \, \overline{u}$ to convert $time$ into $length$. However, the disadvantage of using the zero crossing as an integration limit lies in the fact that the autocorrelation function often exhibits overlying periodical structures ($e.g.$ shedding behind a cylinder) or does not cross 0 at all within a certain range - a frequent issue in non-homogeneous turbulence. Furthermore, when estimating $L$ in inhomogeneous turbulence, it is expected that $L$ becomes dependent on the direction, which does not pose a limitation here, as only the streamwise velocity component is considered.}

\new{In the present work, $L$ is estimated and checked calculating the following three procedures, which are neither directly influenced by superimposed flow structures nor assume any \emph{a priori} turbulence model:}

\begin{enumerate}
\item Integration of the decreasing function $R_{uu}(\tau \, \overline{u})$ until it crosses 1/e \color{blue} $\rightarrow L_e$ \color{black}
\item Zero-crossing method (based on Mora \& Obligado~\cite{mora2020estimating}) \color{blue} $\rightarrow L_z$ \color{black}
\item $L=r$ where the extrapolated fit of $\Lambda^2(r)$ in the inertial range crosses 0 \color{blue} $\rightarrow L_s$ \color{black}
\end{enumerate}

\noindent \new{The first, $L_e$, is determined as}

\begin{equation} \new{
    L_e = \int_{0}^{r_{1/\mathrm{e}}} R_{uu}(\tau \, \overline{u}) \, \mathrm{d} r,}
    \label{equation_L_e}
\end{equation}

\noindent \new{where the normalized autocorrelation function is integrated until it crosses the value of 1/e ($\approx 0.37$, with e as Euler's number). This approach ensures that even pronounced shedding within the autocorrelation function does not contribute to the integral. Due to its robustness we used this method throughout the paper, setting $L = L_{\mathrm{e}}$~\cite{mora2020estimating}.}

\new{Figure~\ref{figure_eight_examples_L} shows $R_{uu}$ as a function of the normalized lag $r/L$ for the eight example VTS listed in table~\ref{table_eight_examples}. The autocorrelation functions in a), b) and d) exhibit the expected smooth decay from 1 to 0. In contrast, c) and e) both remain positive over an extended range rather than decreasing smoothly to 0, despite their otherwise distinct characteristics. The autocorrelation functions in f) and g) display clear periodic structures, producing a quasi-sinusoidal oscillation around 0 after the initial decay. Remarkably, VTS h) exhibits a surprisingly smooth decay albeit it was ultimately discarded.}

\begin{figure}[htbp]
    \centering
    \begin{subfigure}[t]{0.49\textwidth}
        \centering        \includegraphics[width=\linewidth]{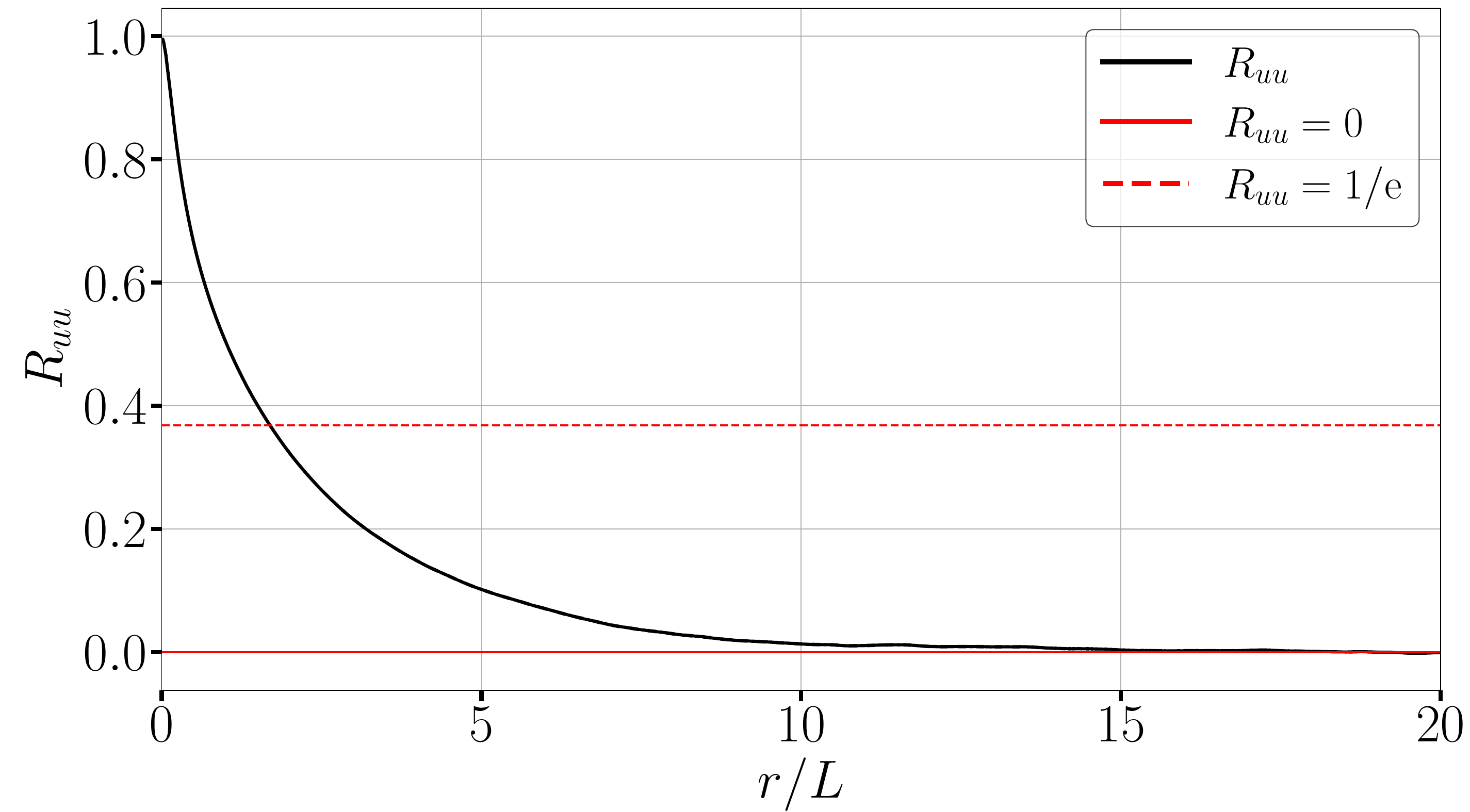}
        \caption{}
    \end{subfigure}
    \hfill
    \begin{subfigure}[t]{0.49\textwidth}
        \centering        \includegraphics[width=\linewidth]{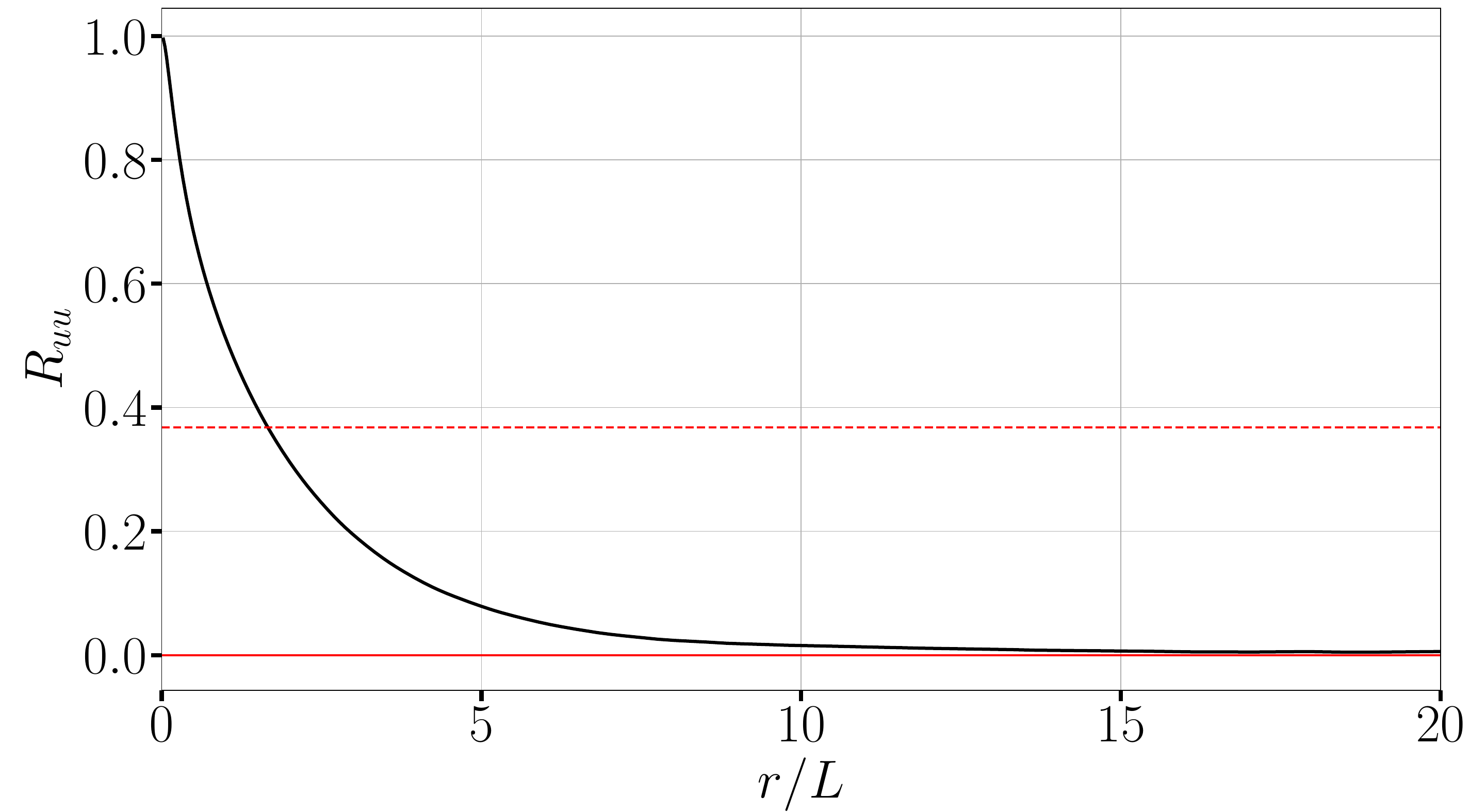}
        \caption{}
    \end{subfigure}
    \begin{subfigure}[t]{0.49\textwidth}
        \centering        \includegraphics[width=\linewidth]{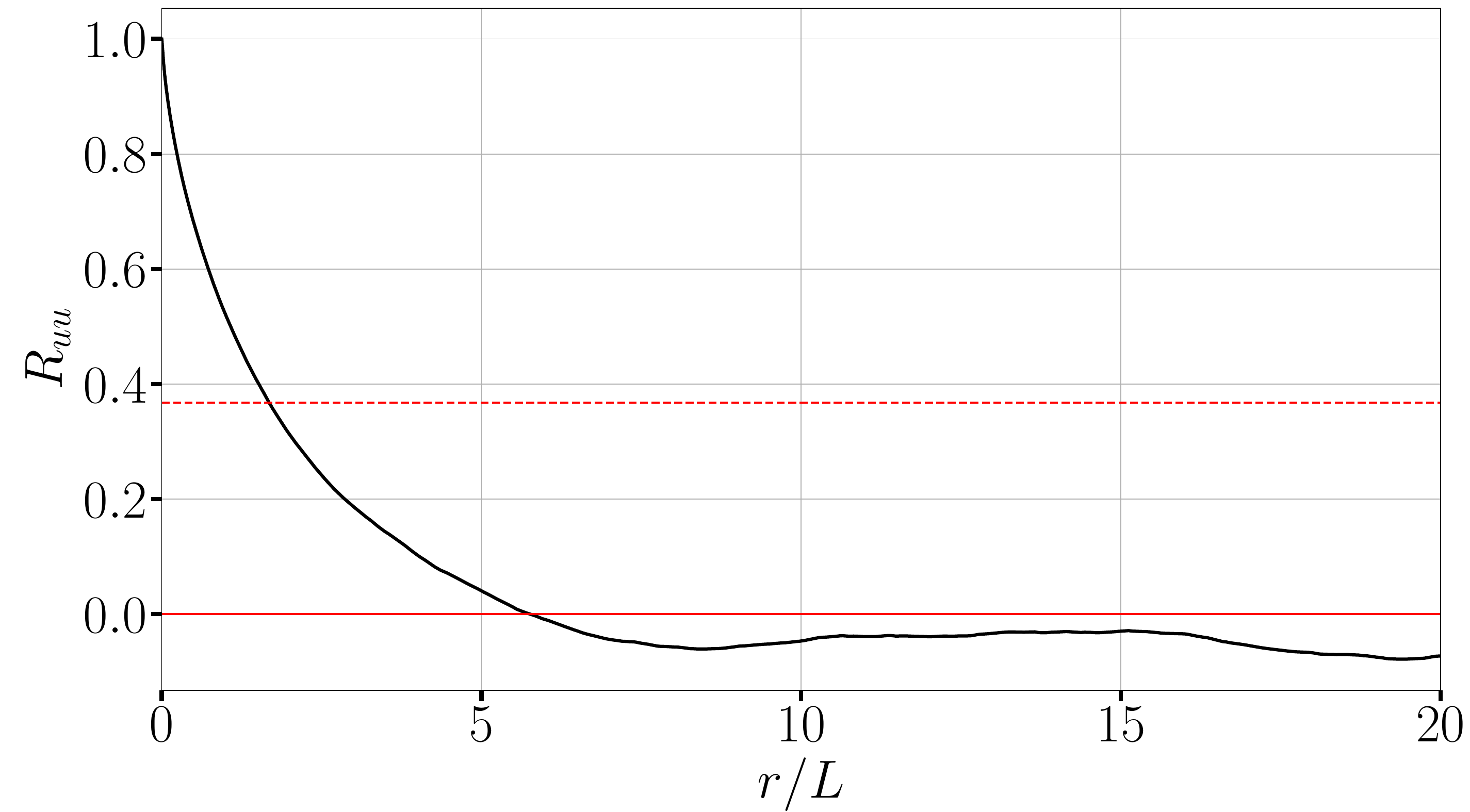}
        \caption{}
    \end{subfigure}
    \hfill
    \begin{subfigure}[t]{0.49\textwidth}
        \centering        \includegraphics[width=\linewidth]{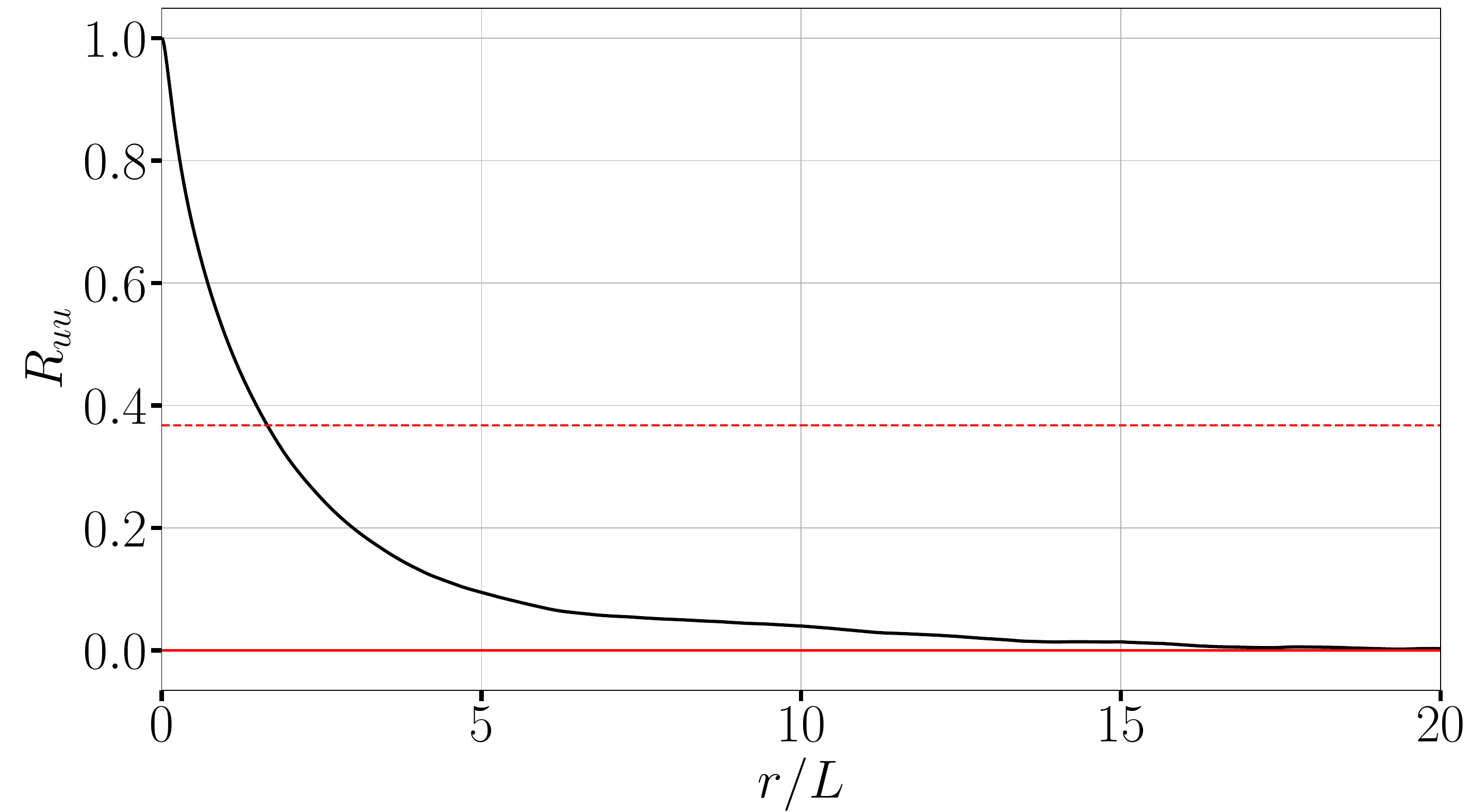}
        \caption{}
    \end{subfigure}
    \begin{subfigure}[t]{0.49\textwidth}
        \centering        \includegraphics[width=\linewidth]{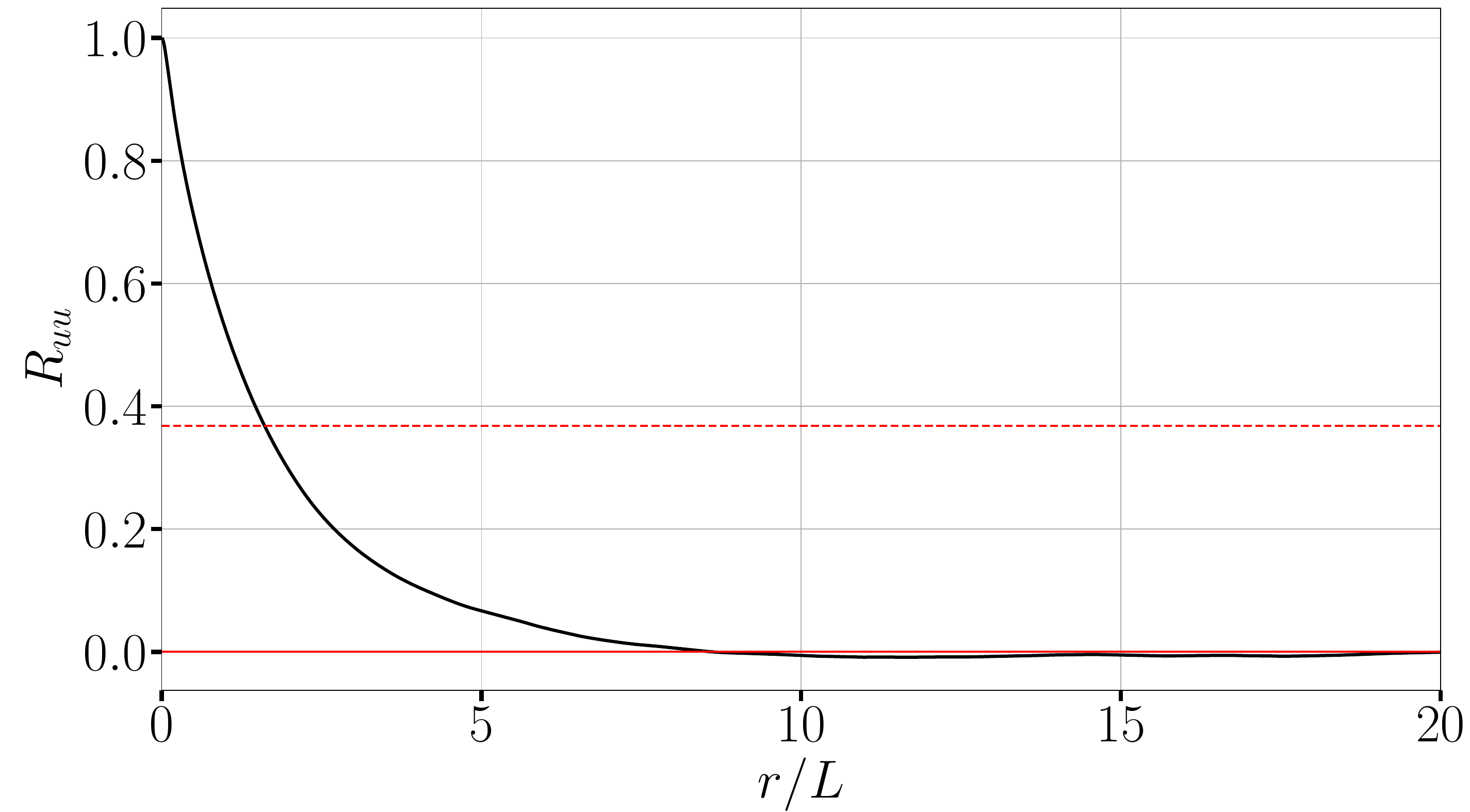}
        \caption{}
    \end{subfigure}
    \hfill
    \begin{subfigure}[t]{0.49\textwidth}
        \centering        \includegraphics[width=\linewidth]{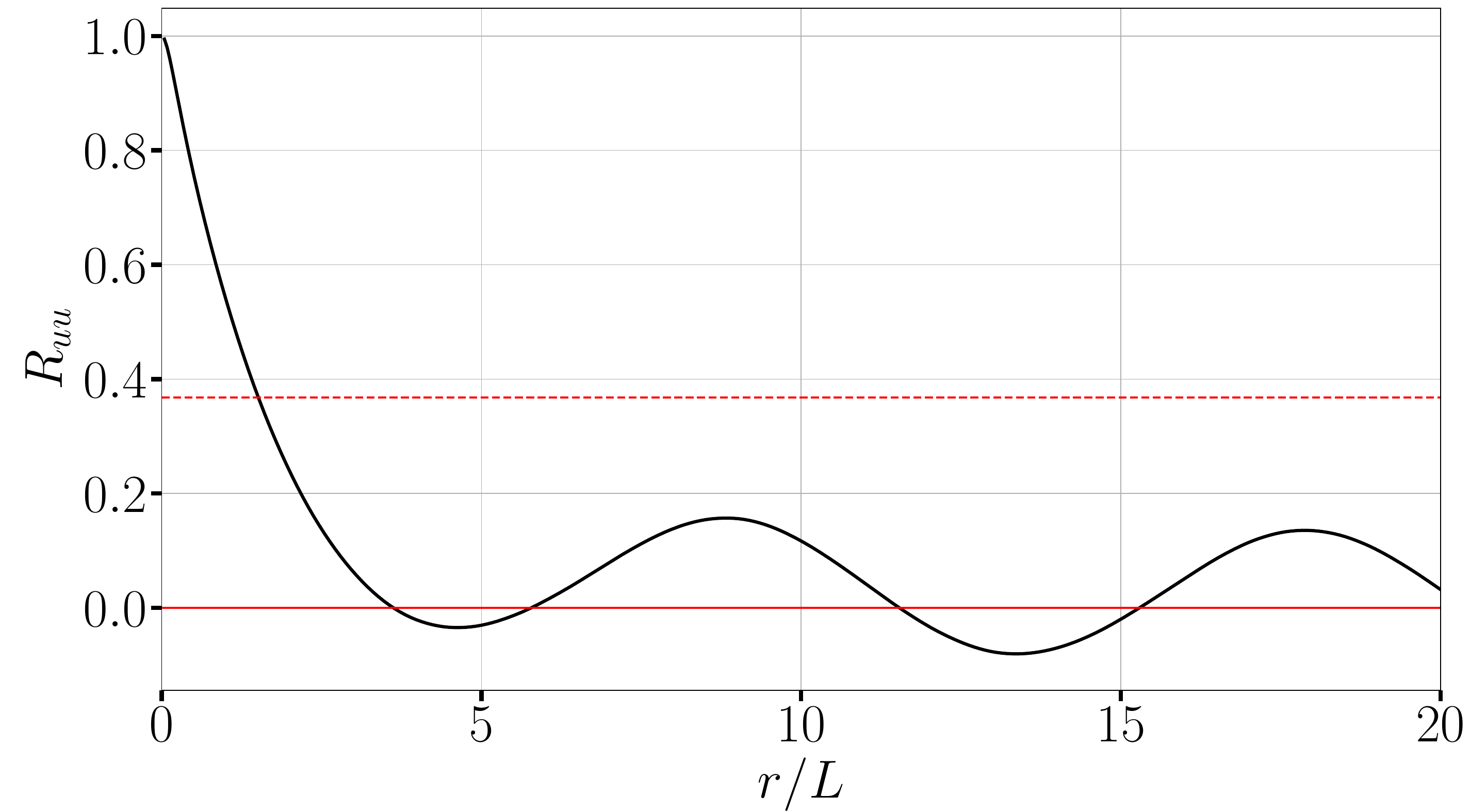}
        \caption{}
    \end{subfigure}
    \begin{subfigure}[t]{0.49\textwidth}
        \centering        \includegraphics[width=\linewidth]{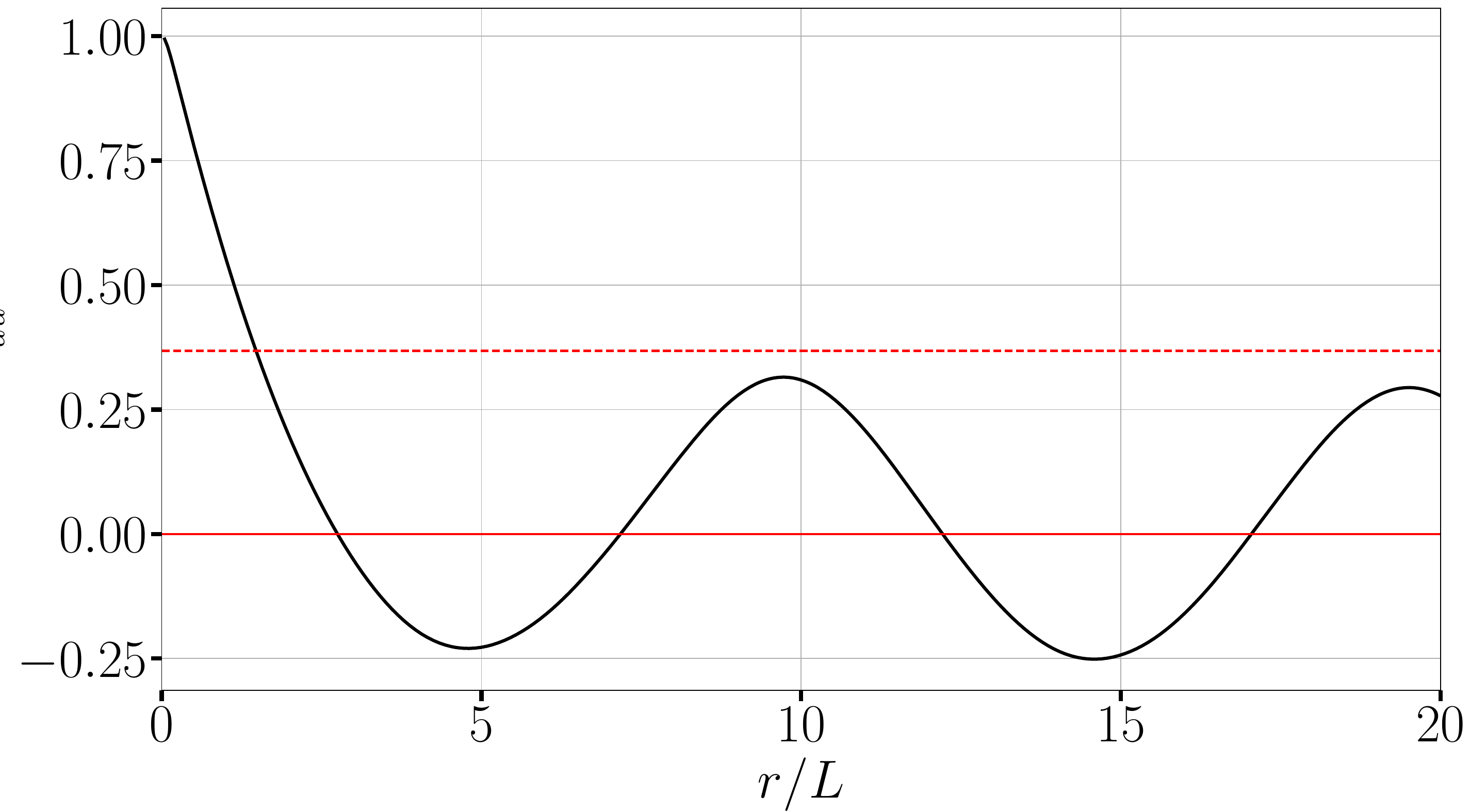}
        \caption{}
    \end{subfigure}
    \hfill
    \begin{subfigure}[t]{0.49\textwidth}
        \centering        \includegraphics[width=\linewidth]{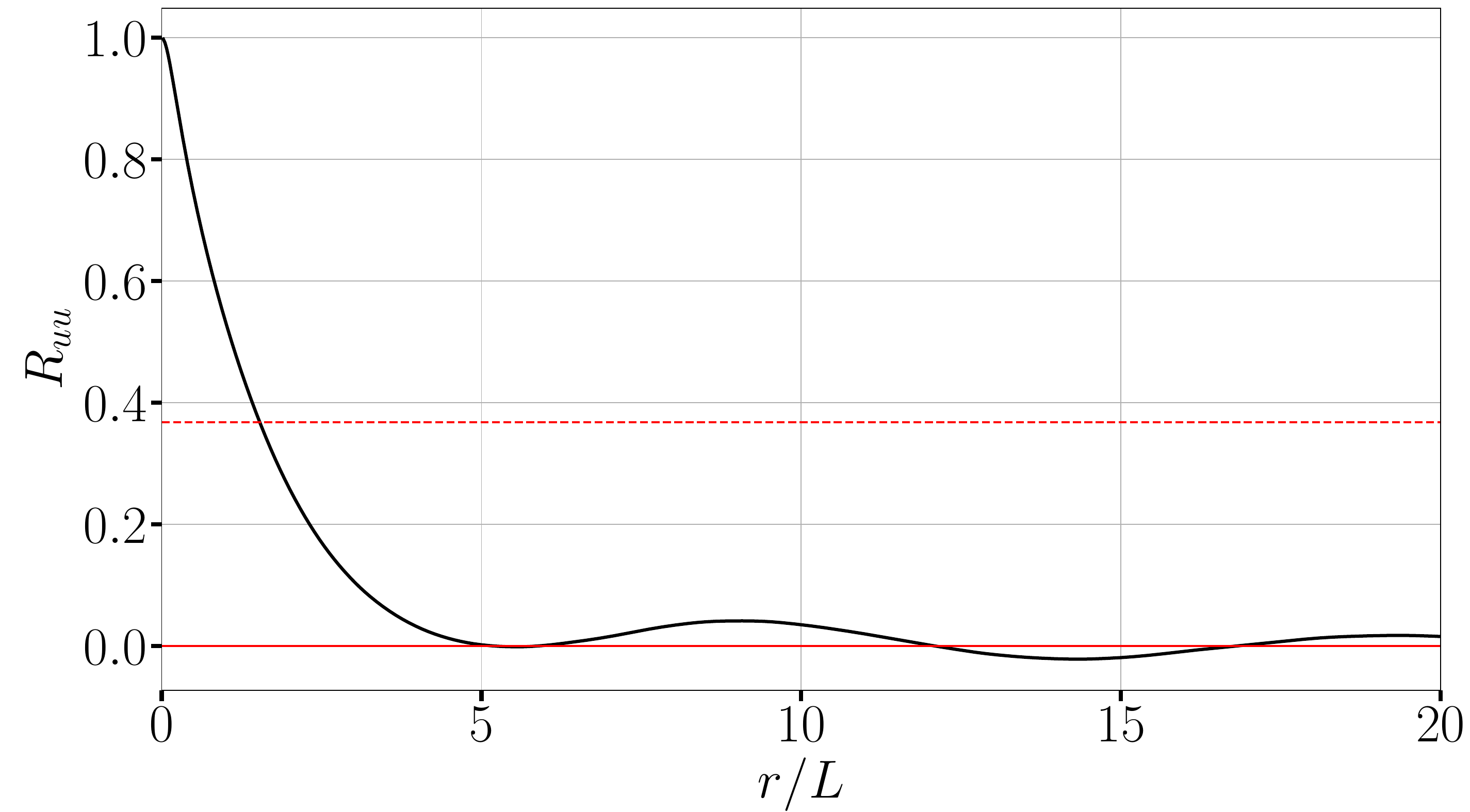}
        \caption{}
    \end{subfigure}
    \caption{\new{The autocorrelation functions (black) are shown for the same eight VTS as in table~\ref{table_eight_examples} (also in the same order of appearance). The solid red line represents the zero level, while the dashed red line represents $R_{uu} = 1/\mathrm{e}$, respectively. a), b), c), e), f) and g) display VTS that satisfy the restriction criteria and taken together, span almost the full range of $\mu$-values. d) also fulfills the restriction criteria and exhibits the same value of $\mu$ as c) but with a $Re_\lambda$ more than three times smaller. h) however does not meet the restriction criteria since $\Lambda_0^2$ shows clear signatures of non-Gaussianity at large scales. The data stem from the cases G20, G24, C8, D11, C6, C1, C1, G23, respectively in order of appearance. For a detailed list of the characteristic values of the VTS and details about their cases see table~\ref{table_eight_examples} and table~\ref{tab:PhD measurements in LEGI 2023} in the appendix~\cite{SM}.}}
    \label{figure_eight_examples_L}
\end{figure}

\new{For wakes and grid-generated turbulence, the integral length scale $L$ is generally expected to increase with streamwise distance~\cite{krogstad2010grid,cafiero2020length}. For $L_e$, this behavior was verified by examining its downstream evolution.}

\new{For the second approach, $L_z$, the McFadden equation~\cite{mcfadden1958axis} has been recently applied to turbulent signals to estimate the longitudinal integral length scale $L$ using the variance and mean of the zero-crossing distances~\cite{mora2020estimating}. Assuming that the zero-crossing distances are statistically independent and, again, that the signal is a Gaussian process with a Gaussian derivative, the McFadden equation can be used as a basis to relate the integral length scale to the statistics of $\Delta Z$ as,}

\begin{equation}
    L_z = \frac{\pi \: \sigma_{\Delta Z}^2}{4 \: \overline{\Delta Z}},
    \label{equation_L_z}
\end{equation}

\noindent \new{where $\sigma_{\Delta Z}$ denotes the standard deviation. Note that the estimation of $L_z$ \emph{a priori} neglects the different manifestations of intermittency discussed above.} 

\new{For the third approach, $L_s$ represents the integral length scale derived from the behavior of the $r$-dependent shape parameter $\Lambda^2(r)$. Within the inertial range, $\Lambda^2(r)$ follows a linear trend in log/lin representation, and $L_s$ is defined as the scale where its linear extrapolation intersects 0. The extrapolation is preferred over the actual zero crossing of $\Lambda^2(r)$ at large scales to minimize the influence of deviations occurring at the transition from linear to constant behavior at the large scales.} 

\new{Figure~\ref{figure_validation_L} a) shows that $L_e$ agrees well with $L_z$. Although the scatter is considerably larger for subplot b), the overall trend observed for $L_e$ and $L_z$ remains clearly evident. This supports that $L_e$ as a reliable estimate of the integral length scale. All this shows that the different estimation methods of $L$ will not change principally the found relation of chapter\ref{new_laws}.}

\begin{figure}[htbp]
    \centering
    \begin{subfigure}[t]{0.49\textwidth}
        \centering        
        \includegraphics[width=\linewidth]{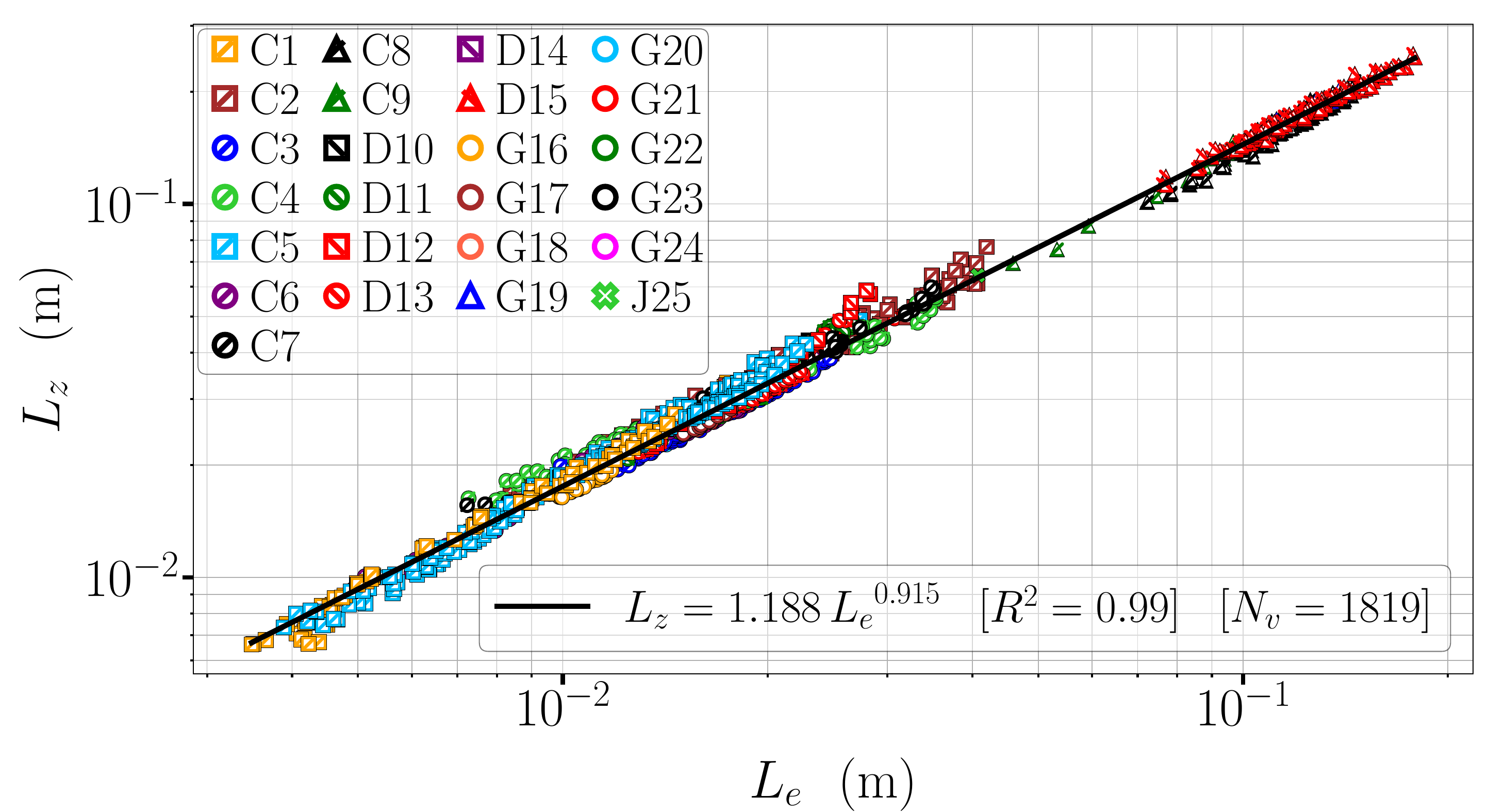}
        \caption{}
    \end{subfigure}
    \hfill
    \begin{subfigure}[t]{0.49\textwidth}
        \centering        
        \includegraphics[width=\linewidth]{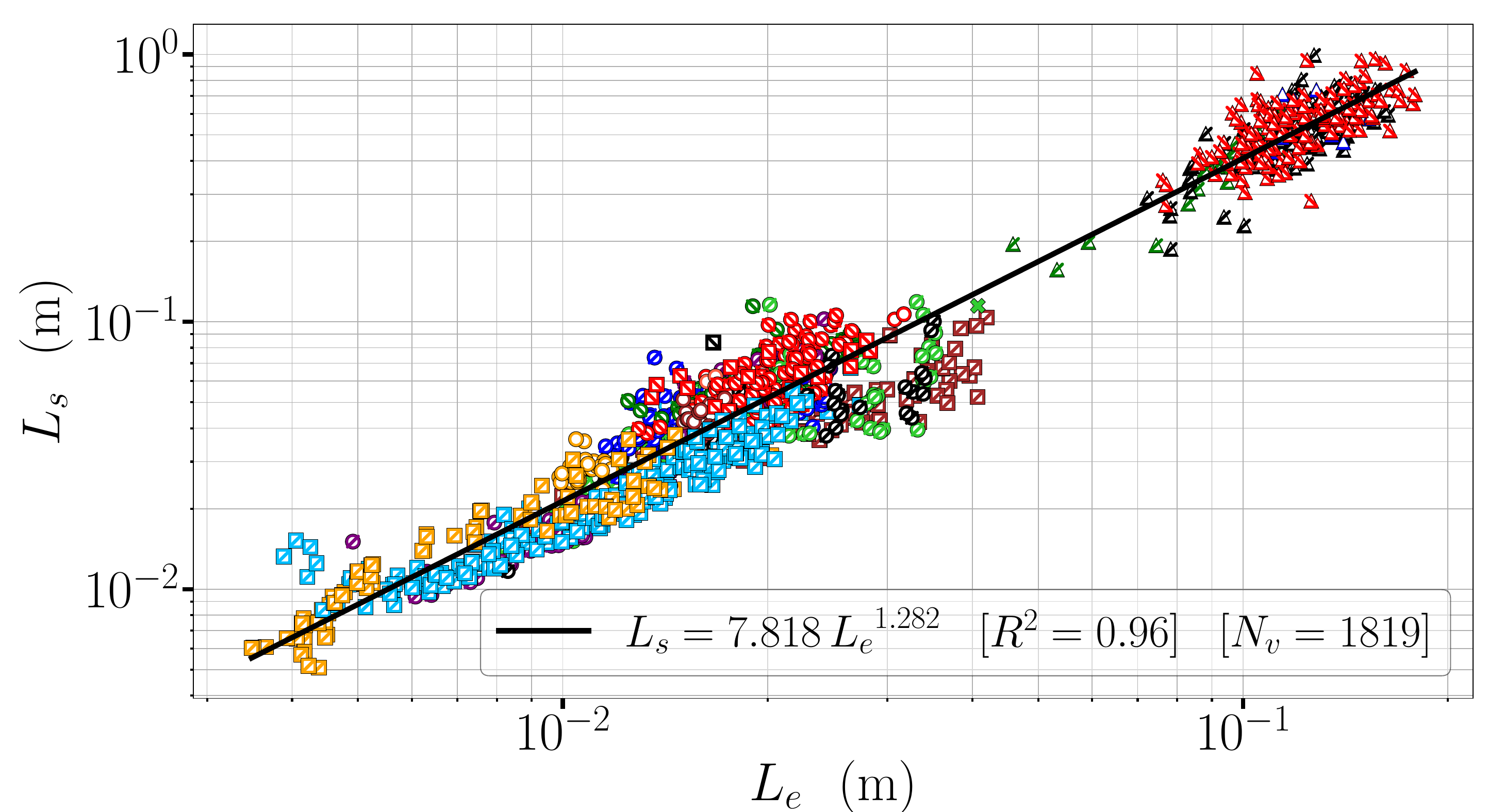}
        \caption{}
    \end{subfigure}
    \caption{\new{a) $L_z$ as a function of $L_e$ for all VTS used. b) $L_s$ as a function of $L_e$ for all VTS used. In general the black lines correspond to a least-squares fit with $R^2$ being the coefficient of determination. $N_v$ indicates the number of VTS shown in the plot. The symbols in the legend are identical for all figures throughout this manuscript and correspond to the configurations in table~\ref{tab:PhD measurements in LEGI 2023} in the appendix~\cite{SM}. For laminar inflow, squared markers are used. For the regular grid and the active grid, markers are shaped as circles and triangles, respectively. For cylinders as generators, the markers contains a ``$/$'' (case names start with ``C'') while for disks a ``$\backslash$'' (case names start with ``D'') is used. Hollow circular markers means no object and is equivalent to grid turbulence (case names start with ``G''). The ``x'' marker indicates a free jet (case names start with ``J'').}}
    \label{figure_validation_L}
\end{figure}

\new{The uncertainty of $L_e$ (used \textit{e.g.} in figures~\ref{law_1} d) and~\ref{law_2} d) is calculated based on the deviation between $L_e$ and $L_z$ in figure~\ref{figure_validation_L} a). The relation obtained from the least-squaress fit between $L_e$ and $L_z$ is used by constructing the difference between the actual value of $L_z$ and $L_z=1.188 \, {L_e}^{0.915}$.}

\FloatBarrier

\subsection{\new{Exponent of the Spectral Law}}

\new{Next, the absolute value of the exponent of the spectral law for non-homogeneous turbulence $\gamma$ is considered, which is introduced in eq.~(\ref{equation_energy_spectrum}). In this context, the selection of the fitting range is crucial for a valid estimation. In the present work, $\gamma$ is estimated and checked by calculating the following two quantities:}

\begin{enumerate}
\item Slope of the energy spectral density within the inertial range \color{blue} $\rightarrow \gamma$ \color{black}
\item Slope of the second order structure function within the inertial range \color{blue} $\rightarrow \zeta_2$ \color{black}
\end{enumerate}

\FloatBarrier

\noindent {The parameter $\gamma$ is estimated as follows: to accurately fit the spectral energy density $E(k)$ within the inertial range, several steps are required. $E(f)$ is computed using Welch's method with a Hann window function, a window length of $2^{15}$ and a 25$\, \%$ overlap~\cite{welch1967use}. Subsequently, $E(f)$ is transformed into $E(k)$ applying Taylor's hypothesis where the angular wavenumber is given by $k=2\pi\,f/\overline{u}$. The wavenumber range is then divided into 50 logarithmically spaced, equidistant windows and a block-wise static mean is calculated for both $E(k)$ and $k$. Starting at the smallest scales, the algorithm identifies the first empty window, typically at the large scales, whose center is denoted as $k_{void}$. Then, a representative, fully smoothed spectral energy density $E(k)$ is obtained by taking the block-wise static mean values until $k_{void}$ and appending the large-scale part from Welch's method. Finally, instrumental noise is removed examining the smoothed derivative of $E(k)$ in log/log representation. The smoothed derivative is computed as the gradient of the fully smoothed spectrum, followed by a moving average with a {window size of 3 bins}. The noise cutoff is set at the local minimum of the smoothed derivative of $E(k)$ in log/log representation, marking the transition from the dissipation range to measurement noise. Note that this cutoff defines $l_c$, the cutoff length introduced above. This yields a smooth spectral energy density $E(k)$ representing only the physics of turbulence, and is well suited for fitting.} 

\new{Next, the limits of the inertial range need to be determined. A cornerstone assumption of turbulence theory is that $E(k)$ follows a power law scaling within the inertial range~\cite{frisch1995turbulence}. This implies that $\mathrm{d \: log}(E(k)) / \mathrm{d \: log}(k)$ should remain constant. We define the inertial range as the region where this derivative forms a plateau. The identification of this plateau is realized examining the second derivative of $E(k)$ in log/log representation, starting at the smallest scales. The algorithm marks the first zero crossing of the second derivative as $k_{i}$ as the initial point of a possible plateau extending toward both smaller and larger scales. A threshold box is then defined with bounds of $\pm 0.25$ around $\mathrm{d \: log}(E(k_{i})) / \mathrm{d \: log}(k_{i})$, which proved robust for most datasets (the test is provided below). From $k_{i}$, the algorithm identifies the first data points on both sides that lie outside this box, defining the inertial range limits $k_{s}$ and $k_{l}$ toward small scales and large scales, respectively. The exponent $\gamma$ is then obtained by performing a linear fit of $E(k)$ in log/log representation. To validate the method, several diagnostic quantities are calculated simultaneously. The extent of the inertial range $\mathrm{log}(k_{s})-\mathrm{log}(k_{l})$ is calculated. In addition, the slope of the derivative of $E(k)$ in log/log representation is computed by a linear fit in log/lin representation. For a perfect power law, this slope equals 0, hence deviations quantify the departure of $E(k)$ from ideal scaling. On average, we found that the slope of the derivative of $E(k)$ in a log/log representation has a slightly negative value for the data used. The value of the slope shows no clear trend with respect to either $\gamma$ or $Re_\lambda$. This confirms that the power-law assumption holds to first order. Experimental evidence shows that, even under approximately homogeneous and isotropic conditions, ideal power-law scaling in the inertial range is only approached and may exhibit systematic deviations~\cite{sinhuber2017dissipative, reinke2018universal}. For both fits $R^2$ is calculated as the coefficient of determination. Ultimately, a lower limit of $\gamma_{min} = 1.25$ is applied as stated in the selection criteria. All main check figures were individually scanned and only VTS with valid $E(k)$ fits were retained.}

\new{Figure~\ref{figure_eight_examples} displays the smoothed $E(k)$ as a function of the wavenumber $k$ for the eight example VTS listed in table~\ref{table_eight_examples}. Note that the spectra in figure~\ref{figure_eight_examples} are shown before noise removal and with an unnormalized x-axis. The black lines indicate the fits of $E(k)$ within the inertial range. Here again, the VTS in f) and g) display clear periodic structures, visible as peaks in $E(k)$ while the VTS h) again shows no apparent indication for exclusion based on the applied criteria. Several spectra exhibited peaks in $E(k)$, caused by energy accumulation in periodic flow structures. The normalized position and extent of each peak were determined, yet no consistent relations emerged among these quantities or with others.}

\begin{figure}[htbp]
    \centering
    \begin{subfigure}[t]{0.49\textwidth}
        \centering        \includegraphics[width=\linewidth]{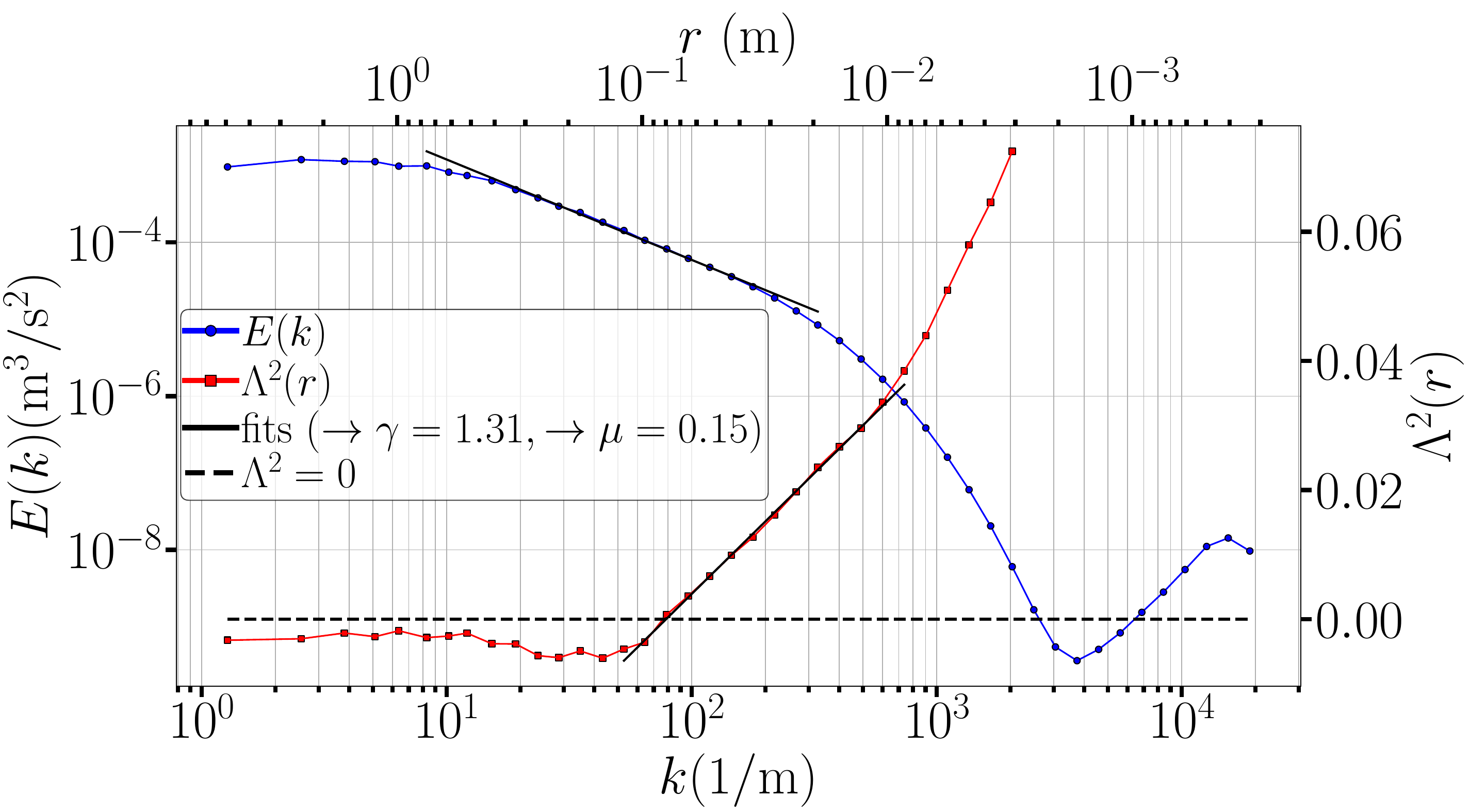}
        \caption{}
    \end{subfigure}
    \hfill
    \begin{subfigure}[t]{0.49\textwidth}
        \centering        \includegraphics[width=\linewidth]{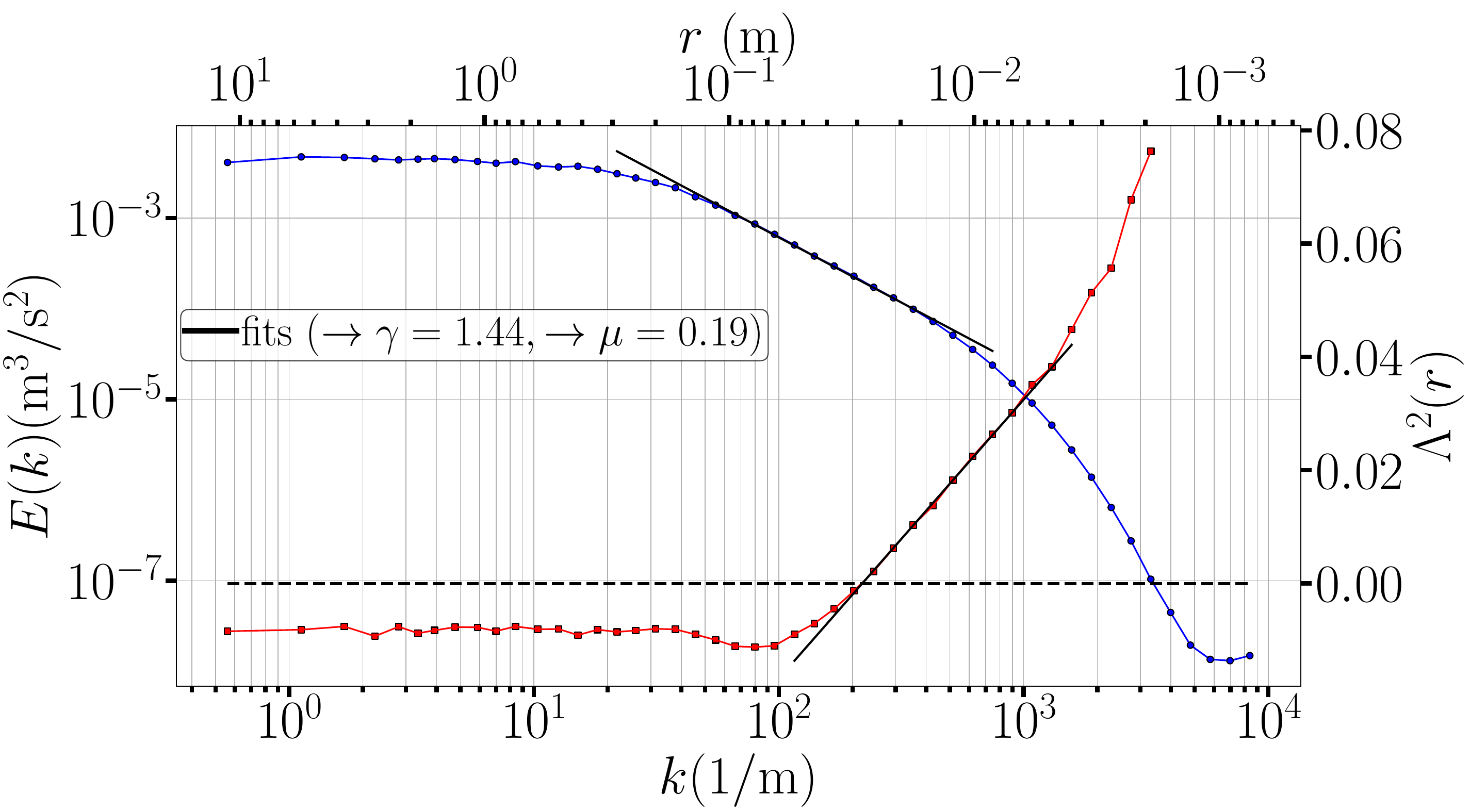}
        \caption{}
    \end{subfigure}
    \begin{subfigure}[t]{0.49\textwidth}
        \centering        \includegraphics[width=\linewidth]{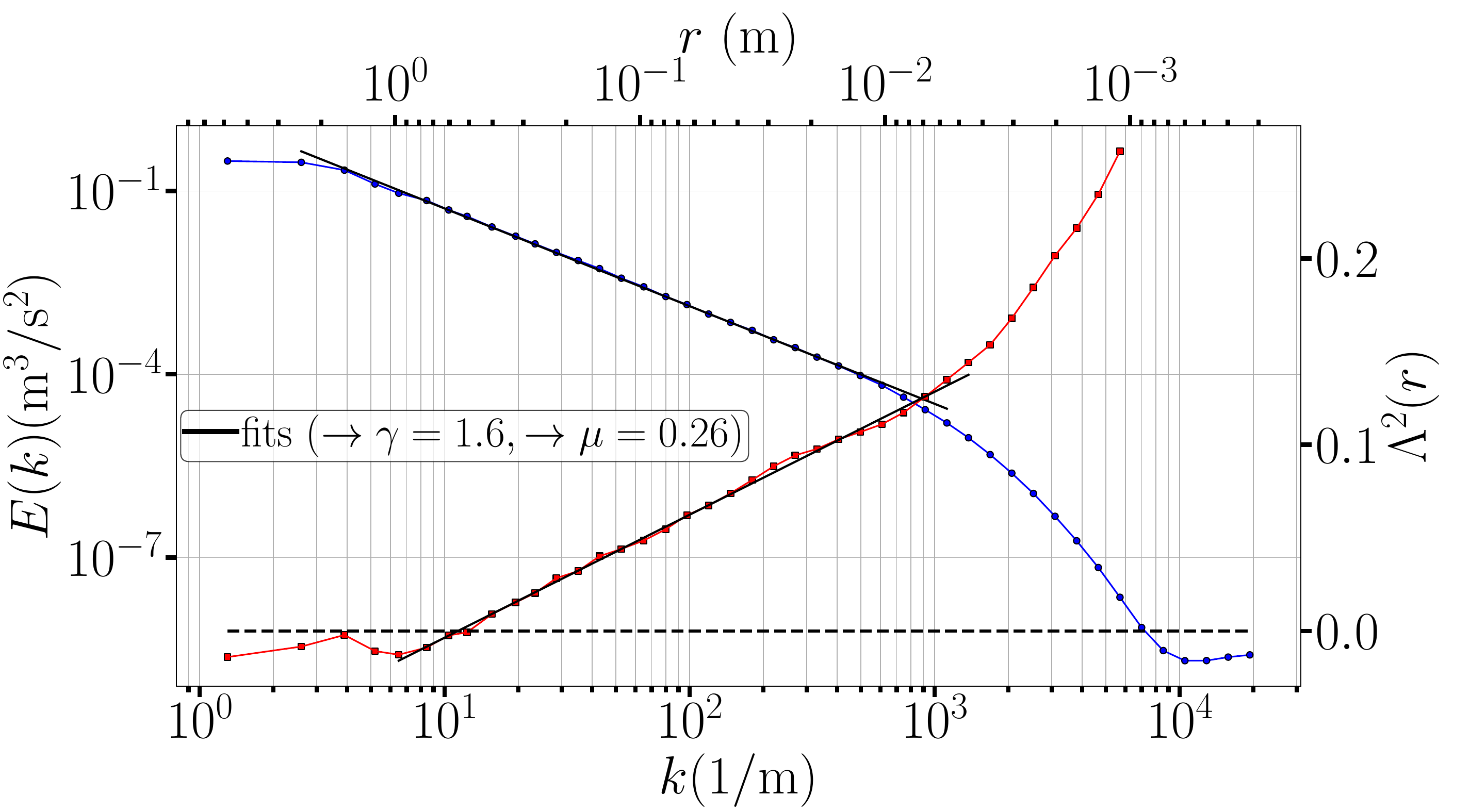}
        \caption{}
    \end{subfigure}
    \hfill
    \begin{subfigure}[t]{0.49\textwidth}
        \centering        \includegraphics[width=\linewidth]{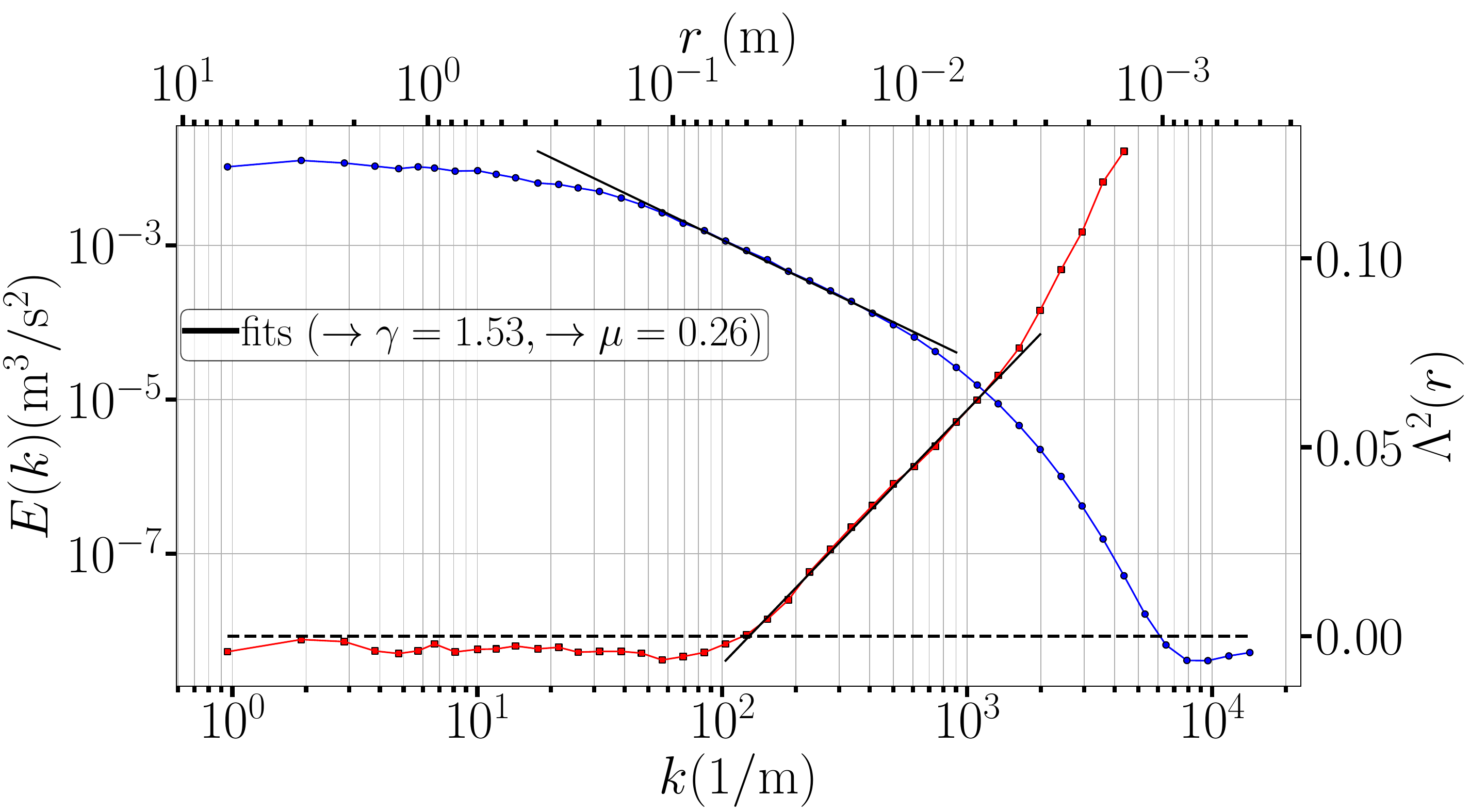}
        \caption{}
    \end{subfigure}
    \begin{subfigure}[t]{0.49\textwidth}
        \centering        \includegraphics[width=\linewidth]{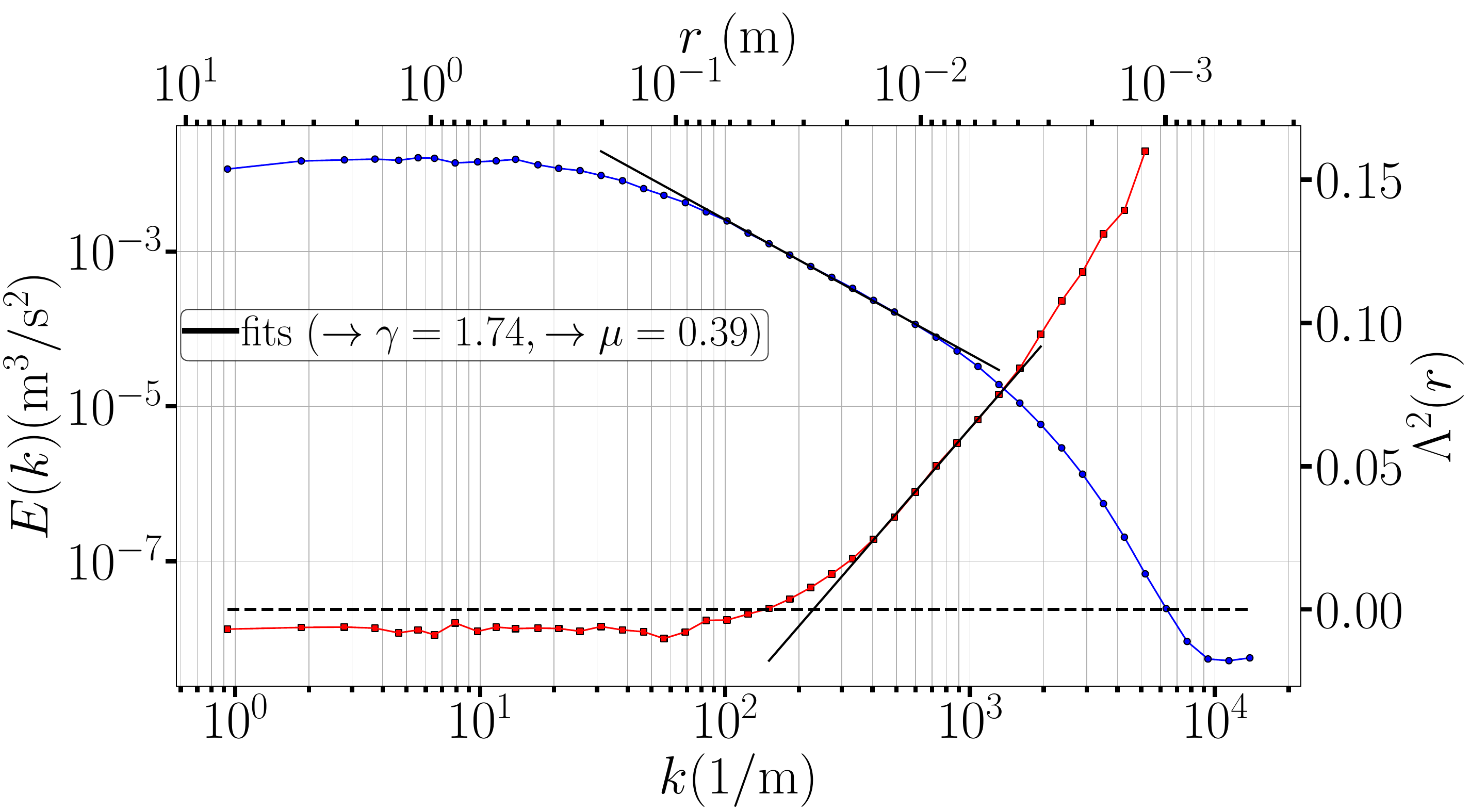}
        \caption{}
    \end{subfigure}
    \hfill
    \begin{subfigure}[t]{0.49\textwidth}
        \centering        \includegraphics[width=\linewidth]{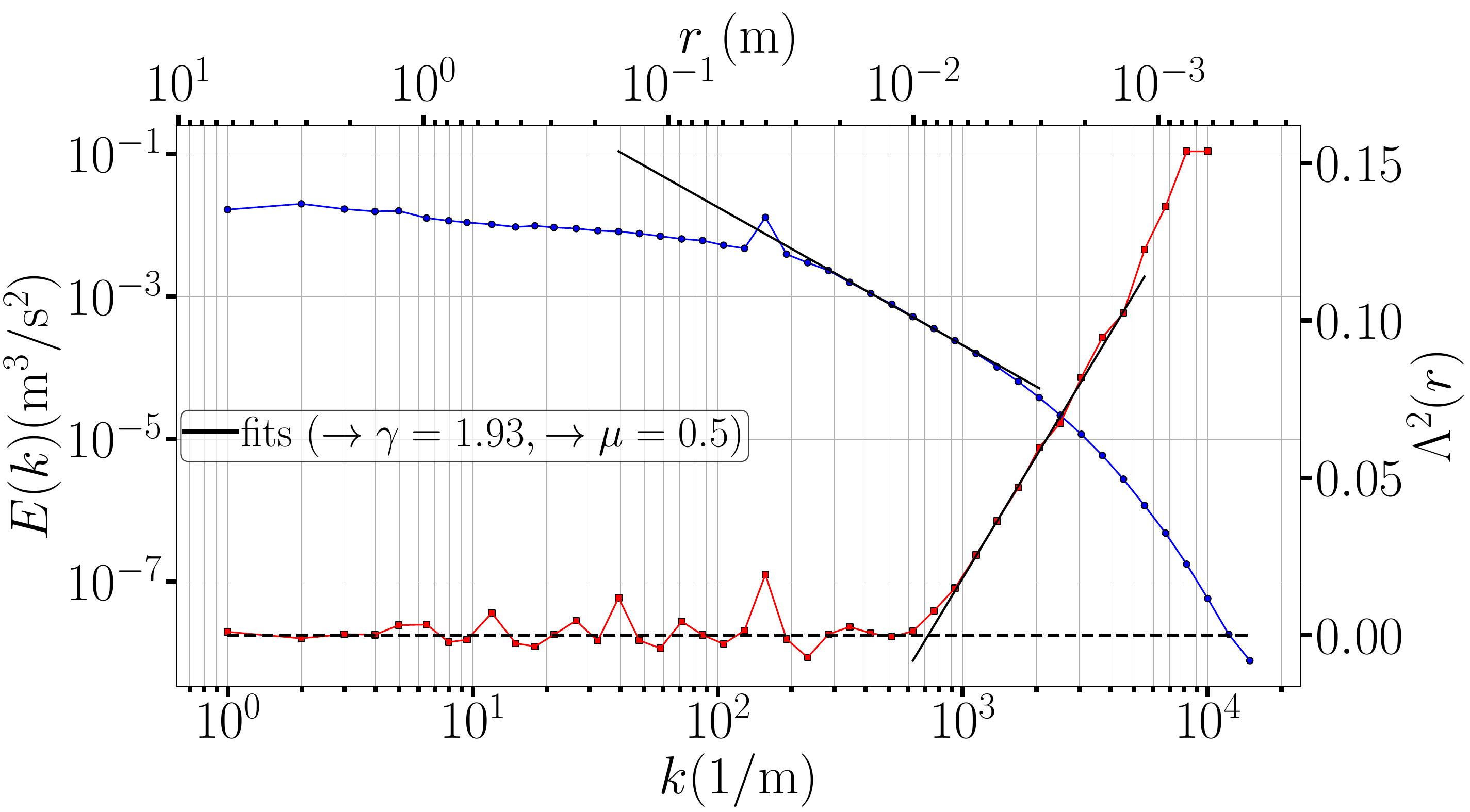}
        \caption{}
    \end{subfigure}
    \begin{subfigure}[t]{0.49\textwidth}
        \centering        \includegraphics[width=\linewidth]{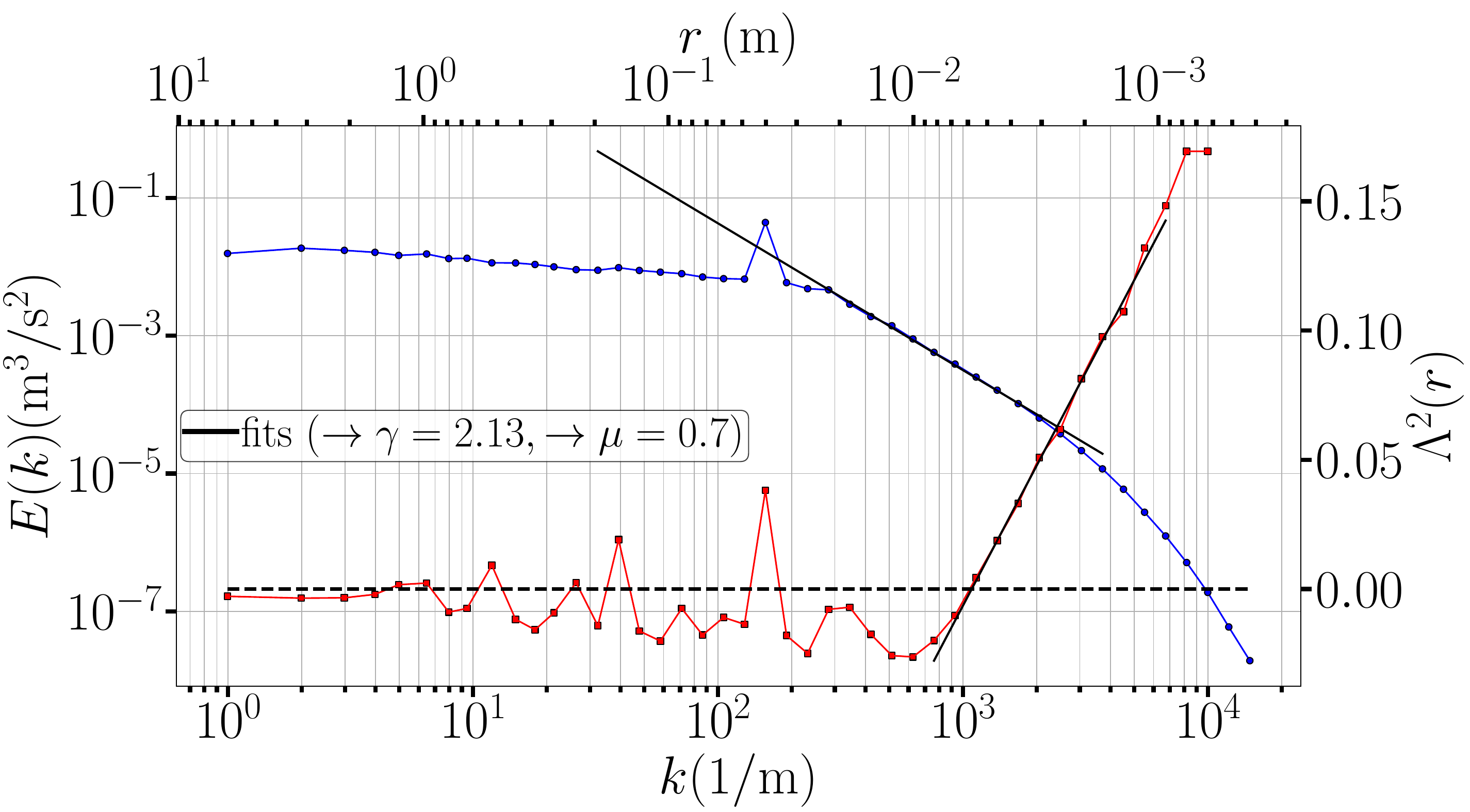}
        \caption{}
    \end{subfigure}
    \hfill
    \begin{subfigure}[t]{0.49\textwidth}
        \centering        \includegraphics[width=\linewidth]{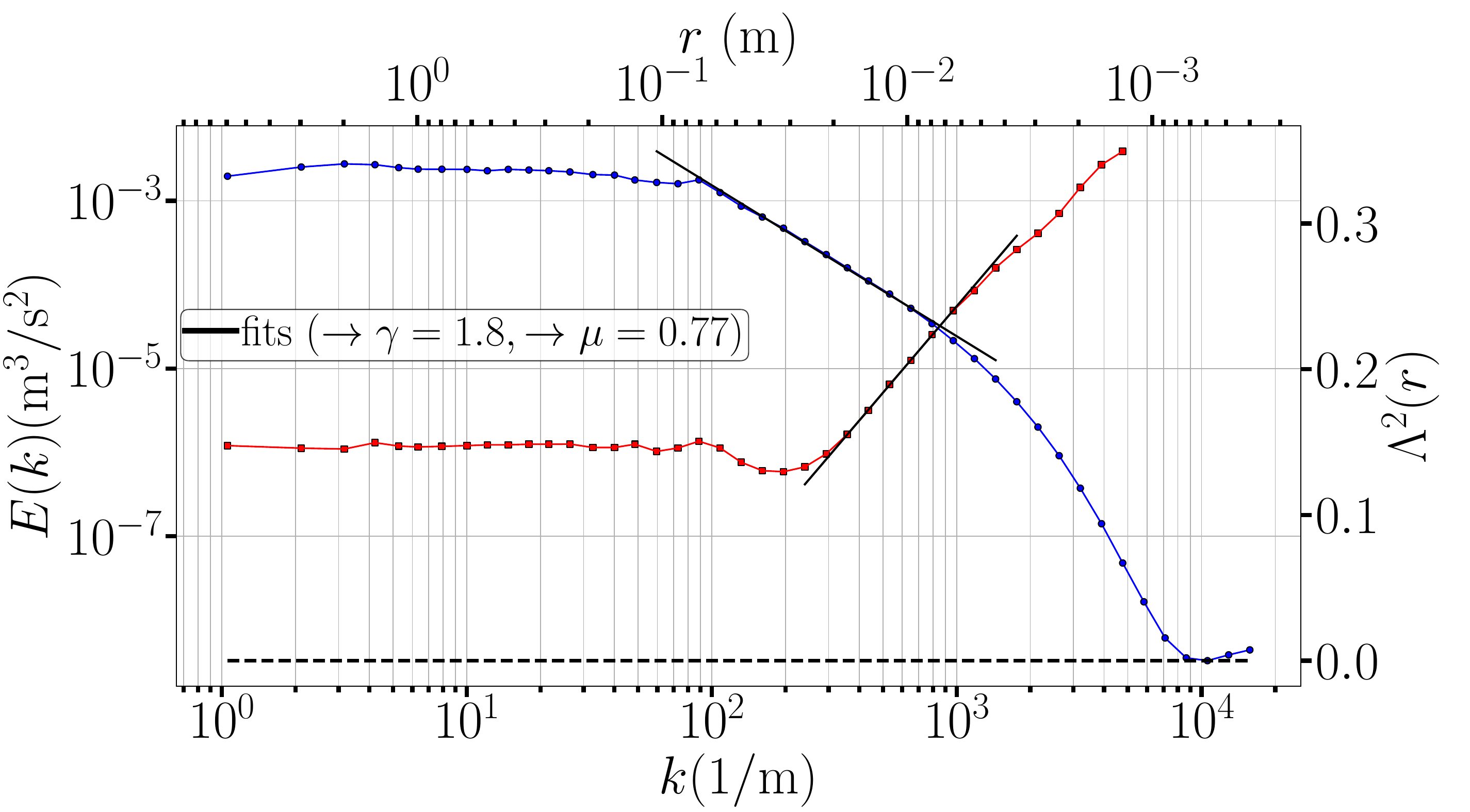}
        \caption{}
    \end{subfigure}
    \caption{\new{The energy spectral density (blue) and the shape parameter curve (red) are shown for the same eight VTS as in table~\ref{table_eight_examples} (also in the same order of appearance). The fits for obtaining $\gamma, C_k$ and $\mu$ are plotted as solid black lines, while the dashed black line represents $\Lambda_0^2 = 0$. a), b), c), e), f) and g) display VTS that satisfy the restriction criteria and taken together, span almost the full range of $\mu$-values. d) also fulfills the restriction criteria and exhibits the same value of $\mu$ as c) but with a $Re_\lambda$ more than three times smaller. h) however does not meet the restriction criteria since $\Lambda_0^2$ shows clear signatures of non-Gaussianity at large scales. The data stem from the cases G20, G24, C8, D11, C6, C1, C1, G23, respectively in order of appearance. For a detailed list of the characteristic values of the VTS and details about their cases see table~\ref{table_eight_examples} and table~\ref{tab:PhD measurements in LEGI 2023} in the appendix~\cite{SM}.}}
    \label{figure_eight_examples}
\end{figure}

\new{Figure~\ref{figure_validation_gamma} a) finally presents all normalized, denoised spectra superimposed for all used VTS. Normalization was performed according to eq. (\ref{equation_energy_spectrum}), affecting only the $x$-axis. To assess further the validity of the estimated $\gamma$-values, the robustness regarding the introduced threshold of $\pm 0.25$ for the derivative in log/log representation was addressed exemplarily. For three representative cases (C1, C4, and D13), $\gamma$ was additionally calculated by changing the value of the threshold. In particular, the threshold was set to $\pm 0.15$ and $\pm 0.35$, resulting in corresponding values of $\gamma^*$. Figure~\ref{figure_validation_gamma} b) shows the results where the blue markers correspond to $\pm 0.15$ and the red markers stand for $\pm 0.35$. It confirms, that the estimation of $\gamma$ is not strongly biased by the choice of the threshold of for the derivative in log/log representation.}

\begin{figure}[htbp]
    \centering
    \begin{subfigure}[t]{0.49\textwidth}
        \centering        \includegraphics[width=\linewidth]{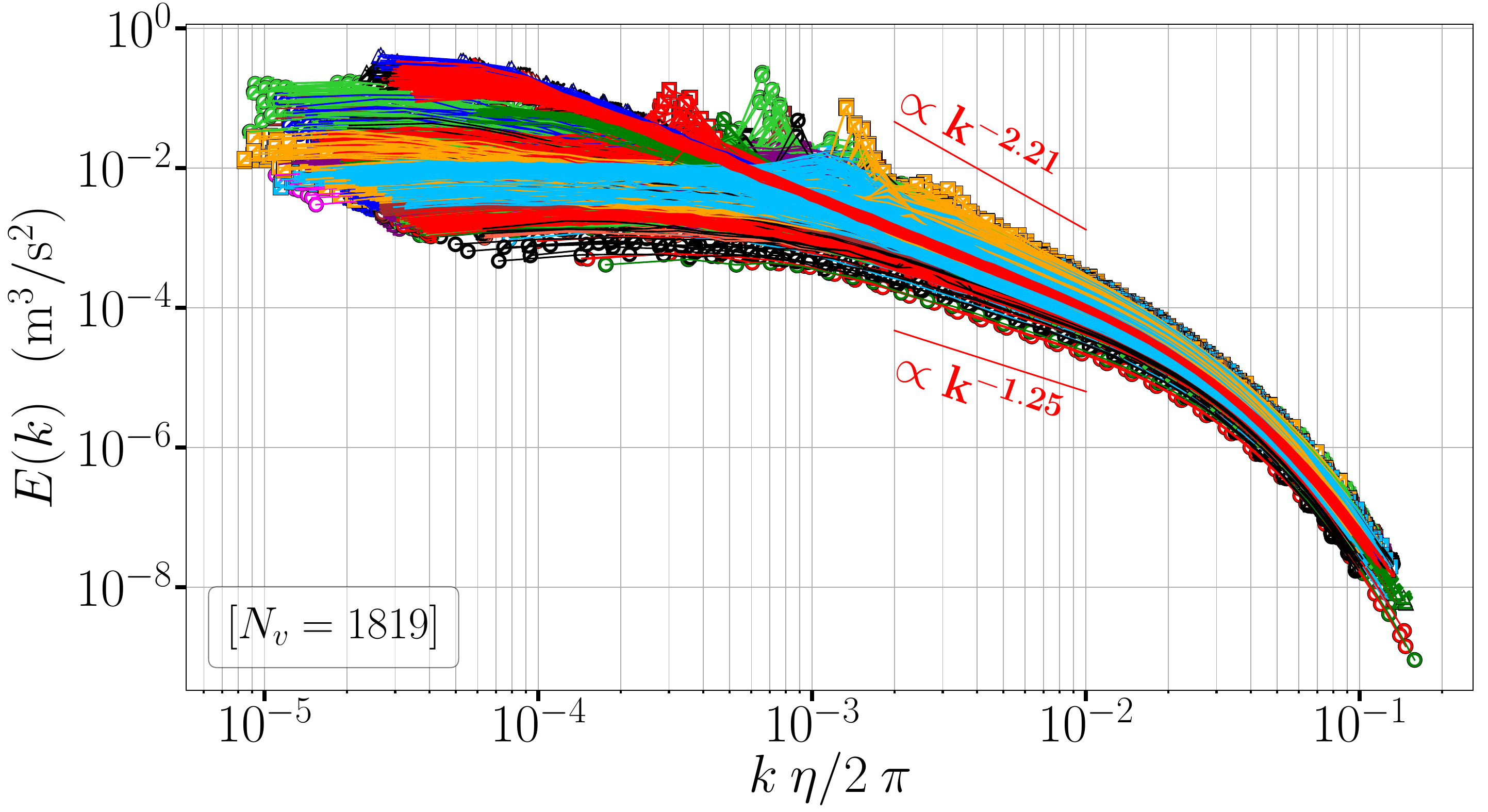}
        \caption{}
    \end{subfigure}
    \hfill
    \begin{subfigure}[t]{0.49\textwidth}
        \centering        \includegraphics[width=\linewidth]{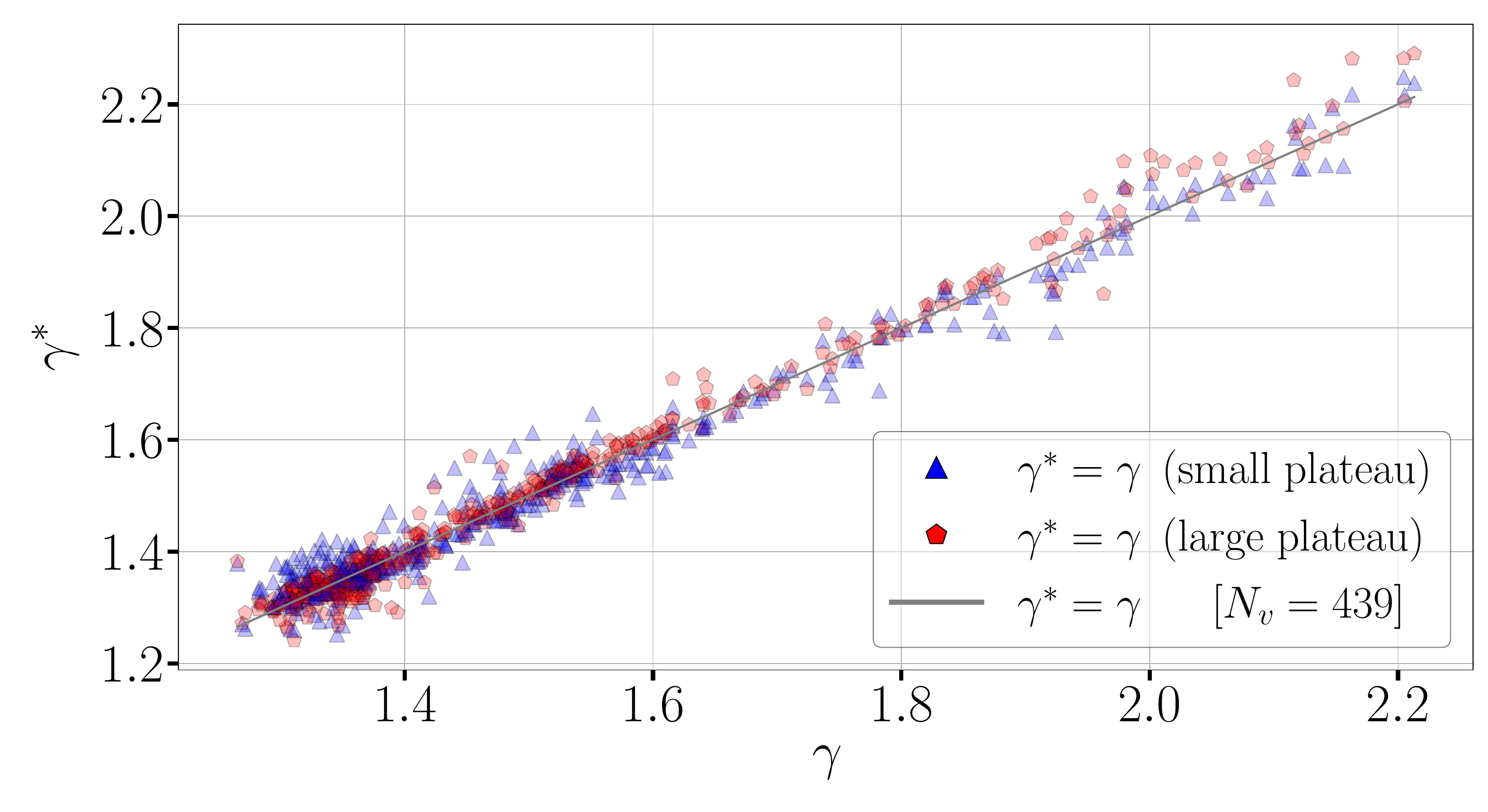}
        \caption{}
    \end{subfigure}
    \begin{subfigure}[t]{0.49\textwidth}
        \centering        \includegraphics[width=\linewidth]{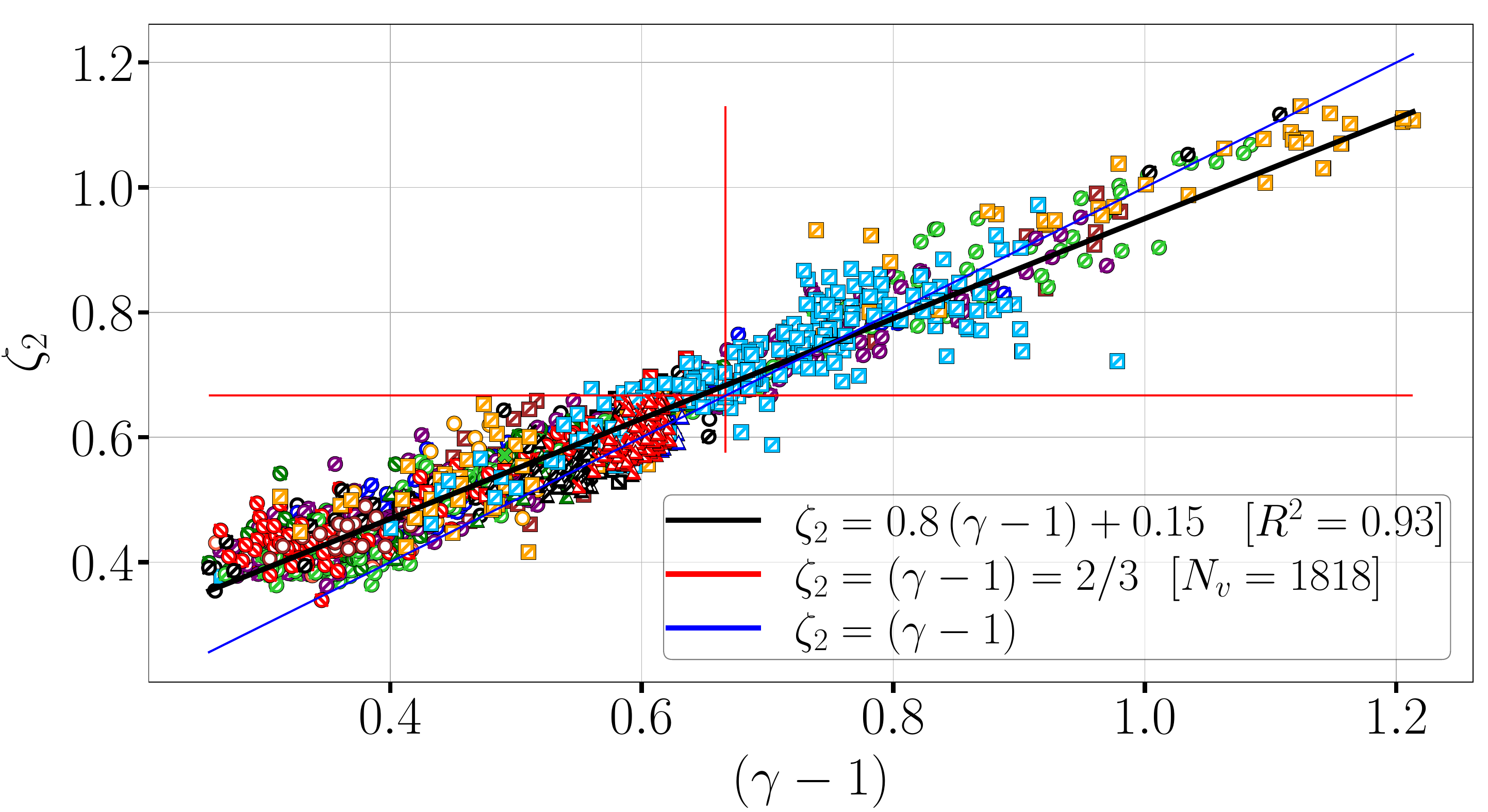}
        \caption{}
    \end{subfigure}
    \hfill
     \begin{subfigure}[t]{0.49\textwidth}
        \centering        \includegraphics[width=\linewidth]{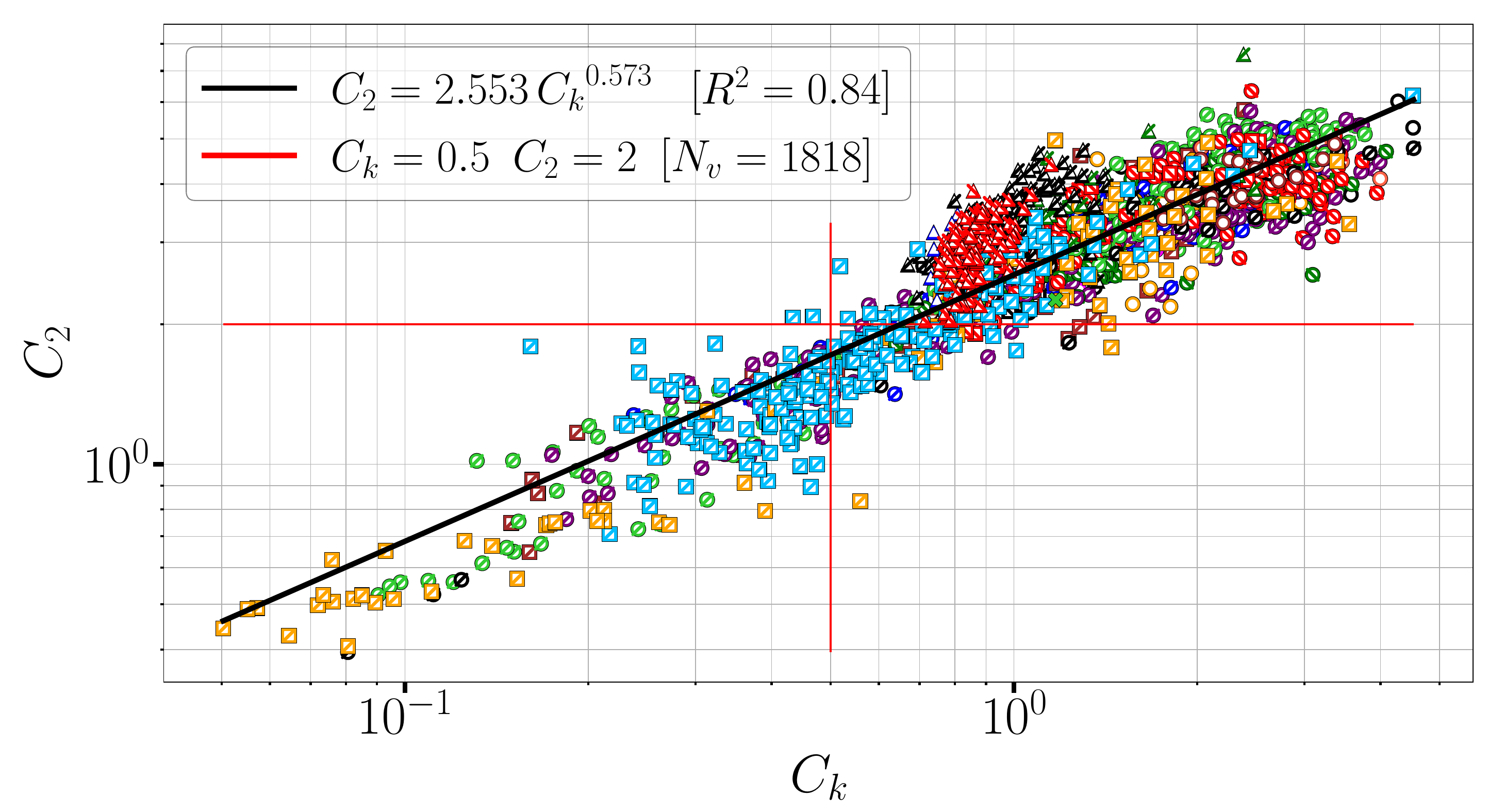}
        \caption{}
    \end{subfigure}
    \caption{\new{a) $E(k)$ as a function of $k \, \eta / 2\,\pi$ for all VTS used. The red lines indicate the highest and the lowest slope present within the data. b) $\gamma^*$ as a function of $\gamma$ for all VTS used from three representative cases (C1, C4 and D13). While the gray solid line represents identity, the blue markers indicate the values of $\gamma$ obtained by using a smaller threshold for the derivative plateau and the red markers indicate the values of $\gamma$ obtained by using a larger threshold for the derivative plateau. c) $\zeta_2$ as a function of $\gamma - 1$ for all VTS used. The blue line indicates where $\zeta_2 = \gamma - 1$. d) $C_2$ as a function of $C_k$ for all VTS used. In general the red lines indicate the commonly accepted values for $\zeta_2$, $C_2$, $\gamma$ and $C_k$ for HIT~\cite{sreenivasan1995universality}. The black lines correspond to a least-squares fit with $R^2$ being the coefficient of determination. $N_v$ indicates the number of VTS shown in the plot. The symbols and corresponding configurations are shown and explained in table~\ref{tab:PhD measurements in LEGI 2023} in the appendix~\cite{SM}.}}
    \label{figure_validation_gamma}
\end{figure}

\new{Moreover, $\gamma$ is checked by calculating $\zeta_2$. As explained above, the Wiener-Khintchine theorem~\cite{wiener1930generalized,khintchine1934korrelationstheorie} allows a direct comparison between $\gamma - 1$ and $\zeta_2$. Equivalent to eq.~(\ref{equation_energy_spectrum}), the second order structure function can be defined as,}

\begin{equation}
    S_2(r) = C_2 \: \varepsilon^{2/3} \: r^{2/3} \: \bigg(\frac{r}{\eta}\bigg)^{\zeta_2 - {2/3}},
    \label{equation_second_order_SF}
\end{equation}

\noindent \new{within the inertial range, accounting for inhomogeneity. Fitting $S_2$ requires appropriate limits for the fitting range, determined using two approaches. The first approach mirrors the determination of $\gamma$, identifying plateaus in the derivative of $S_2(r)$ in log/log representation. The second approach adopts the fitting limits from $\Lambda^2(r)$, which shares the same scale parameter $r$. The first method proved more valid, hence, $\zeta_2$ was derived from the behavior of the derivative of $S_2$ in log/log representation. Figure~\ref{figure_validation_gamma} c) confirms that $\gamma-1$ provides a valid estimate of the power-law exponent of $E(k)$ in the inertial range showing good agreement with $\zeta_2$.}

\new{The uncertainties of $\gamma$ are determined according to the uncertainties shown in figure~\ref{figure_validation_gamma} b). Accordingly, the largest deviation of $\gamma^*$ based on the corresponding value of $\gamma$ is taken as the error bound of $\gamma$. Overall we conclude again that the relations found in chapter \ref{new_laws} will not be changed by different estimations methods of $\gamma$. }

\FloatBarrier

\subsection{\new{Kolmogorov Parameter}}

\new{Next, the Kolmogorov parameter $C_k$ is considered, which is introduced in eq.~(\ref{equation_energy_spectrum}) in which it sets the level of turbulent kinetic energy as a prefactor. In the present work, $C_k$ is estimated and checked calculating the following two quantities:}

\begin{enumerate}
\item prefactor of the energy spectral density within the inertial range \color{blue} $\rightarrow C_k$ \color{black}
\item prefactor of the second order structure function within the inertial range \color{blue} $\rightarrow C_2$ \color{black}
\end{enumerate}

\FloatBarrier

\noindent \new{Although $C_k$ is derived using the same power-law fit as $\gamma$, the chosen normalization of $k$ crucially affects its value outside HIT. We therefore explored several normalization approaches, including scaling with $2\pi/L$ and $2\pi/\lambda$. In addition, we considered a normalization based also on $\varepsilon$, in which $C_\varepsilon$ and an additional length scale ($L$, $\lambda$, or $\eta$) are introduced to ensure the non-dimensionality of $C_k$. Among them, the normalization used in eq.~(\ref{equation_energy_spectrum}) proved to be the most robust and valid estimate.}

\new{To assess the validity of $C_k$, we computed $C_2$, the prefactor of the second order structure function within the inertial range, according to eq. (\ref{equation_second_order_SF}). $C_2$ is related to $C_k$ and should be around a value of 2 in HIT~\cite{frisch1995turbulence}. Figure~\ref{figure_validation_gamma} d) confirms that $C_k$ provides an estimate of the Kolmogorov parameter showing in average good agreement with $C_2$.}

\new{The uncertainties of $\mathrm{log}(C_k)$ were calculated similar to the uncertainties of $\gamma$, since $\gamma$ and $\mathrm{log}(C_k)$ are estimated through the same fit.}

\FloatBarrier

\subsection{\new{Mean Dissipation Rate}}

\new{Next, the mean dissipation rate $\varepsilon$ is considered, representing the amount of energy dissipated per unit mass and time. In the present work, $\varepsilon$ is estimated and checked calculating the following three quantities:}

\begin{enumerate}
\item Variance of the derivative of the velocity fluctuations\color{blue} $\rightarrow \varepsilon_g$ \color{black}
\item Integration of the dissipation spectrum \color{blue} $\rightarrow \varepsilon_d$ \color{black}
\item Zero-crossing method \color{blue} $\rightarrow \varepsilon_z$ \color{black}
\end{enumerate}

\FloatBarrier

\noindent \new{The mean dissipation rate for 1$\,$D data assuming small scale isotropy is defined as~\cite{davidson2015turbulence},} 

\begin{equation}
    \varepsilon_g = 15 \, \nu \overline{\bigg(\frac{\partial {u}}{\partial x}\bigg)^2}.
    \label{equation_dissipation_rate_2}
\end{equation}

\noindent \new{Following our dissipation assumption, we further allow eq.~(\ref{equation_dissipation_rate_2}) to be applied to inhomogeneous turbulence by interpreting $\varepsilon_g$ as a one-dimensional surrogate. Moreover, we deliberately avoid any direct or implied scaling laws such as K41. Schröder \emph{et al.}~\cite{schroder2024estimating} provide an overview of the accuracy of different methods based on 1$\,$D data, showing that any approach relying on scaling laws yields major deviations, while the most accurate method employs the gradient of the velocity fluctuations. { The gradient was performed using a central difference. The VTS was filtered before using a Butterworth lowpass filter to cut off the measurement noise at the same frequency as explained above, corresponding to the cutoff length of $l_c$. Note that the calculation of $\varepsilon_r$ using eq.~(\ref{equation_epsilon_r}) was treated the same way.} The disadvantage of this method lies in the finite sampling frequency, which prevents resolving the full mean dissipation rate.}

\new{Alternatively to $\varepsilon_g$, $\varepsilon$ can also be obtained by integrating the dissipation spectrum~\cite{davidson2015turbulence}. Although this method is mathematically equivalent to eq.~(\ref{equation_dissipation_rate_2}) by invoking Parseval's theorem, the limitations imposed by the finite sampling frequency can be mitigated in a straightforward manner. We define,}

\begin{equation}
    \varepsilon_d = \int_{0}^\infty  15 \, \nu \, k^2 \, E(k) \, \mathrm{d} k.
    \label{equation_dissipation_rate_1}
\end{equation}

\new{To represent the smallest scales, the dissipation range of the spectrum was extended before. After removing instrumental noise at $k=2\pi/l_c$ (the high-frequency tails are shown in figure~\ref{figure_eight_examples}), each spectrum was extended by fitting a quadratic function in log/log representation, starting at the end of the inertial range toward small scales. Other fits were tested but the quadratic fit proved robust and straightforward to automate, yielding equivalent results. Several validation parameters were computed to assess the quality of the fits. The coefficient of determination $R^2$ was determined to quantify how well the quadratic fit overlapped with measured dissipation range. As stated before, each main check figure was individually scanned and only VTS with consistent dissipation fits were retained.}

\new{To quantify the effect of the extension of the dissipation spectrum, we define $\Delta \varepsilon$ as the added portion, resulting that the ratio $\Delta \varepsilon / \varepsilon$ remained in the order of $5\,\%$, shown in figure~\ref{figure_validation_epsilon} a). Note that the quantization of the data are due to the wavenumber at which the noise was cut off. As mentioned above, these cutoffs were determined from one of the 50 windows into which the spectrum was divided. Figure~\ref{figure_validation_epsilon} b) presents $\Delta \varepsilon / \varepsilon$ as a function of the normalized cutoff length $l_c/\eta$. As expected, $\Delta \varepsilon / \varepsilon$ scales approximately linear with the temporal resolution of the measurement. } 

\new{Finally, the third approach provides,}

\begin{equation}
    \varepsilon_z = 15 \, \nu {\bigg(\frac{{u}^\prime}{ \lambda_z}\bigg)^2} = 15 \, \nu {\bigg(\frac{C_z \: \pi \:{u}^\prime}{\overline{\Delta Z}}\bigg)^2},
    \label{equation_dissipation_rate_3}
\end{equation}

\noindent \new{using the zero-crossing method and the general definition of $\lambda = (15 \: \nu \: u^{\prime 2}/\varepsilon)^{1/2}$~\cite{mora2020estimating}. Following Mora~\emph{et al.}~\cite{mora2020estimating}, the Taylor length scale $\lambda_z = \overline{\Delta Z} / (C_z \: \pi)$ is derived via the zero-crossing method, providing an independent estimate of $\varepsilon$. Note that $C_z$ is set to 1.}

\new{Figure~\ref{figure_validation_epsilon} confirms that $\varepsilon_d$ provides a valid estimate of the mean dissipation rate showing good agreement with both $\varepsilon_g$ and $\varepsilon_z$. The deviation between $\varepsilon_d$ and $\varepsilon_z$ toward higher dissipation originates from the departure of $C_z$ from unity, which is caused by a stronger departure from Gaussianity at high dissipation. Throughout this paper, we use $\varepsilon = \varepsilon_d$.}

\begin{figure}[htbp]
    \centering
    \begin{subfigure}[t]{0.49\textwidth}
        \centering        \includegraphics[width=\linewidth]{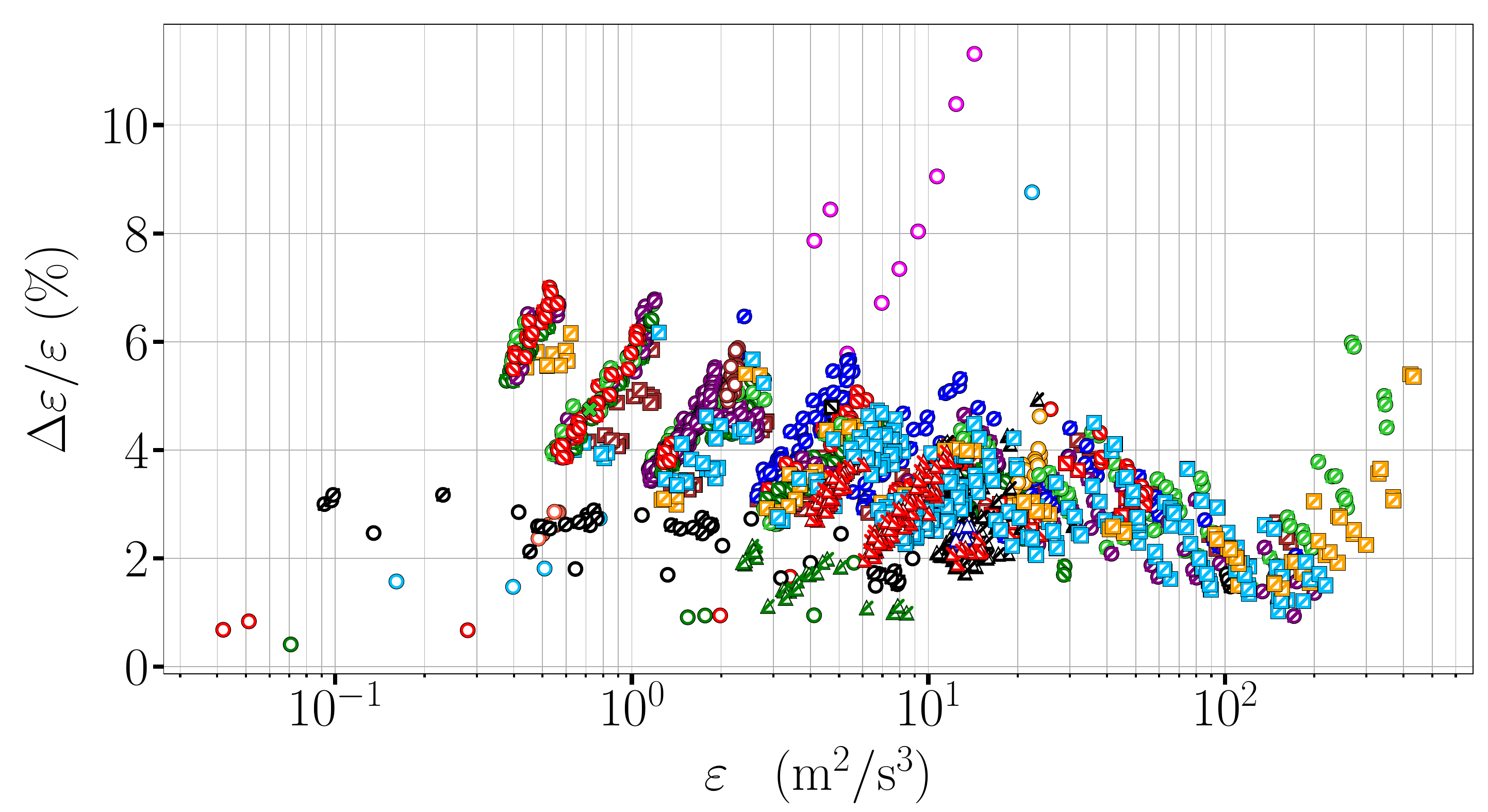}
        \caption{}
    \end{subfigure}
    \hfill
    \begin{subfigure}[t]{0.49\textwidth}
        \centering        \includegraphics[width=\linewidth]{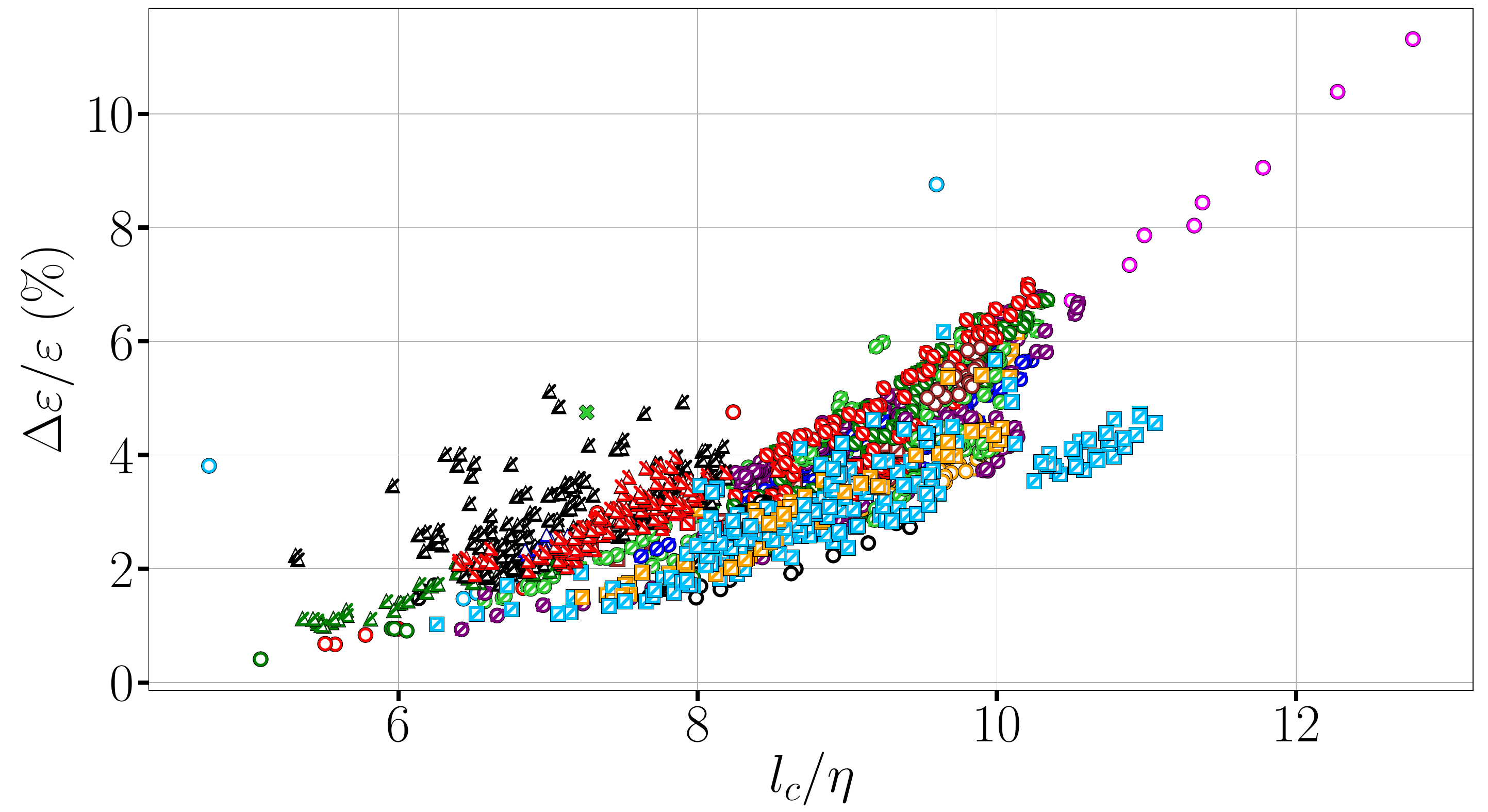}
        \caption{}
    \end{subfigure}
    \begin{subfigure}[t]{0.49\textwidth}
        \centering        \includegraphics[width=\linewidth]{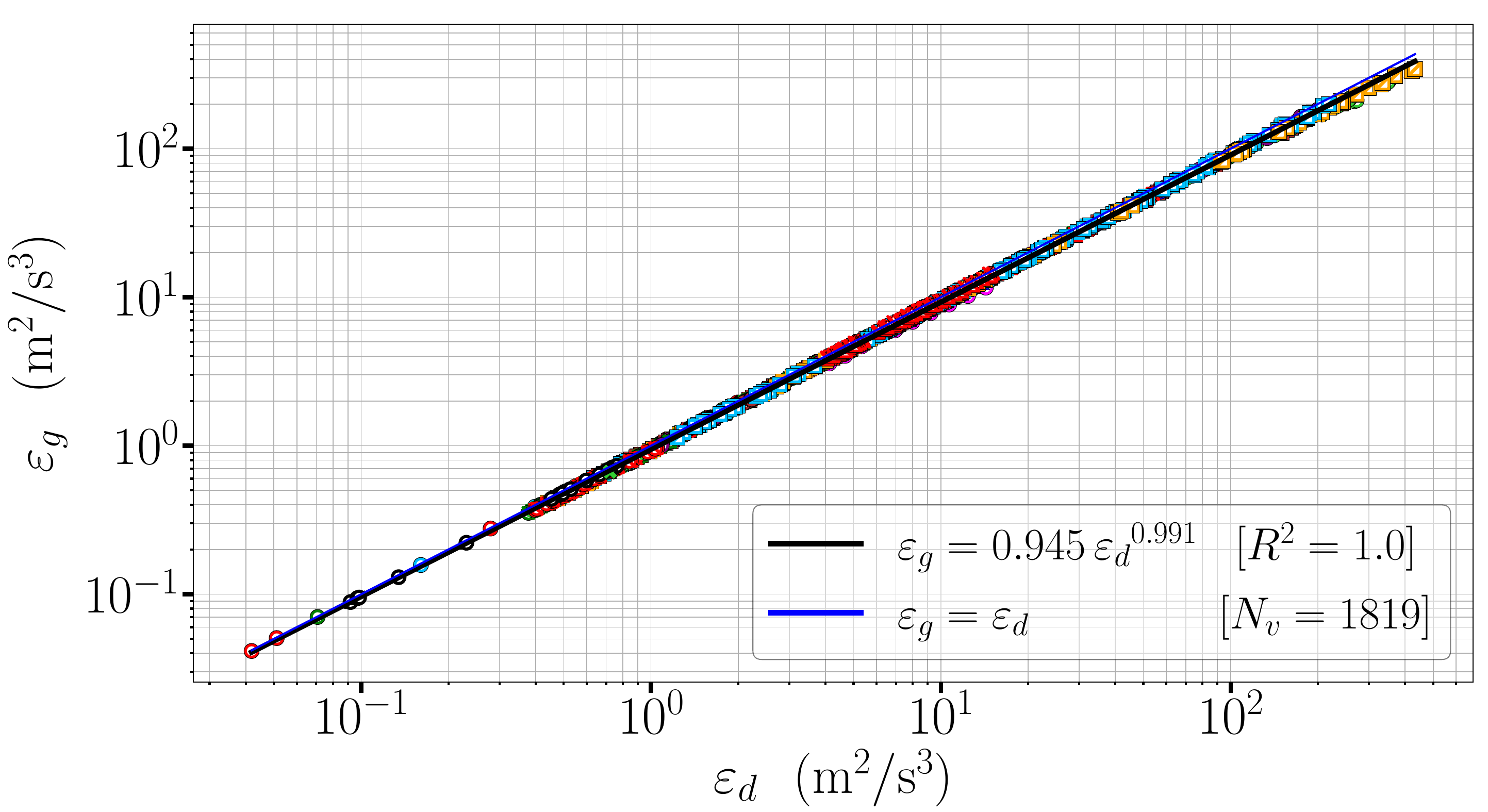}
        \caption{}
    \end{subfigure}
    \hfill
    \begin{subfigure}[t]{0.49\textwidth}
        \centering        \includegraphics[width=\linewidth]{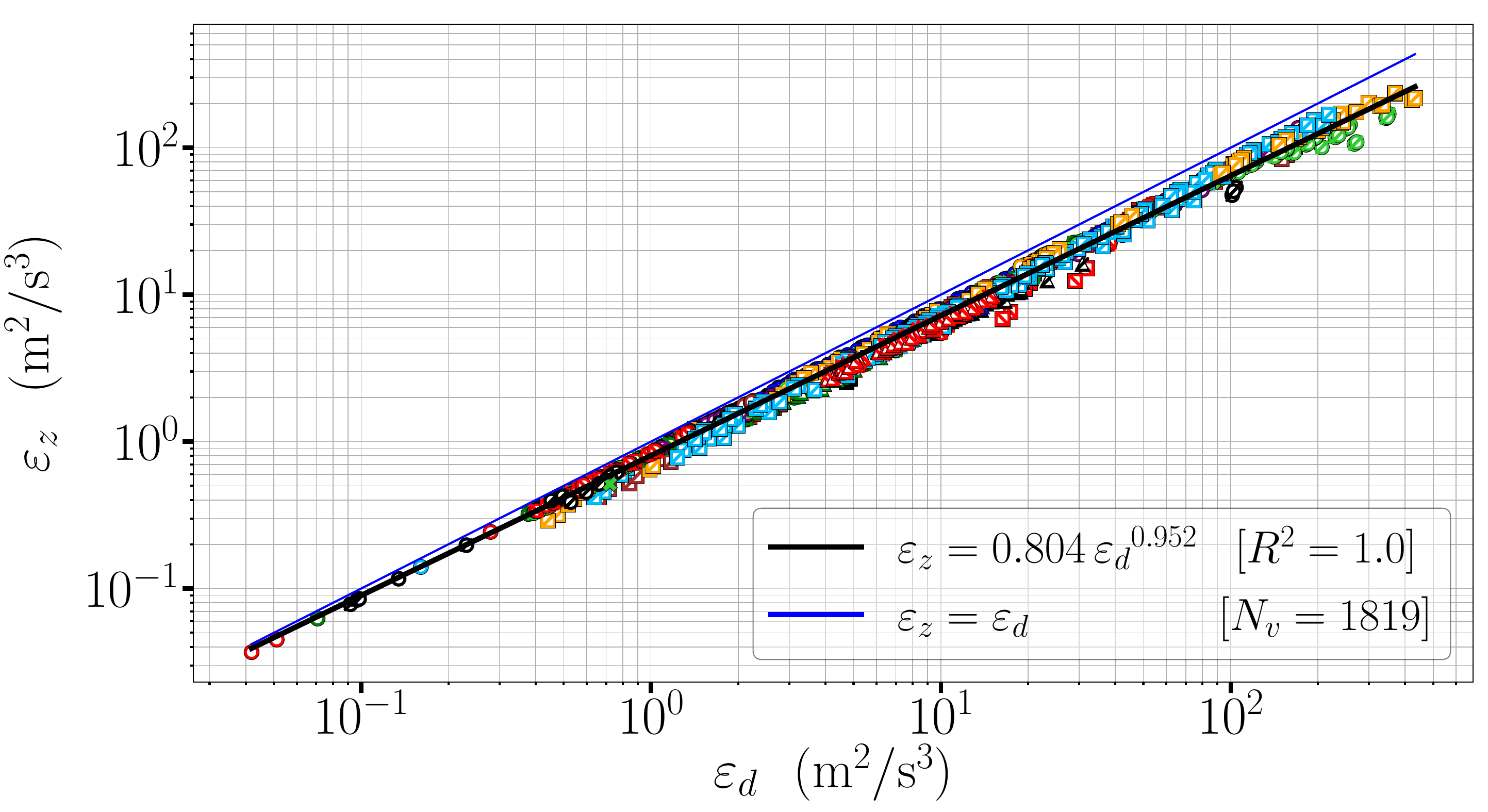}
        \caption{}
    \end{subfigure}
    \caption{\new{a) $\Delta \varepsilon / \varepsilon$ as a function of $\varepsilon$ with $\varepsilon = \varepsilon_d$. b) $\Delta \varepsilon / \varepsilon$ as a function of the normalized cutoff length $l_c/\eta$. c) $\varepsilon_g$ as a function of $\varepsilon_d$ for all VTS used. The blue line indicates $\varepsilon_g = \varepsilon_d$. d) $\varepsilon_z$ as a function of $\varepsilon_d$ for all VTS used. The blue line indicates $\varepsilon_z = \varepsilon_d$. In general the black lines correspond to a least-squares fit with $R^2$ being the coefficient of determination. $N_v$ indicates the number of VTS shown in the plot. The symbols and corresponding configurations are shown and explained in table~\ref{tab:PhD measurements in LEGI 2023} in the appendix~\cite{SM}.}}
    \label{figure_validation_epsilon}
\end{figure}

\new{The uncertainties in $\varepsilon$ were determined by taking the contribution of $\varepsilon$ that results from extrapolating the dissipation spectrum. Regarding the uncertainties in $C_\varepsilon$, we see that $\varepsilon$ and $L$ as the dominant contributors to the uncertainties. Since these two uncertainties are independent, the uncertainty of $C_\varepsilon$ is estimated using Gaussian error propagation.}

\FloatBarrier

\subsection{\new{Intermittency Parameter}}
\label{Intermittency parameter}

\new{Next, the intermittency parameter $\mu$ is considered, which is introduced in eq.~(\ref{equation_lambda_two}). In the present work, $\mu$ is estimated and checked calculating the following two quantities:}

\begin{enumerate}
\item Slope of the scale-dependent shape parameter in the inertial range \color{blue} $\rightarrow  \mu_{\Lambda^2}$ \color{black}
\item Zero-crossing method \color{blue} $\rightarrow \mu_z$ \color{black}
\end{enumerate}

\FloatBarrier

\noindent \new{For the estimation of $\mu_{\Lambda^2}$ we use, as stated above, an estimation method that directly analyzes the trend of the normalized increment PDFs ${p(u_r(x)/\sigma_{r})}$ across the scales $r$. The procedure is described in detail below.}

\new{The scales $r$ at which eq.~(\ref{equation_lambda_one}) is determined are chosen according to the points of $E(k)$, with $r = 2 \pi / k$. $\mu$ is then determined by fitting eq.~(\ref{equation_lambda_two}). Figure~\ref{figure_eight_examples} displays the smoothed $\Lambda^2(r)$ as a function of the scale $r$ for the eight example VTS listed in table~\ref{table_eight_examples}. Note that the shape parameter curves in figure~\ref{figure_eight_examples} are shown with an unnormalized x-axis. The black lines indicate the fits of $\Lambda^2(r)$ within the inertial range.}

{The fitting follows the same procedure as for $\gamma$, with some key differences. $\Lambda^2(r)$ is analyzed in log/lin representation, the derivative of $\Lambda^2(r)$ in log/lin representation is smoothed with a {window size of 5 bins}, and the threshold of the maximum box height is here not fixed but individually adjusted. This adjustment is necessary because the limits of the visible plateau in $\mathrm{d} \: \Lambda^2(r)/ \mathrm{d} \: \mathrm{log}(r)$ occur with a different maximum box height which depends approx. on the value of $\Lambda^2(\lambda)$. Therefore, it was reasonable to define the individual threshold of the maximum box height as a fraction of the value of $\Lambda^2(\lambda)$. The used fraction was 0.15 (the test is provided below).}

{Finally, in contrast to the determination of $\gamma$, no lower limit was imposed for the possible values of $\mu$ (note that through the lack of a comprehensive theory regarding $\mu$, there are no lower or upper limits known). As a result, we also obtain the extent of the inertial range $\mathrm{log}(r_{s})-\mathrm{log}(r_{l})$, the slope of the derivative of $\Lambda^2(r)$ {in log/lin representation computed} by a linear fit in log/lin representation, and the corresponding $R^2$ values of the aforementioned fit and the original fit to compute $\mu$. Only VTS with $R^2 > 0.99$ for the original fit were retained, as stated in the selection criteria. Similar to the determination of $\gamma$, the slope of the derivative of $\Lambda^2(r)$ in log/lin representation has on average a slightly positive value for the data used. The value of the slope shows no clear trend with respect to either $\mu$ or $Re_\lambda$.} 

\new{All main check figures were individually scanned and only VTS with valid $\Lambda^2(r)$ fits were retained. Figure~\ref{figure_validation_mu} a) presents all normalized shape parameter curves superimposed for all used VTS. Normalization was performed according to eq. (\ref{equation_lambda_two}), affecting only the x-axis. The resulting value of $\mu = \mu_{\Lambda^2}$ is considered the true value of $\mu$ and is used throughout this paper.} 

\begin{figure}[htbp]
    \centering
    \begin{subfigure}[t]{0.49\textwidth}
        \centering        \includegraphics[width=\linewidth]{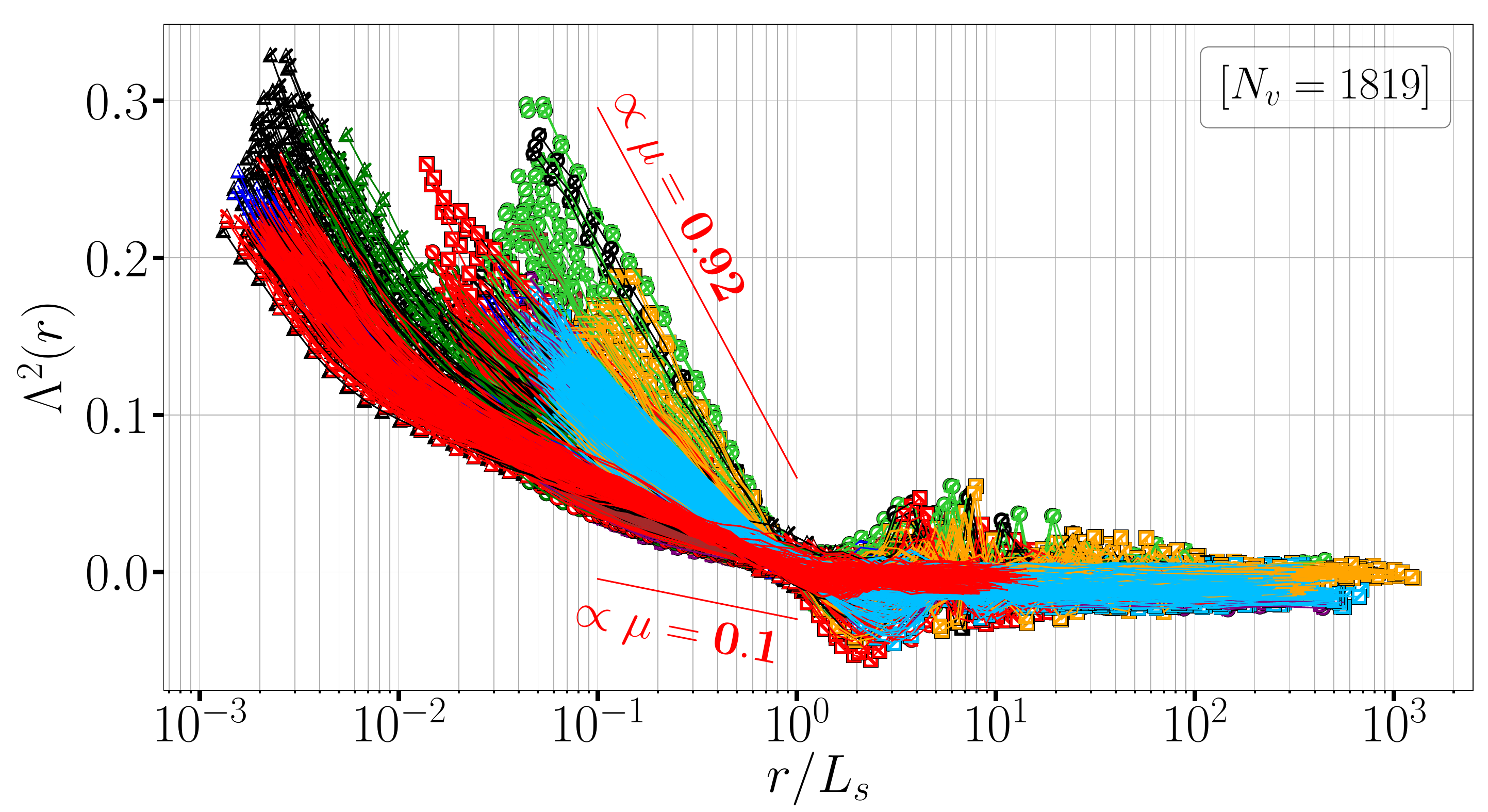}
        \caption{}
    \end{subfigure}
    \hfill
    \begin{subfigure}[t]{0.49\textwidth}
        \centering        \includegraphics[width=\linewidth]{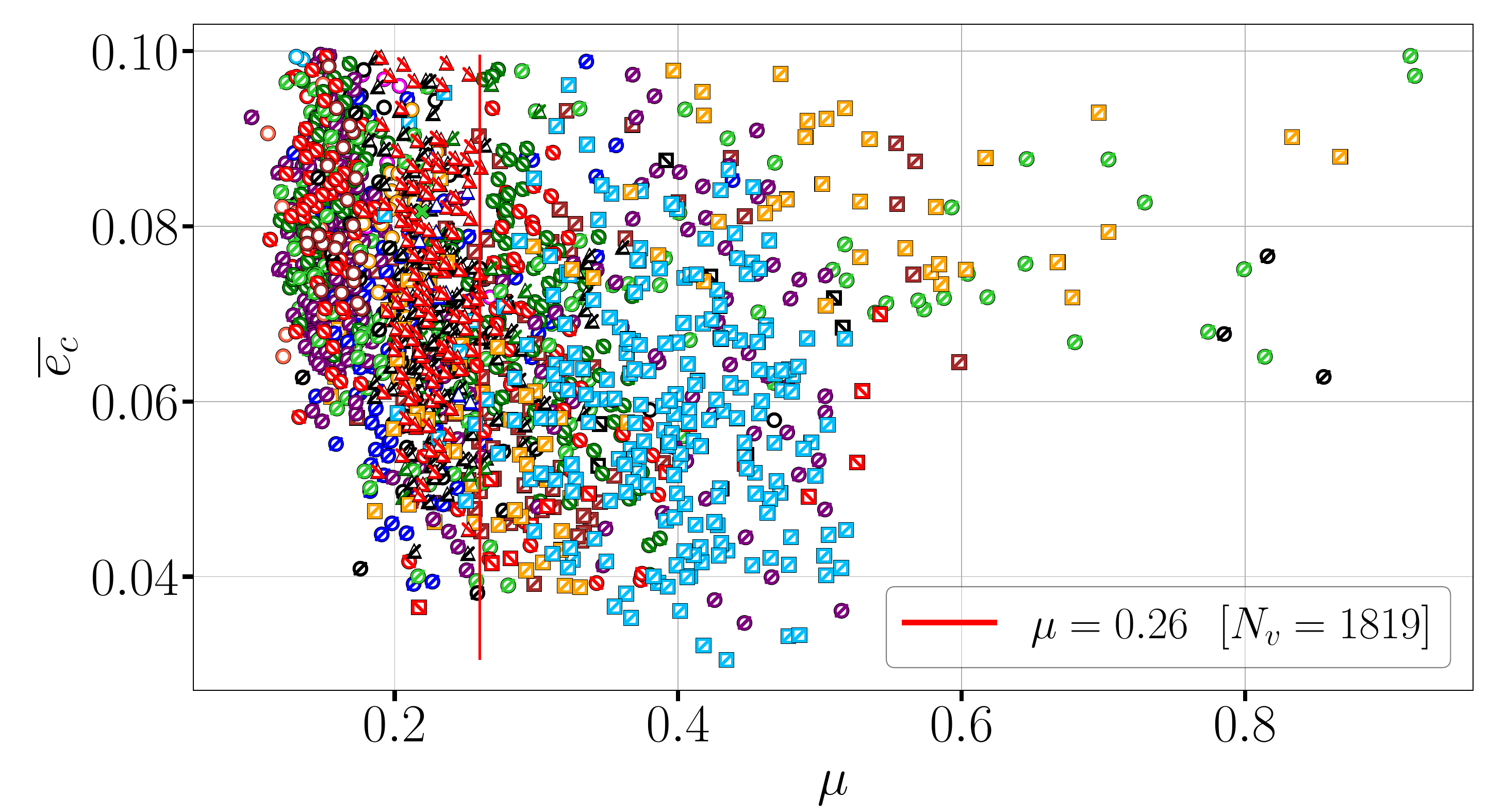}
        \caption{}
    \end{subfigure}
    \begin{subfigure}[t]{0.49\textwidth}
        \centering        \includegraphics[width=\linewidth]{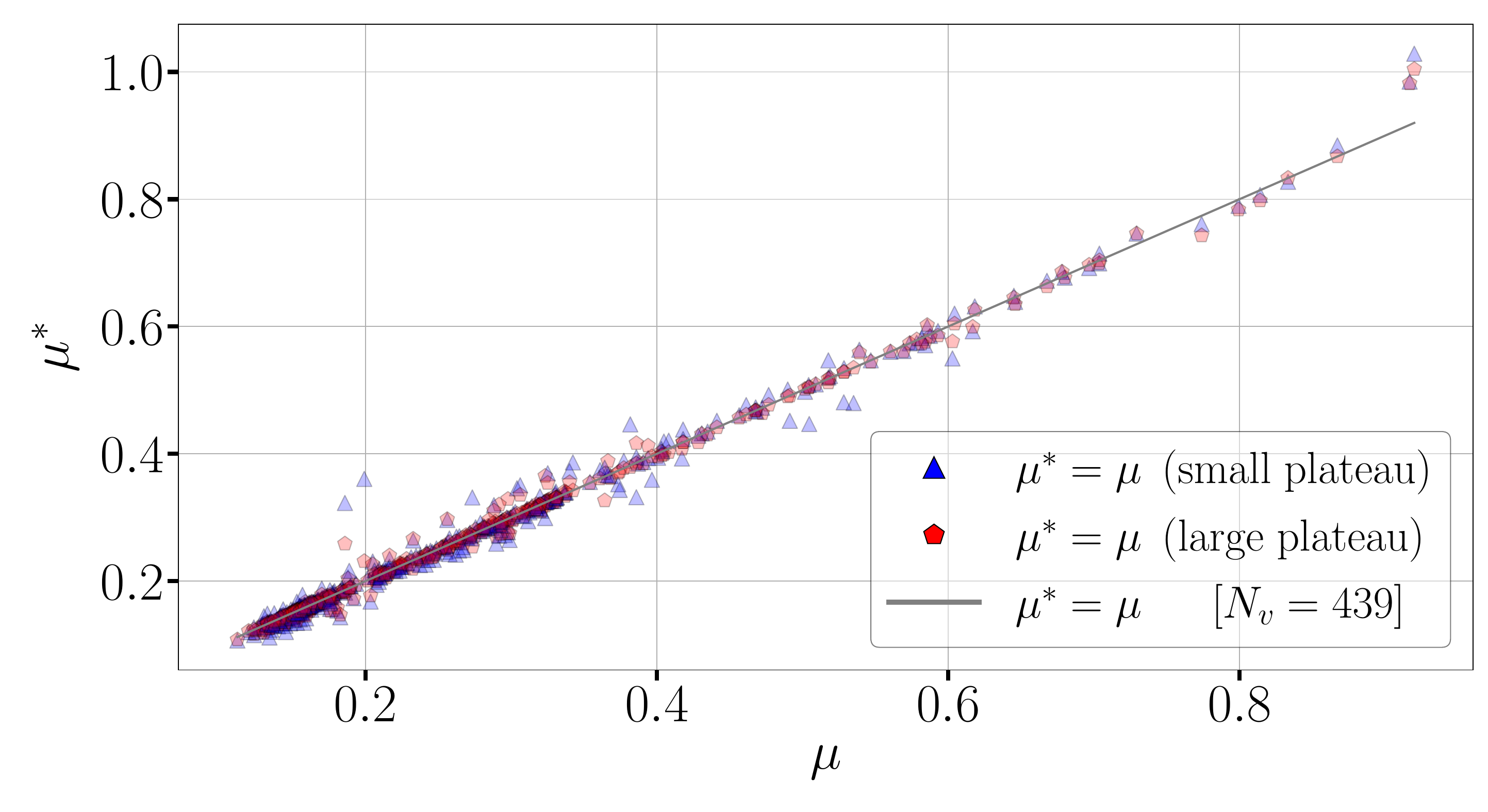}
        \caption{}
    \end{subfigure}
    \hfill
    \begin{subfigure}[t]{0.49\textwidth}
        \centering        \includegraphics[width=\linewidth]{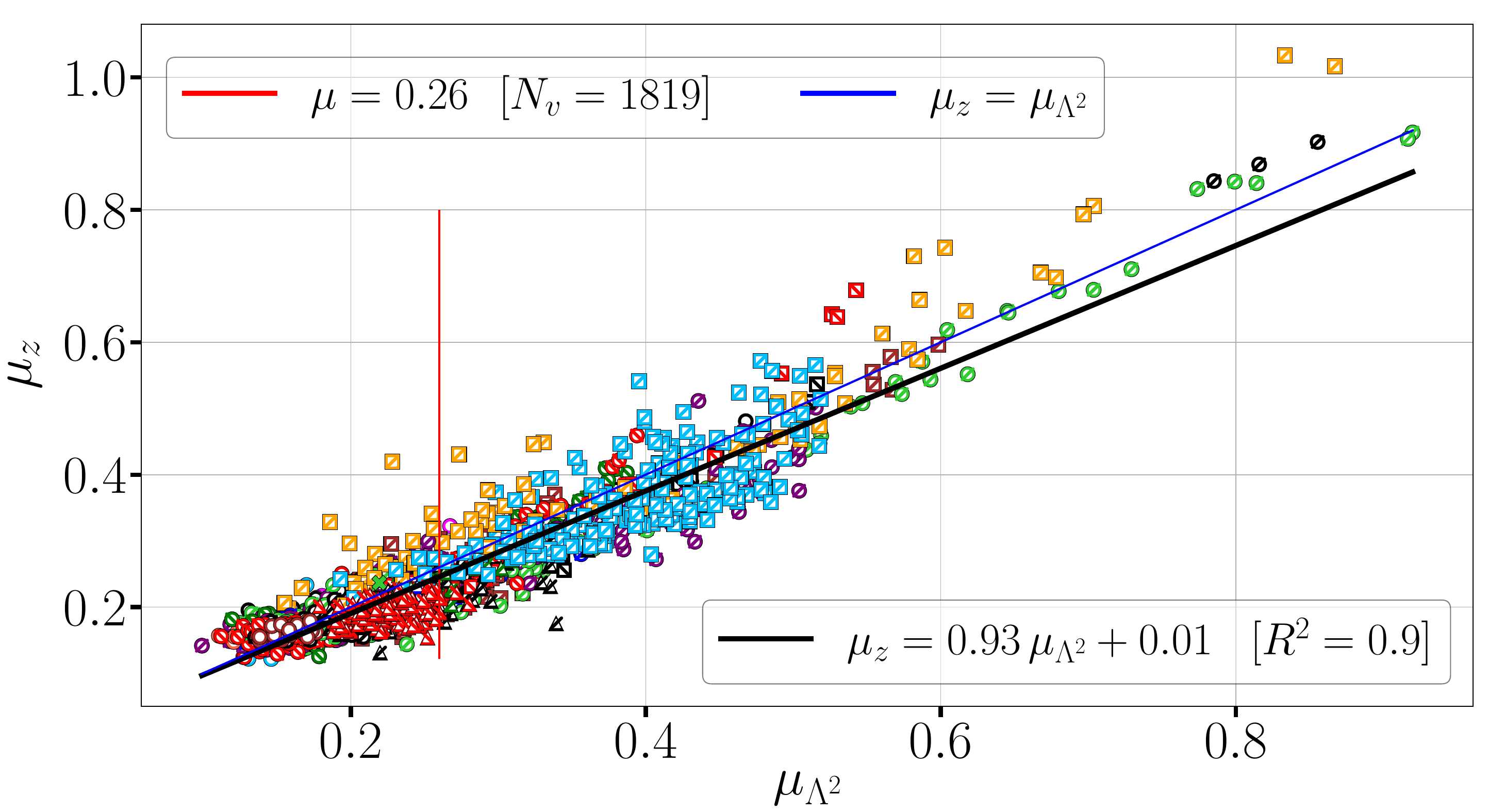}
        \caption{}
    \end{subfigure}
    \caption{\new{a) $\Lambda^2(r)$ as a function of $r / L_s$ for all VTS used. The red lines indicate the highest and the lowest slope present within the data (they are shifted vertically for better visibility). b) $\overline{e_c}$ as a function of $\mu$ for all VTS used. c) $\mu^*$ as a function of $\mu$ for all VTS used from three representative cases (C1, C4 and D13). While the gray solid line represents identity, the blue markers indicate the values of $\mu$ obtained by using a smaller threshold for the derivative plateau and the red markers indicate the values of $\mu$ obtained by using a larger threshold for the derivative plateau. d) $\mu_z$ as a function of $\mu_{\Lambda^2}$ for all VTS used. The blue line indicates $\mu_z = \mu_{\Lambda^2}$. In general the red lines indicate both the commonly accepted values for $\mu$ for HIT~\cite{arneodo1996structure}. The black lines correspond to a least-squares fit with $R^2$ being the coefficient of determination. $N_v$ indicates the number of VTS shown in the plot. The symbols and corresponding configurations are shown and explained in table~\ref{tab:PhD measurements in LEGI 2023} in the appendix~\cite{SM}.}}
    \label{figure_validation_mu}
\end{figure}

\new{In the following, further details, intrinsic consistency checks, and an alternative approach for determining $\mu$ are presented. With regard to figure~\ref{figure_eight_examples}, it is striking that similar to the spectra and the autocorrelation functions, the VTS in f) and g) display again clear periodic structures in the shape parameter curve, visible as peaks in $\Lambda^2(r)$. For both VTS, the highest peak of $\Lambda^2(r)$ coincides closely with the peak in $E(k)$ regarding the scale, respectively. This behavior is consistent with the expected response of a sinusoidal structure if both x-axis of $E(k)$ and $\Lambda^2(r)$ are consistent respectively, since a sinusoidal structure yields the largest increments for spatial lags equal to half the wavelength. This consistency supports the validity of the transformation $k = 2\pi/r$ between the angular wavenumber $k$ and the spatial lags $r$. Within this rationale, a systematic shift was observed between the inertial range of $E(k)$ and $\Lambda^2(r)$, with the latter always located toward smaller scales. The ratio of their extents and their overlap was quantified but no relation neither between them nor between other quantities was found. In contrast to $E(k)$ and $R_{uu}$, the VTS h) deviates from the others - its shape parameter at large scales remains far from 0, explaining its exclusion.} 

\new{The first intrinsic check concerns the validity of the $\Lambda^2$ values determined using eq.~(\ref{equation_lambda_one}). In contrast to $E(k)$, determining $\Lambda^2(r)$ is not straightforward, since it must be ensured that $\Lambda^2(r)$ really captures the shape of the PDF of the normalized spatial velocity increments. Therefore, the normalized increment PDFs for both $r=L$ and $r=\lambda$ are computed exemplarily with a bin size of 100 {(which results in 5-10 bins per $\sigma_r$)}. Next, for these two PDFs, the corresponding Castaing-curves~\cite{castaing1990velocity},}

\begin{equation}
{\hat p(u_r(x)/\sigma_{r})} = 
\frac{1}{2 \pi \Lambda(r)} \:
\int_{0}^{\infty} \: 
\frac{1}{{\sigma}^2} \:
\exp \left[- \frac{{(u_r(x)/\sigma_{r})}^2}{{2\sigma}^2} \right] \:
\exp \left[ -\frac{\mathrm{\ln}^2 \left({\sigma/\sigma_0}\right)}{2\Lambda^{2}(r)} \right] \: \mathrm{d}\sigma, 
\label{equation_castaing_curve_one}
\end{equation}

\noindent \new{where the {parameter $\sigma_0$} is implicitly (by normalization) defined as,}

\begin{equation}
\sigma_{0}^2 = \overline{(u_r(x)/\sigma_{r})^2} \:
\exp \left[- 2\Lambda^2(r) \right],
\label{equation_castaing_curve_two}
\end{equation}

\noindent \new{are calculated. Note that eq.~(\ref{equation_castaing_curve_one}) does not include skewness corrections. Moreover, note that $\sigma_{r}$ is directly related to $E(k)$. Eq. (\ref{equation_castaing_curve_one}) can be interpreted within the framework of superstatistics~\cite{beck2004superstatistics}, where the standard deviation $\sigma$ of a Gaussian distribution is modulated based by a superimposed log-normal distribution of $\sigma$. The log-normal distribution is characterized by the parameter $\sigma_0$ and its variance $\Lambda^2$. As $\Lambda^2$ approaches $0$, the distribution (\ref{equation_castaing_curve_one}) converges to a normal distribution~\cite{castaing1990velocity,morales2012characterization}.} 

\new{In theory, the Castaing curve ${\hat p(u_r(x)/\sigma_{r})}$ should coincide with the normalized PDF ${p(u_r(x)/\sigma_{r})}$ from which the shape parameter was obtained~\cite{castaing1990velocity}. This comparison serves as a validation of $\Lambda^2$. Figure~\ref{figure_eight_examples_PDF} presents the PDFs $p(u_r(x)/\sigma_{r})$ for $r=\lambda$ and $r=L$, along with the corresponding Castaing distributions $\hat p(u_r(x)/\sigma_{r})$ are superimposed for $r=\lambda$ and $r=L$. For the VTS a)-g), the empirical PDFs (red) show the expected behavior and match the Castaing PDFs well. In contrast, figure~\ref{figure_eight_examples_PDF} h) illustrates why it was discarded - both empirical PDFs $p(u_r(x)/\sigma_{r})$ display overlapping distributions with distinct variances.}

\begin{figure}[htbp]
    \centering
    \begin{subfigure}[t]{0.49\textwidth}
        \centering        \includegraphics[width=\linewidth]{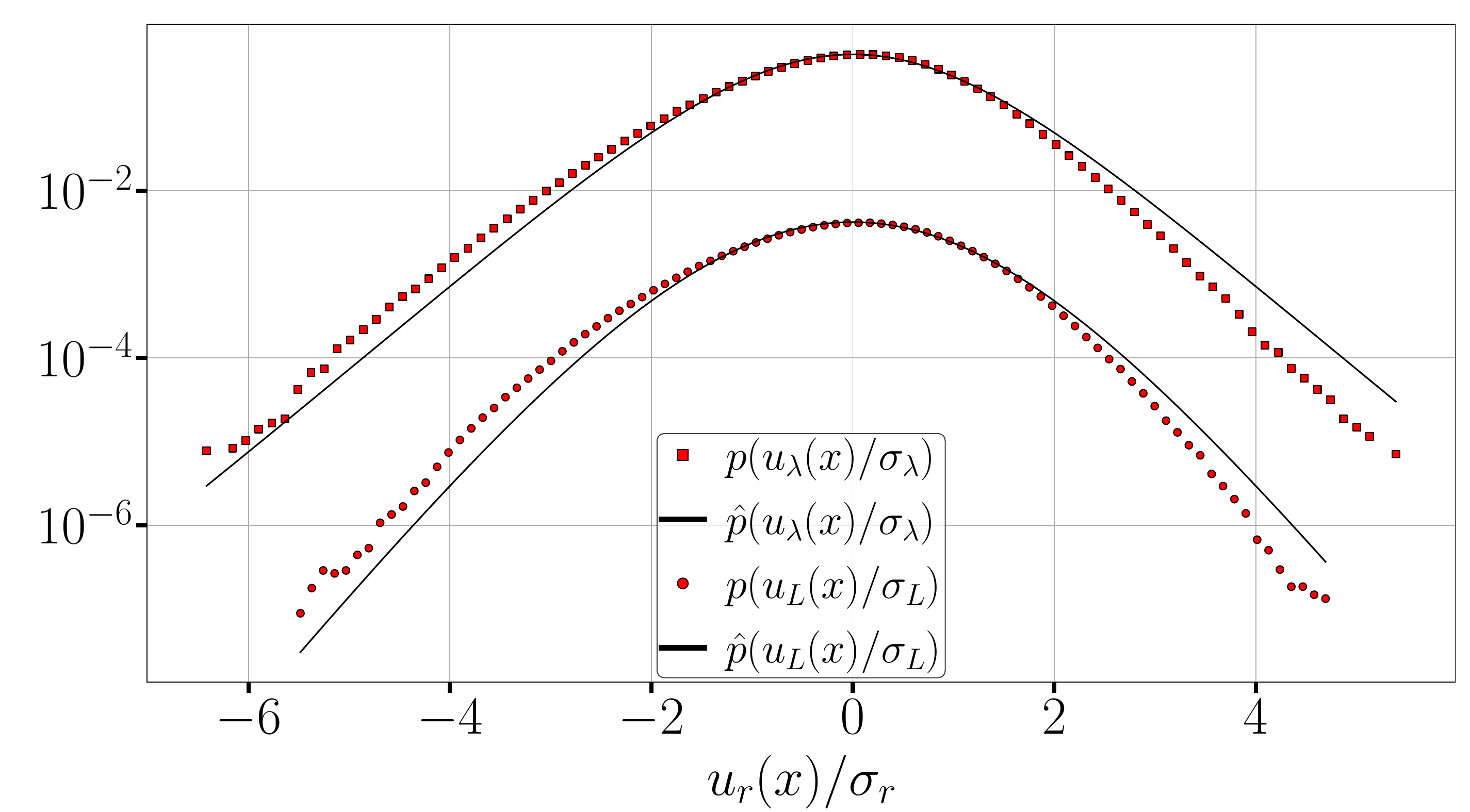}
        \caption{}
    \end{subfigure}
    \hfill
    \begin{subfigure}[t]{0.49\textwidth}
        \centering        \includegraphics[width=\linewidth]{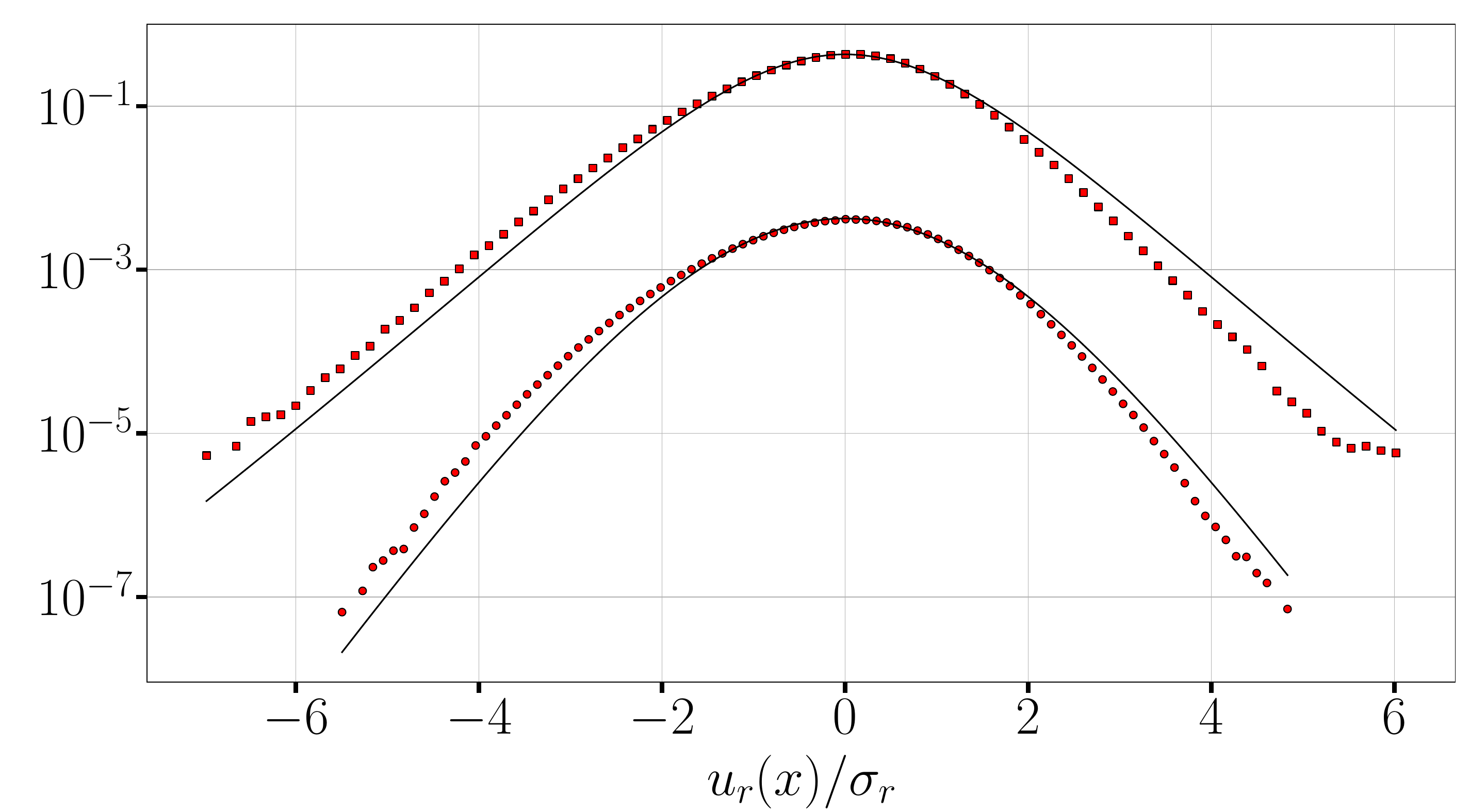}
        \caption{}
    \end{subfigure}
    \begin{subfigure}[t]{0.49\textwidth}
        \centering        \includegraphics[width=\linewidth]{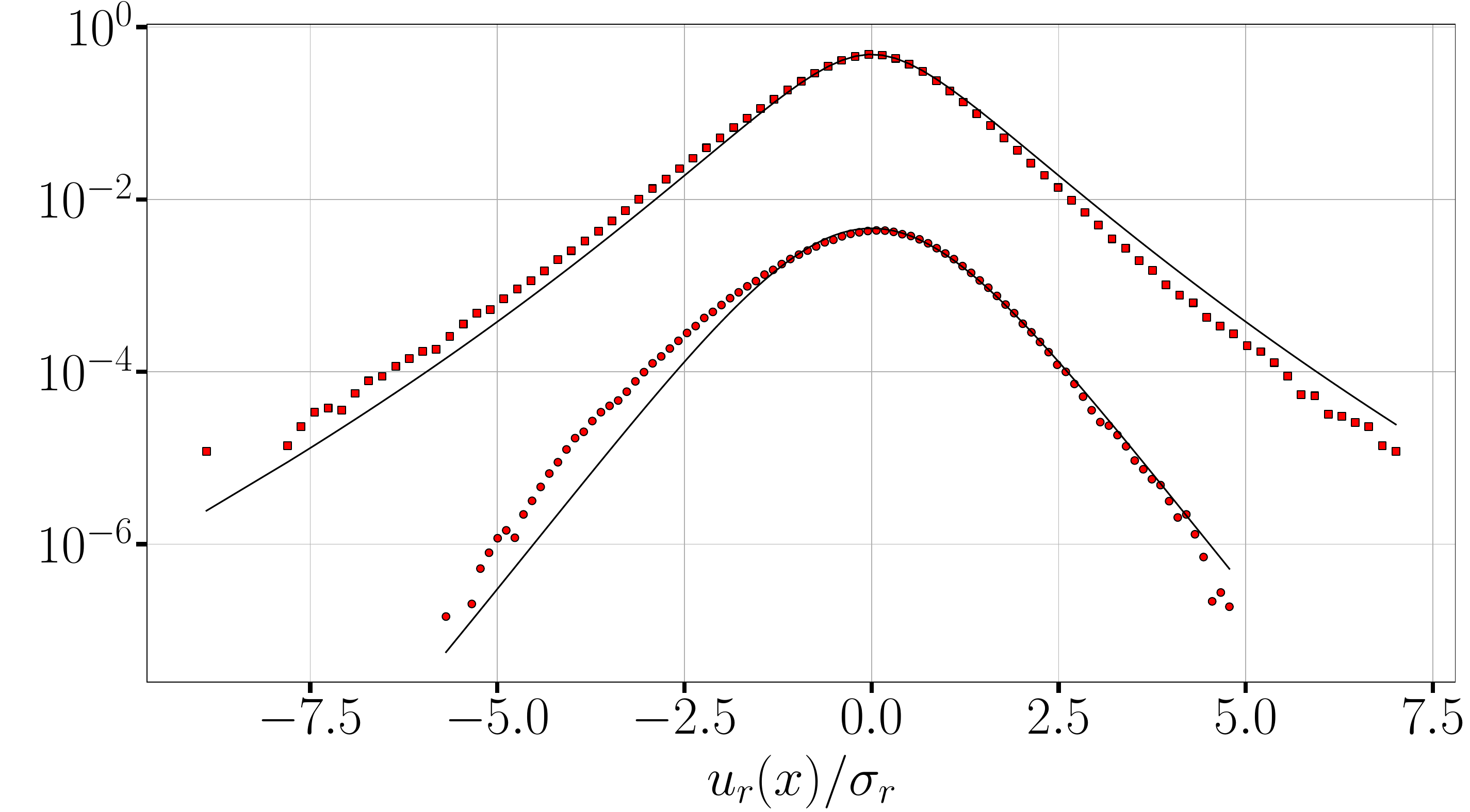}
        \caption{}
    \end{subfigure}
    \hfill
    \begin{subfigure}[t]{0.49\textwidth}
        \centering        \includegraphics[width=\linewidth]{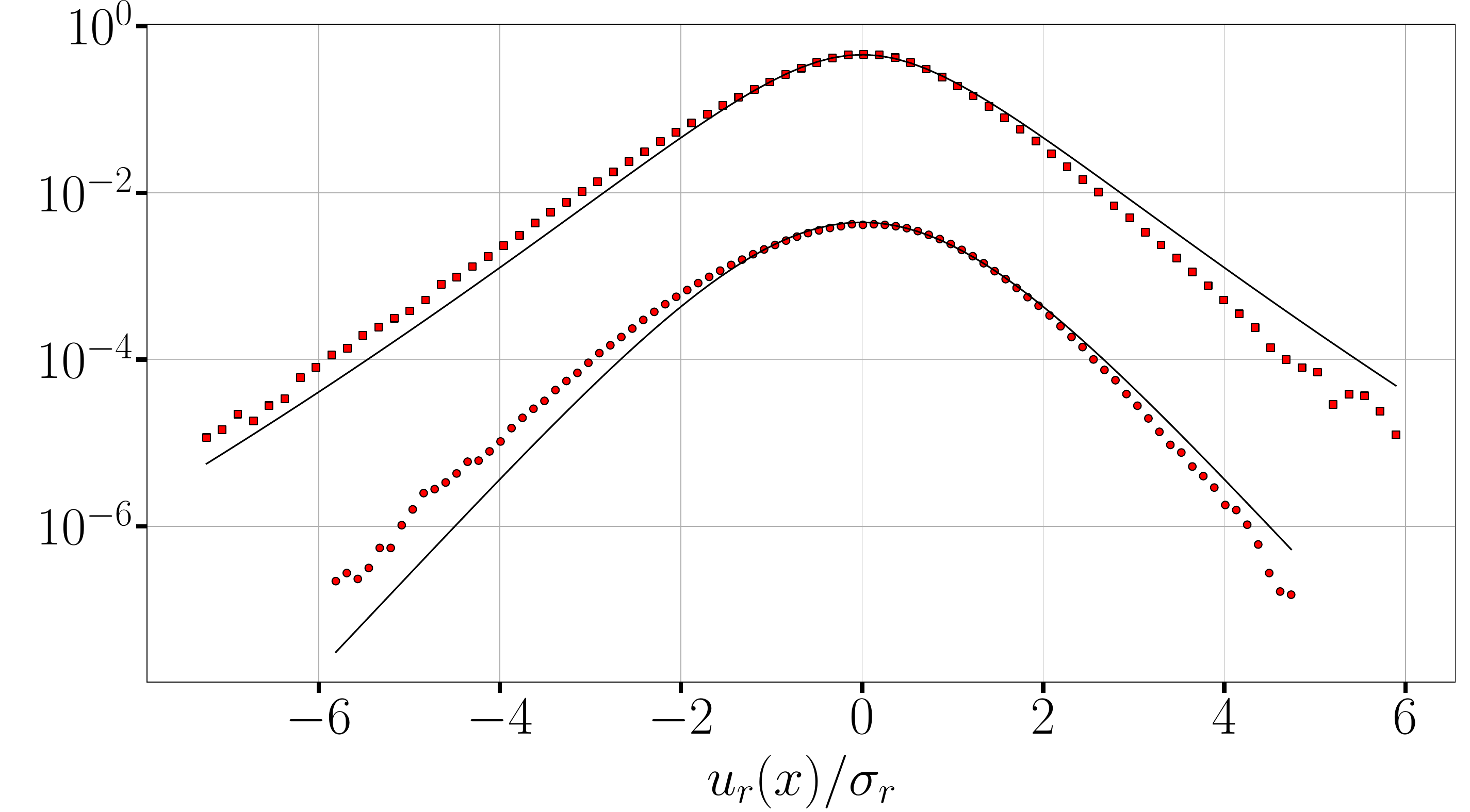}
        \caption{}
    \end{subfigure}
    \begin{subfigure}[t]{0.49\textwidth}
        \centering        \includegraphics[width=\linewidth]{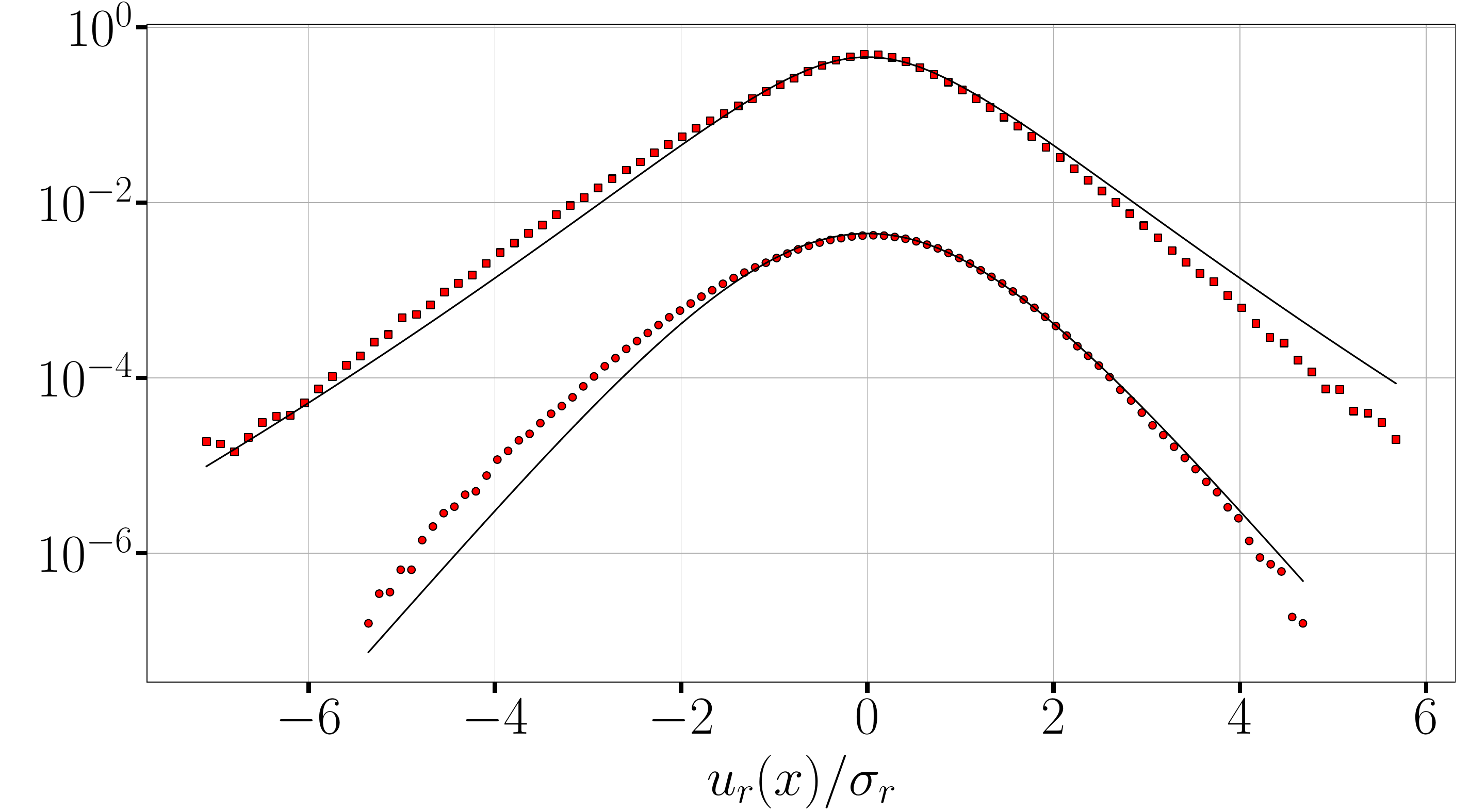}
        \caption{}
    \end{subfigure}
    \hfill
    \begin{subfigure}[t]{0.49\textwidth}
        \centering        \includegraphics[width=\linewidth]{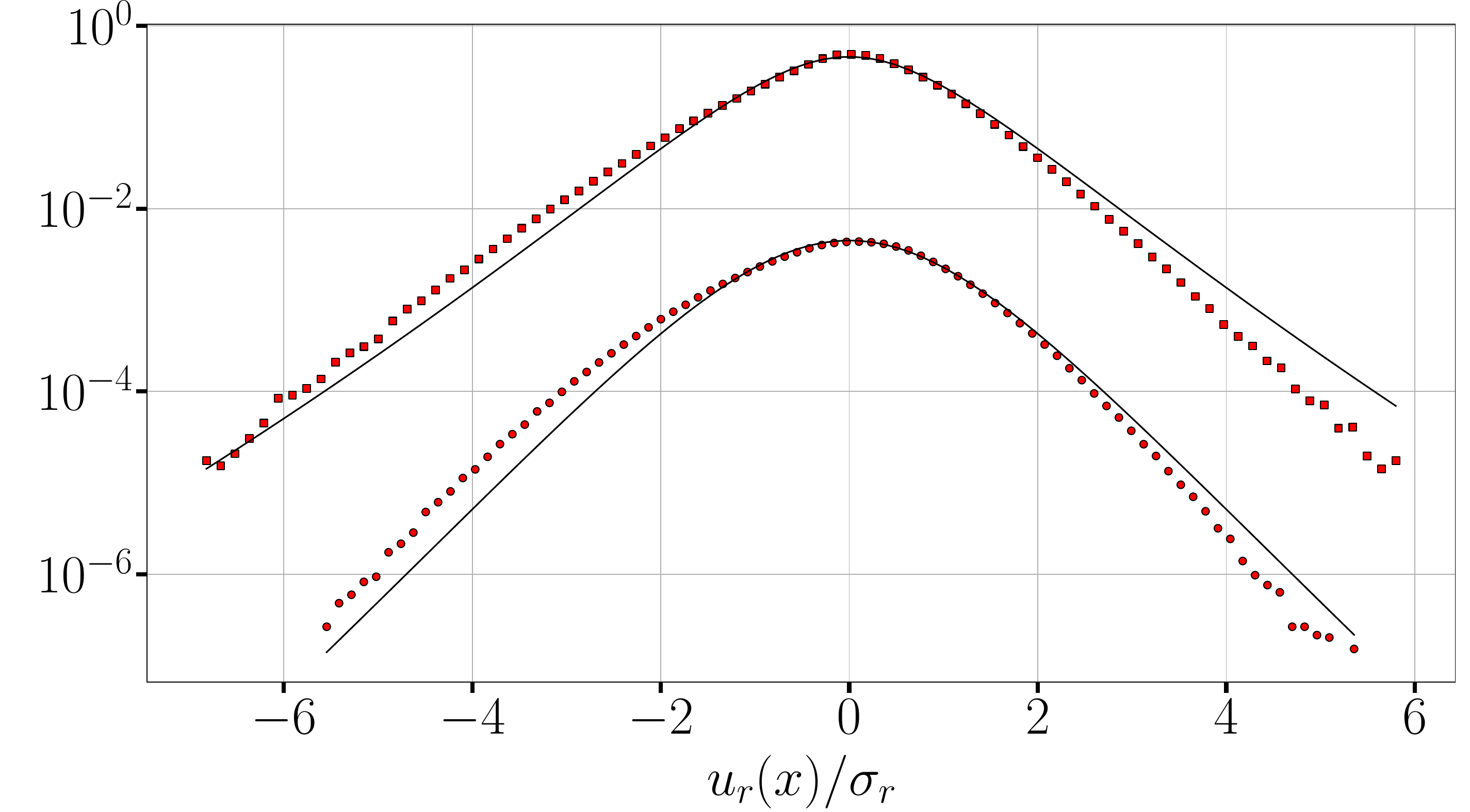}
        \caption{}
    \end{subfigure}
    \begin{subfigure}[t]{0.49\textwidth}
        \centering        \includegraphics[width=\linewidth]{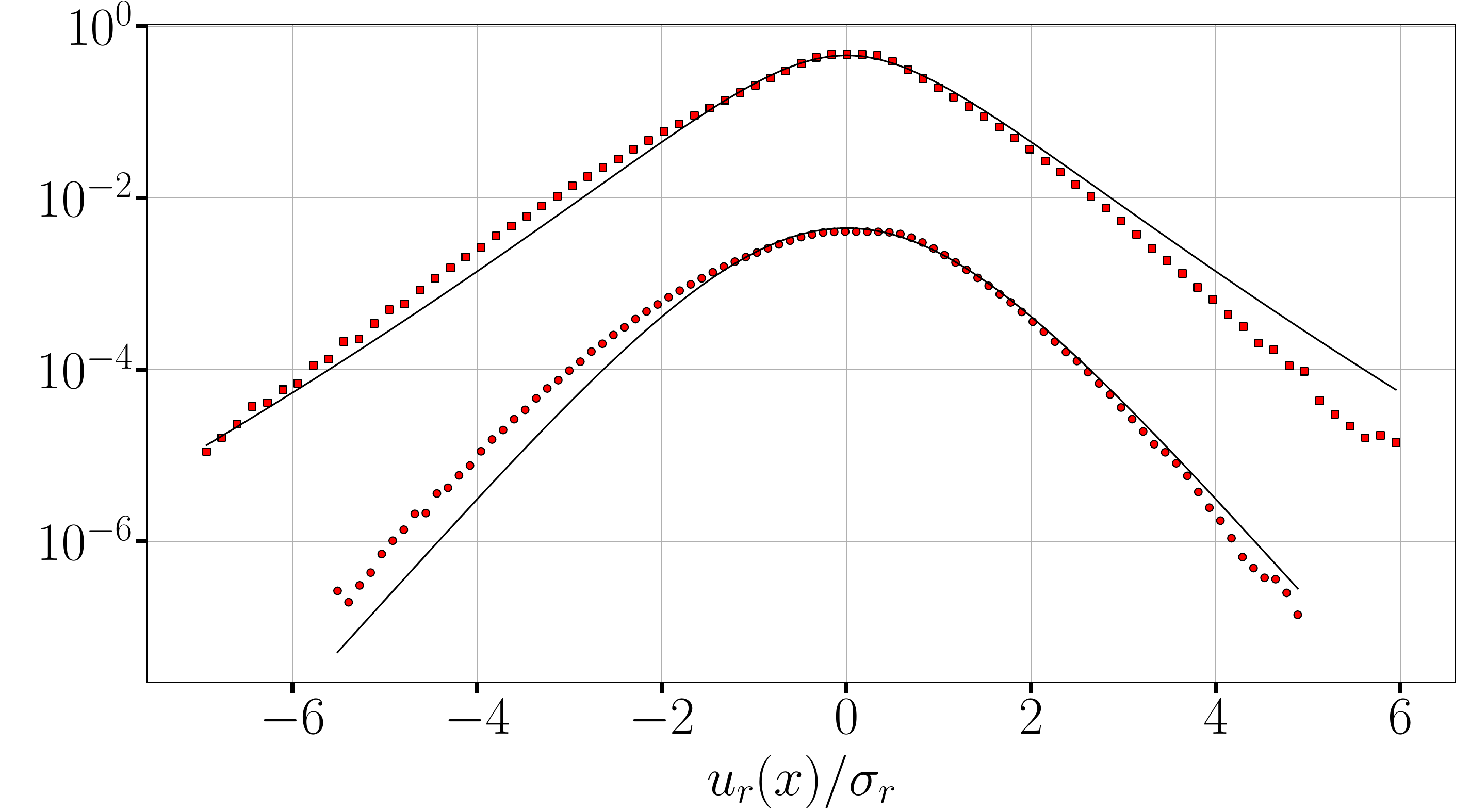}
        \caption{}
    \end{subfigure}
    \hfill
    \begin{subfigure}[t]{0.49\textwidth}
        \centering        \includegraphics[width=\linewidth]{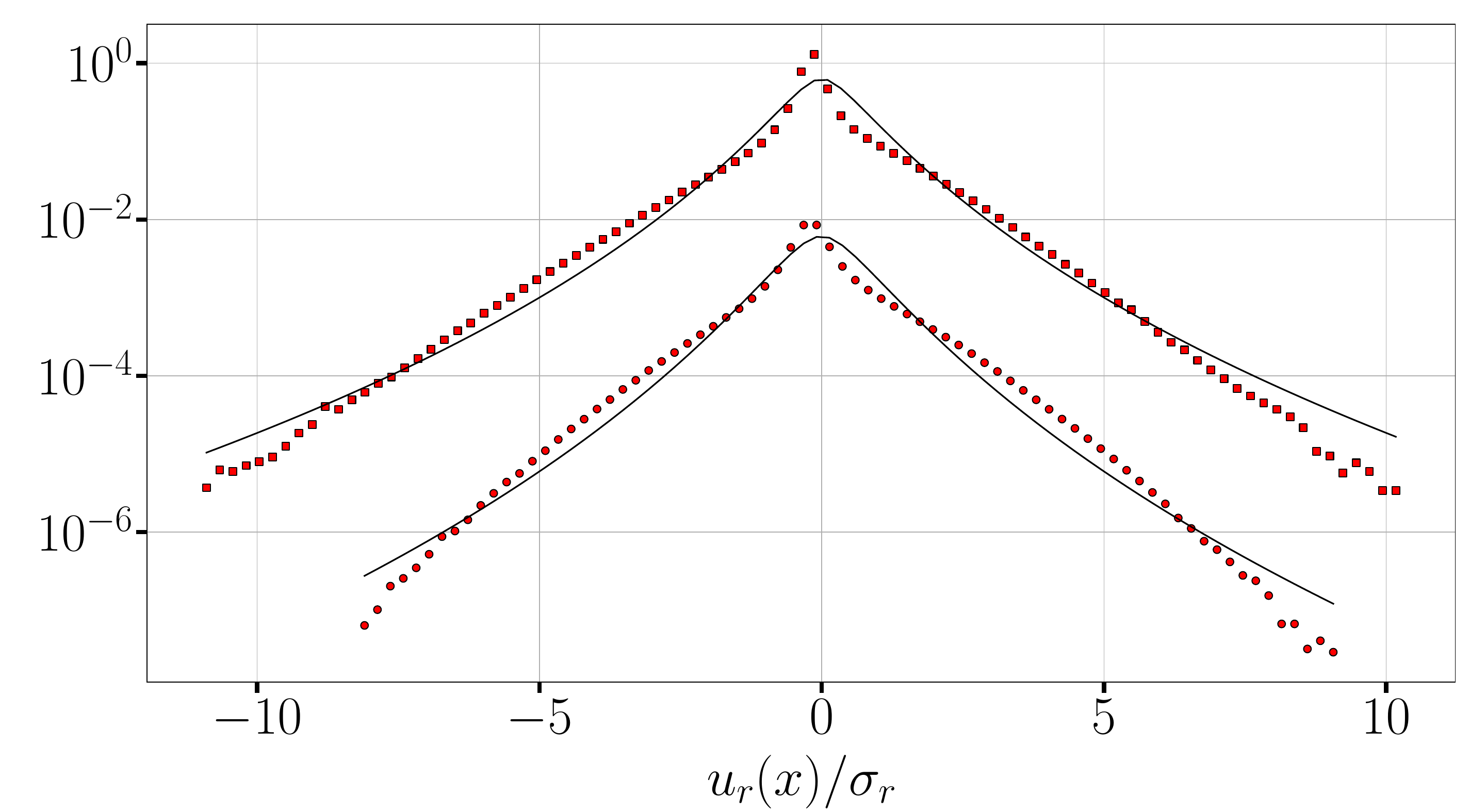}
        \caption{}
    \end{subfigure}
    \caption{\new{The normalized spatial velocity increment PDFs (red) are shown for the same eight VTS as in figures~\ref{figure_eight_examples} and \ref{figure_eight_examples_L} (also in the same order). For each VTS, two PDFs are presented. While the lower PDF represents the normalized spatial velocity increments for the scale $r=L$ (circles as markers), the upper PDF represents the normalized spatial velocity increments for the scale $r=\lambda$ (squares as markers). Note that due to a better visibility, the PDFs are shifted vertically and that only bins are shown that have at least 10 values. The black curves show the corresponding Castaing curve for the two scales, respectively. a), b), c), e), f) and g) display VTS that satisfy the restriction criteria and taken together, span almost the full range of $\mu$-values. d) also fulfills the restriction criteria and exhibits the same value of $\mu$ as c) but with a $Re_\lambda$ more than three times smaller. h) however does not meet the restriction criteria since $\Lambda_0^2$ shows clear signatures of non-Gaussianity at large scales. The data stem from the cases G20, G24, C8, D11, C6, C1, C1, G23, respectively in order of appearance. For a detailed list of the characteristic values of the VTS and details about their cases see table~\ref{table_eight_examples} and table~\ref{tab:PhD measurements in LEGI 2023} in the appendix~\cite{SM}.}}
    \label{figure_eight_examples_PDF}
\end{figure}

\new{To quantify the deviation, the Castaing error $\overline{e_c}$ is computed as the average of ${e_c(L)}$ and ${e_c(\lambda)}$, where the mean squared deviation between the empirical PDF and the modeled PDF from eq. (\ref{equation_castaing_curve_one}) is defined as,}

\begin{equation}
    e_c = \overline{\big(\mathrm{log}({p(u_r(x)/\sigma_{r}))} - \mathrm{log}({\hat p(u_r(x)/\sigma_{r})})\big)^2}.
    \label{equation_castaing_error}
\end{equation}

\noindent \new{{Note that, prior to computing the Castaing error, only bins of the empirical PDF containing at least 10 samples were retained.} Only VTS with $\overline{e_c} < 0.1$ were retained, as stated in the selection criteria. Figure~\ref{figure_validation_mu} b) shows the Castaing error $\overline{e_c}$ as a function of $\mu$ for all used VTS, showing no distinct clustering. Note that a valid scaling of $\Lambda^2(r) \propto \mu \, \mathrm{ln}(L/r)$ closes the theoretical framework by justifying the use of the Castaing curves to describe the empirical increment PDFs \emph{c.f.}~\cite{castaing1990velocity}. By normalizing $p(u_r(x))$ with $\sigma_{r}$, the information of the variance (and $E(k)$) is eliminated and thus, $p(u_r(x)/\sigma_{r})$ only contains the information of the form of $p(u_r(x))$. Hence, a negligible error in $\overline{e_c}$ covers all higher moments of $u_r(x)$.}

\new{A second intrinsic validity check concerns robustness of their previously introduced threshold fraction of $0.15$ for the derivative in lin/log representation. For three representative cases (C1, C4, and D13), $\mu$ was in addition calculated by changing the value of the threshold fraction. In particular, the threshold fraction was set to $0.1$ and $0.2$, resulting in corresponding values of $\mu^*$. Figure~\ref{figure_validation_mu} c) shows the results where the blue markers correspond to $0.1$ and the red markers stand for $0.2$. It confirms, that the estimation of $\mu$ is not strongly biased by the choice of the threshold fraction for the derivative in lin/log representation.}

\new{Furthermore, the value of $\mu$ was validated independently using the zero-crossing method by determining} 

\begin{equation}
    \mu_z = \frac{\big(C_z - 1\big)}{\mathrm{log}(L_e/\lambda_d)} = \frac{\big(\frac{\lambda_z}{\lambda_d} - 1\big)}{\mathrm{log}(L_e/\lambda_d)}. 
    \label{equation_dissipation_rate_mu_z}
\end{equation}

\noindent {The aforementioned departure of $C_z$ from unity is exploited, as this deviation provides a measure of the non-Gaussianity of the smallest (measurable) scales~\cite{ferran2023characterising}. The deviation of $C_z$ from unity can be expressed by the ratio of $\lambda_z$ to $\lambda_d$, since $\lambda_z$ is estimated using eq. (\ref{equation_rice_theorem}) with $C_z = 1$ and we assume that $\lambda_d = C_z \, \lambda_z$ with a variable $C_z \geq 1$. Subtracting 1 from this ratio yields a measure of non-Gaussianity that, similar to $\mu$, vanishes for a Gaussian distribution. However, the non-Gaussianity of the smallest (measurable) scales does not directly represent intermittency in the sense of Kolmogorov~\cite{kolmogorov1962refinement} and Castaing~\cite{castaing1990velocity}. Although intermittency involves more than quantifying non-Gaussianity at a single (small) scale, plotting $(\lambda_z / \lambda_d) - 1$ versus the shape parameter at the Taylor length scale $\Lambda^2(\lambda_d)$ reveals a clear proportionality {(not shown here)}, confirming the reliability of this measure toward non-Gaussianity itself. Intermittency in the sense of Kolmogorov and Castaing is defined as the evolution of non-Gaussianity across scales within the inertial range. Therefore, $\mu_z$, a reliable parameter characterizing intermittency, is constructed by normalizing $(\lambda_z / \lambda_d) - 1$ by the extent of the inertial range, $\mathrm{log}(L_e/\lambda_d)$. Figure~\ref{figure_validation_mu} d) shows that $\mu_{z}$ is in good agreement with $\mu_{\Lambda^2}$, confirming that $\mu_{\Lambda^2}$ is a valid estimate of the intermittency parameter. Note additionally, that although this work does not go beyond two-point quantities, recent work by Schmitt~\emph{et al.}~\cite{schmitt2026small} demonstrates that intermittency estimated using multi-point statistics also aligns with the values of $\mu_{\Lambda^2}$.}

\new{The uncertainty in $\mu$, \textit{e.g.} used in figure~\ref{figure_law_1} d) is calculated according to the uncertainties shown in figure~\ref{figure_validation_mu} c). Accordingly, the largest deviation of $\mu^*$ based on the corresponding value of $\mu$ is taken as the error bound of $\mu$.}

\FloatBarrier

\subsection{\new{Large-Scale Non-Gaussianity}}

\new{Finally, we consider $\Lambda_0^2$, a parameter of the non-Gaussianity of the large scales, which is introduced in eq.~(\ref{equation_lambda_two}). In the present work, $\Lambda_0^2$ is estimated and checked calculating the following two quantities:}

\begin{enumerate}
\item Averaging the values of the scale-dependent shape parameter at large scales \color{blue} $\rightarrow \Lambda_0^2$ \color{black}
\item Flatness of the VTS \color{blue} $\rightarrow F_u$ \color{black}
\end{enumerate}

\FloatBarrier

\noindent \new{We define which we define $\Lambda_0^2$ empirically as the average of the first 5 $\Lambda^2$ values starting at the largest scales (compare figure~\ref{figure_eight_examples}). Only VTS with $\Lambda_0^2 < 0.005$ were retained, as stated in the selection criteria. To check the consistency of $\Lambda_0^2$, $F_u$ is computed as the flatness of the VTS, which should be directly related. Figure~\ref{figure_validation_shape_zero} a) shows $\Lambda_0^2$ as a function of $F_u$ for all measured VTS where $\Lambda_0^2$ could be computed. The black rectangle includes the retained data. Figure~\ref{figure_validation_shape_zero} b) presents the same quantities but only the VTS used for our study. Figure~\ref{figure_validation_shape_zero} confirms that $\Lambda_0^2$ provides a valid estimate of the degree of Gaussianity at the large scales showing good agreement with $F_u$. The distinction between Gaussian and non-Gaussian behaviour is based solely on the threshold $\Lambda_0^2 < 0.005$. Although negative values of $\Lambda_0^2$ also indicate non-Gaussianity, they are classified as Gaussian here because they exhibit a clear lower bound, and the trends of the investigated quantities remain unchanged whether $\Lambda_0^2$ is set to 0.005, 0, or $-0.02$.} 

\begin{figure}[htbp]
    \centering
    \begin{subfigure}[t]{0.49\textwidth}
        \centering
        \includegraphics[width=\linewidth]{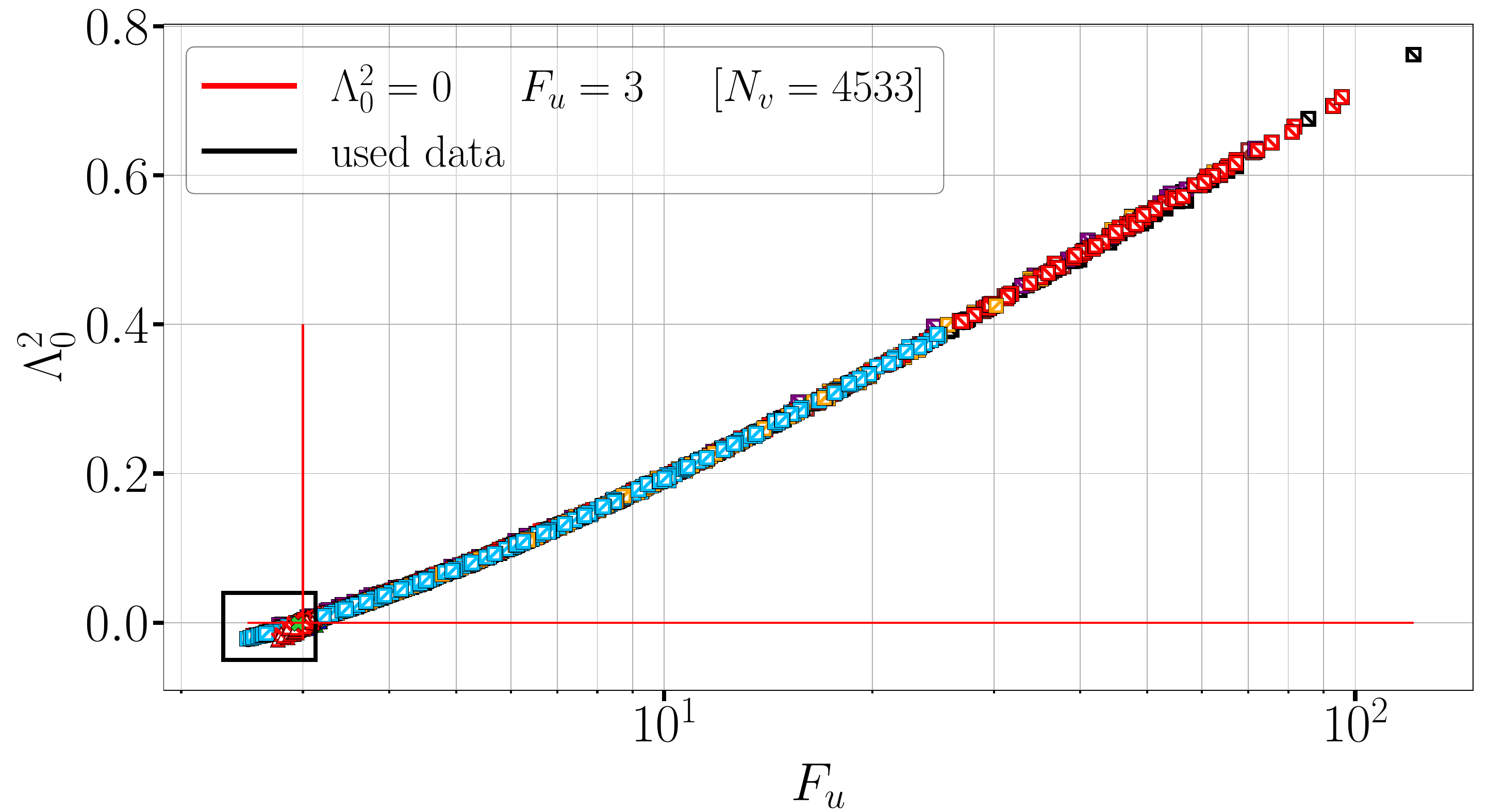}
        \caption{}
    \end{subfigure}
    \hfill
    \begin{subfigure}[t]{0.49\textwidth}
        \centering        \includegraphics[width=\linewidth]{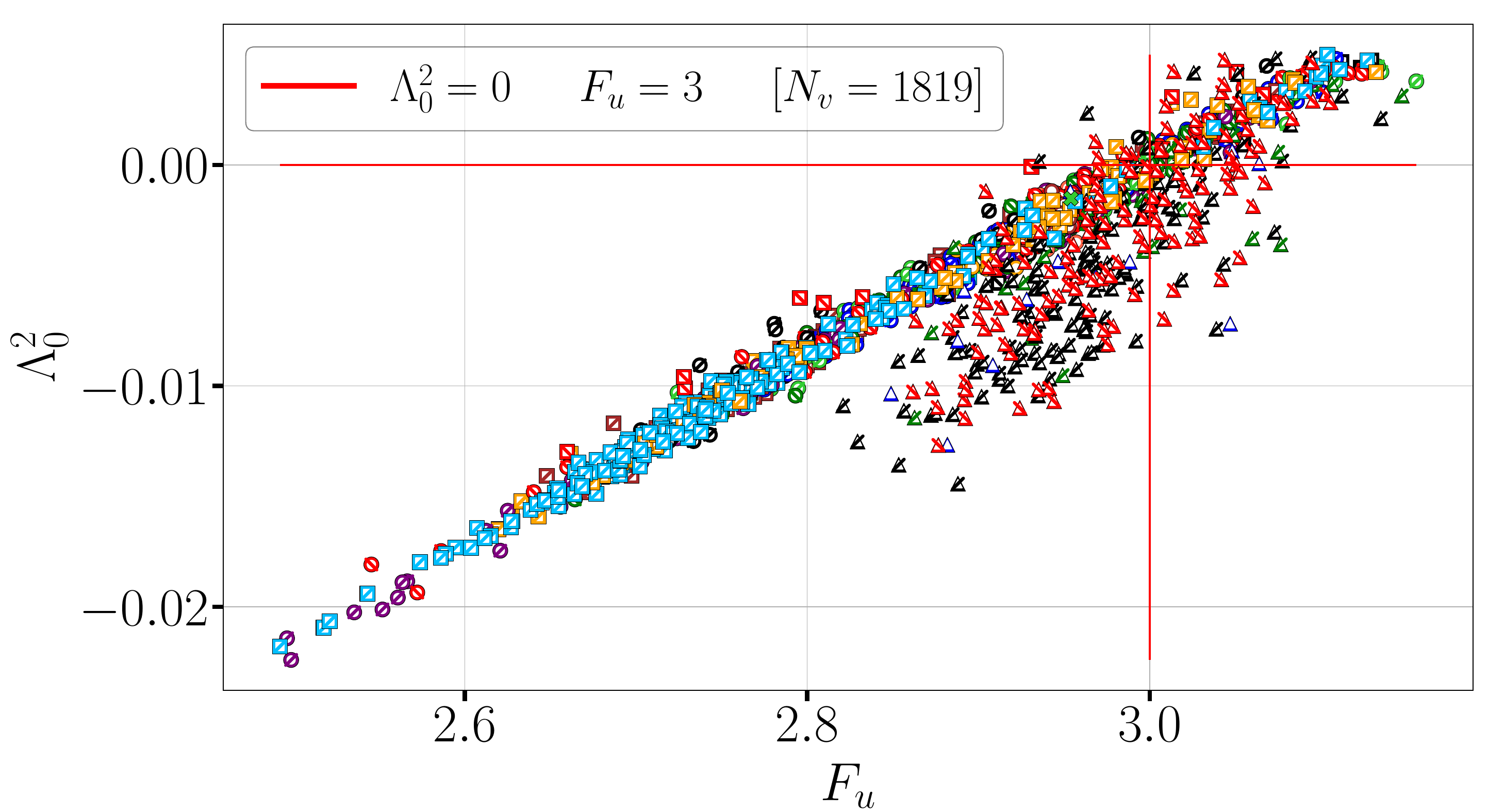}
        \caption{}
    \end{subfigure}
    \caption{\new{a) $\Lambda_0^2$ as a function of $F_u$ for all VTS measured that allowed a determination of $\Lambda_0^2$. The black rectangle includes the used data. b) $\Lambda_0^2$ as a function of $F_u$ for all VTS used. Note that according to eq. (\ref{equation_lambda_one}), $\Lambda_0^2$ takes negative values if $F(u_r(x))$ has values smaller than 3. In general the red lines indicate both $\Lambda^2_0=0$ and $F_u=3$. $N_v$ indicates the number of VTS shown in the plot. The symbols and corresponding configurations are shown and explained in table~\ref{tab:PhD measurements in LEGI 2023} in the appendix~\cite{SM}.}}
    \label{figure_validation_shape_zero}
\end{figure}

\FloatBarrier

\new{To conclude this chapter, all quantities that cannot be determined straightforwardly are estimated using methods that do not rely on any specific turbulence model and are validated against at least one independent alternative method. Thus, the results presented in chapter~\ref{new_laws} can also be reproduced using the alternative estimation methods introduced for the individual underlying quantities. However, the estimation methods ultimately adopted for the main analysis, and therefore used in chapter~\ref{new_laws}, yield the lowest scatter. From this we conclude that our found relations are not
a spurious results of a special method.}


\section{\new{Dependence on Conventional Control Parameters}}
\label{control}

\new{Many aspects of turbulence research commonly involve studying how features develop as the Reynolds number and/or turbulence intensity increase. In this chapter, we thus investigate the according dependencies of  $\mu$, $C_\varepsilon$, $\gamma$, and $C_k$ first, followed by the corresponding relations involving $\alpha$, $\beta$, and $\phi$.}

\subsection{\new{Dependence on $Re_\lambda$}}

\new{Figure~\ref{figure_four_quantities_reynolds} a), b), c), and d) shows the dependence of $C_\varepsilon$, $\mu$, $\gamma$ and $C_k$ on $Re_\lambda$, respectively. No clear trend is observed in any of the subplots. The data exhibit substantial scatter.  In the subplots b), c), and d), a group of values are distributed around the well-established HIT reference values for $\mu$, $\gamma$, and $C_k$, respectively. Overall no universal or predictable dependence on $Re_\lambda$ is obvious, but the observed scatter is not entirely unstructured. The data remain organized according to the individual cases, marked by the same symbols and following common curves or clusters. While most cases, corresponding to low to moderate $Re_\lambda$, exhibit such case-specific trends, we find a group os data for the highest $Re_\lambda$ which are  close to the well-established reference values for HIT. Notably, these asymptotic values are all coming from active grid flows. We interpret the behavior shown in figure~\ref{figure_four_quantities_reynolds} as evidence that, within SST, $Re_\lambda$ does not act as a universal control parameter, but rather as a case-dependent one. }

\new{Considering the curves and clusters from a global perspective and comparing the four subplots reveals an additional feature. For Fig 13 a) and b) as well as for b) and c) similar pattern are present. This recurring pattern suggests that the observed behavior is not coincidental, but may reflect a deeper connection among $C_\varepsilon$, and $C_k$, as well as between $\mu$ and $\gamma$.}

\begin{figure}[htbp]
    \centering
    \begin{subfigure}[t]{0.49\textwidth}
        \centering        \includegraphics[width=\linewidth]{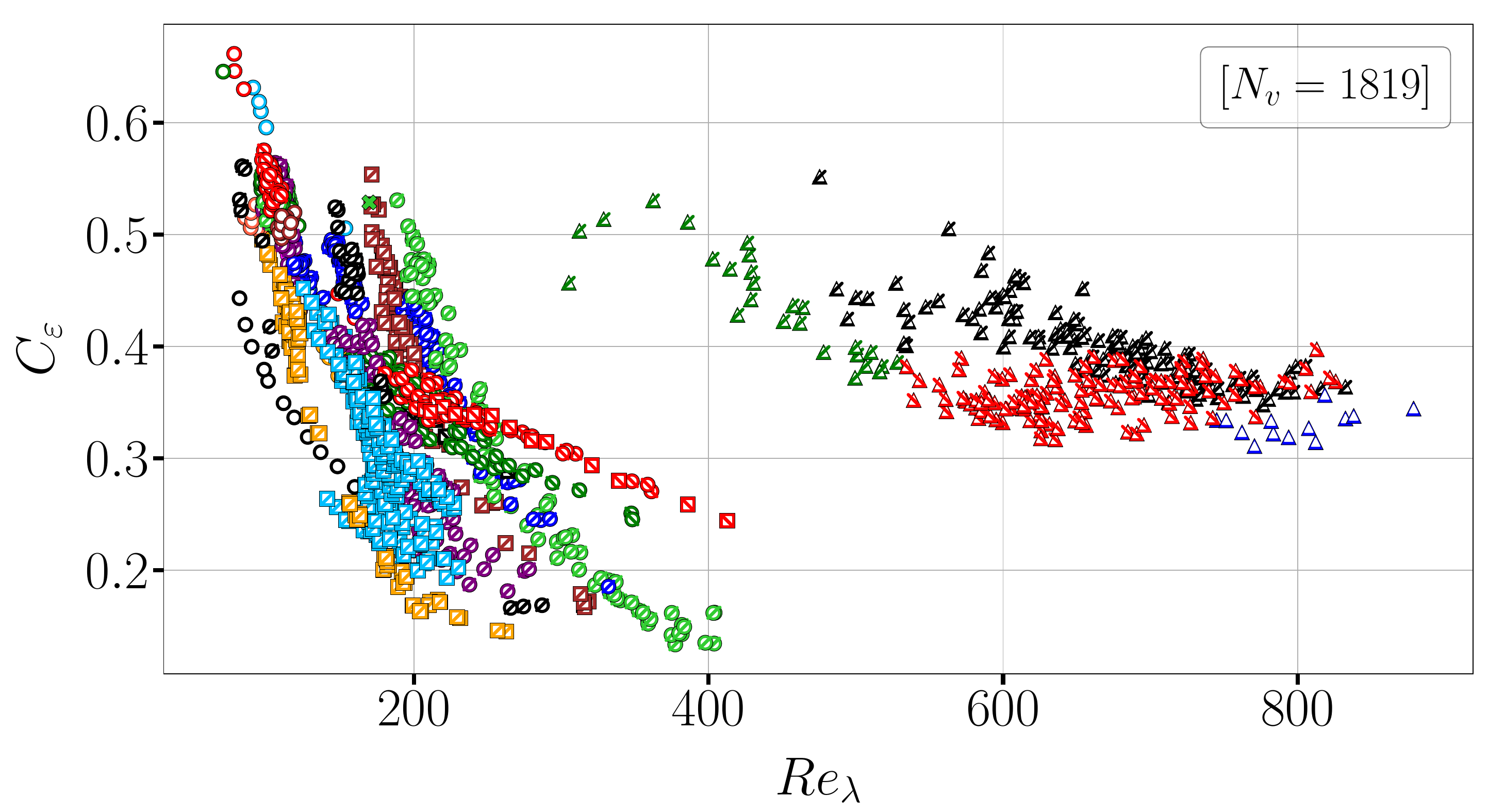}
        \caption{}
    \end{subfigure}
    \hfill
    \begin{subfigure}[t]{0.49\textwidth}
        \centering        \includegraphics[width=\linewidth]{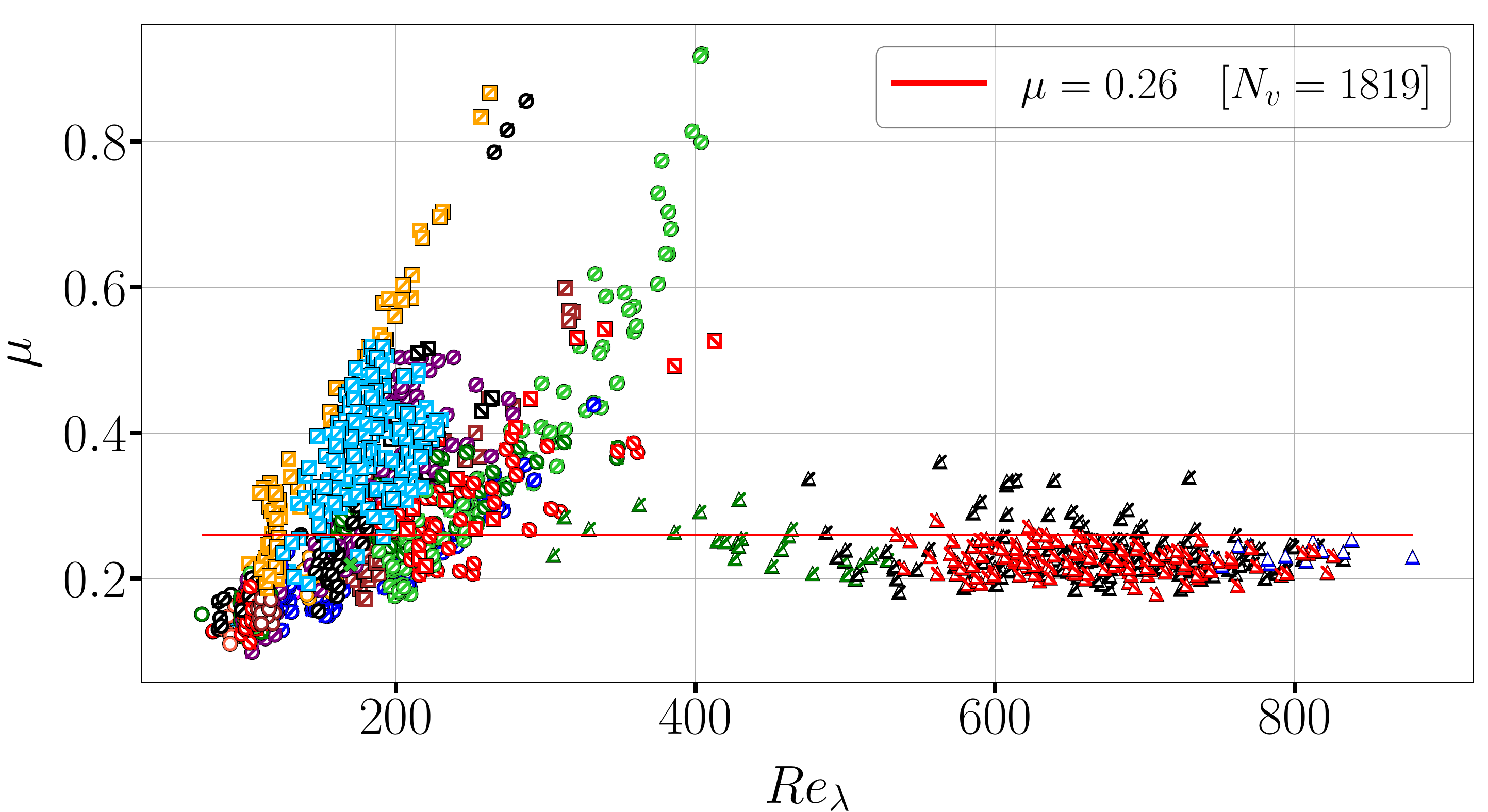}
        \caption{}
    \end{subfigure}
    \begin{subfigure}[t]{0.49\textwidth}
        \centering        \includegraphics[width=\linewidth]{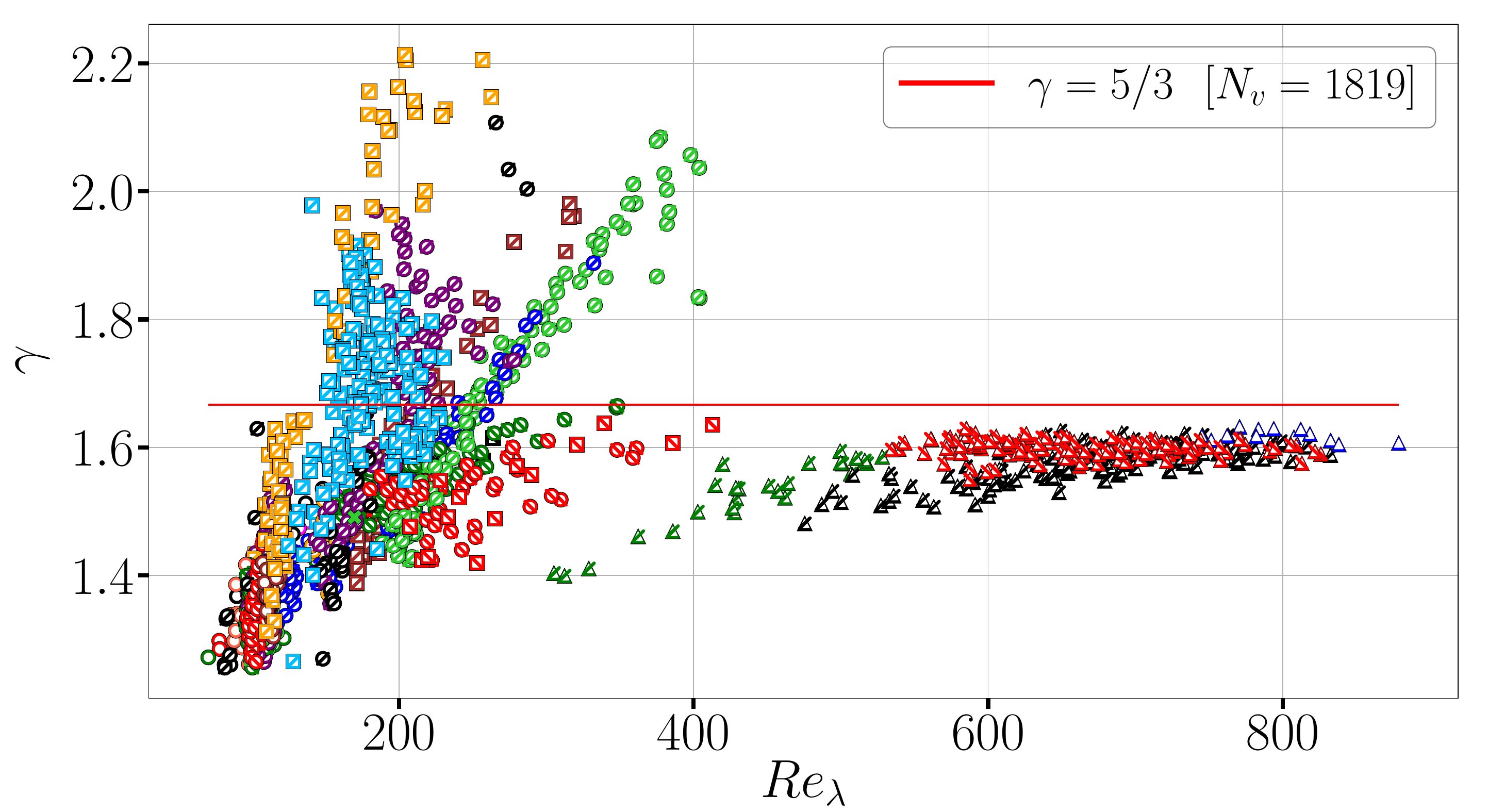}
        \caption{}
    \end{subfigure}
    \hfill
    \begin{subfigure}[t]{0.49\textwidth}
        \centering        \includegraphics[width=\linewidth]{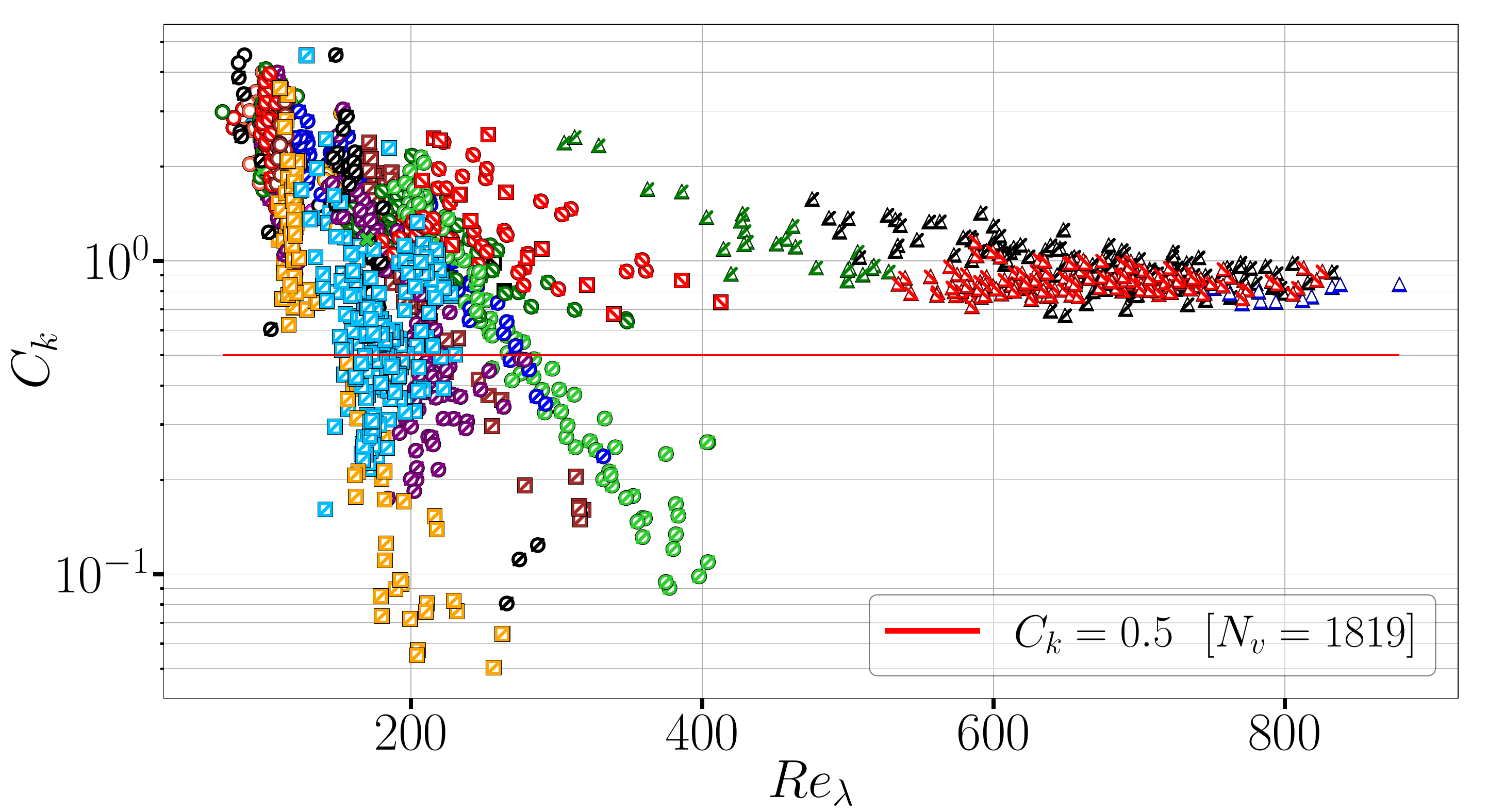}
        \caption{}
    \end{subfigure}
    \caption{\new{a) $C_\varepsilon$, b) $\mu$, c) $\gamma$, and d) $C_k$ as a function of $Re_\lambda$ for all VTS used, respectively. In general the red lines indicate the commonly accepted values for $\gamma$, $C_k$ and $\mu$ for HIT~\cite{sreenivasan1995universality, arneodo1996structure}. $N_v$ indicates the number of VTS shown in the plot. The symbols and corresponding configurations are shown and explained in table~\ref{tab:PhD measurements in LEGI 2023} in the appendix~\cite{SM}.}}
    \label{figure_four_quantities_reynolds}
\end{figure}

\FloatBarrier

\new{Figures~\ref{figure_alpha_beta_phi_reynolds} a), c), and e) show $\alpha$, $\beta$, and $\phi$ as functions of $Re_\lambda$, respectively. Compared with the corresponding results in figure~\ref{figure_four_quantities_reynolds}, the overall scatter is strongly reduced, while the distinct case-specific trajectories largely disappear or, as in the case of $\beta$, become considerably less pronounced. For $\alpha$ and $\beta$, subplots b) and d) reproduce the relations shown in a) and c), respectively, for three representative cases (C1, C4, and D13), with the corresponding uncertainties additionally indicated by error bars. The uncertainties in $C_\varepsilon$, $\mu$, and $\gamma$ are adopted from figures~\ref{law_1} d) and~\ref{law_2} d), while those associated with $\alpha$ and $\beta$ are obtained from the uncertainties of their underlying quantities using Gaussian error propagation. It can be seen that the uncertainties in $\alpha$ and $\beta$ remain within an acceptable range and that the magnitude of the error bars does not exhibit any systematic dependence on $Re_\lambda$.}

\begin{figure}[htbp]
    \centering
    \begin{subfigure}[t]{0.49\textwidth}
        \centering        \includegraphics[width=\linewidth]{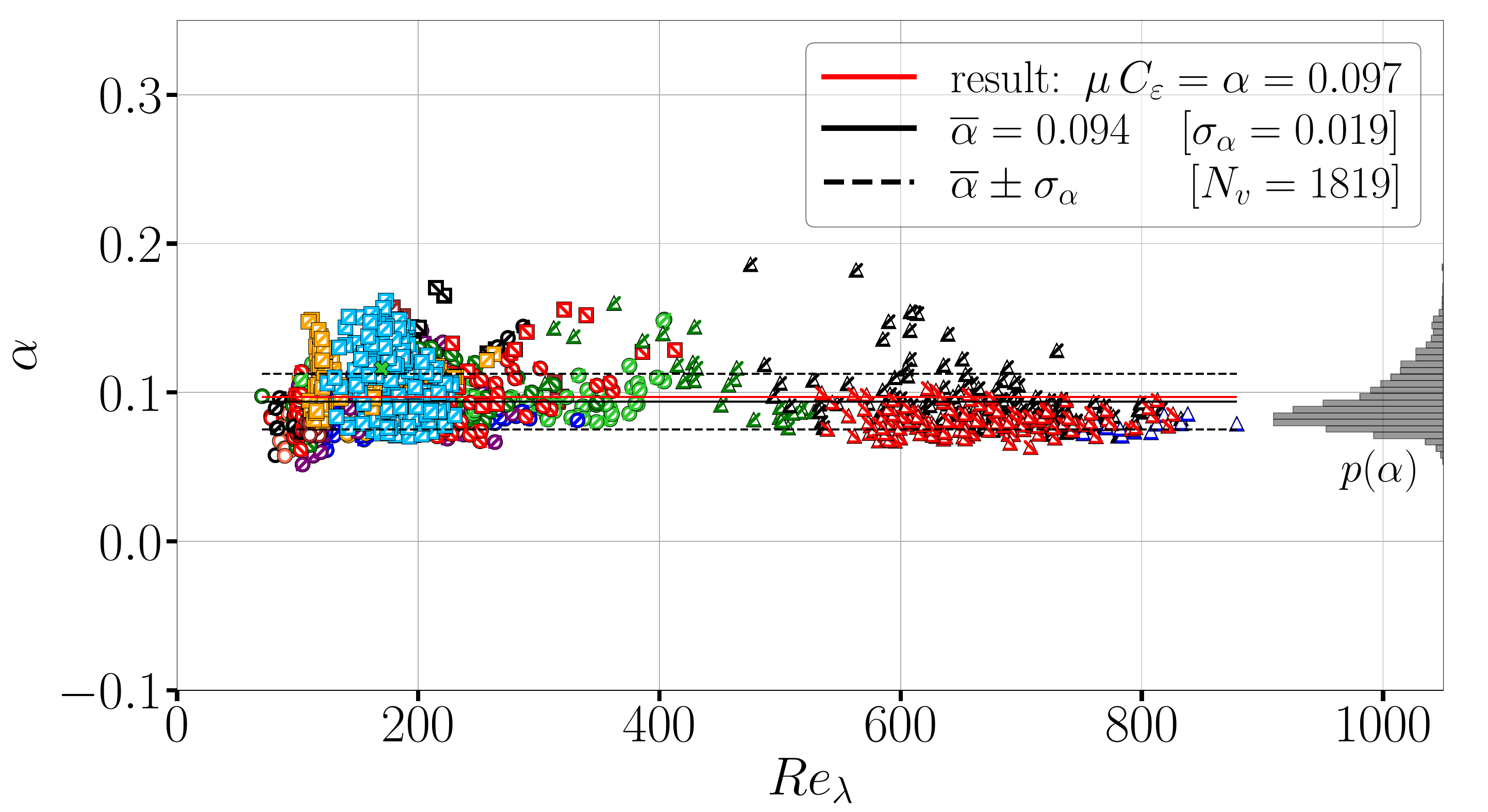}
        \caption{}
    \end{subfigure}
    \hfill
    \begin{subfigure}[t]{0.49\textwidth}
        \centering        \includegraphics[width=\linewidth]{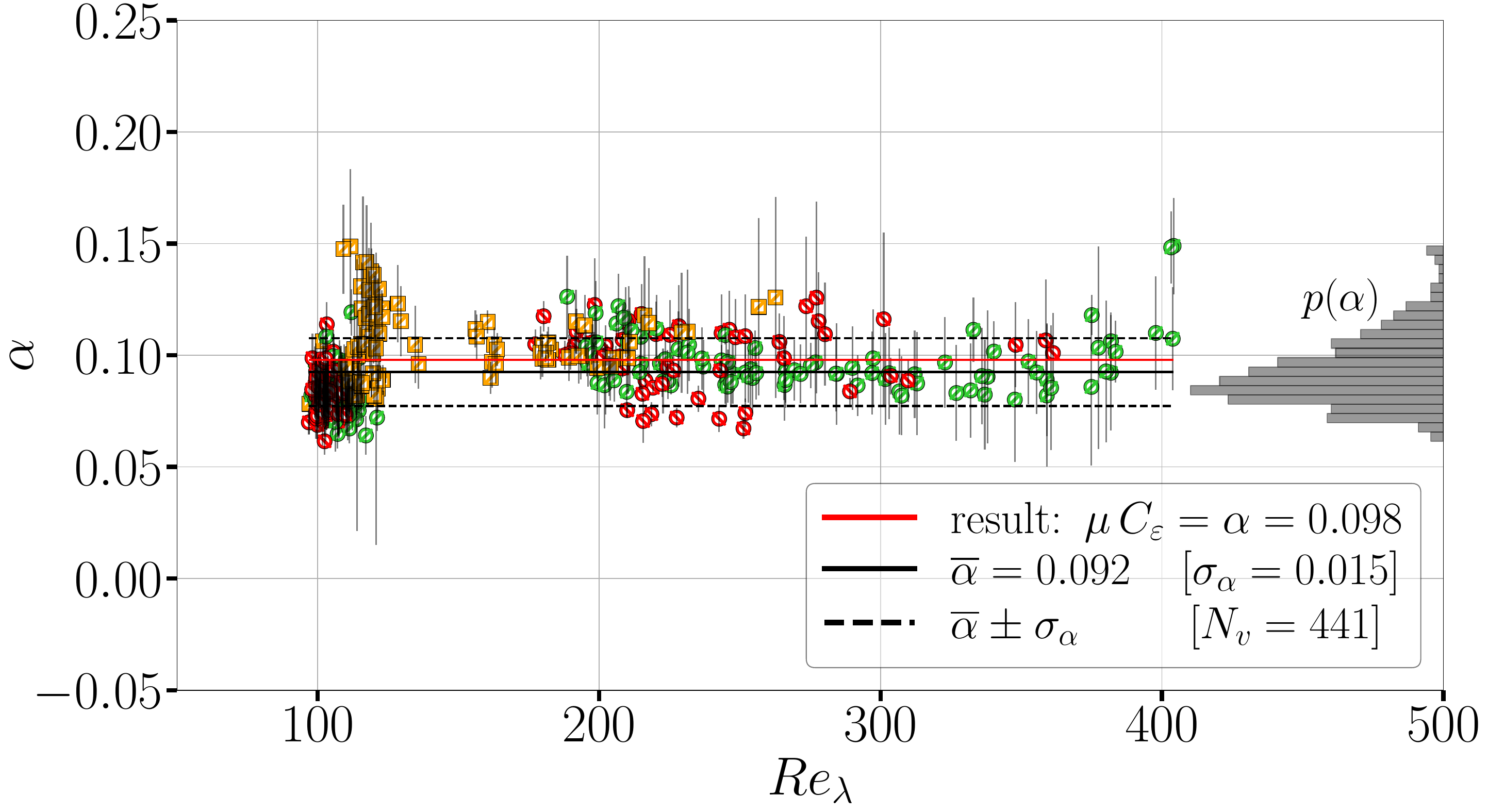}
        \caption{}
    \end{subfigure}
    \begin{subfigure}[t]{0.49\textwidth}
        \centering        \includegraphics[width=\linewidth]{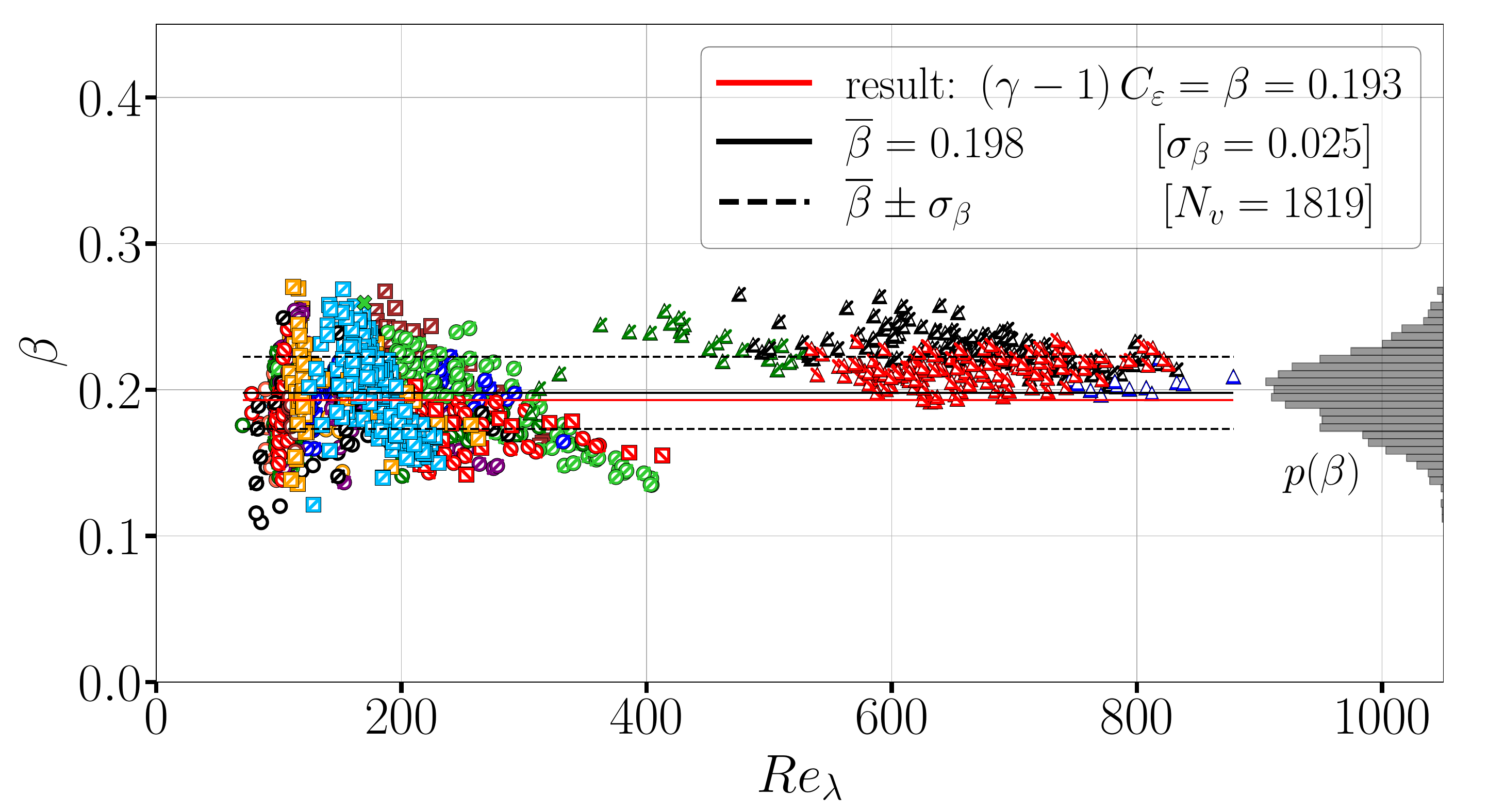}
        \caption{}
    \end{subfigure}
    \hfill
    \begin{subfigure}[t]{0.49\textwidth}
        \centering        \includegraphics[width=\linewidth]{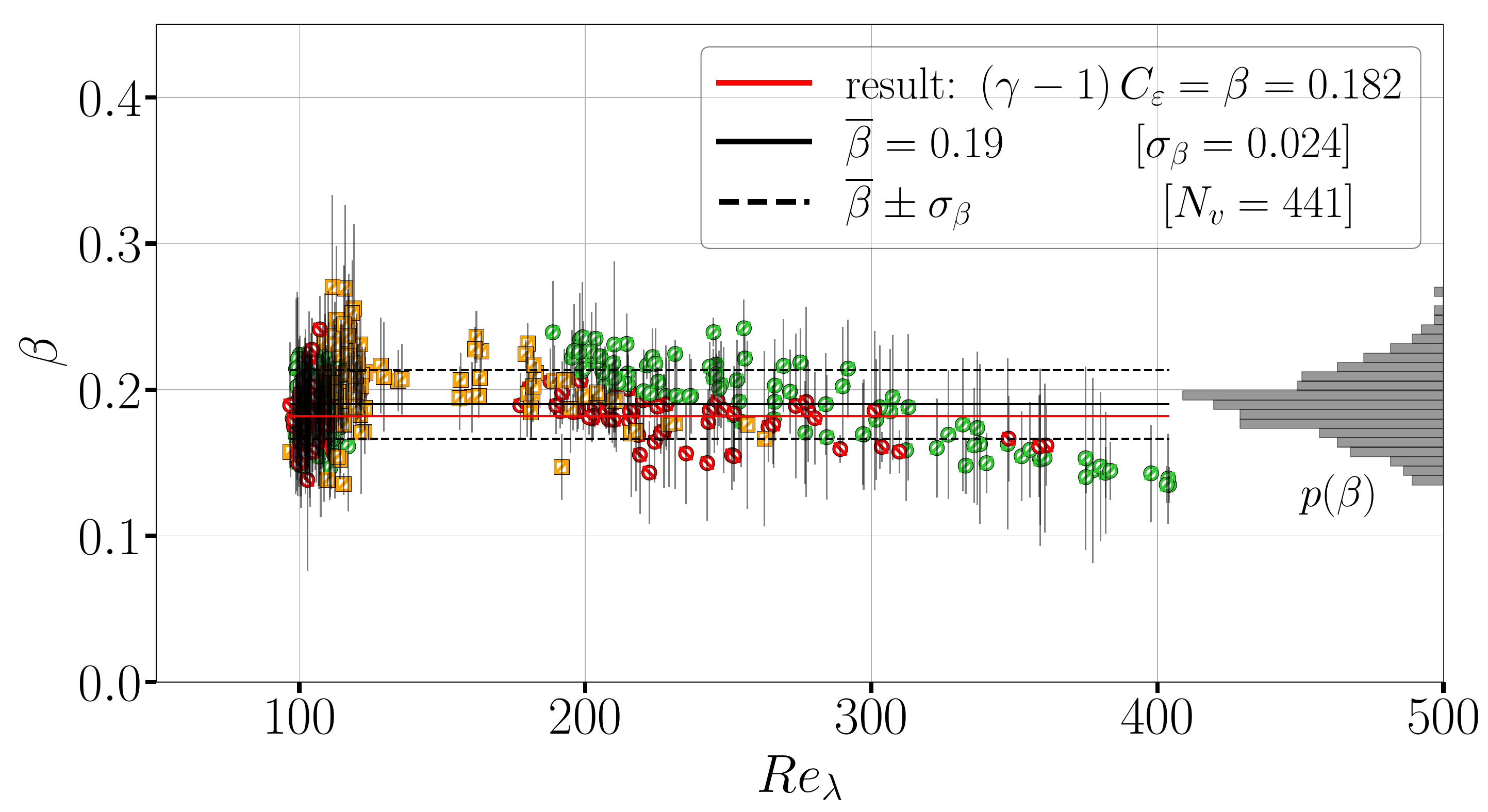}
        \caption{}
    \end{subfigure}
    \begin{subfigure}[t]{0.49\textwidth}
        \centering        \includegraphics[width=\linewidth]{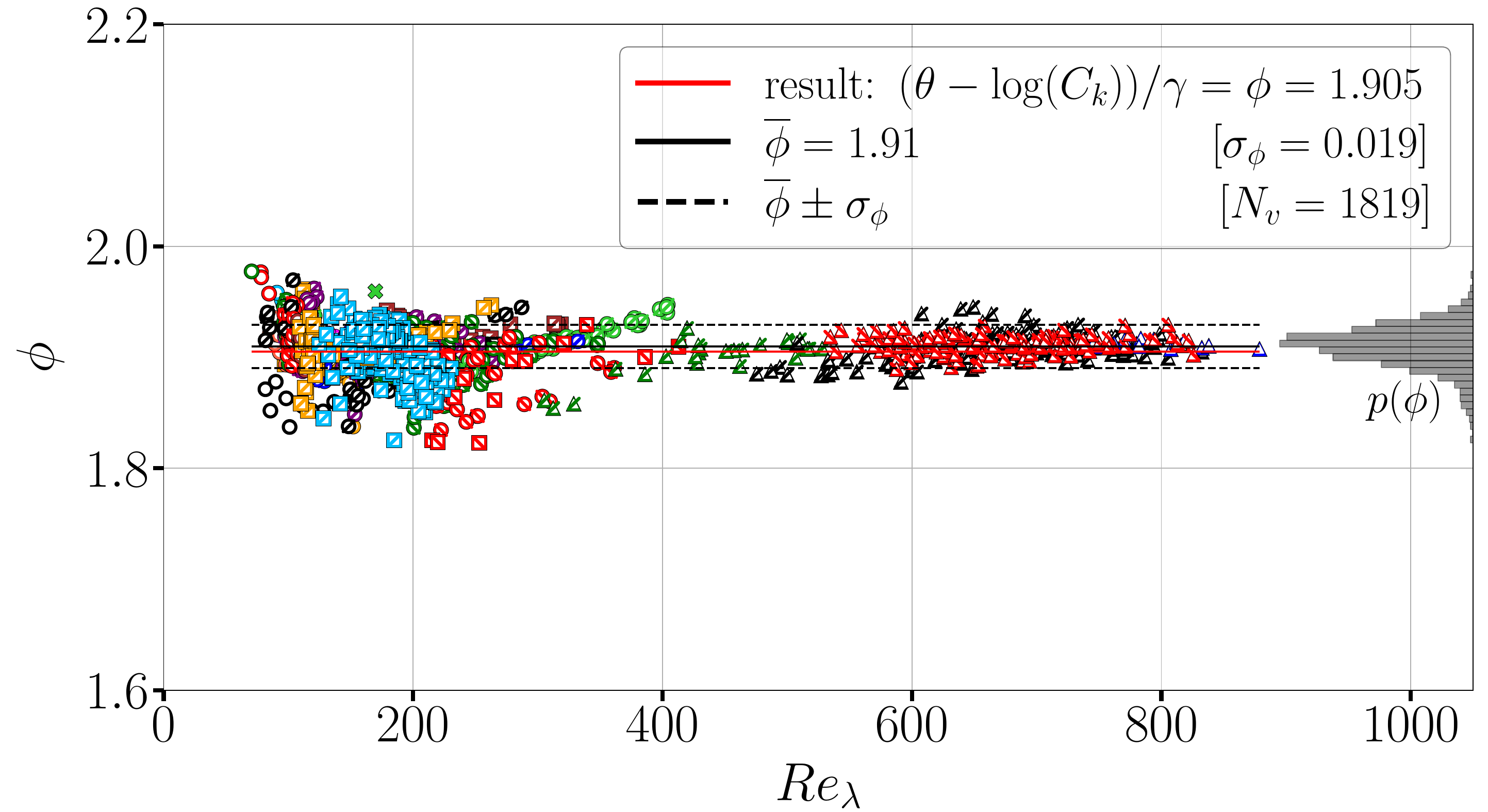}
        \caption{}
    \end{subfigure}
    \caption{\new{$\alpha=\mu \, C_\varepsilon$ as a function of $Re_\lambda$ for a) all VTS used and for b) only the used VTS from the representative cases C1, C4, and D13, respectively. The red line indicates the result for $\alpha$ from the fit shown in figure~\ref{figure_law_1}, while the black solid line represents the actual mean value of the ensemble of $\alpha$ values. The two black dashed lines denote the corresponding standard deviation from the mean. Furthermore, the PDF of $\alpha$ values $p(\alpha)$ is presented. Additionally, b) shows the errors in $\alpha$ based on the used methods. The corresponding results for $\beta=(\gamma-1)\,C_\varepsilon$ are presented in c) and d), based on the fit shown in figure~\ref{figure_law_2}. In addition, e) shows $\phi=(\theta-\log(C_k))/\gamma$ as a function of $Re_\lambda$ for all VTS considered, using the same notation as in a) and c). In general, $N_v$ indicates the number of VTS shown in the plot. The symbols and corresponding configurations are shown and explained in table~\ref{tab:PhD measurements in LEGI 2023} in the appendix~\cite{SM}.}}
    \label{figure_alpha_beta_phi_reynolds}
\end{figure}

\FloatBarrier

\subsection{\new{Dependence on $TI$}}

\new{Next, the same analysis is performed using the second conventional control parameter, namely $TI$. Figure~\ref{figure_four_quantities_TI} presents the same quantities as figure~\ref{figure_four_quantities_reynolds}, but as functions of $TI$. Likewise, figure~\ref{figure_alpha_beta_phi_TI} shows the dependence of $\alpha$, $\beta$, and $\phi$ on $TI$. The overall behavior closely resembles that observed for $Re_\lambda$, suggesting that $TI$ likewise does not act as a universal control parameter within SST, but rather as a case-dependent one.}

\begin{figure}[htbp]
    \centering
    \begin{subfigure}[t]{0.49\textwidth}
        \centering        \includegraphics[width=\linewidth]{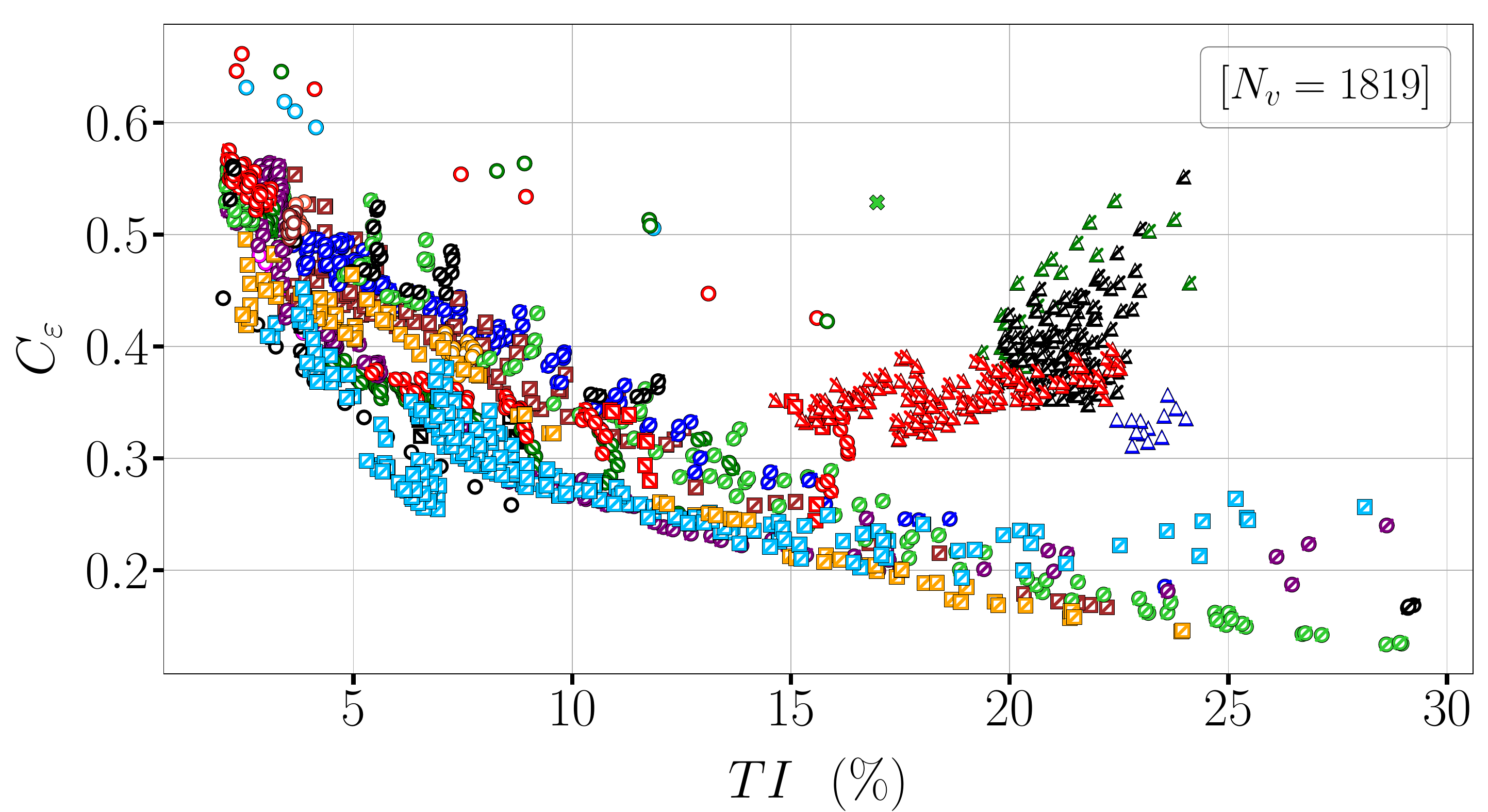}
        \caption{}
    \end{subfigure}
    \hfill
    \begin{subfigure}[t]{0.49\textwidth}
        \centering        \includegraphics[width=\linewidth]{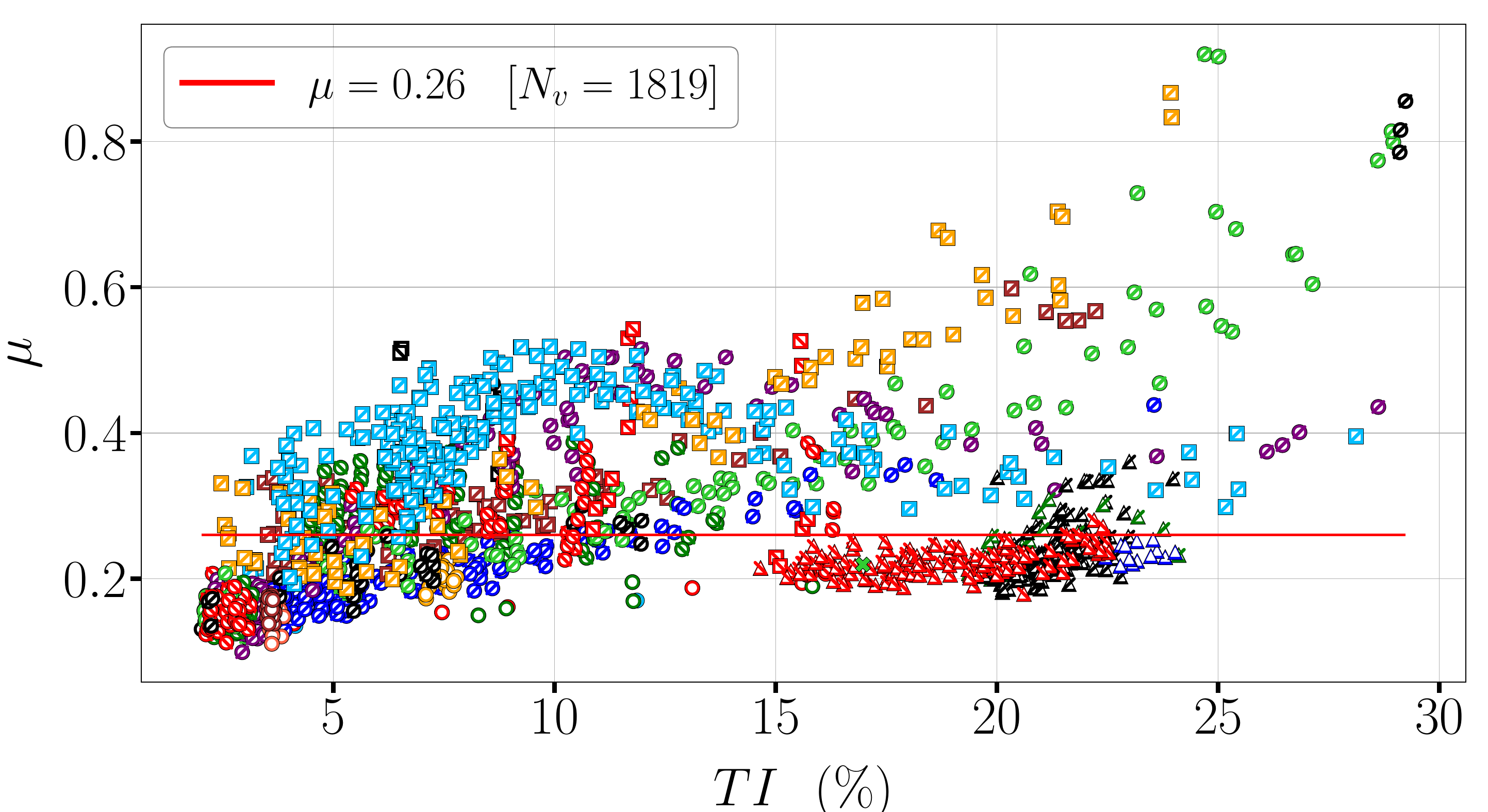}
        \caption{}
    \end{subfigure}
    \begin{subfigure}[t]{0.49\textwidth}
        \centering        \includegraphics[width=\linewidth]{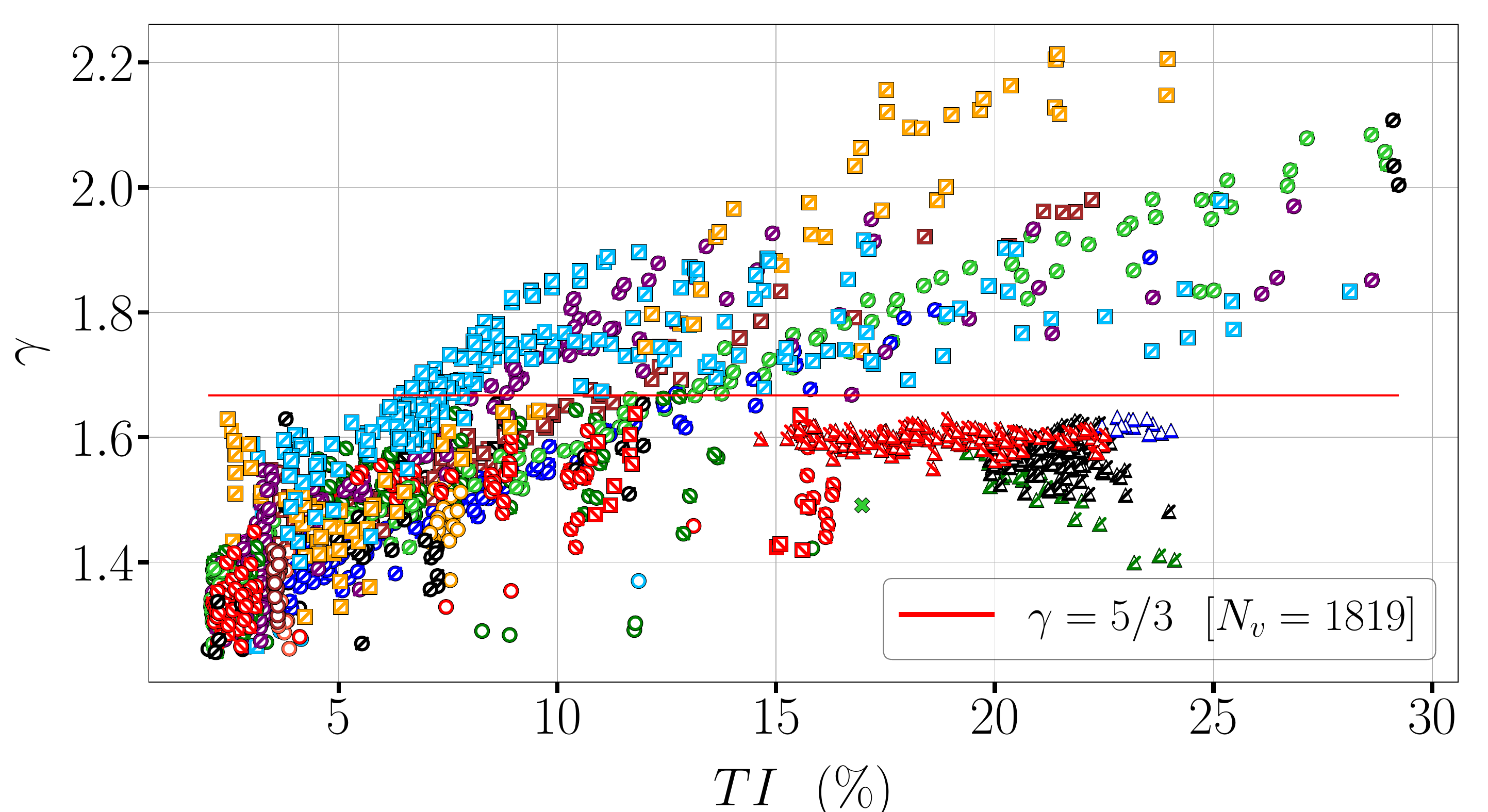}
        \caption{}
    \end{subfigure}
    \hfill
    \begin{subfigure}[t]{0.49\textwidth}
        \centering        \includegraphics[width=\linewidth]{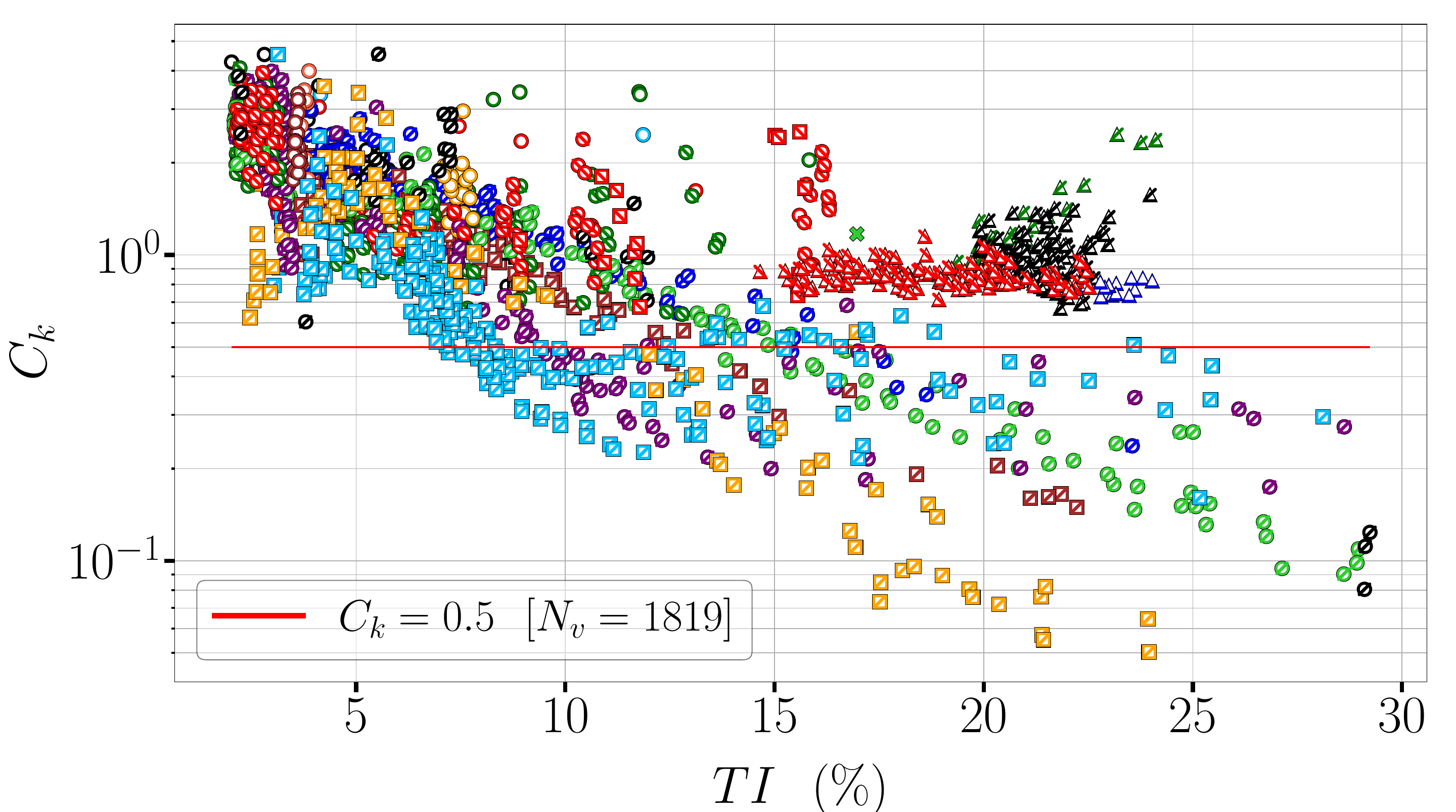}
        \caption{}
    \end{subfigure}
    \caption{\new{a) $C_\varepsilon$, b) $\mu$, c) $\gamma$, and d) $C_k$ as a function of $TI$ for all VTS used, respectively. In general the red lines indicate the commonly accepted values for $\gamma$, $C_k$ and $\mu$ for HIT~\cite{sreenivasan1995universality, arneodo1996structure}. $N_v$ indicates the number of VTS shown in the plot. The symbols and corresponding configurations are shown and explained in table~\ref{tab:PhD measurements in LEGI 2023} in the appendix~\cite{SM}.}}
    \label{figure_four_quantities_TI}
\end{figure}

\FloatBarrier

\begin{figure}[htbp]
    \centering
    \begin{subfigure}[t]{0.49\textwidth}
        \centering        \includegraphics[width=\linewidth]{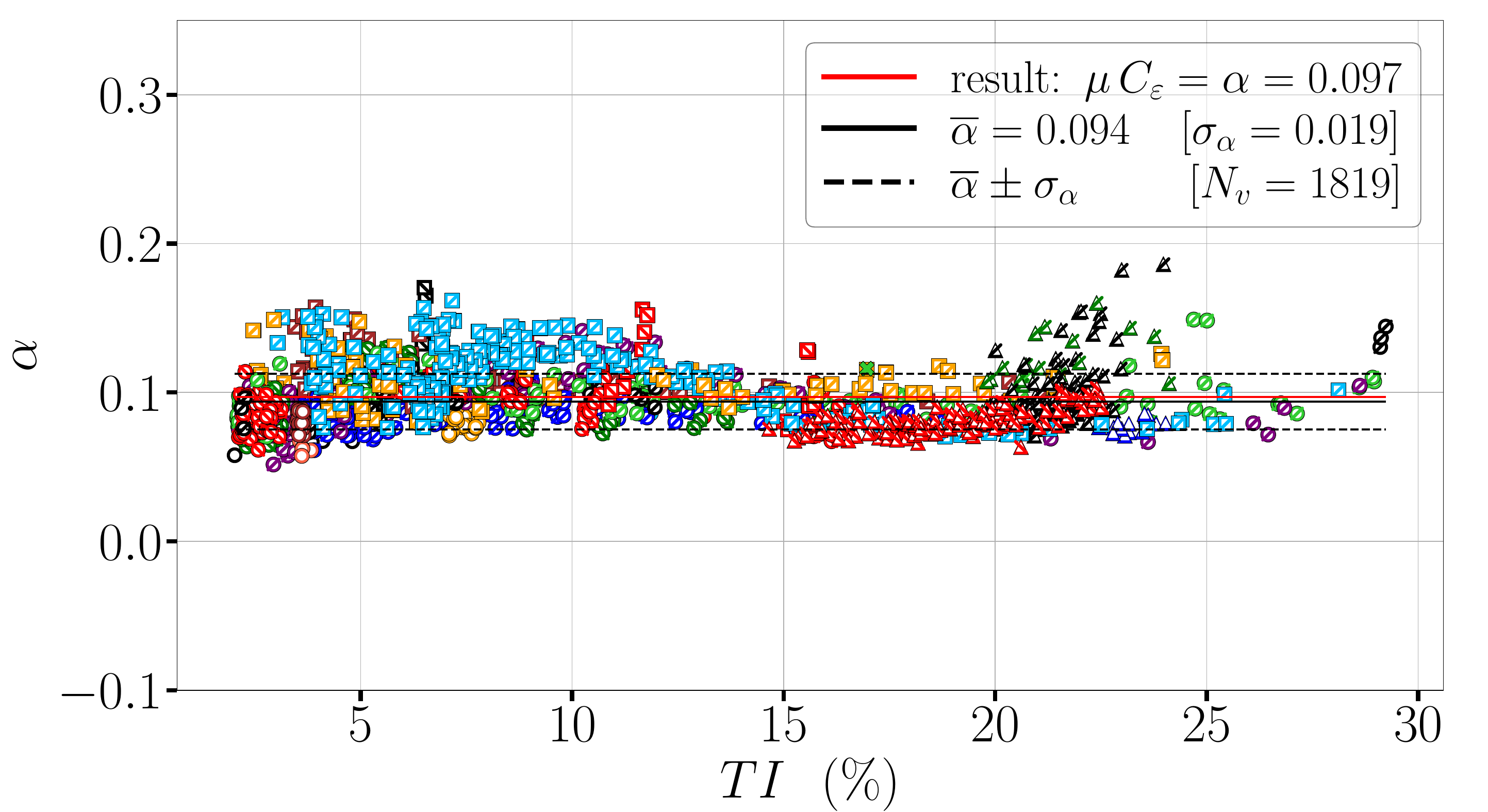}
        \caption{}
    \end{subfigure}
    \hfill
    \begin{subfigure}[t]{0.49\textwidth}
        \centering        \includegraphics[width=\linewidth]{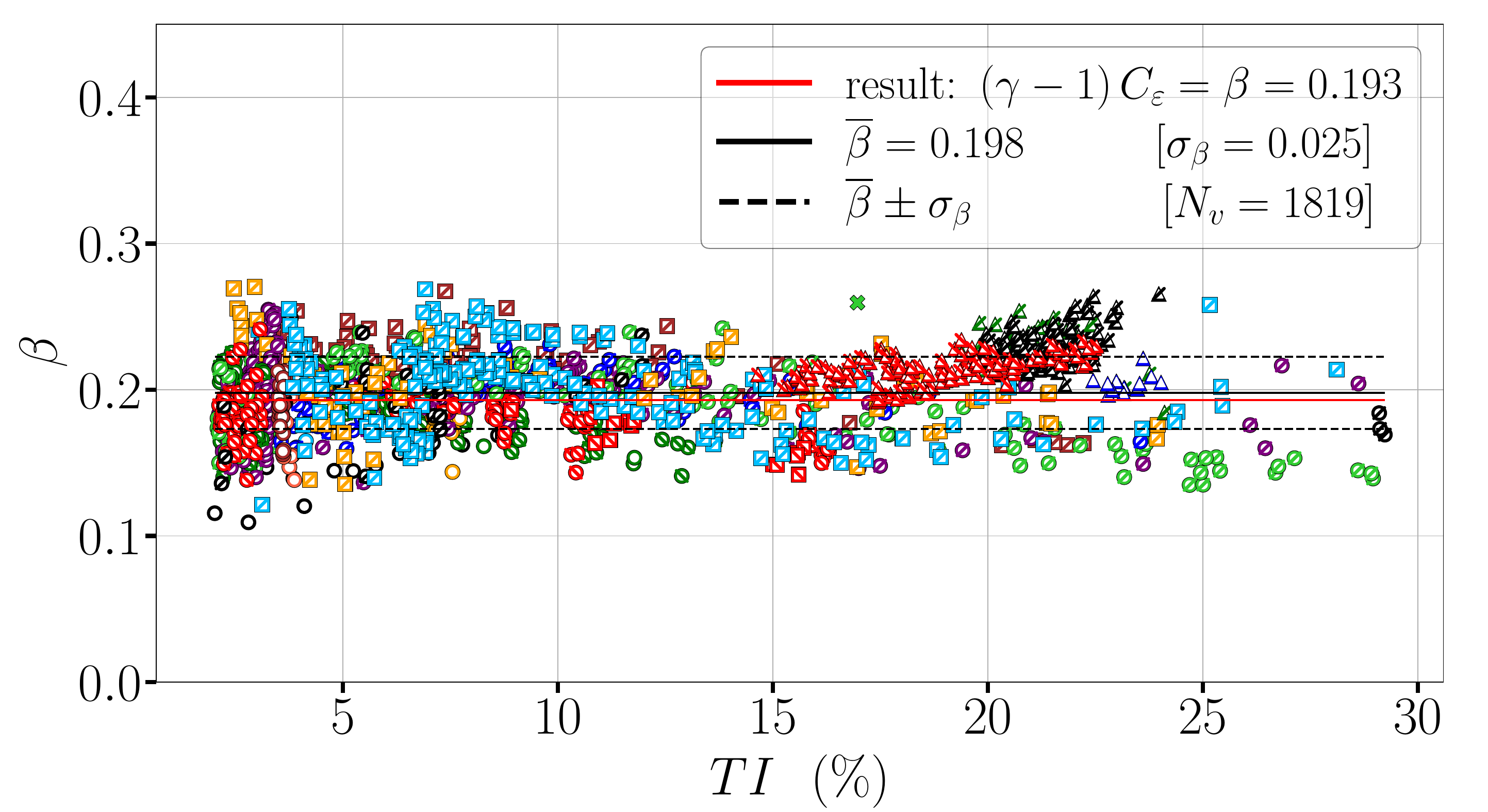}
        \caption{}
    \end{subfigure}
    \begin{subfigure}[t]{0.49\textwidth}
        \centering        \includegraphics[width=\linewidth]{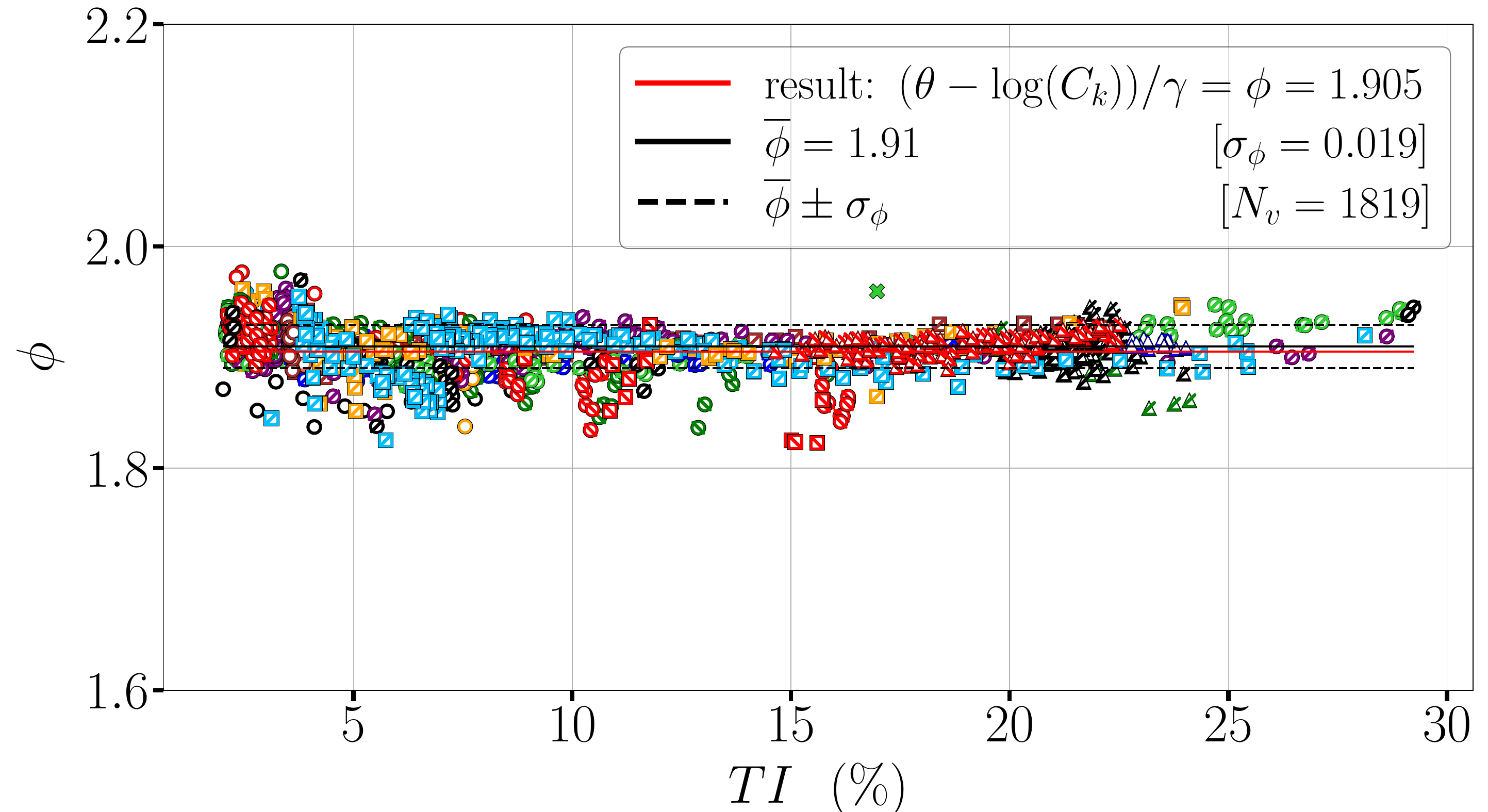}
        \caption{}
    \end{subfigure}
    \caption{\new{a) $\alpha=\mu \, C_\varepsilon$, b) $\beta=(\gamma-1)\,C_\varepsilon$, and c) $\phi=(\theta-\log(C_k))/\gamma$ as a function of $TI$ for all VTS used. The notation follows that introduced in figure~\ref{figure_alpha_beta_phi_reynolds}. In general, $N_v$ indicates the number of VTS shown in the plot. The symbols and corresponding configurations are shown and explained in table~\ref{tab:PhD measurements in LEGI 2023} in the appendix~\cite{SM}.}}
    \label{figure_alpha_beta_phi_TI}
\end{figure}

\FloatBarrier

\subsection{\new{Error Analysis}}

\new{So far, it has been shown that the distributions of $\alpha$, $\beta$, and $\phi$ are well centered and that their scatter appears visually smaller than that of the underlying quantities $C_\varepsilon$, $\mu$, $\gamma$, and $C_k$. However, this assessment has so far been based only on visual inspection. To quantify whether $\alpha$, $\beta$, and $\phi$ indeed provide a more compact description, their relative scatter is evaluated using the coefficient of variation (ratio of standard deviation and mean) and compared directly with that of the underlying quantities. Consequently, tables~\ref{table_law_1}, \ref{table_law_2}, and \ref{table_law_3} summarize the statistical properties of the distributions of $\alpha$, $\beta$, and $\phi$, respectively, together with those of their respective underlying quantities. Across all three tables, the coefficient of variation of the respective composite quantity, highlighted in orange, is consistently and substantially smaller than the coefficients of variation of both underlying quantities, highlighted in gray.}

\begin{table} [h!]
	\centering
	\begin{tabular}{lccccccccc}
		\toprule
		    fit result ($\alpha$)  & $\overline{\alpha}$ & $\tilde{\alpha}$ &  $\sigma_{\alpha}$ & $\sigma_{\alpha} / \overline{\alpha}$ [$\%$]  & $\sigma_{C_\varepsilon}$ & $\sigma_{C_\varepsilon} / \overline{C_\varepsilon}$ [$\%$] & $\sigma_{\mu}$ & $\sigma_{\mu} / \overline{\mu}$ [$\%$] &\\
		\midrule
		    0.0971 $\pm$ 0.0004 & 0.094 & 0.089 & 0.019 & \cellcolor{orange!50} 20.2 & 0.107 & \cellcolor{darkgray!30}26.2 & 0.115 & \cellcolor{darkgray!30}44.9 & \\
		\bottomrule
	\end{tabular}
	\caption{\new{Location parameters of the distribution of $\alpha$ from figure~\ref{figure_law_1} c) and of the distribution of both $\mu$ and $C_\varepsilon$ from figure~\ref{figure_law_1} a) are presented. The fit result refers to a least-squares fit shown in~\ref{figure_law_1} a). An overline denotes the mean, a tilde denotes the median and $\sigma$ denotes the standard deviation of a distribution. The relative spread of the newly identified parameter $\alpha$ is shown in orange.}}
	\label{table_law_1}
\end{table}

\begin{table} [h!]
	\centering
	\begin{tabular}{lccccccccc}
		\toprule
		    fit result ($\beta$)  & $\overline{\beta}$ & $\tilde{\beta}$ &  $\sigma_{\beta}$ & $\sigma_{\beta} / \overline{\beta}$ [$\%$]  & $\sigma_{C_\varepsilon}$ & $\sigma_{C_\varepsilon} / \overline{C_\varepsilon}$ [$\%$] & $\sigma_{\gamma - 1}$ & $\sigma_{\gamma - 1} / \overline{\gamma - 1}$ [$\%$] &\\
		\midrule
		   0.1933 $\pm$ 0.0006 & 0.198 & 0.199 & 0.025 & \cellcolor{orange!50} 12.6 & 0.107 & \cellcolor{darkgray!30}26.2 & 0.169 & \cellcolor{darkgray!30}32.1 & \\
		\bottomrule
	\end{tabular}
	\caption{\new{Location parameters of the distribution of $\beta$ from figure~\ref{figure_law_2} c) and of the distribution of both $\gamma - 1$ and $C_\varepsilon$ from figure~\ref{figure_law_2} a) are presented. The fit result refers to a least-squares fit shown in~\ref{figure_law_2} a). An overline denotes the mean, a tilde denotes the median and $\sigma$ denotes the standard deviation of a distribution.The relative spread of the newly identified parameter $\beta$ is shown in orange.}}
	\label{table_law_2}
\end{table}

\begin{table} [h!]
	\centering
	\begin{tabular}{lccccccccc}
		\toprule
		    fit result ($\phi$) & $\overline{\phi}$ & $\tilde{\phi}$ &  $\sigma_{\phi}$ & $\sigma_{\phi} / \overline{\phi}$ [$\%$]  & $\sigma_{\gamma}$ & $\sigma_{\gamma} / \overline{\gamma}$ [$\%$] & $\sigma_{\mathrm{log}(C_k)}$ & $\sigma_{\mathrm{log}(C_k)} / \overline{\mathrm{log}(C_k)}$ [$\%$] &\\
		\midrule
		   1.905 $\pm$ 0.004 & 1.91 & 1.912 & 0.019 & \cellcolor{orange!50} 1.0 & 0.169 & \cellcolor{darkgray!30}11.1 & 0.323 & \cellcolor{darkgray!30}417.2 & \\
		\bottomrule
	\end{tabular}
	\caption{\new{Location parameters of the distribution of $\phi$ from figure~\ref{figure_law_3} c) and of the distribution of both $\mathrm{log}(C_k)$ and $\gamma$ from figure~\ref{figure_law_3} a) are presented. The fit result refers to a least-squares fit shown in~\ref{figure_law_3} a). An overline denotes the mean, a tilde denotes the median and $\sigma$ denotes the standard deviation of a distribution. The relative spread of the newly identified parameter $\phi$ is shown in orange.}}
	\label{table_law_3}
\end{table}

\new{To conclude this chapter, it has been demonstrated that, within SST, the composite quantities $\alpha$, $\beta$, and $\phi$ substantially reduce the scatter of their underlying quantities. Moreover, their behavior cannot be explained solely by conventional control parameters such as $Re_\lambda$ and $TI$. The following chapter examines the dependence on spatial position within the flow for one representative case and additionally illustrates which data are excluded by the SST selection criteria.}

\FloatBarrier

\section{\new{Dependence on Spatial Variation}}
\label{spatial}

\new{As a last point, we investigate whether systematic trends emerge in the downstream evolution of the turbulence. It is well known that many turbulent flows show transitions to a developed state. Thus, we examine the spatial behavior of several turbulence quantities in detail, including which data are retained or rejected by the SST selection criteria. We select VTS from a wake flow, case C4, as a representative example. This is the same case shown in figure~\ref{data_selection} b) and covers nearly the full range of values observed for $C_\varepsilon$, $\mu$, $\gamma$, and $C_k$. Figure~\ref{figure_support_color_along_x} shows the progressions of $Re_\lambda$, $TI$, $C_\varepsilon$, $\mu$, $\gamma$, and $C_k$ along the streamwise direction.}

\new{In contrast to the preceding figures, with the exception of figure~\ref{data_selection} b), all subplots in Figure~\ref{figure_support_color_along_x} include the complete dataset without applying any selection criteria. The individual data points are instead distinguished according to their respective category. VTS satisfying all SST criteria ($\approx46\,\%$) are indicated by the established C4 marker in green. Red markers ($\approx46\,\%$) denote VTS excluded based on the value of $\Lambda_0^2 > 0.005$, whereas black markers represent VTS rejected by any of the other selection criteria introduced above. The curves connect data points sharing the same spanwise coordinate, and the color of each curve indicates the absolute distance from the centerline. } 

\new{In general, for all subplots, three distinct data categories can be identified.  There is the inner part of the wake (green markers and light-green curves), the outer part of the flow (green markers and blue curves) and there is a part given by the discarded red markers, which correspond to the mixing zone of the inner and the outer regions in figure~\ref{data_selection} b).}  

\new{As shown in figure~\ref{figure_support_color_along_x}, in the outer region of the flow (blue curves), all six parameters vary only weakly along the streamwise direction. In contrast, and as expected, the corresponding values in the inner region (green curves) change much more rapidly with streamwise position. The discarded data occupy different regions of the parameter space, some values remain bounded by those of the inner and outer flow regions, whereas others extend beyond this range. This behavior is observed for all parameters except $TI$ and $\mu$. For $TI$, all discarded values remain entirely bounded by those of the inner and outer flow regions. In contrast, for $\mu$, nearly all discarded data lie outside the range spanned by the inner and outer flow regions. 
Surprisingly, for all six parameters, the values associated with the three data categories appear to converge toward a common level as the streamwise position increases. At first sight, this behavior might be interpreted as an indication of fully developed turbulence. However, fully developed turbulence is generally associated with the limit of large $Re_\lambda$, whereas the most downstream positions considered here correspond to the lowest values of $Re_\lambda$. The observed convergence is therefore more plausibly interpreted as a progressive homogenization of the overall flow. This interpretation is supported by the large distance downstream of the cylinder; a transition to developed turbulence typically occurs earlier. Furthermore, in this experiment, the cylinder wake is embedded in a turbulent environment (inflow), which becomes increasingly mixed into the wake as the flow propagates downstream.
}

\begin{figure}[htbp]
    \centering
    \begin{subfigure}[t]{0.49\textwidth}
        \centering        \includegraphics[width=\linewidth]{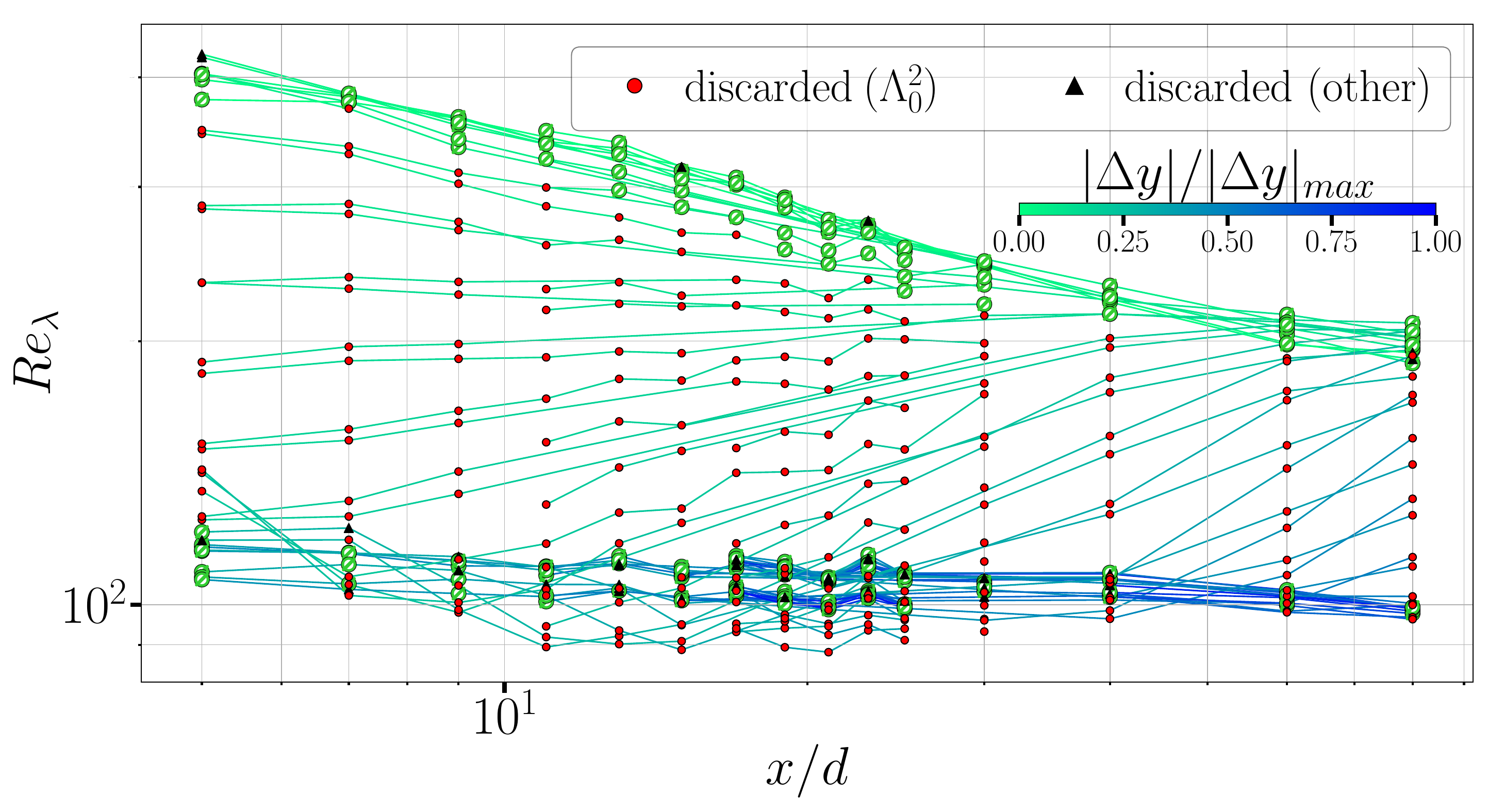}
        \caption{}
    \end{subfigure}
    \hfill
    \begin{subfigure}[t]{0.49\textwidth}
        \centering        \includegraphics[width=\linewidth]{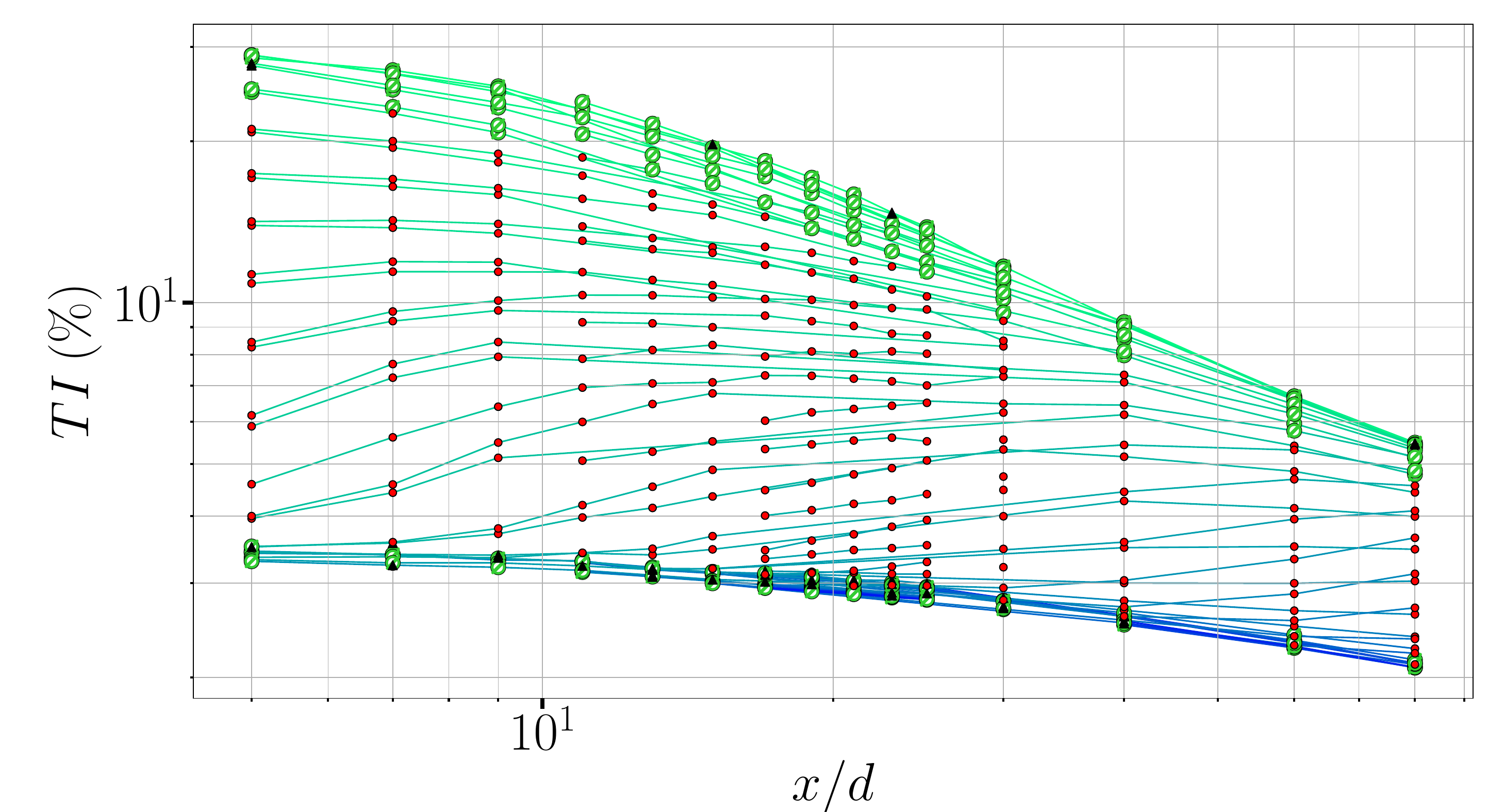}
        \caption{}
    \end{subfigure}
    \begin{subfigure}[t]{0.49\textwidth}
        \centering        \includegraphics[width=\linewidth]{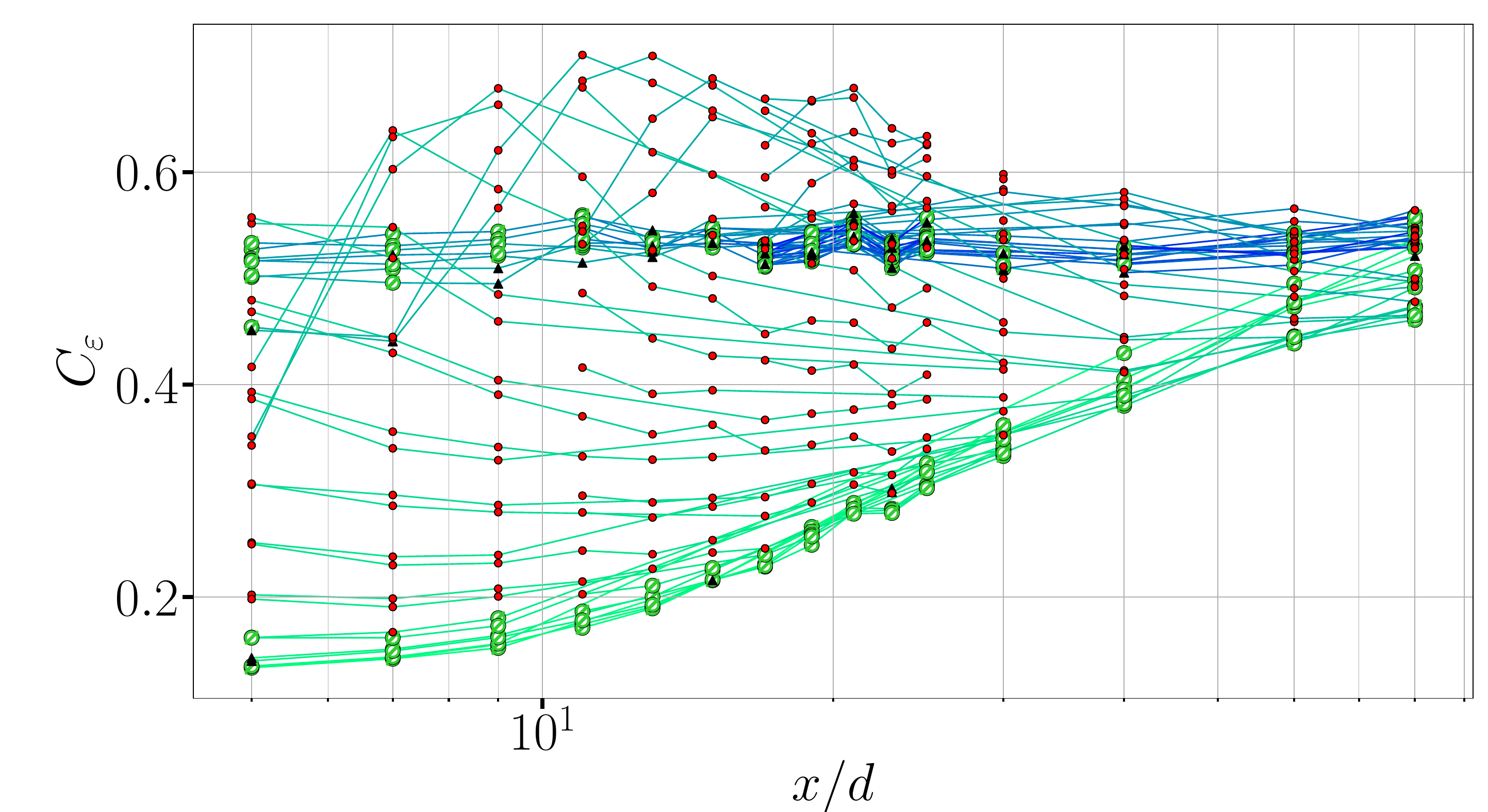}
        \caption{}
    \end{subfigure}
    \hfill
    \begin{subfigure}[t]{0.49\textwidth}
        \centering        \includegraphics[width=\linewidth]{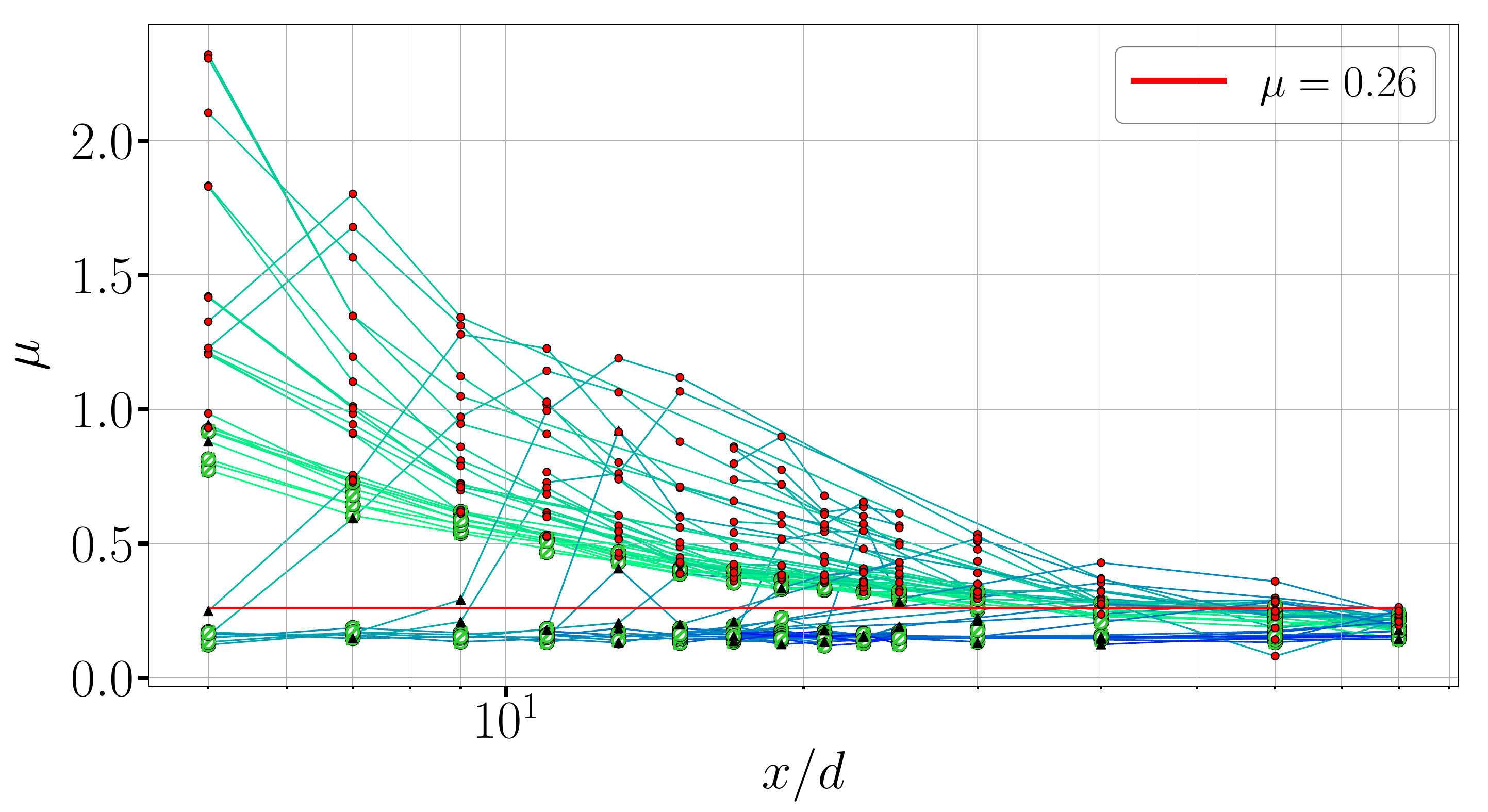}
        \caption{}
    \end{subfigure}
    \begin{subfigure}[t]{0.49\textwidth}
        \centering        \includegraphics[width=\linewidth]{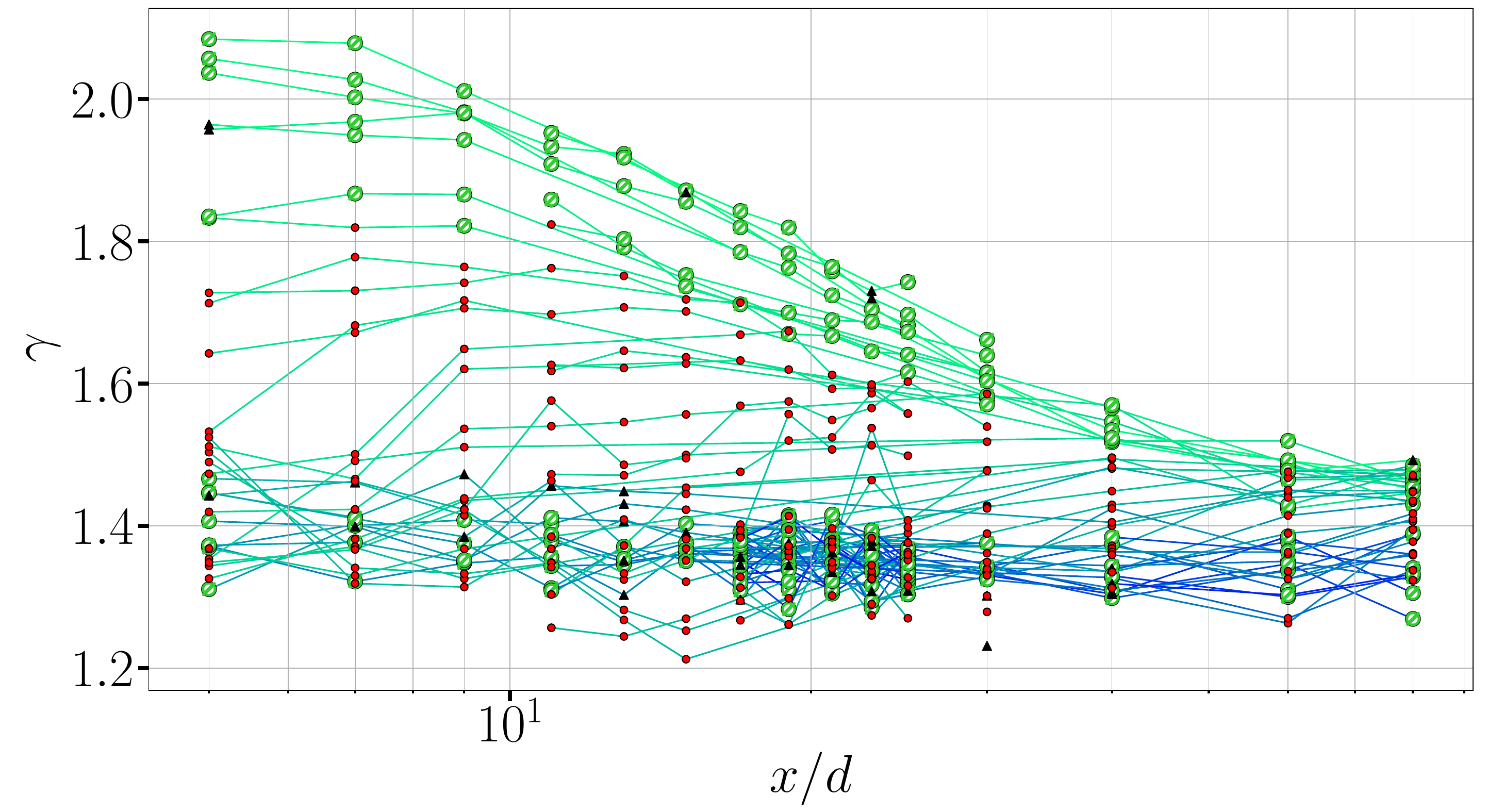}
        \caption{}
    \end{subfigure}
    \hfill
    \begin{subfigure}[t]{0.49\textwidth}
        \centering        \includegraphics[width=\linewidth]{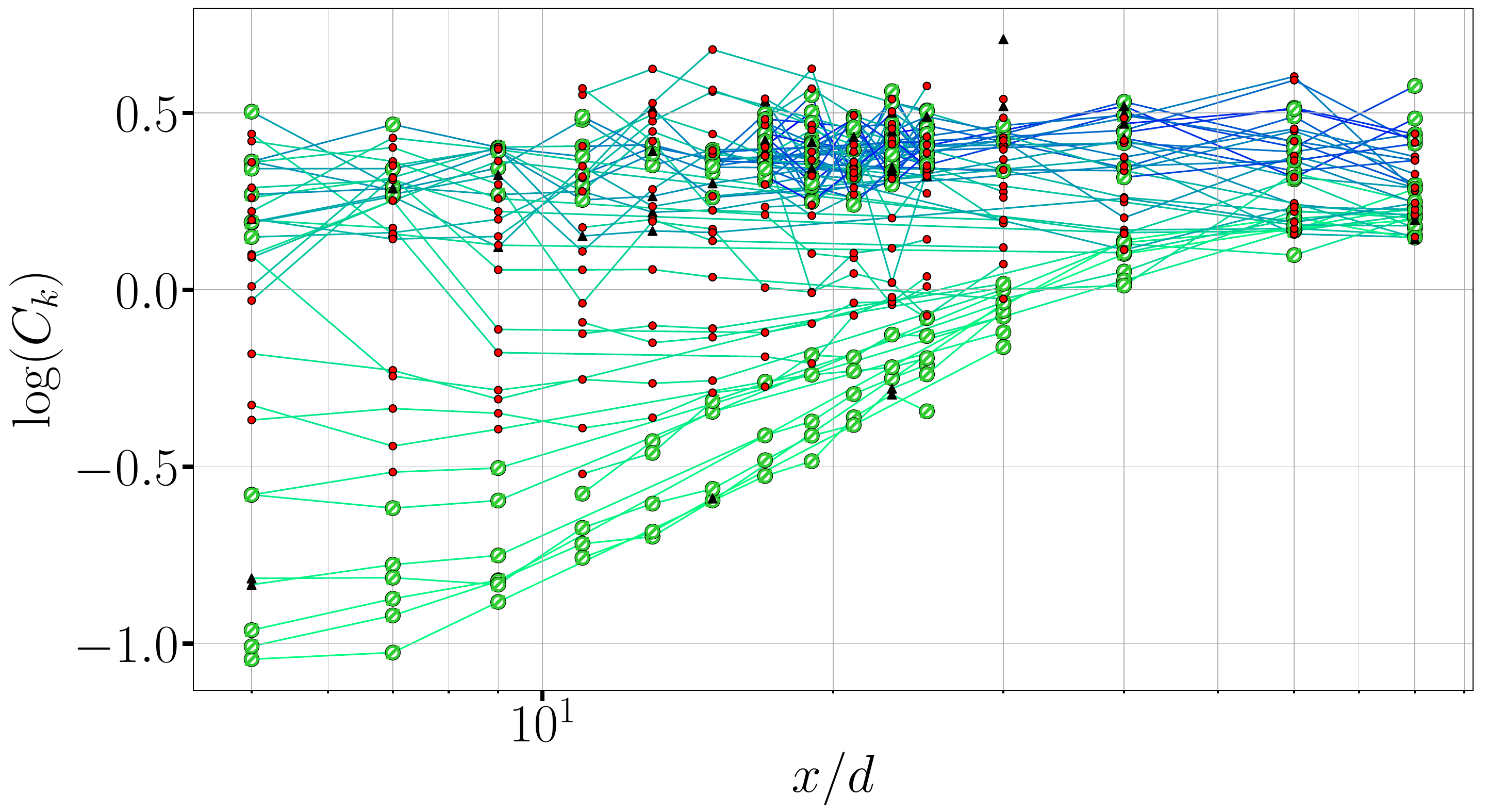}
        \caption{}
    \end{subfigure}
    \caption{{$Re_\lambda$, $TI$, $C_\varepsilon$, $\mu$, $\gamma$, and $C_k$ are shown as a function of $x/d$ for one representative case (C4), compare figure~\ref{data_selection} b). The green markers (the markers are identical to all other representations of the case C4) indicate the VTS that were retained for further analysis, whereas the red markers correspond to VTS discarded based on the value of $\Lambda^2_0$. The black markers correspond to VTS discarded based on other selection criteria outlined above. Note that the values of $Re_\lambda$, $C_\varepsilon$, $\mu$, $\gamma$, and $C_k$ associated with the discarded (red and black) markers were not individually examined for methodological reliability. The curves connecting the points correspond to distinct spanwise positions, and the color denotes the normalized absolute distance from the centerline, $|\Delta y|/|\Delta y|_{max}$. In d) the red line indicates a commonly accepted value of $\mu$ for HIT~\cite{arneodo1996structure}. Further details regarding case C4 are provided in table~\ref{tab:PhD measurements in LEGI 2023} in the appendix~\cite{SM}.}}
    \label{figure_support_color_along_x}
\end{figure}

\new{To examine which data are retained or discarded for the composite quantities, figure~\ref{figure_support_alpha_beta_phi_color_along_x} a), b), and c) presents the streamwise evolution of $\alpha$, $\beta$, and $\phi$, respectively, in the same manner as figure~\ref{figure_support_color_along_x}.}

\new{For $\beta$ and $\phi$, the discarded values largely remain within the range spanned by the retained data, particularly for VTS excluded because of their $\Lambda_0^2$ values. A markedly different behavior is observed for $\alpha$. Almost all data discarded based on $\Lambda_0^2$ lie above the retained values, in some cases by a considerable margin. Note that the axis for $\alpha$ is logarithmic, emphasizing the magnitude of these deviations. This is a remarkable result, as it indicates that the selection criterion $\Lambda_0^2 > 0.005$ has a pronounced effect primarily on $\alpha$, whereas $\beta$ and $\phi$ remain comparatively unaffected. The strong deviations in $\alpha$ can be attributed mainly to the elevated values of the intermittency parameter $\mu$ associated with these discarded VTS.}

\begin{figure}[htbp]
    \centering
    \begin{subfigure}[t]{0.49\textwidth}
        \centering        \includegraphics[width=\linewidth]{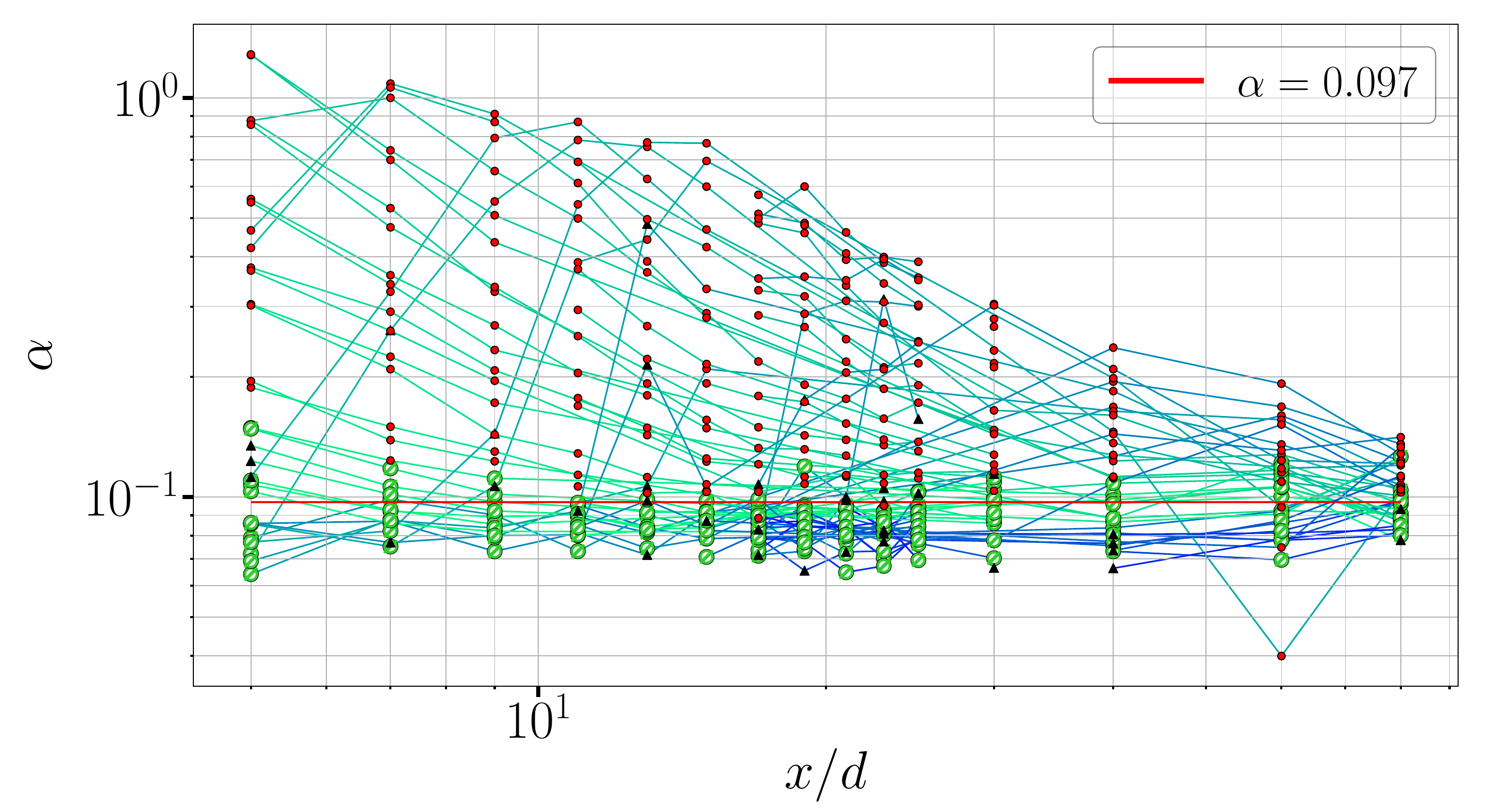}
        \caption{}
    \end{subfigure}
    \hfill
    \begin{subfigure}[t]{0.49\textwidth}
        \centering        \includegraphics[width=\linewidth]{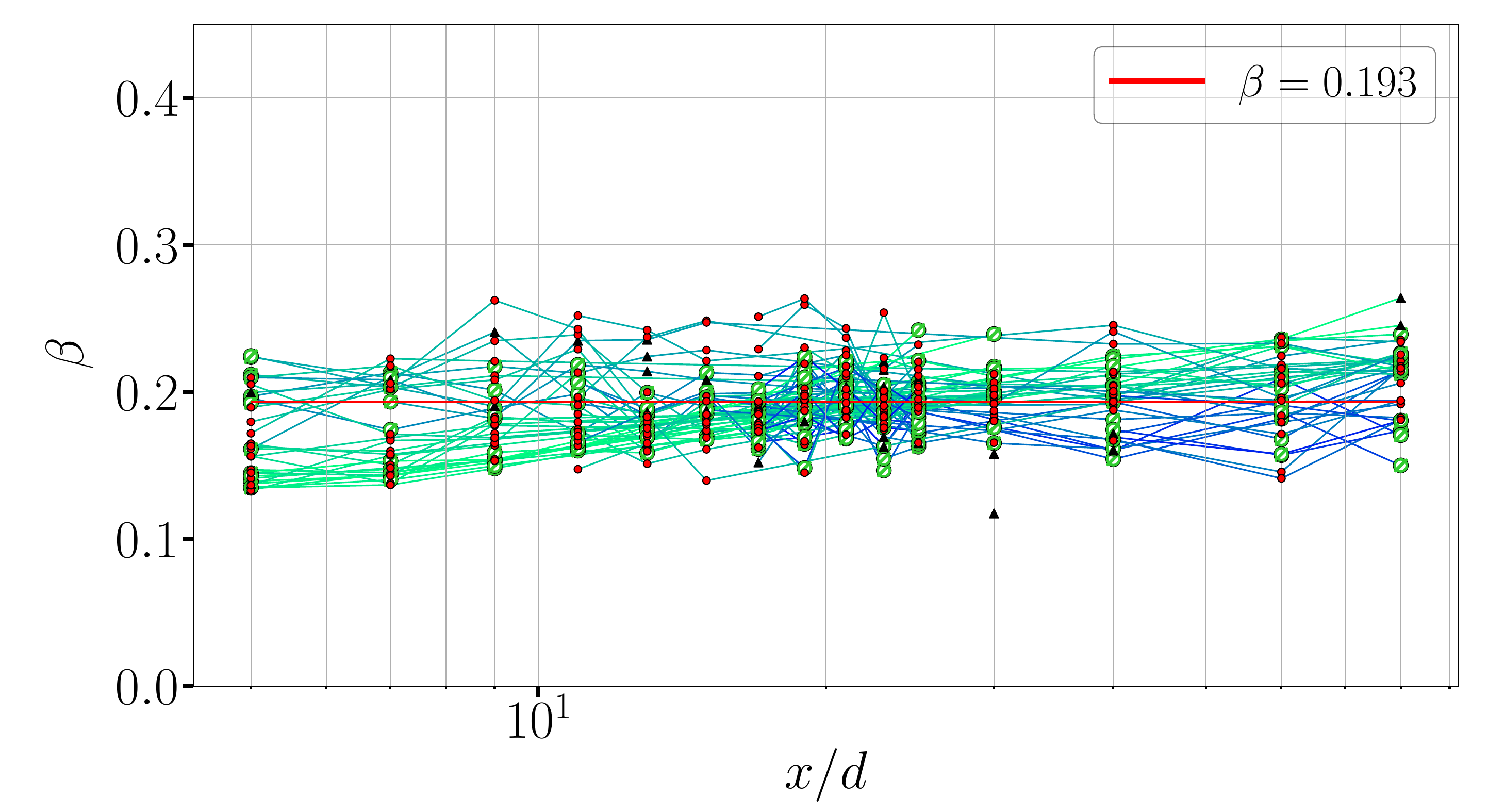}
        \caption{}
    \end{subfigure}
    \begin{subfigure}[t]{0.49\textwidth}
        \centering        \includegraphics[width=\linewidth]{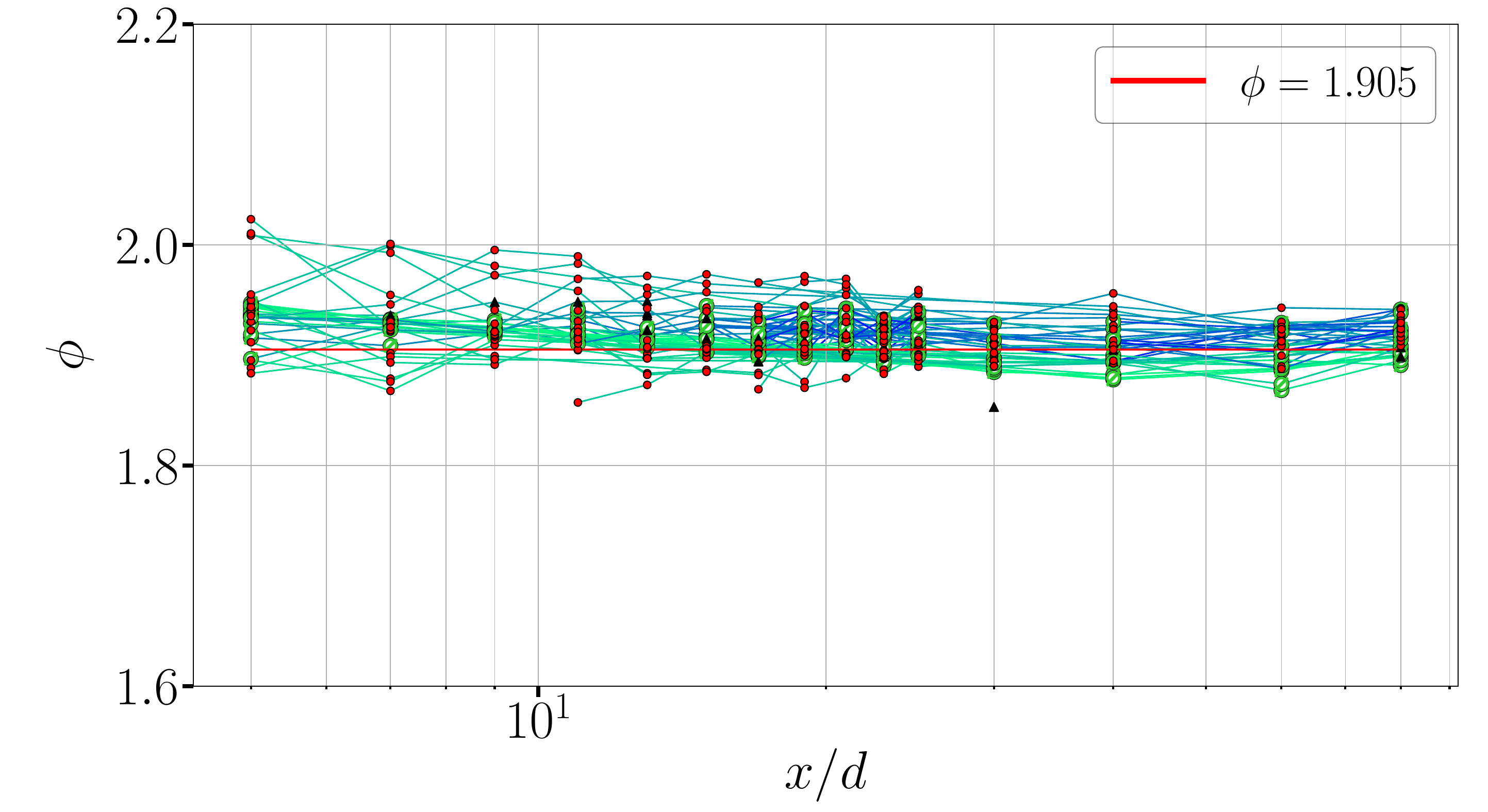}
        \caption{}
    \end{subfigure}
    \caption{\new{$\alpha$, $\beta$, and $\phi$ are shown as a function of $x/d$ for one representative case (C4). Data and presentation style like in figure~\ref{figure_support_color_along_x}.}}
    \label{figure_support_alpha_beta_phi_color_along_x}
\end{figure}

\FloatBarrier

\new{Up to this point, the overall spatial evolution of the flow has been illustrated using a single case. For this case, the retained data are characterized by decreasing $Re_\lambda$ and $TI$ with increasing streamwise position and therefore correspond to decaying turbulence. Overall, most cases exhibit qualitatively similar streamwise behavior. Remarkably, however, several cases deviate noticeably from this general trend, although the data retained from these cases still satisfy relations~(\ref{law_1}, \ref{law_2}, and \ref{law_3}). We therefore provide a table in the appendix~\cite{SM}, where the streamwise evolution of $Re_\lambda$, $TI$, $\overline{u}$, $u'$, $\varepsilon$, $L$, $\Lambda_0^2$, $\mu$, $C_\varepsilon$, and $\gamma$ of the retained data is presented for all cases. Additionally, the table provides for each case the portion of data that was retained, that was discarded due to the value of $\Lambda^2_0$, and that was discarded due to the fact that the VTS did not exhibit all signatures of turbulence.}

\new{If one considers only the results for the SST data (green symbols), we see again that also the downstream position does not have an influence on the values  $\alpha$, $\beta$, and $\phi$.  }

\FloatBarrier

\section{\new{Discussion}}
\label{discussion}

\new{This chapter provides a more detailed discussion of several aspects introduced in the preceding chapters, with particular emphasis on the newly obtained results. First, the role of the conventional control parameters is examined in further depth. The robustness of the identified relations is then summarized and critically assessed. Based on these relations, a forward-estimation approach is subsequently introduced, demonstrating how several turbulence parameters can be inferred from knowledge of $C_\varepsilon$ alone.}

\new{Possible physical interpretations of the observed relations are discussed next. This discussion is accompanied by a justification of why the conceptual framework developed for HIT can, from an experimental perspective, be extended to the SST turbulent states considered here. Finally, these considerations lead to the Kolmogorov--Castaing--Beck model, which brings together the overall approach and the principal findings of this work within a unified framework and provides a common perspective.} 

\FloatBarrier

\subsection{\new{Conventional Flow Parameters}}

\new{The centerline data for each individual case followed the relation proposed by Vassilicos~\cite{vassilicos2015dissipation} (shown already in~\cite{schmitt2024universal}), exhibiting a case-dependent linear relationship between $C_\varepsilon$ and the normalized Reynolds number $\sqrt{Re_G}/Re_\lambda$, where $Re_G$ denotes a Reynolds number based on the inflow velocity and the characteristic width of the turbulence generator.}

\new{Figure~\ref{christos_law} a) shows $C_\varepsilon$ as a function of the normalized Reynolds number $\sqrt{Re_G}/Re_\lambda$ for all VTS used, where the characteristic width of the turbulence generator is given by $d^*$, the characteristic width of the turbulence generator closest to the measurement location. In particular, $d^*$ is taken as $d$, the diameter of the disk, cylinder or nozzle, for cases where one of these elements was used. For grid turbulence cases, $d^*$ corresponds to the mesh size $M$.}

\begin{figure}[htbp]
    \centering
    \begin{subfigure}[t]{0.49\textwidth}
        \centering        \includegraphics[width=\linewidth]{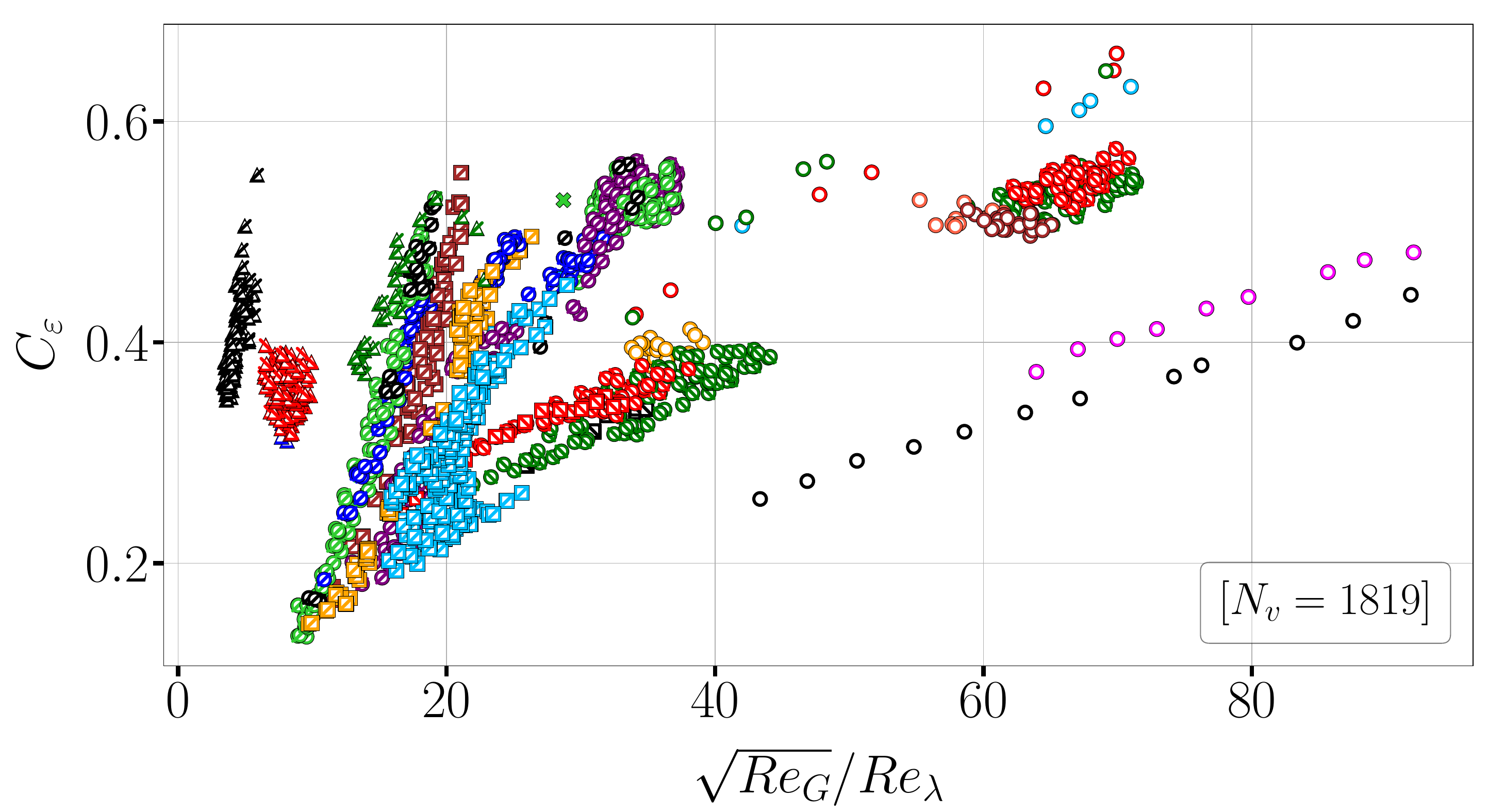}
        \caption{}
    \end{subfigure}
    \hfill
     \begin{subfigure}[t]{0.49\textwidth}
        \centering        \includegraphics[width=\linewidth]{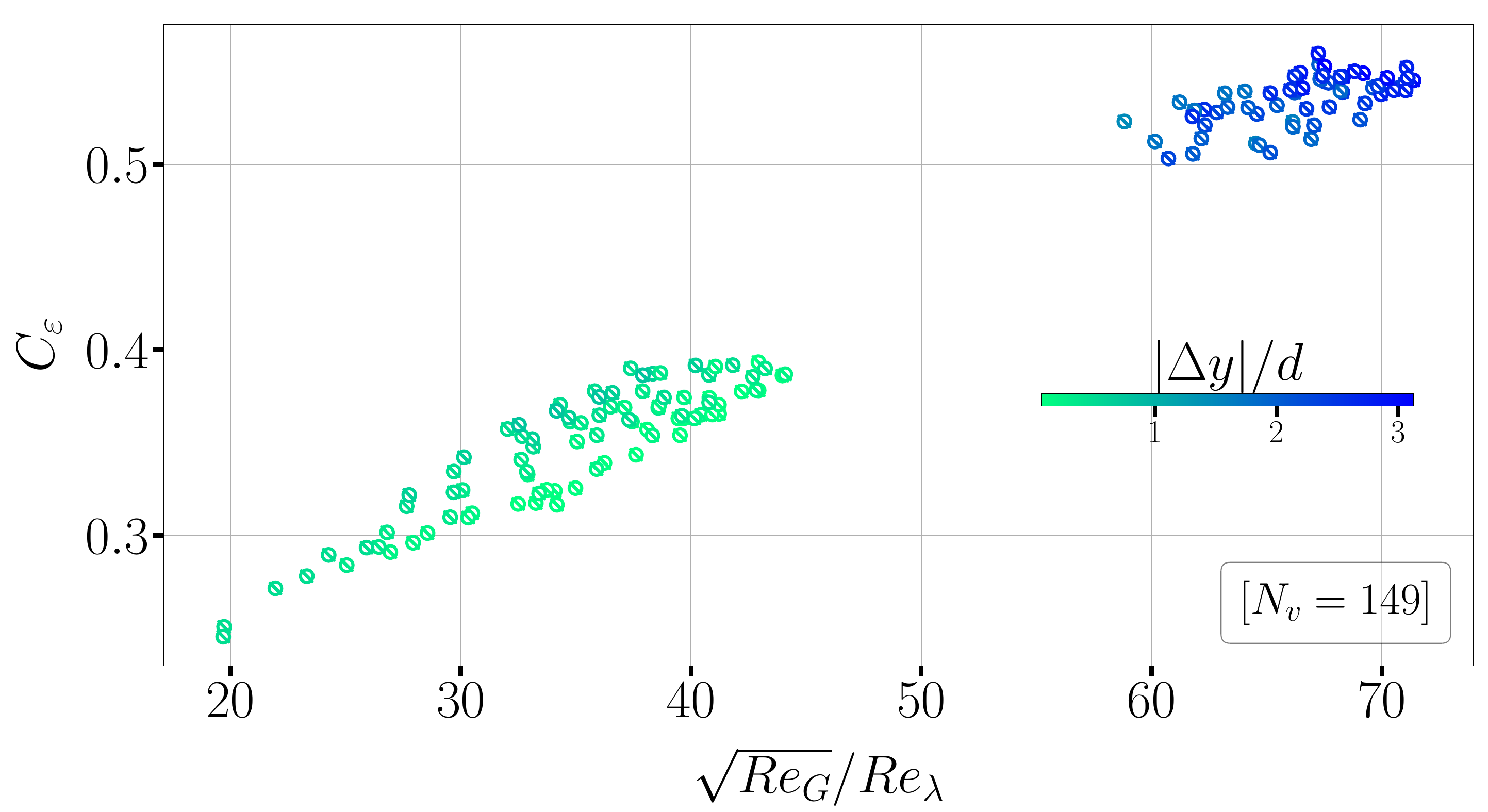}
        \caption{}
    \end{subfigure}
    \caption{\new{a) $C_\varepsilon$ as a function of $\sqrt{Re_G}/Re_\lambda$ for all VTS used. b) $C_\varepsilon$ as a function of $\sqrt{Re_G}/Re_\lambda$ for all VTS of case D11. The color indicates the absolute spanwise distance from the centerline normalized by the diameter of the wake generating object. In general, $N_v$ indicates the number of VTS shown in the plot. The symbols and corresponding configurations are shown and explained in table~\ref{tab:PhD measurements in LEGI 2023} in the appendix~\cite{SM}.}}
    \label{christos_law}
\end{figure}

\new{According to Vassilicos~\cite{vassilicos2015dissipation}, the dissipation parameter $C_{\varepsilon}$ should have a linear relation to ${{Re_G}^{m_1}}/{{Re_\lambda}^{m_2}}$, where the free exponents $m_1$ and $m_2$ take the values $1/2$ and $1$, respectively. These suggested values were used. Notably, the proposed scaling not only works for centerline data but also for out-of-centerline data. The presence of two integral length scales (for instance inflow and cylinder) is not considered in Vassilicos' formalism and thus, further work is needed to for a proper extension of the formalism.}

\new{Interestingly, figure~\ref{christos_law} a) also reveals a clear clustering within the individual flow cases, marked by common symbols. This arises from situations where background turbulence (from a grid) and a wake (from an object) coexist, each characterized by distinct boundary conditions. To emphasize this, figure~\ref{christos_law} b) presents the same plot as a), however for a single case (D11) and supplemented with a color map indicating $|\Delta y|/d$, the spanwise distance from the centerline, normalized with the diameter of the wake generating object. $|\Delta y|/d$ explains the clustering, confirming that varying boundary conditions lead to different behaviors with respect to the linear relation between $C_{\varepsilon}$ and $\sqrt{Re_G}/Re_\lambda$ according to Vassilicos~\cite{vassilicos2015dissipation}.  From this we conclude that our results are confirming results on turbulent flows, which were worked out well. We see that these hold for single flow cases but do not collapse on a simple common behavior for our SST data of several different flow cases. This is a clear difference to our new relations reported here.}

\subsection{\new{Robustness of the New Relations}}

\new{It has been shown that, to first-order approximation, the three introduced constants $\alpha$, $\beta$, and $\phi$ are independent of the Reynolds number , turbulence intensity and the downstream location. Furthermore the results are independent of the method used for their estimation. This is consistent with their intended purpose of providing a coarse characterization of SST. Nevertheless, several systematic second-order effects (referring to the scattering in the range of 10 to 20~$\%$) remain visible and are examined in more detail in this chapter.}

{Starting with the overall distributions of the data in figure~\ref{figure_law_1} a) and figure~\ref{figure_law_2} a) exhibit slightly different shapes, even though both are described by the same functional dependence, namely an inverse proportionality of the form $1/x$. Moreover, the distribution $p(\alpha)$ shown in figure~\ref{figure_law_1} c) appears closer to a lognormal than to a Gaussian distribution. If the deviations in $\mu$ and $C_\varepsilon$ were dominated by purely random fluctuations, a more Gaussian distribution would be expected. The observed lognormal-like shape therefore points instead toward systematic or structurally organized deviations. Furthermore, figure~\ref{figure_alpha_beta_phi_reynolds} c) and d) indicate that, although $\beta$ remains approximately constant with respect to $Re_\lambda$, individual subsets exhibit a weak tendency for $\beta$ to decrease as $Re_\lambda$ increases. Overall, these second-order effects become particularly apparent whenever either $\mu$ or $\gamma$ is involved. The following analysis therefore provides further evidence for the possible origins of these residual systematic dependencies.

\new{To illustrate how systematic errors in the estimation of $\gamma$ may contribute to second-order variations in $\beta$, figure~\ref{figure_support_second_order_effects_gamma} a) shows $\beta$ as a function of $\gamma-1$ for all VTS considered. Although the leading-order behavior is described by $\beta=\mathrm{const.}$, panel a) reveals additional systematic deviations from this relation. The most pronounced behavior is observed for case D13, corresponding to a disk wake subjected to passive-grid-generated inflow.}

\begin{figure}[htbp]
    \centering
    \begin{subfigure}[t]{0.49\textwidth}
        \centering        \includegraphics[width=\linewidth]{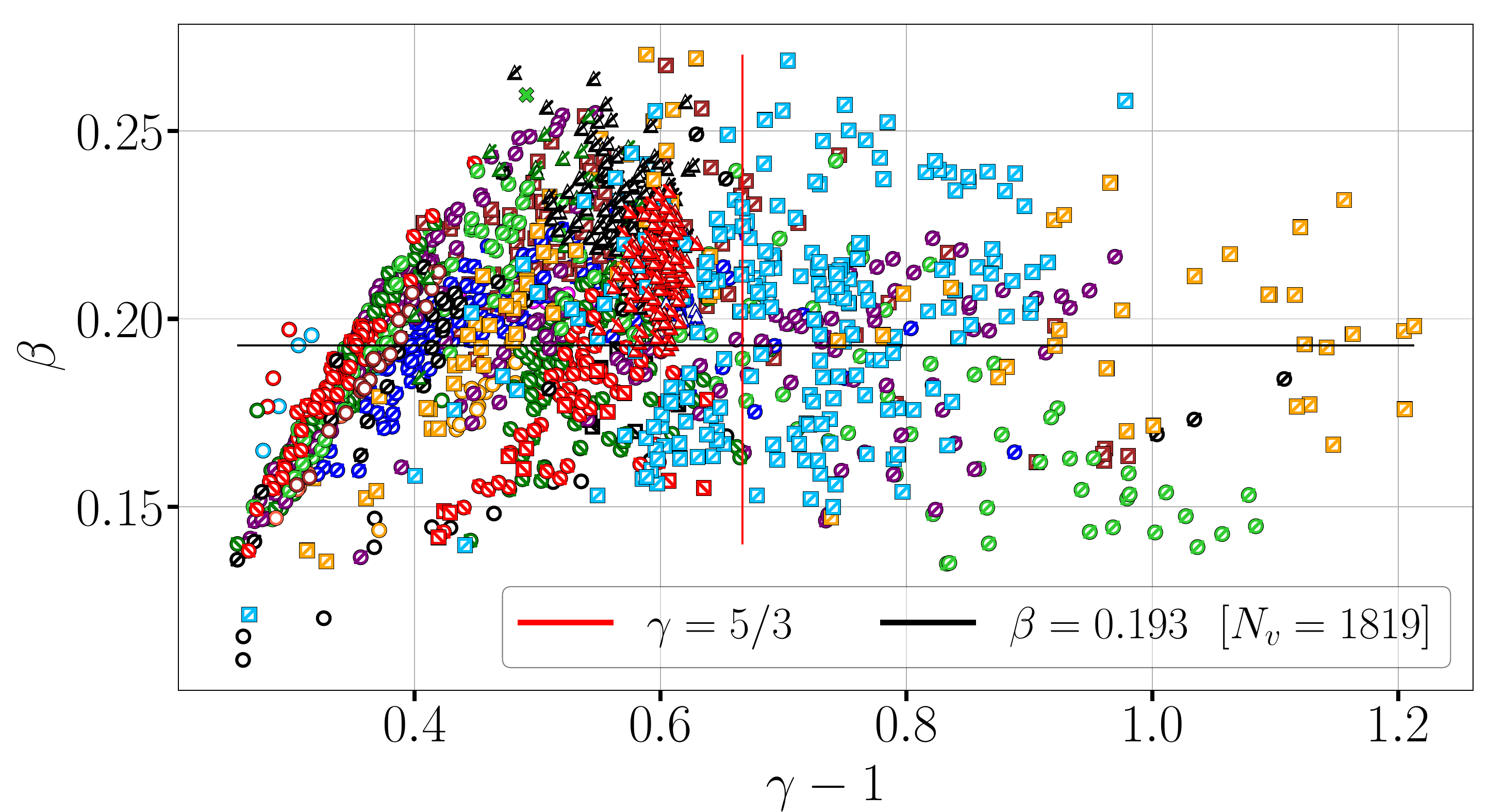}
        \caption{}
    \end{subfigure}
    \hfill
    \begin{subfigure}[t]{0.49\textwidth}
        \centering        \includegraphics[width=\linewidth]{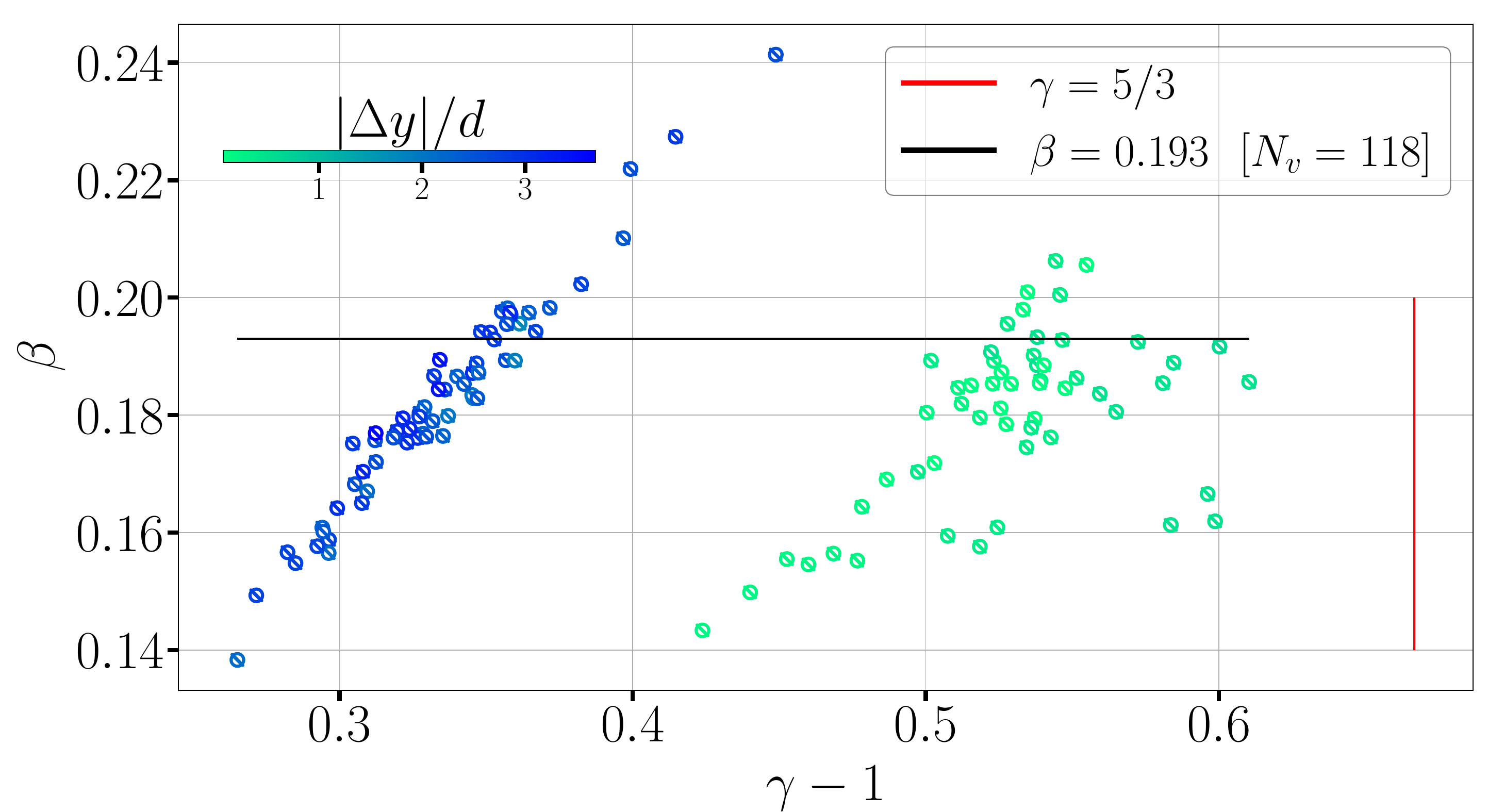}
        \caption{}
    \end{subfigure}
    \begin{subfigure}[t]{0.49\textwidth}
        \centering        \includegraphics[width=\linewidth]{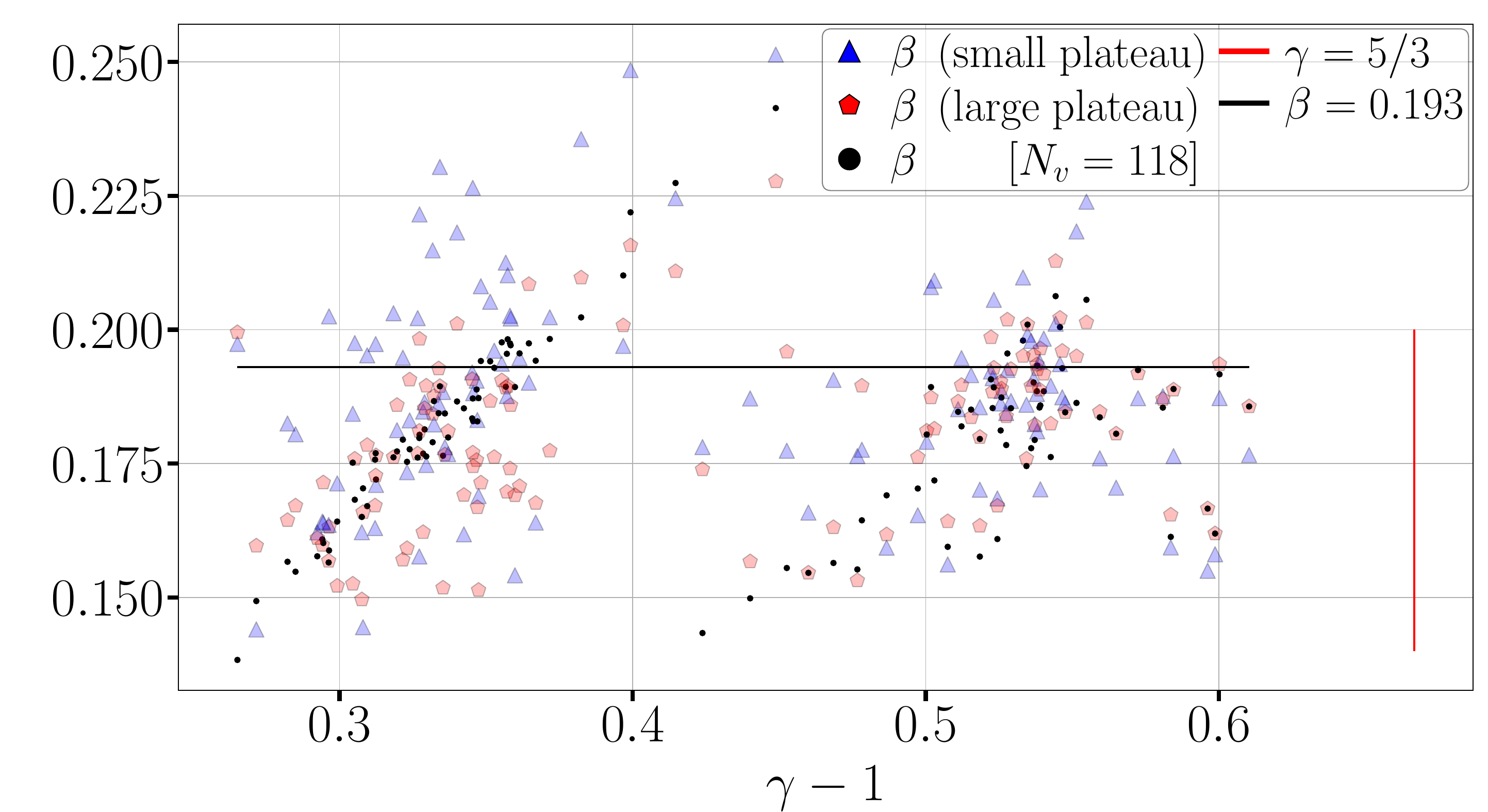}
        \caption{}
    \end{subfigure}
    \caption{\new{a) $\beta$ as a function of $\gamma-1$ for all velocity time series used. b) $\beta$ as a function of $\gamma - 1$ for the used VTS of one representative case (D13). The color indicates $|\Delta y| / d$, the absolute spanwise distance from the centerline normalized by the diameter of the wake generating object. c) different representations of $\beta$ as a function of $\gamma - 1$ for the used VTS of one representative case (D13). While the black dots represent the actual $\beta$-values shown in panel b), the blue markers show $\beta$ built by a version of $\gamma$ where the plateau threshold was smaller than usual (while value of $C_\varepsilon$ remains unchanged). Similar, the red markers represent $\beta$ built by a version of $\gamma$ where the plateau threshold was larger than usual. In general, the red lines indicate commonly accepted values for $\gamma$~\cite{sreenivasan1995universality} for HIT, and the black solid line indicates the obtained values from the least-squares fits shown in figure~\ref{figure_law_2} a). $N_v$ indicates the number of VTS shown in the plot. The symbols and corresponding configurations are shown and explained in table~\ref{tab:PhD measurements in LEGI 2023} in the appendix~\cite{SM}.}}
    \label{figure_support_second_order_effects_gamma}
\end{figure}

\new{Case D13 is therefore examined separately in figure~\ref{figure_support_second_order_effects_gamma} b), where the same quantities are displayed exclusively for this configuration. In addition, the data are color-coded according to the absolute spanwise distance from the centerline. The two distinct clusters can clearly be associated with the inner and outer regions of the flow. Importantly, both regions exhibit the same systematic deviation, indicating that the observed second-order behavior is not restricted to a particular spanwise region of the flow.}

\new{Systematic uncertainties in the estimation of $\gamma$ associated with the applied method were already discussed in figure~\ref{figure_validation_gamma} c). These uncertainties arise from the threshold used to identify the plateau in the derivative of $E(k)$ in log--log representation and, consequently, from the resulting definition of the inertial range. Figure~\ref{figure_support_second_order_effects_gamma} c) reproduces the analysis of figure~\ref{figure_support_second_order_effects_gamma} b), but now incorporates the uncertainty in $\gamma$ associated with the inertial-range definition into the corresponding values of $\beta$.}

\new{The black markers represent the original estimates shown in figure~\ref{figure_support_second_order_effects_gamma} b), whereas the blue and red markers indicate the values of $\beta$ that would result if $\gamma$ were estimated using alternative threshold values. The resulting spread shows that variations in the inertial-range definition can partially blur the previously observed systematic deviations. It therefore remains difficult to determine unambiguously whether these second-order effects are entirely of physical origin or partly introduced by the estimation procedure. Nevertheless, the analysis indicates that at least a fraction of the observed deviations can plausibly be attributed to methodological influences.}

\FloatBarrier

\new{As seen in figure~\ref{figure_support_second_order_effects_gamma} a), the method-related systematic effects discussed above primarily concern the lower values of $\gamma$. In contrast, the discrepancy between the overall shapes of the data in figure~\ref{figure_law_1} a) and figure~\ref{figure_law_2} a) appears to originate mainly from the upper range of $\gamma$, and correspondingly from larger values of $\mu$. To investigate this behavior in greater detail, figure~\ref{figure_support_detrended_gamma} presents $\gamma-1$ as a function of $\mu$.}

\begin{figure}[htbp]
    \centering
    \begin{subfigure}[t]{0.49\textwidth}
        \centering        \includegraphics[width=\linewidth]{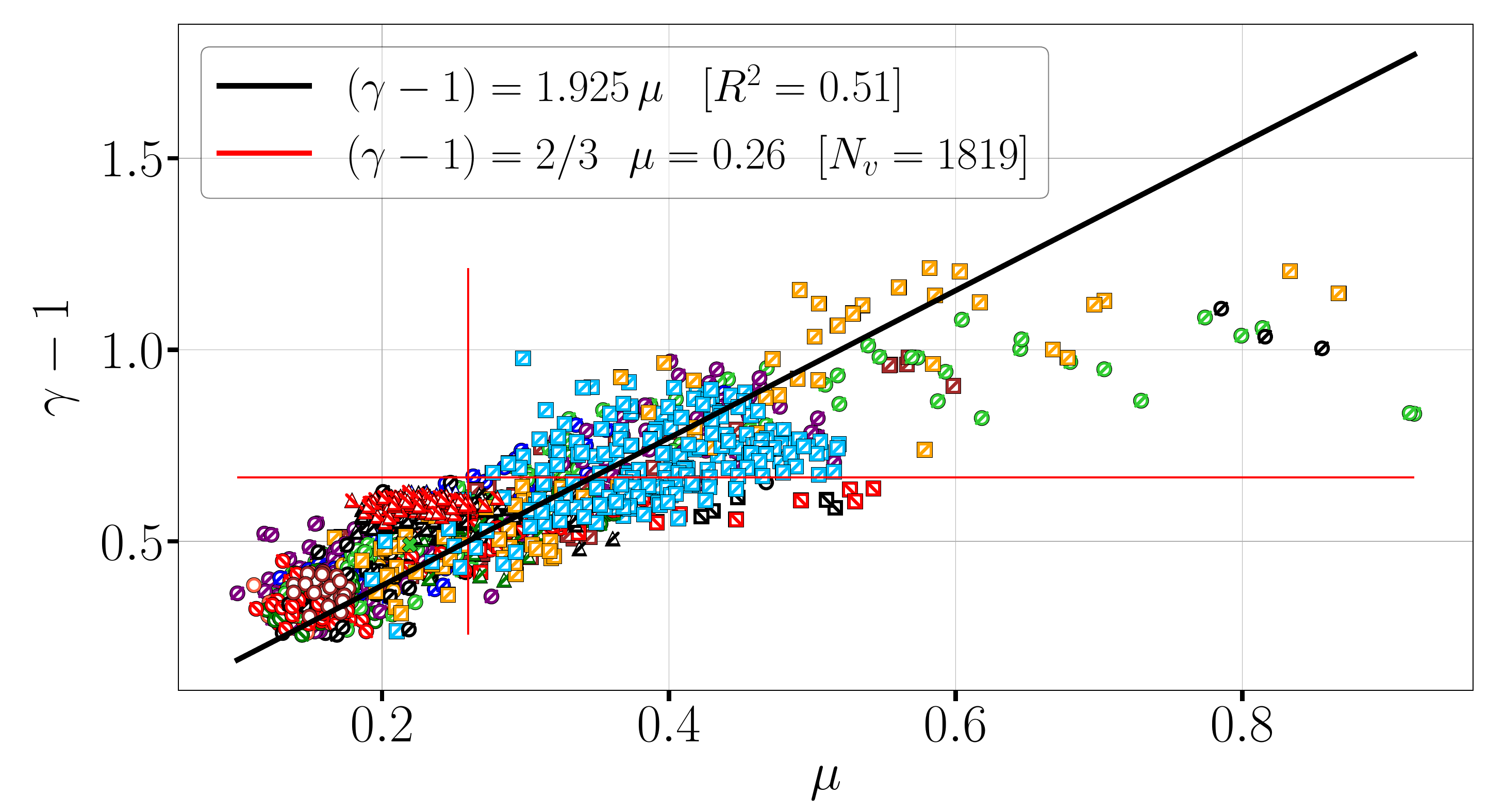}
        \caption{}
    \end{subfigure}
    \caption{\new{$\gamma - 1$ as a function of $\mu$ for all VTS used. The black line corresponds to a least-squares fit. In general the red lines indicate the commonly accepted values for $\gamma$ and $\mu$ for HIT~\cite{sreenivasan1995universality, arneodo1996structure}, while $R^2$ being the coefficient of determination. $N_v$ indicates the number of VTS shown in the plot. The symbols and corresponding configurations are shown and explained in figure~\ref{figure_validation_L} a) and table~\ref{tab:PhD measurements in LEGI 2023} in the appendix~\cite{SM}.}}
    \label{figure_support_detrended_gamma}
\end{figure}

\new{Eqs.~(\ref{equation_law_1}) and~(\ref{equation_law_2}) together imply
$\gamma - 1 = (\beta/\alpha) \cdot \mu$, with $\beta/\alpha \approx 2$. Accordingly, the data in figure~\ref{figure_support_detrended_gamma} are fitted by a linear relation. For $\mu<0.5$, the data follow this relation closely, with a fitted slope near the expected value of 2. In contrast, the data points for $\mu>0.5$ exhibit a coherent deviation from the linear trend. This systematic departure accounts for the previously observed difference in the overall data shapes between figure~\ref{figure_law_1} a) and figure~\ref{figure_law_2} a).}

\new{Interestingly, the data points that deviate from the linear trend in figure~\ref{figure_support_detrended_gamma} a) are precisely those for which coherent structures can be identified. These structures appear as noticable peaks at large scales in $E(k)$ and $\Lambda^2(r)$ in figures~\ref{figure_eight_examples} f) and g), as well as through oscillatory behavior of $R_{uu}(r)$ in figures~\ref{figure_eight_examples_L} f) and g). Moreover, we have shown that the level of $\Lambda_0^2$ plays an important role in determining whether SST is established. In particular, figures~\ref{figure_support_color_along_x} d) and e) showed that data discarded because of $\Lambda_0^2>0.005$ tend toward larger values of $\mu$ and smaller values of $\gamma$.}

\new{If individual coherent structures produce a qualitatively similar, but considerably weaker, influence than a general elevation of $\Lambda^2(r)$ at the large scales, this could provide a possible explanation for the systematic behavior observed in figure~\ref{figure_support_detrended_gamma}. The correspondence between the occurrence of coherent structures and the observed deviations is therefore suggestive of such a connection. However, the present analysis does not provide sufficient evidence to establish a causal relation, and this interpretation should consequently be regarded as a hypothesis rather than a demonstrated mechanism.}

\FloatBarrier

\new{The systematic deviations of $\alpha$, $\beta$, and $\phi$ have now been discussed, with particular attention to the systematic uncertainties associated with the estimation of $\gamma$ and $\mu$. To conclude the analysis of second-order effects, we compare in figures~\ref{figure_support_alpha_beta_phi} a), b), and c) the values of $\alpha$, $\beta$, and $\phi$ pairwise.} 

\new{The corresponding data are represented as contour plots, with the contour levels indicating the local data density and thereby making the consequences of the systematic deviations discussed above visible. Figures~\ref{figure_support_alpha_beta_phi} a) and b) exhibit approximately circular contour patterns, (respectively,  symmetic contour with respect to the x and y axis,)
which is consistent with pairwise independence to first-order approximation. Nevertheless, small distortions from this idealized shape remain visible. In both panels, the quantity represented on the horizontal axis, \textit{e.g.} $\alpha$, involves $\mu$, whereas that on the vertical axis involves $\gamma$, such that the previously identified systematic effects associated with these parameters can propagate into the contour shapes.}

\new{Figure~\ref{figure_support_alpha_beta_phi} c) exhibits a somewhat more pronounced distortion with a double structure tilted from the horizontal axis. Interestingly, in this case both quantities, $\beta$ and $\phi$, contain $\gamma$, which is consistent with the second-order deviations, associated with its estimation, as shown in figure~\ref{figure_support_second_order_effects_gamma}. Thus, despite the observable second-order distortions, $\alpha$, $\beta$, and $\phi$ remain approximately independent of one another at leading order.}

\begin{figure}[htbp]
    \centering
    \begin{subfigure}[t]{0.49\textwidth}
        \centering        \includegraphics[width=\linewidth]{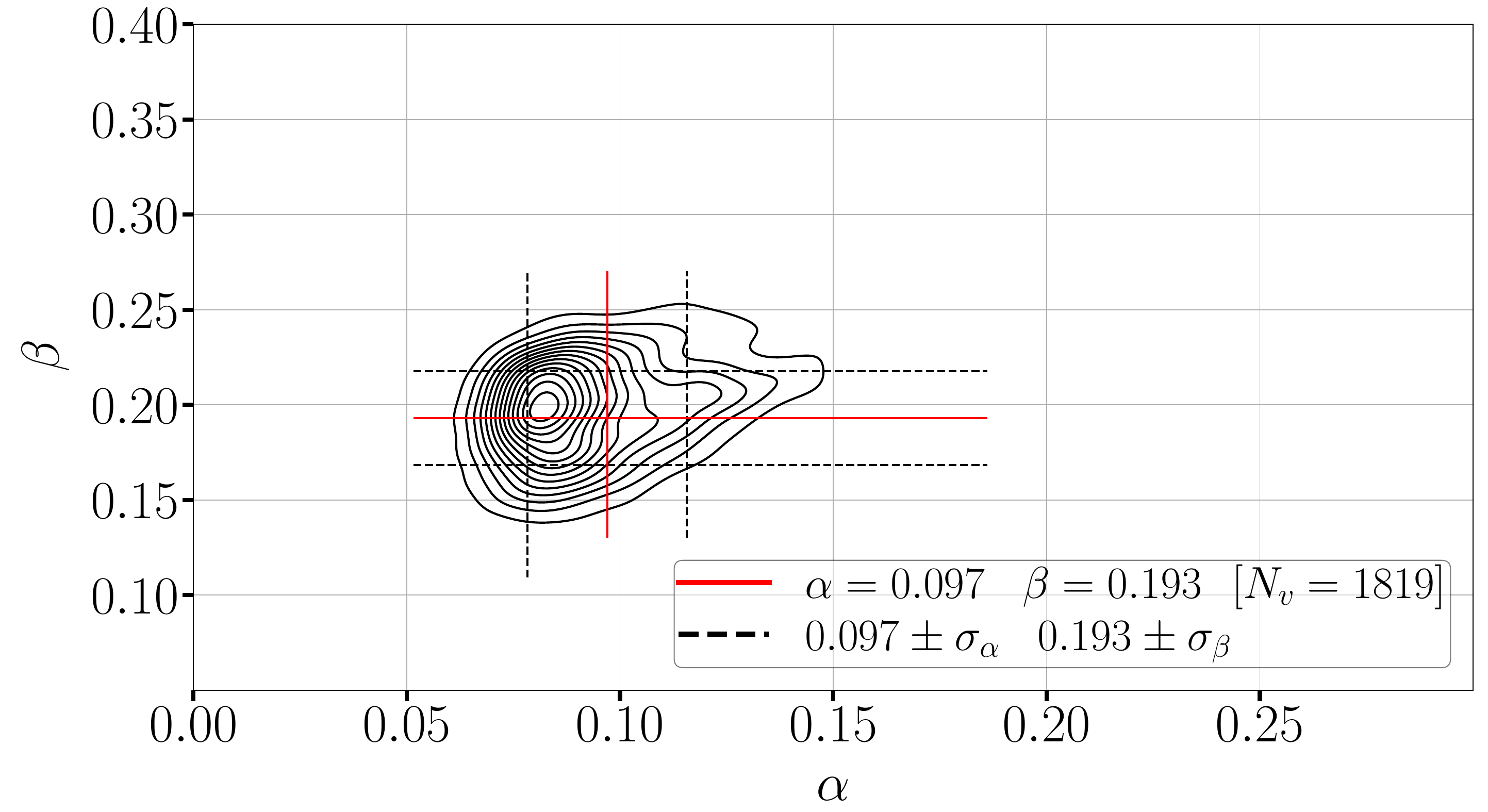}
        \caption{}
    \end{subfigure}
    \hfill
    \begin{subfigure}[t]{0.49\textwidth}
        \centering        \includegraphics[width=\linewidth]{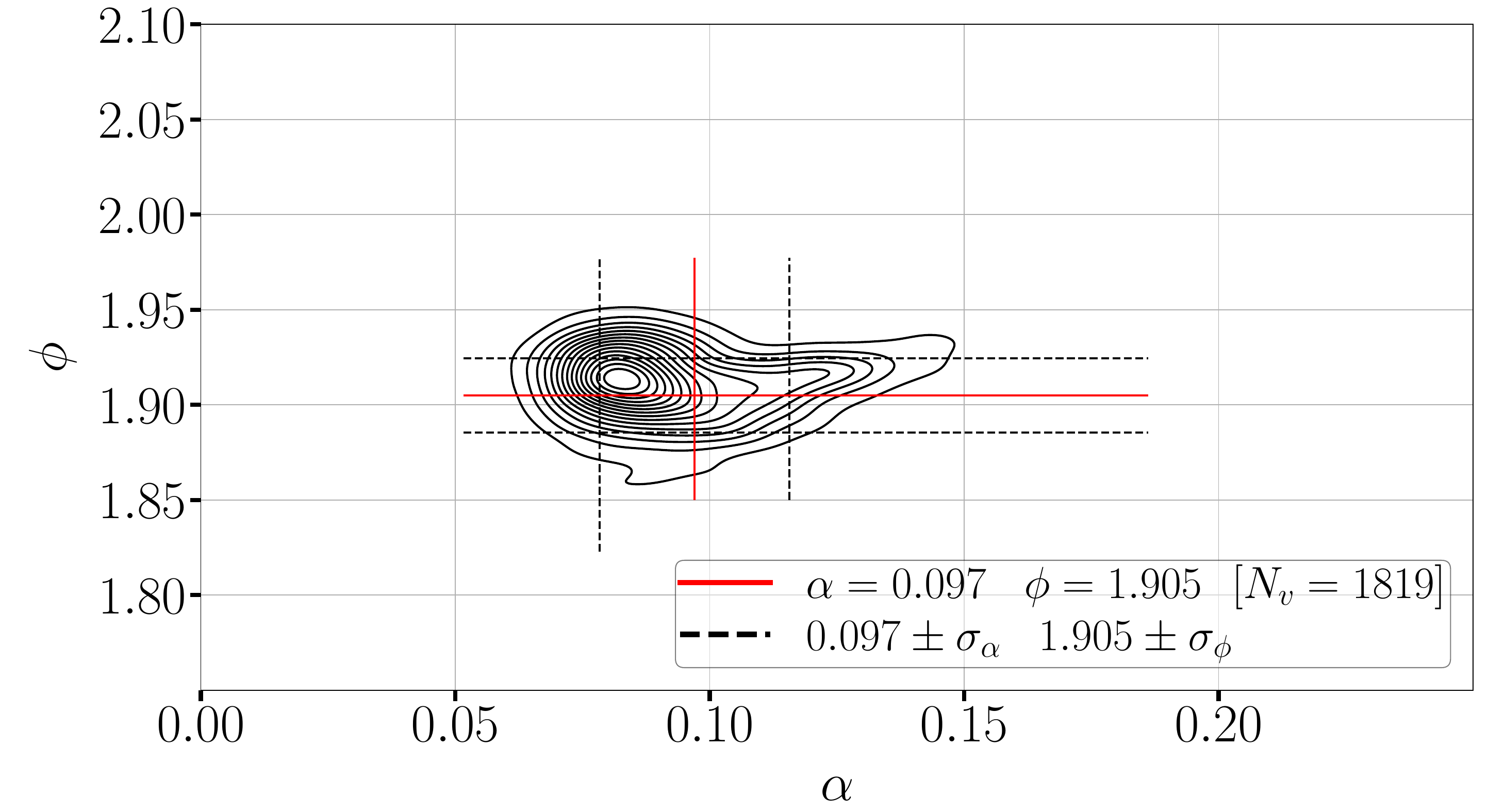}
        \caption{}
    \end{subfigure}
    \begin{subfigure}[t]{0.49\textwidth}
        \centering        \includegraphics[width=\linewidth]{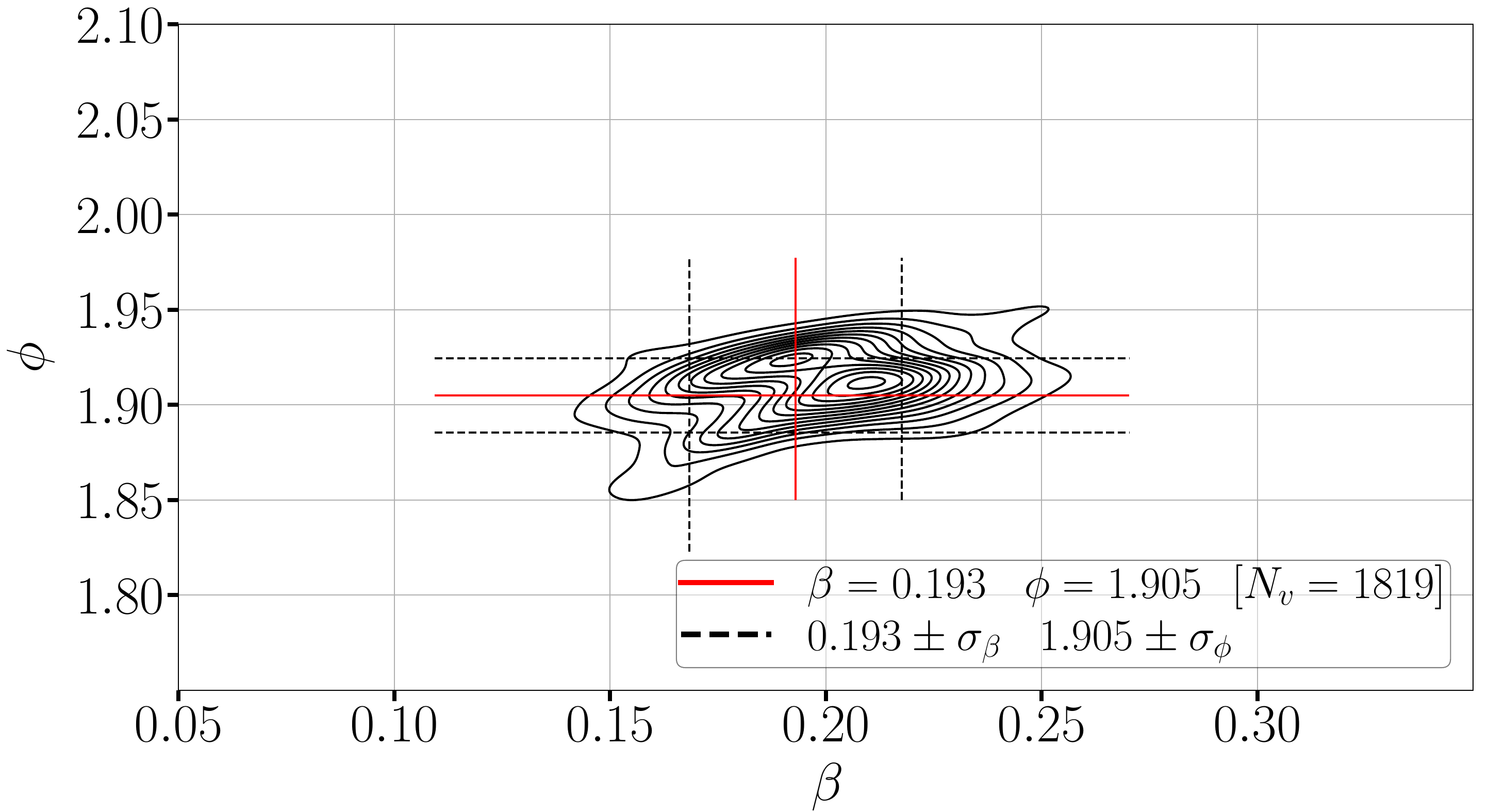}
        \caption{}
    \end{subfigure}
    \caption{\new{Contour plots showing the probability density of a) $\beta$ as a function of $\alpha$, b) $\phi$ as a function of $\alpha$, and c) $\phi$ as a function of $\beta$ for all VTS used. In general, the red lines indicate the obtained values from the least-squares fits shown in figures~\ref{figure_law_1} a), \ref{figure_law_2} a), and \ref{figure_law_3} a). The black dashed lines represent the standard deviation from the fit result and the $N_v$ indicates the number of VTS shown in the plot.}}
    \label{figure_support_alpha_beta_phi}
\end{figure}

\FloatBarrier

\subsection{\new{Estimation of $\mu$, $\gamma$, and $C_k$ from $C_\varepsilon$ Using New Relations}} 
\label{forward}

\new{Next, we will consider some more practical applications. It is important to bear in mind that the dimensionless coefficients  $\mu$, $\gamma$,  $C_k$, and  $C_\varepsilon$  provide a detailed characterisation of a turbulent VTS. These coefficients determine the power spectrum, intermittency, including velocity increment statistics, and energy dissipation. Based on the presented relations between these coefficients, we showed that they are all interchangeable with knowledge of the parameters $\alpha$, $\beta$, $\theta$ and $\phi$,  which appear to have constant universal values. In the following part, we demonstrate how accurate our estimations are using the empirically determined values in table~\ref{table_estimated_values}. We proceed under the assumption that only $C_\varepsilon$ is known.
}

\begin{table}[h]
    \centering
    \begin{tabular}{lcccc}
        \toprule
        parameter & $\alpha$ & $\beta$ & $\theta$ & $\phi$ \\
       \midrule
        estimated value & 0.097 & 0.193 & 2.99 & 1.905 \\
        \bottomrule
    \end{tabular}
    \caption{\new{The values of the parameters $\alpha$, $\beta$, $\theta$ and $\phi$, obtained by least-squares fits, are presented. For uncertainties, see~figure~\ref{figure_alpha_beta_phi_reynolds}}}.
    \label{table_estimated_values}
\end{table}

\new{We will validate our relations by applying a forward computation, which quantifies the differences between measured and model-predicted values, applying the found relations. The forward computation is solely based on the estimated value of $C_\varepsilon$. 
Based on $C_\varepsilon$, the estimated values of $\mu$, $\gamma$, and $C_k$ are compared with the predictions based on eqs.~(\ref{equation_law_1})–(\ref{equation_law_3}). This model validation is done for the eight example VTS from figure~\ref{figure_eight_examples} and table~\ref{table_eight_examples} in detail and additionally for all used VTS.}

\new{Table~\ref{table_eight_examples_forward_computing} shows the results for the eight VTS. Regarding the relative error that is made by comparing the measured value to the predicted value, we see that both for $\mu$ and $\gamma$ the relative error is smaller than $12.9\,\%$ while the average relative error is $7.8\,\%$ for $\mu$ and $2.0\,\%$ for $\gamma$ for the SST data (highlighted in green). For the VTS that do not fulfill the criteria both relative errors of $\mu$ and $\gamma$ are in comparison quite high with $63.6\,\%$ and $13.9\,\%$, respectively (highlighted in red).}

\begin{table} [h!]
	\centering
        \setlength{\tabcolsep}{2pt} 
        \scriptsize
	\begin{tabular}{lcccccccccccc}
		\toprule
		      & $F_u$ & ${\Lambda}^2_0$ & \cellcolor{blue!30} $C_\varepsilon$ (meas.) & $\mu$ (pred.) & $\mu$ (meas.) & $|\Delta \mu| / \mu$ [$\%$] & $\gamma$ (pred.) & $\gamma$ (meas.) & $|\Delta \gamma| / \gamma$ [$\%$] & $C_k$ (pred.) & $C_k$ (meas.) & $|\Delta C_k| / C_k$ [$\%$]\\
		\midrule
		   a)  & \cellcolor{green!50} 2.93 & \cellcolor{green!50} -0.003 & 0.63 & 0.15 & 0.15 & \cellcolor{green!50} 0.0 & 1.31 & 1.31 & \cellcolor{green!50} 0.0 & 3.17 & 2.71 & \cellcolor{green!50} 17.0\\
           \midrule
		   b) & \cellcolor{green!50} 2.81 & \cellcolor{green!50} -0.008 & 0.46 & 0.21 & 0.19 & \cellcolor{green!50} 10.5 & 1.42 & 1.44 & \cellcolor{green!50} 1.4 & 1.93 & 1.71 & \cellcolor{green!50} 12.9\\
           \midrule
		   c)  & \cellcolor{green!50} 2.94 & \cellcolor{green!50} -0.01 & 0.36 & 0.27 & 0.26 & \cellcolor{green!50} 3.8 & 1.54 & 1.6 & \cellcolor{green!50} 3.8 & 1.16 & 0.84 & \cellcolor{green!50} 38.1\\
           \midrule
		   d)  & \cellcolor{green!50} 2.93 & \cellcolor{green!50} -0.003 & 0.34 & 0.29 & 0.26 & \cellcolor{green!50} 11.5 & 1.57 & 1.53 & \cellcolor{green!50} 2.6 & 1.01 & 1.3 & \cellcolor{green!50} 22.3\\
           \midrule
		   e) & \cellcolor{green!50} 2.84 & \cellcolor{green!50} -0.007 & 0.27 & 0.36 & 0.39 & \cellcolor{green!50} 7.7 & 1.71 & 1.74 & \cellcolor{green!50} 1.7 & 0.53 & 0.46 & \cellcolor{green!50} 15.2 \\
           \midrule
		   f) & \cellcolor{green!50} 3.02 & \cellcolor{green!50} 0.001 & 0.21 & 0.46 & 0.5 & \cellcolor{green!50} 8 & 1.92 & 1.93 & \cellcolor{green!50} 0.5 & 0.22 & 0.36 & \cellcolor{green!50} 38.9\\
           \midrule
		   g) & \cellcolor{green!50} 2.95 & \cellcolor{green!50} -0.002 & 0.16 & 0.61 & 0.7 & \cellcolor{green!50} 12.9 & 2.21 & 2.13 & \cellcolor{green!50} 3.8 & 0.06 & 0.07 & \cellcolor{green!50} 14.3\\
           \midrule
		   h) & \cellcolor{red!50} 7.81 & \cellcolor{red!50} 0.147 & 0.35 & 0.28 & 0.77 &  \cellcolor{red!50} 63.6 & 1.55 & 1.8 &  \cellcolor{red!50} 13.9 & 1.08 & 0.3 &  \cellcolor{red!50} 260.0\\
		\bottomrule
	\end{tabular}
	\caption{\new{Model validation of the proposed equations. Based on eqs.~(\ref{equation_law_1})–(\ref{equation_law_3}) and the constant values in table~\ref{table_estimated_values}, the parameters of $\mu$, $\gamma$ and $C_k$ are forward-computed for the eight example VTS used in figure~\ref{figure_eight_examples} and table~\ref{table_eight_examples}. The order of appearance of the VTS remains unchanged. The forward computation is solely based on the measured value of $C_\varepsilon$ (highlighted in blue). ``(pred.)'' denotes the predicted values while ``(meas.)'' refers to the measured ones. In addition, the relative error made by using the prediction with respect to the measured values of $\mu$, $\gamma$ and $C_k$ is presented. The data stem from the cases G20, G24, C8, D11, C6, C1, C1, G23, respectively in order of appearance. For details about their cases see table~\ref{tab:PhD measurements in LEGI 2023} in the appendix~\cite{SM}. Values of $F_u$ and $\Lambda_0^2$ highlighted in green indicate Gaussianity at the large scales and are therefore retained in the analysis. In contrast, values of $F_u$ and $\Lambda_0^2$ highlighted in red denote non-Gaussianity at the large scales and are consequently discarded.}}
	\label{table_eight_examples_forward_computing}
\end{table}

\new{ For $C_k$ the results for the SST data are similar, although definitely higher. The relative errors are between $12.9\,\%$ and $38.9\,\%$ while the average relative error is $22.7\,\%$. This behavior is expected, since the forward computation relies solely on the value of $C_\varepsilon$. Regarding eqs.~(\ref{equation_law_1})–(\ref{equation_law_3}), for the prediction of $\mu$ and $\gamma$ only one equation is needed. However, for predicting $C_k$, two equations are needed. The relative error for the VTS that does not fulfill the criteria is with $260\,\%$ also very high.} 

\new{Figure~\ref{figure_forward_computing} presents the relative errors made by the same prediction for all used VTS as a function of the $Re_\lambda$. The average relative prediction error is $16.8\,\%$, $3.5\,\%$ and $28.7\,\%$ for $\mu$, $\gamma$ and $C_k$, respectively. Thus, $\gamma$ can be predicted most accurately, the estimation of $\mu$ gives the second best result, while the computation of $C_k$ is the most unpredictable under these circumstances. Remarkably, there are no trends observable.} 

\new{Overall, reliable values of $\mu$, $\gamma$, and $C_k$ can be obtained solely from knowledge of $C_\varepsilon$, further strengthening the validity of the proposed relations. As a last comment we want to speculate on a further simplification. Looking at the obtained values of $\alpha$, $\beta$, $\theta$ and $\phi$ we see that $\beta/\alpha \approx 2$, which is equal to the value of $\phi$ within $10 \, \%$ and even more accurately to $\theta-1$. A relation of the form}

\begin{equation}
\frac{\beta}{\alpha} = \phi = \theta-1
\end{equation}
 
would further simplify the overall framework.

\FloatBarrier

\begin{figure}[htbp]
    \centering
    \begin{subfigure}[t]{0.49\textwidth}
        \centering        \includegraphics[width=\linewidth]{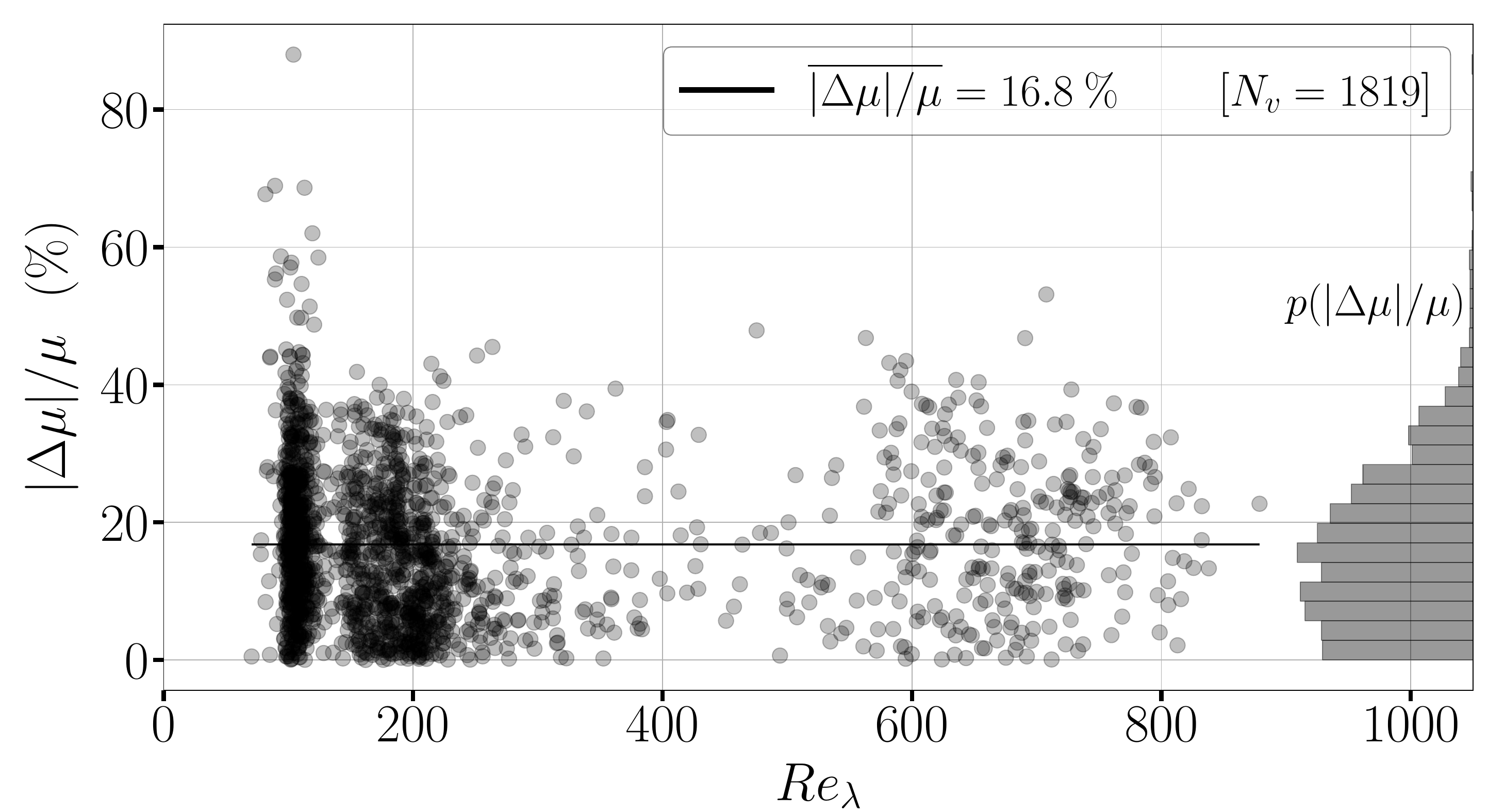}
        \caption{}
    \end{subfigure}
    \hfill
    \begin{subfigure}[t]{0.49\textwidth}
        \centering        \includegraphics[width=\linewidth]{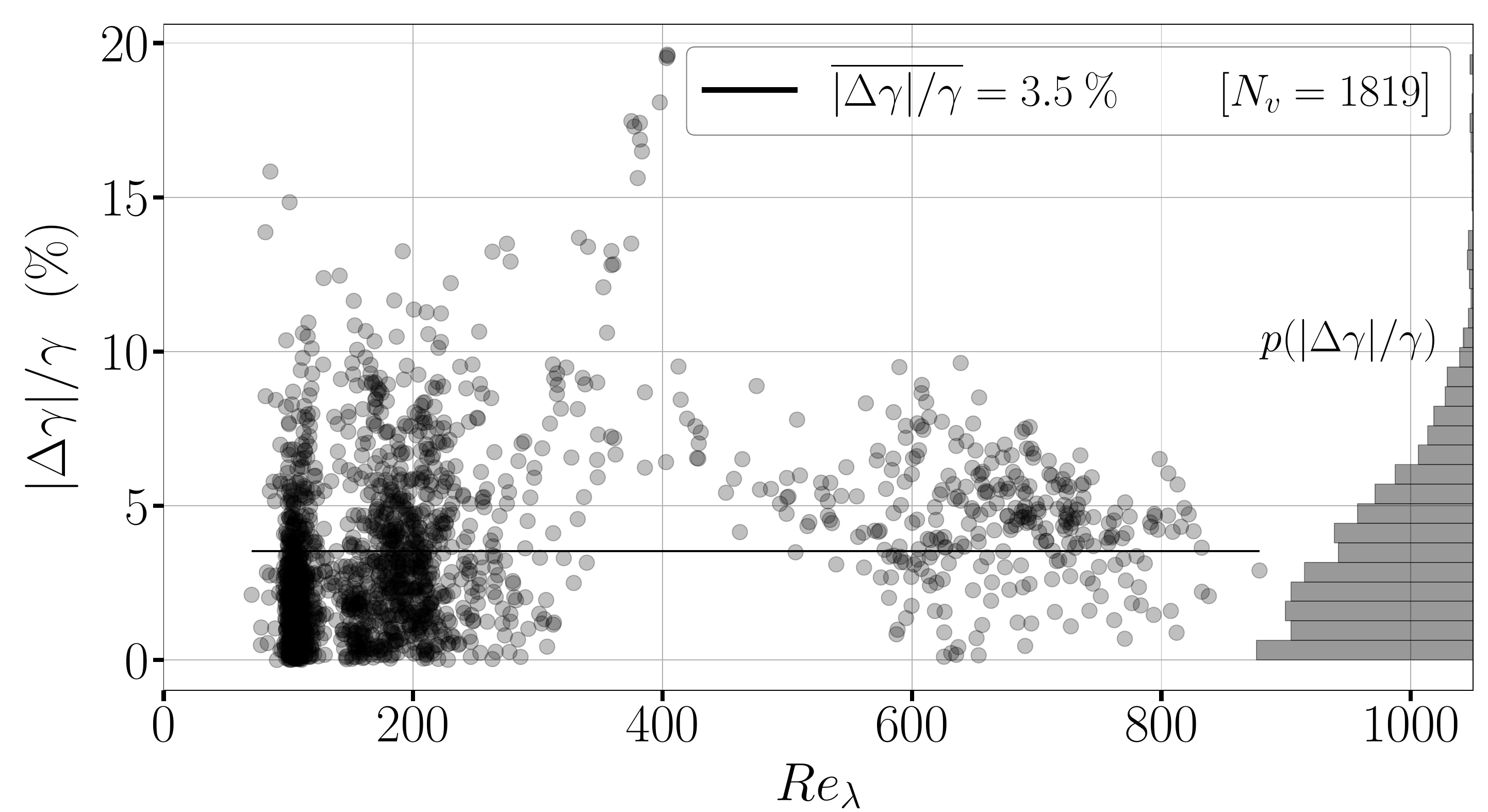}
        \caption{}
    \end{subfigure}
    \hfill
    \begin{subfigure}[t]{0.49\textwidth}
        \centering        \includegraphics[width=\linewidth]{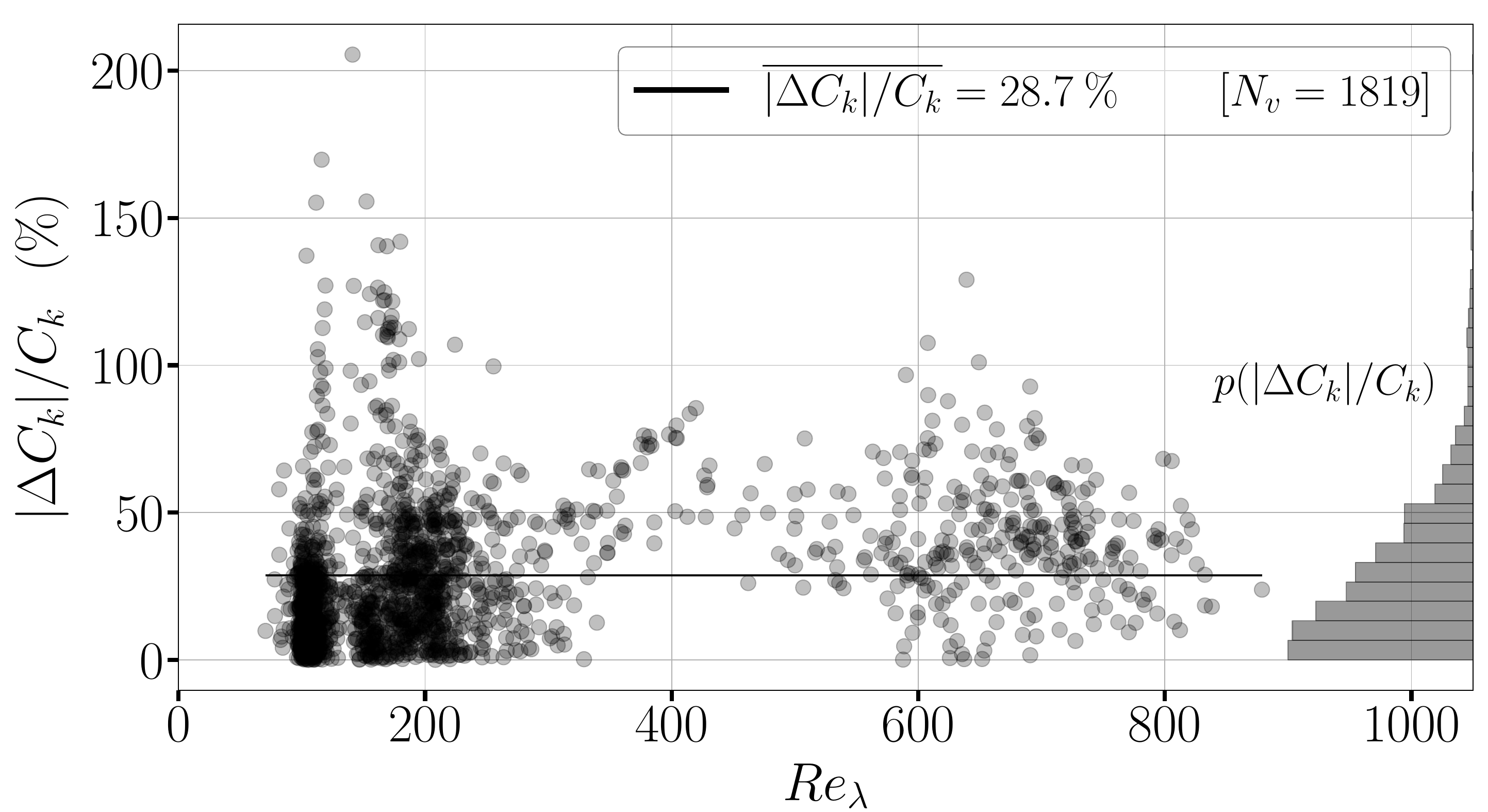}
        \caption{}
    \end{subfigure}
    \caption{\new{a) $|\Delta \mu| / \mu$ as a function of $Re_\lambda$ for all VTS used. b) $|\Delta \gamma| / \gamma$ as a function of $Re_\lambda$ for all VTS used. c) $|\Delta C_k| / C_k$ as a function of $Re_\lambda$ for all VTS used. In general the black solid lines represent the mean and the gray histogram shows the PDF of the deviations. All markers have the same shape and the same amount of transparency. Differences in color are solely due to overlapping markers. The $N_v$ indicates the number of VTS shown in the plot. Note that all markers have the same color and differences in color are solely due to overlapping markers.}}
    \label{figure_forward_computing}
\end{figure}

\FloatBarrier

\subsection{\new{Interpretation Approaches}}

\new{Within this chapter, possible explanations for the origin of the identified relations are explored, together with their physical interpretation.}

{Following Siefert $\&$ Peinke~\cite{siefert2004different}, $\mu$ can be associated with the speed of the cascade~(\textit{cf.}~Schmitt \emph{et al.}~\cite{schmitt2024universal}) as}

\begin{equation}
\mu \propto \frac{\mathrm{d}\ \Lambda^2(r)}{\mathrm{d}\log(r)}.
\label{equation_mu_deriv}
\end{equation}

{Furthermore, following Apostolidis~\emph{et al.}~\cite{apostolidis2022scalings}, $C_\varepsilon$ can be interpreted as the ratio between the energy dissipated at the small scales and the energy supplied to the cascade at the large scales~(\textit{cf.}~Schmitt \emph{et al.}~\cite{schmitt2024universal}). Within the same picture, $\gamma$ may likewise be regarded as a measure related to the cascade speed, since}

\begin{equation}
\gamma \propto \frac{\mathrm{d}\log(E(k))}{\mathrm{d}\log(k)}.
\label{equation_gamma_deriv}
\end{equation}

\noindent \new{Different values of $\gamma$ therefore correspond to different intervals in $k$-space over which a given variation in energy density occurs as energy is transferred through the cascade from one $k$-value to another one. A steeper or shallower spectral slope thus reflects how rapidly the energy distribution changes across scales. Yin \emph{et al.}~\cite{yin2024dynamics} interpreted $Re_\lambda$ as the ratio between the energy contained in the large scales and that contained in the small scales. As $\gamma$ is also consistent with this interpretation, it should be noted that eq.~(\ref{equation_law_2}}) is compatible with their reported linear relation between $C_\varepsilon$ and $1/Re_\lambda$.}

\new{Consequently, figure~\ref{figure_relation_second_law} extends the scheme introduced in~\cite{schmitt2024universal} by including $\gamma$ next to $\mu$. Moreover, the scheme is further broadened by adding two cascade models, one representing a fast cascade and the other one a slow cascade where the blue arrows indicate the extent of the inertial range. While the lower model in figure~\ref{figure_relation_second_law} illustrates a n extended cascade, the upper model bypasses energy toward smaller scales, thus representing a faster cascade. As proposed by Velte and Buchhave~\cite{velte2021dynamic}, this can be visualized as two larger vortices counter-rotating in the same plane so that they generate and transfer energy to smaller vortices that differ in size substantially. 
}  

\begin{figure}[h]
    \centering
    \newcommand{\block}[2]{%
        \begin{array}{c}
            \text{#1} \\#2
        \end{array}    }
    \newsavebox{\topbox}
    \sbox{\topbox}{%
        \(
        \block{fast process}{\mu \uparrow, \: \gamma \uparrow}
        \quad \widehat{=}\quad
        \block{high dissipation}{\varepsilon \: (\mathrm{case \; specific \; + \; for \; decaying \; turb.}) \uparrow}
        \quad \widehat{=}\quad
        \block{low energy ratio}{ C_\varepsilon \downarrow}
        \)   }
    \newsavebox{\bottombox}
    \sbox{\bottombox}{%
        \(
        \block{slow process}{\mu \downarrow, \: \gamma \downarrow}
        \quad \widehat{=}\quad
        \block{low dissipation}{\varepsilon \: (\mathrm{case \; specific \; + \; for \; decaying \; turb.}) \downarrow}
        \quad \widehat{=}\quad
        \block{high energy ratio}{ C_\varepsilon \uparrow}
        \)}
    \usebox{\topbox}
    \vspace{0.5em}
    \makebox[\wd\topbox]{%
        \includegraphics[width=\wd\topbox, height=8.5\ht\topbox]{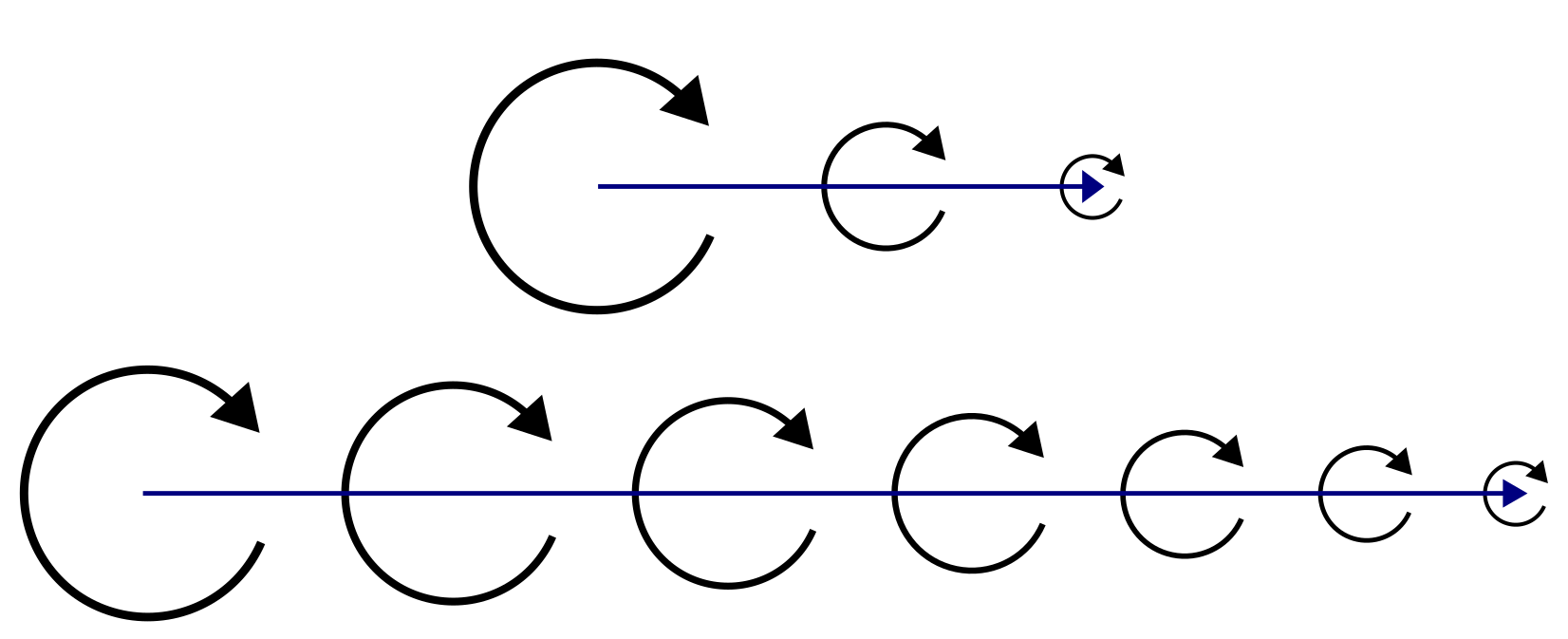}}
    \vspace{0.5em}
    \usebox{\bottombox}
\caption{\new{Scheme of the relations between the intermittency parameter $\mu$, the exponent of the spectral law $\gamma$, the mean dissipation rate $\varepsilon$ (in $\mathrm{m}^2 /\mathrm{s}^{3}$), and the dissipation parameter $C_\varepsilon$. The blue arrow indicates the depth of the cascade in $r$ or $k$ from larger to smaller scales. The lower part of the scheme represents a slow and continuous cascade, where energy is transferred smoothly across scales, consistent with the classical picture of Richardson, Batchelor, and Kolmogorov~\cite{kolmogorov1941local}. In contrast, the upper part illustrates a fast and discontinuous cascade, in which energy is transferred between vortices that differ substantially in size, as proposed by Velte and Buchhave~\cite{velte2021dynamic}.}}
    \label{figure_relation_second_law}
\end{figure}

\FloatBarrier

\new{In chapter~\ref{spatial}, it was noted that not all investigated cases exhibited an immediate streamwise decay of the turbulence. This distinction becomes relevant here and is considered in the context of the conceptual scheme shown in figure~\ref{figure_relation_second_law}. Although a general relation between $\mu$ and $C_\varepsilon$ was identified, the case-specific dependence of either quantity on $\varepsilon$ does not always follow the general tendency indicated in figure~\ref{figure_relation_second_law}. In particular, a decrease in $\mu$ is not invariably accompanied by an increase in $\varepsilon$ throughout the entire dataset.}

\new{Five configurations were identified for which the relation of $\varepsilon$ to $\mu$ and $C_\varepsilon$ deviates, at least over part of the measured streamwise range, from the behavior observed for the remaining cases. These configurations are C5, C6, C8, C9, and D15, as also listed in the streamwise-evolution table in the appendix~\cite{SM}. Importantly, these are precisely the cases for which $Re_\lambda$ does not begin to decrease immediately in the streamwise direction. They correspond to flows generated by the porous cylinder and the active grid.}

\new{The behavior of these cases is interpreted as a build-up stage of the cascade, during which additional mechanisms may still contribute to the flow evolution. This interpretation is consistent with the fact that both the porous-cylinder wake and the active grid operated in triple-random mode represent stronger departures from classical reference flows than solid-cylinder wakes, disk wakes, or conventional grid-generated turbulence. Once a streamwise position is reached beyond which $Re_\lambda$ begins to decrease, corresponding to a decaying regime, the behavior of $\varepsilon$ becomes consistent with the general picture: for a fixed set of boundary conditions, $\varepsilon$ decreases as $\mu$ decreases.}

\new{The notable result is therefore not that decaying and non-decaying turbulence exhibit different streamwise behavior, but rather that the found relations~(\ref{equation_law_1})–(\ref{equation_law_3}) remain largely unaffected by this distinction. Excluding the non-decaying data from the set of VTS leads only to a slight reduction in the scatter around relation~(\ref{equation_law_1}), while its overall form remains unchanged. The approximately lognormal shape of $p(\alpha)$ is likewise preserved. For this reason, all VTS were retained in the present analysis, irrespective of whether the corresponding flow could be classified as decaying over the investigated range. }

\new{To conclude the interpretation of the identified relations, the possible constraints imposed by relations~(\ref{equation_law_1})–(\ref{equation_law_3}) on the underlying quantities $\mu$, $C_\varepsilon$, $\gamma$, and $C_k$ are discussed. As we have seen, the eqs.~(\ref{equation_law_1})–(\ref{equation_law_3}) come with implicit restrictions on the range of values for $C_\varepsilon$, $\mu$, $\gamma - 1$ and $C_k$ that all four quantities need to be bigger than 0. As physical parameters, $C_\varepsilon$, $\mu$ and $C_k$ must be positive. However, for $\gamma - 1$ the range of values is less trivial. The integration of the spectral law gives}

\begin{equation} 
    \int E(k) \: \mathrm{d}k \propto \int k^{-\gamma} \: \mathrm{d}k = - \frac{k^{-(\gamma - 1)}}{(\gamma - 1)} + A^\prime,
    \label{equation_energy_spectrum_deriv}
\end{equation}

\noindent which provides the area under the energy spectral density curve. Considering eq. (\ref{equation_energy_spectrum_deriv}) and eq. (\ref{equation_second_order_SF}), $\gamma$ needs to be larger than 1. Simultaneously, a mathematically motivated upper limit for $\gamma$ is given by 3~\cite{vigneron2019wiener}. This limit is physically also the absolute slope for 2$\,$D-turbulence. Note that the computed values for $\gamma$ in this work fit within both limits. However, since $\gamma$ is bounded by $1<\gamma<3$, the identified relations also impose corresponding constraints on the remaining parameters. In this context, the implications for $\mu$ and $C_\varepsilon$ are of particular interest. This yields the bounds $0<\mu<2\,\alpha/\beta \approx 1$ and $\beta/2<C_\varepsilon<\infty$, which represents a notable consequence of the identified relations. In particular, $\mu$ is not only subject to a finite upper bound. Even more importantly, the relations imply the existence of a finite lower bound for $C_\varepsilon$. The finite lower bound of \(C_\varepsilon\) is consistent with the concept of the dissipation anomaly, according to which the dimensionless dissipation rate remains finite and non-zero in the high-Reynolds-number limit. In the following chapter, a justification of our SST-approach is provided.

\FloatBarrier

\subsection{\newnew{Justification of SST Analysis}}
\label{justification}

\newnew{The application of methods originally derived for HIT to data that explicitly include non-HIT conditions requires careful justification. Beyond the application of general statistical methods to the measured time series, two major assumptions are required for extending the present analysis to SST, neither of which is trivial. The first concerns the transformation from the temporal to the spatial domain through Taylor's hypothesis of frozen turbulence. This assumption directly affects, among others, the estimation of $C_\varepsilon$ and $C_k$, the latter through the spatial normalization of the spectral law. The second concerns the multiscale nature of turbulence and the assumption that the relevant scaling relations can be extended beyond HIT. This particularly affects the estimation and interpretation of $\mu$, $\gamma$, and again $C_k$. In the following, both assumptions are examined in detail, and their applicability within SST is assessed directly on the basis of the present data.}

\newnew{The Taylor assumption was tested by recalculating $C_\varepsilon$, $\mu$, $\gamma$, and $C_k$ from VTS while systematically varying the mean velocity used in the post-processing. In this procedure, $\mu$ and $\gamma$ were only slightly affected due to minor changes in the fitting bounds resulting from the modified mean velocity. This observation is consistent with the findings reported in the literature~\cite{wyngaard1977taylor}.} 

\newnew{Furthermore, Taylor's frozen-turbulence hypothesis is often considered to be applicable only up to a certain turbulence intensity, with commonly cited upper limits typically not exceeding approximately $20\,\%$. This is particularly relevant here because the use of Taylor's hypothesis has been generalized beyond its conventional interpretation. Although it is treated in the present work primarily as a coordinate transformation, it remains important to assess whether the identified trends are affected once $TI$ exceeds $20\,\%$. Figure~\ref{figure_alpha_beta_phi_TI} reveals no systematic threshold in turbulence intensity beyond which the observed relations break down or exhibit a qualitative change. This absence of a visible transition supports the use of the Taylor assumption within the range of turbulence intensities considered here.}

Next we discuss a further justification of the scaling relations, which no longer relies on fitting estimated data, but instead establishes a direct connection to the relevant theoretical concepts originally introduced within the framework of HIT.

\newnew{As explained in section~\ref{Intermittency parameter}, the derivation of $\mu$ relies on the estimation of $\Lambda^2$ and its scaling with $\mathrm{ln}(r)$. The estimation of $\Lambda^2$, in turn, is based on a superposition of Gaussian conditional distributions $p(u_r(x)|\varepsilon_r(x))$, whose contributions are weighted according to the distribution $p(\varepsilon_r(x))$, which is commonly assumed to be lognormal within the Kolmogorov--Castaing framework (\textit{cf.}~eq.~(\ref{equation_castaing_curve_one})). The resulting velocity-increment distribution is therefore given by}

\begin{equation} \newnew{
p(u_r) = \int p(u_r|\varepsilon_r) \, p(\varepsilon_r) \, \mathrm{d}\varepsilon_r,}
\label{equation_superstatistik}
\end{equation}

\noindent This approach originates from HIT frameworks developed by Kolmogorov~\cite{kolmogorov1962refinement} and Castaing~\cite{castaing1990velocity} and was later generalized by Beck~\cite{beck2004superstatistics} to a broader class of complex systems. {Beck replaced $\varepsilon_r$ by a general hidden quantity. More recently, superstatistical approaches have also been applied to turbulent circulation fluctuations, although the corresponding analysis remained restricted to HIT~\cite{lima2026superstatistics}.}

\newnew{In the following, the requirements and assumptions underlying both $p(u_r(x)|\varepsilon_r(x))$ and $p(\varepsilon_r(x))$, as introduced by Kolmogorov and Castaing, are examined in greater detail and compared with the present data. Within the HIT-based framework considered here, the relevant assumptions and expected properties can be summarized as follows:}

\newnew{
\begin{itemize}
    \item \newnew{shape of $p(\varepsilon_r(x))$ $\rightarrow \:$ self-similar \textit{e.g.} log-normal or log-Poisson or $\chi^2$ distributed~\cite{beck2004superstatistics}}
    \item  \newnew{shape of all $p(u_r(x)|\varepsilon_r(x))$$\rightarrow \:$ similarly shaped distributed, \textit{e.g.} normal distributed}
    \item  \newnew{variance of all $p(u_r(x)|\varepsilon_r(x))$ $\rightarrow \:$ unique function of $\varepsilon_r(x)$}
\end{itemize}}

\newnew{Here, self-similarity is understood in the sense that, for a given VTS, the distribution $p(\varepsilon_r(x))$ retains the same functional form across scales $r$, with its scale dependence being captured solely by changes in its variance. The itemized points represent the essential preconditions required for the higher-order moments of $u_r(x)$ to exhibit systematic scaling behavior. It should be noted  that $\varepsilon_r$ is the central quantity for the energy cascade of turbulence.
}

\newnew{Together, these properties define the statistical assumptions underlying the multiscale description adopted here and provide a set of criteria that can be tested individually against the present SST data.}

\newnew{Figure~\ref{figure_eight_examples_epsilon_r} shows $p(\varepsilon_{2L}(x))$ and $p(\varepsilon_{2\lambda}(x))$, (here not normalised to their 
 $\sigma_{r}$) for the eight VTS listed in table~\ref{table_eight_examples}. The scales $r=2L$ and $r=2\lambda$ were selected to represent two distinct scales within the inertial range. Note that, due to our definition for $L = L_e$, the scale $r=2L_e$ is still smaller than using the conventional definitions of $L$.}

\FloatBarrier

\begin{figure}[htbp]
    \centering
    \begin{subfigure}[t]{0.49\textwidth}
        \centering        \includegraphics[width=\linewidth]{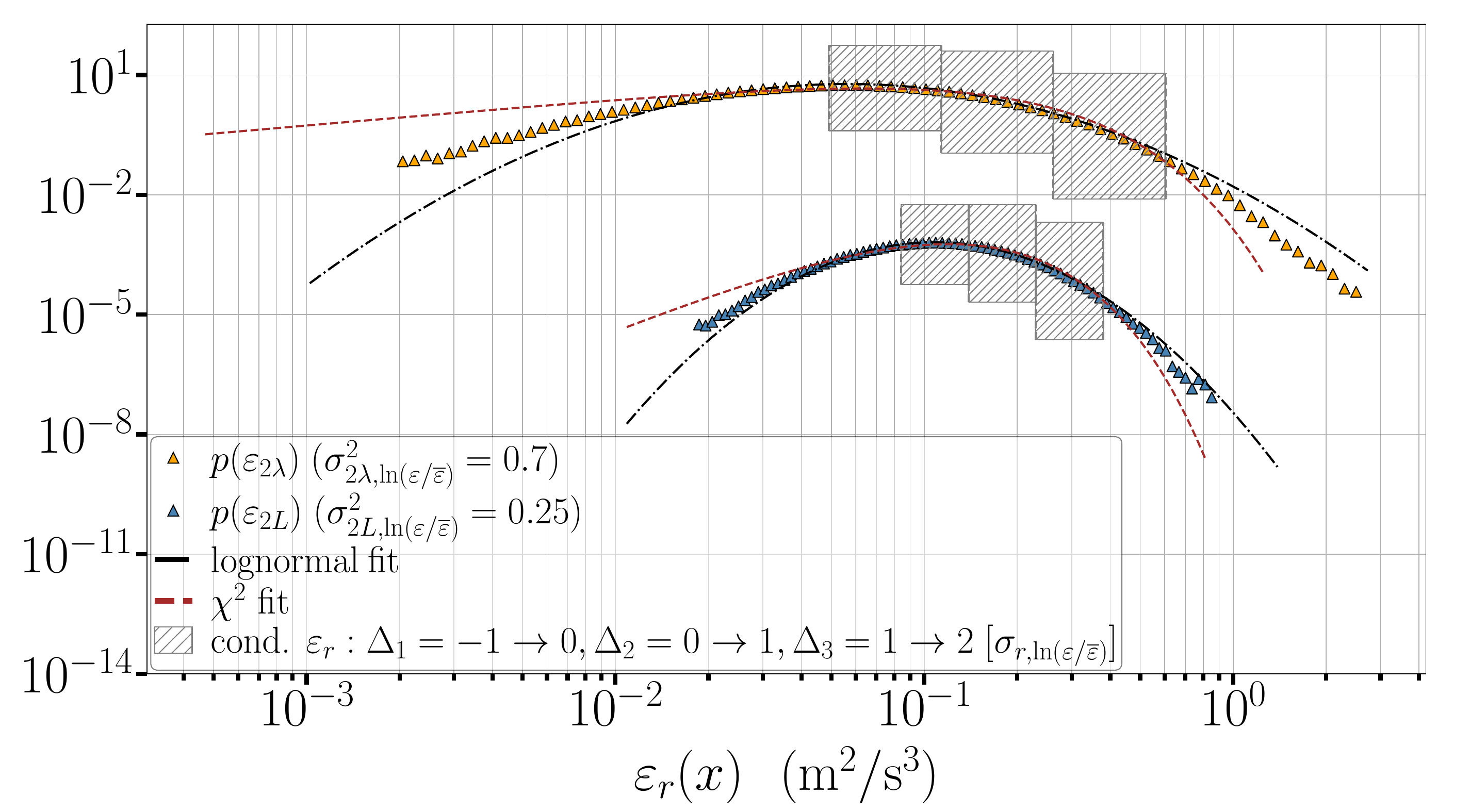}
        \caption{}
    \end{subfigure}
    \hfill
    \begin{subfigure}[t]{0.49\textwidth}
        \centering        \includegraphics[width=\linewidth]{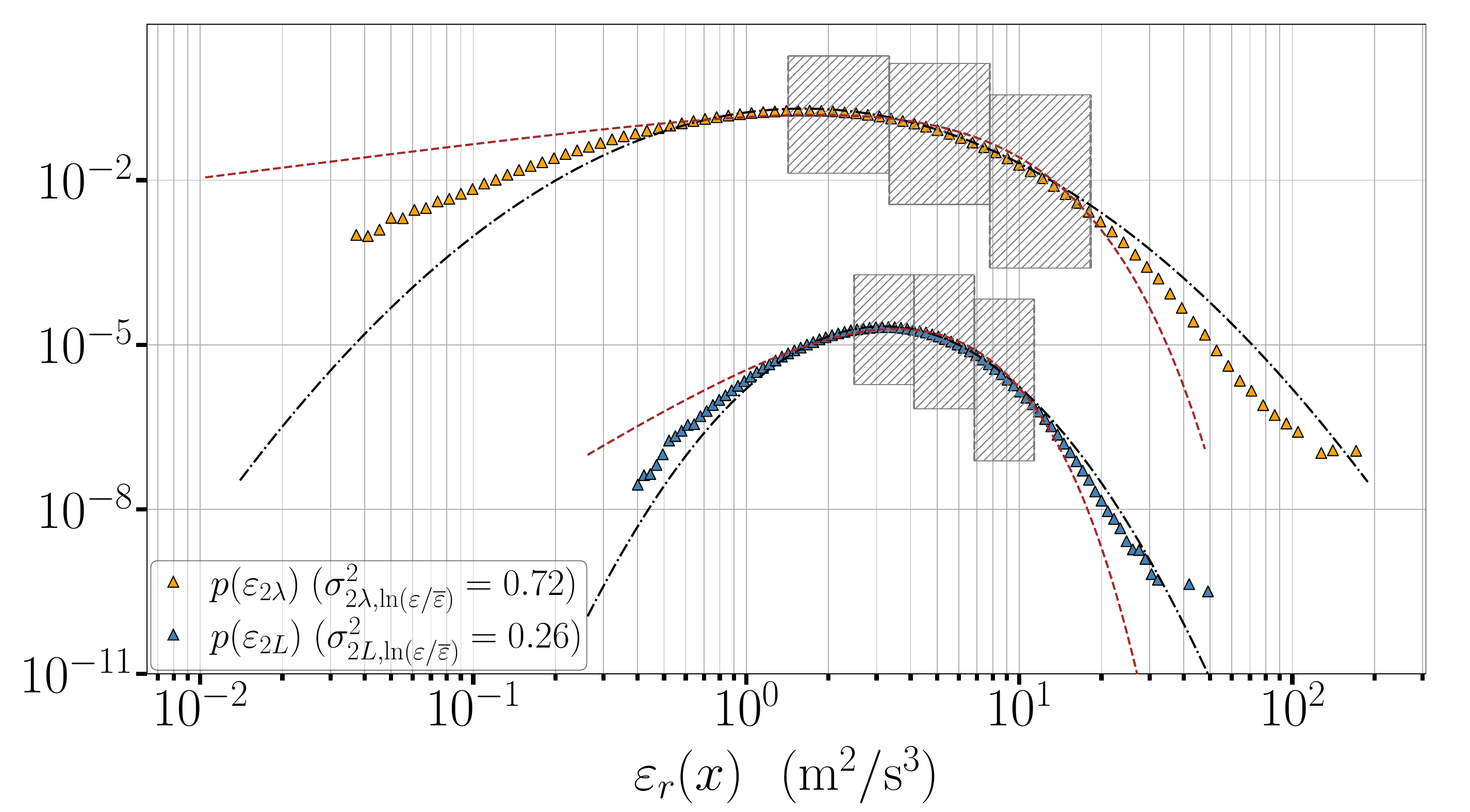}
        \caption{}
    \end{subfigure}
    \begin{subfigure}[t]{0.49\textwidth}
        \centering        \includegraphics[width=\linewidth]{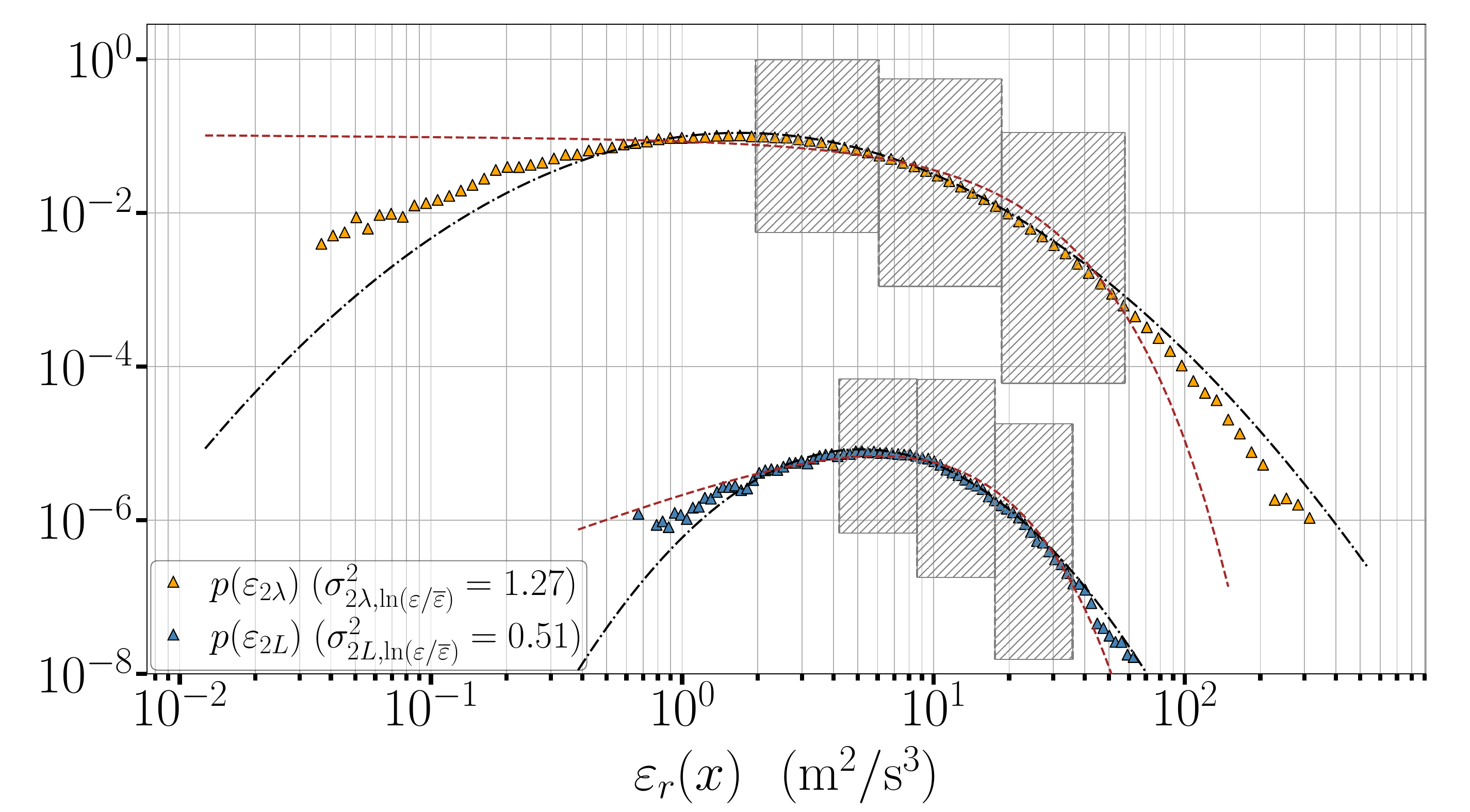}
        \caption{}
    \end{subfigure}
    \hfill
    \begin{subfigure}[t]{0.49\textwidth}
        \centering        \includegraphics[width=\linewidth]{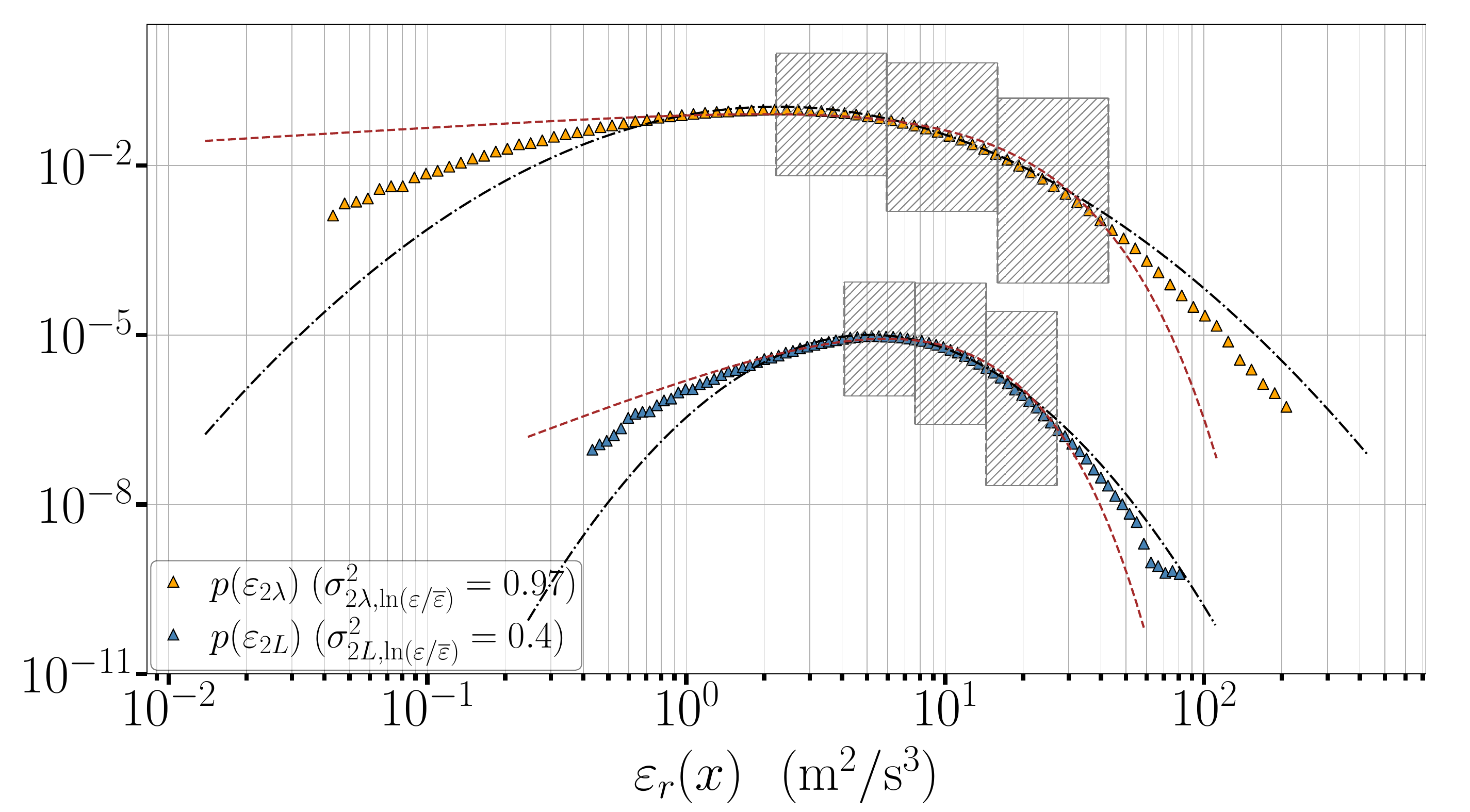}
        \caption{}
    \end{subfigure}
    \begin{subfigure}[t]{0.49\textwidth}
        \centering        \includegraphics[width=\linewidth]{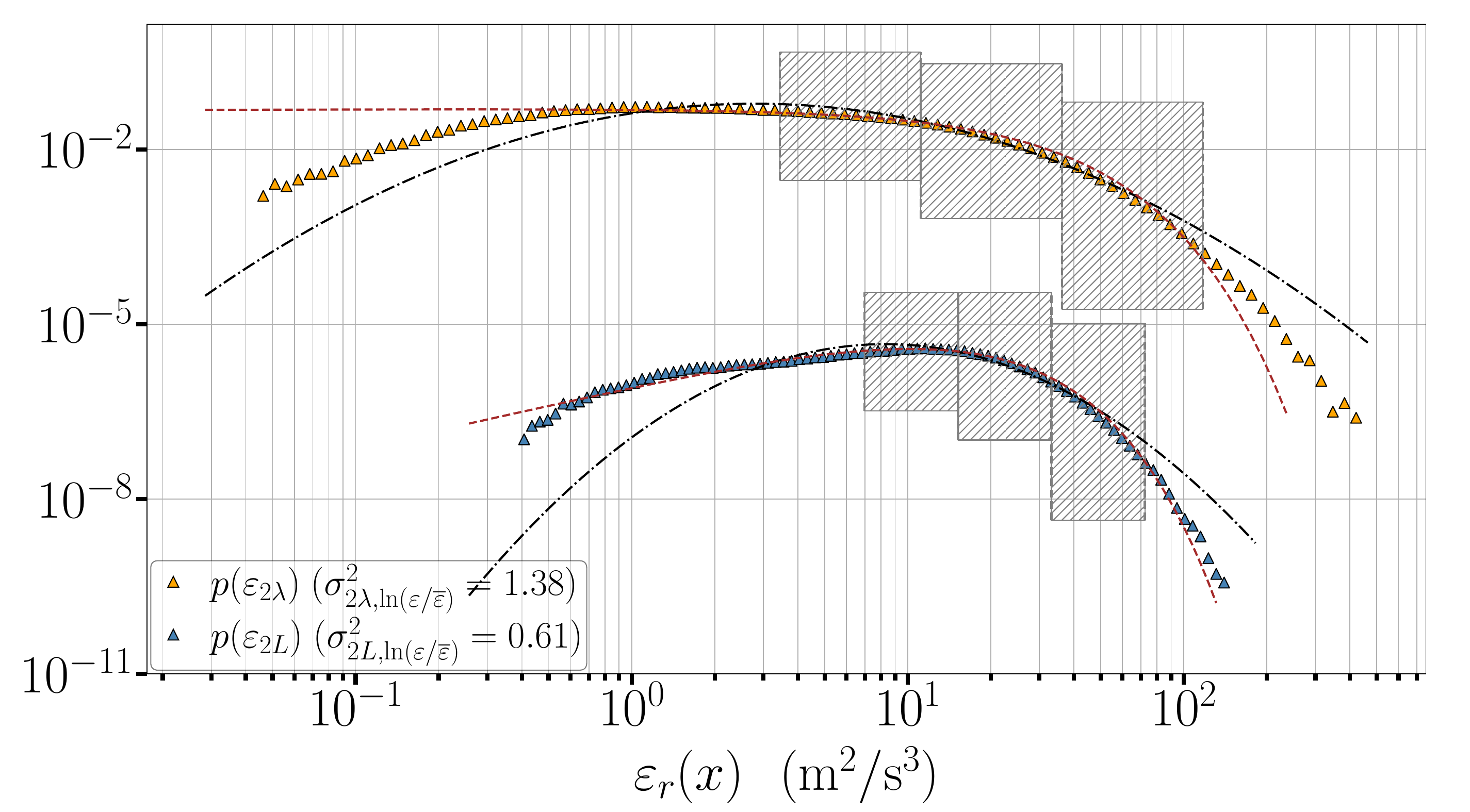}
        \caption{}
    \end{subfigure}
    \hfill
    \begin{subfigure}[t]{0.49\textwidth}
        \centering        \includegraphics[width=\linewidth]{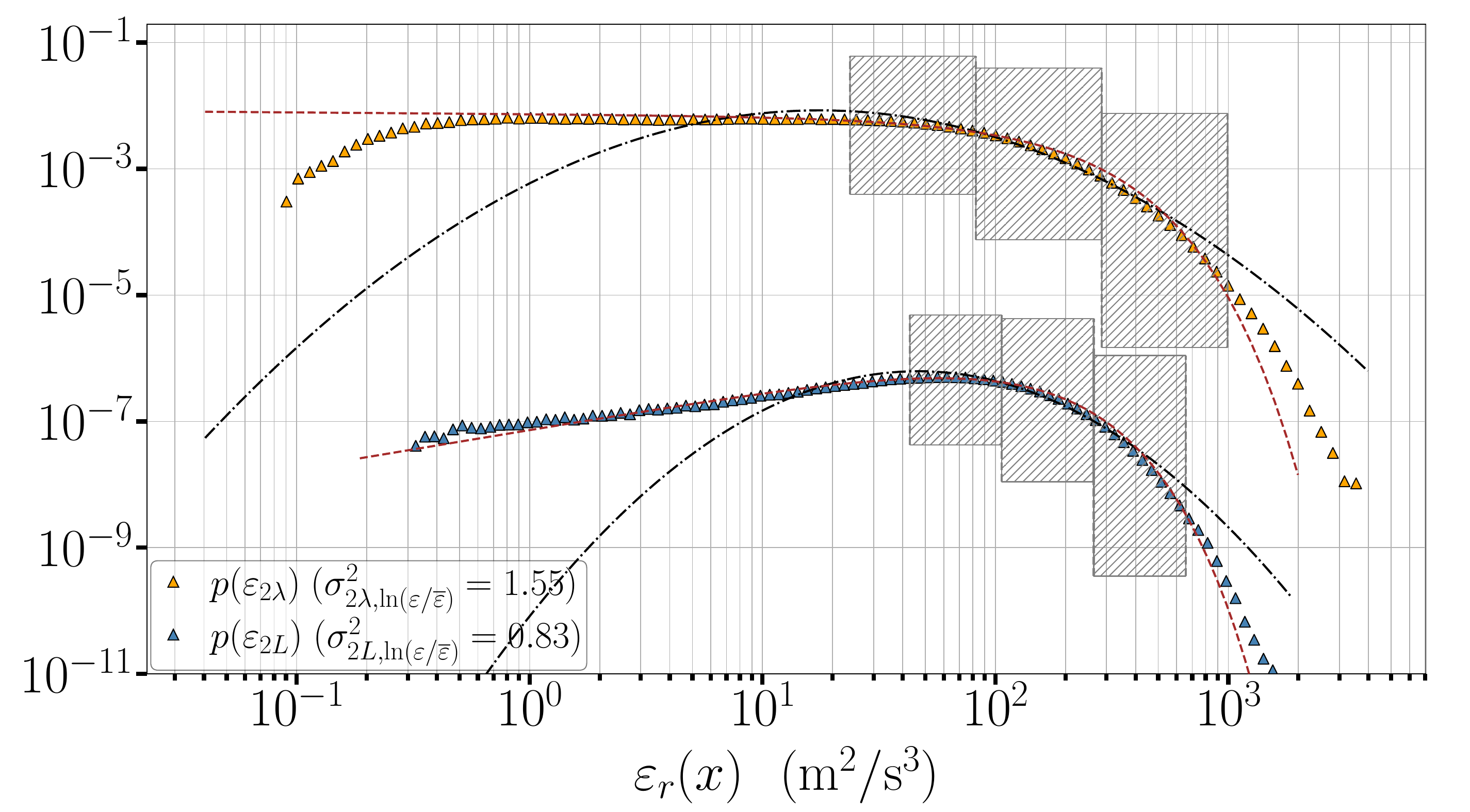}
        \caption{}
    \end{subfigure}
    \begin{subfigure}[t]{0.49\textwidth}
        \centering        \includegraphics[width=\linewidth]{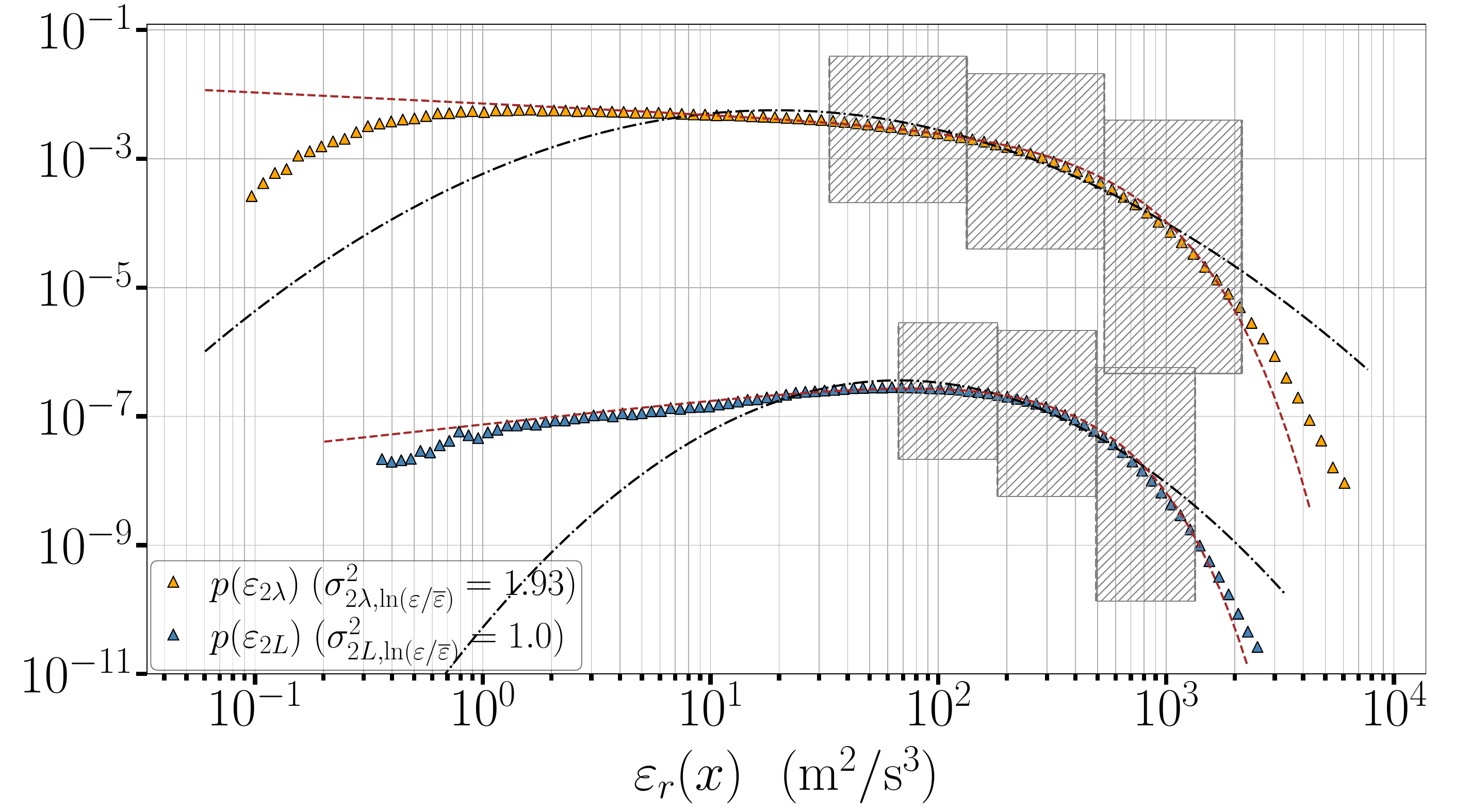}
        \caption{}
    \end{subfigure}
    \hfill
    \begin{subfigure}[t]{0.49\textwidth}
        \centering        \includegraphics[width=\linewidth]{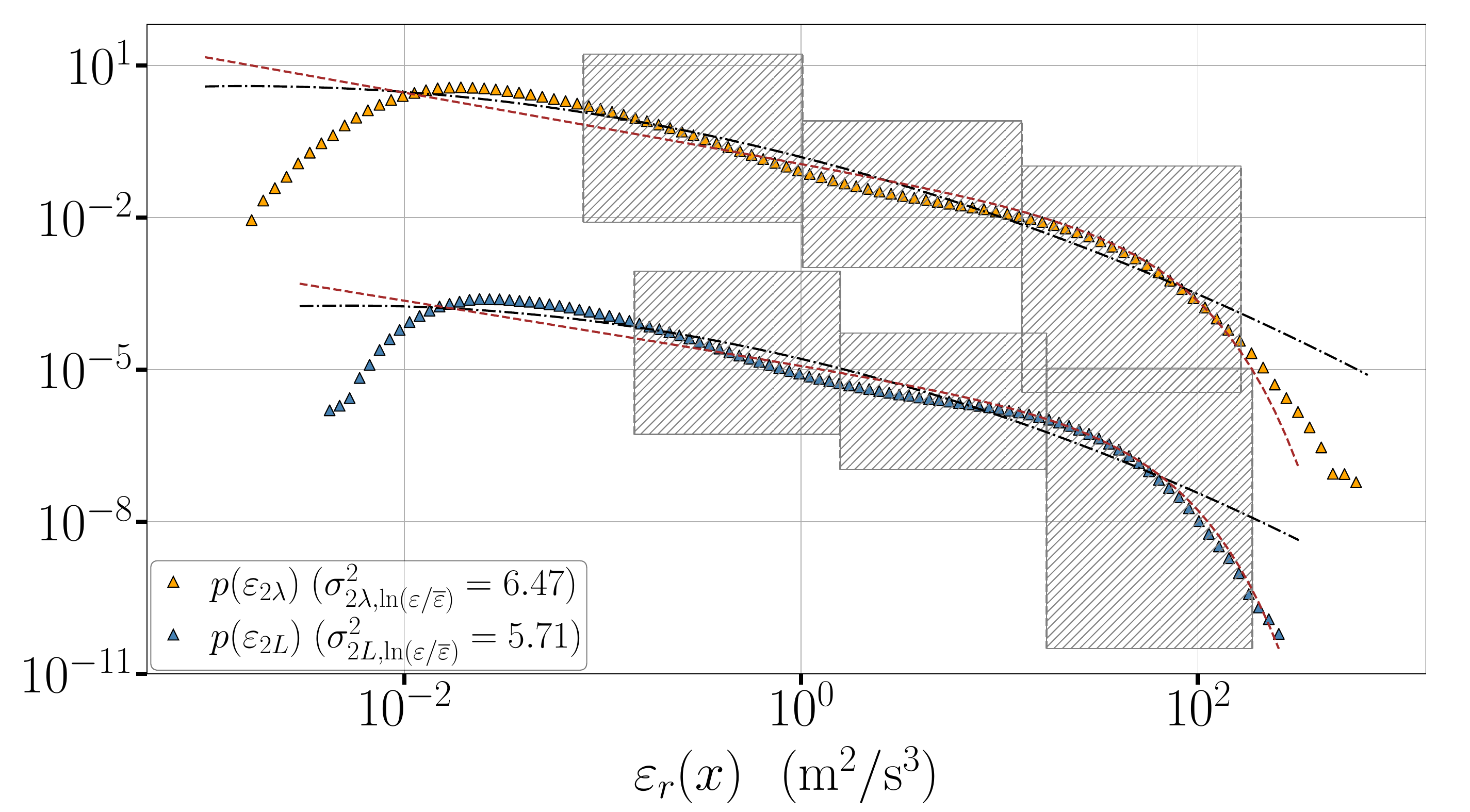}
        \caption{}
    \end{subfigure}
    \caption{\newnew{The PDFs of $\varepsilon_{r}(x)$ with $r=2L$ and $r=2\lambda$ are shown for eight representative VTS. The black and brown curves correspond to log-normal and $\chi^2$ fits, respectively. The hatched regions denote the intervals of $\varepsilon_r$ used for the computation of the conditioned PDFs $p(u_r|\varepsilon_r)$. Note that due to a better visibility, the PDFs are shifted vertically and that only bins are shown that have at least 10 values. a), b), c), e), f) and g) display VTS that satisfy the restriction criteria and taken together, span almost the full range of $\mu$-values. d) also fulfills the restriction criteria and exhibits the same value of $\mu$ as c) but with a $Re_\lambda$ more than three times smaller. h) however does not meet the restriction criteria since $\Lambda_0^2$ shows clear signatures of non-Gaussianity at large scales. The data stem from the cases G20, G24, C8, D11, C6, C1, C1, G23, respectively in order of appearance. For a detailed list of the characteristic values of the VTS and details about their cases see table~\ref{table_eight_examples} and table~\ref{tab:PhD measurements in LEGI 2023} in the appendix~\cite{SM}.}}
    \label{figure_eight_examples_epsilon_r}
\end{figure}

{It can be seen that for a given VTS, the distributions $p(\varepsilon_r(x))$ across different scales $r$ exhibit self-similar behavior, irrespective of the values of $\mu$ and $Re_\lambda$.} 


Moreover, the distributions $p(\varepsilon_r(x))$ are single distributions generally bounded by log-normal and $\chi^2$ distributions, providing reasonable lower and upper approximations for the observed shapes. In detail, it appears that a log-normal distribution provides a better fit at lower values of $\mu$ whereas a $\chi^2$ distribution provides a better fit at higher values of $\mu$. We therefore assume that $p(\varepsilon_r(x))$ is a superposition of these two distributions, with the relative weighting depending on $\mu$, the scale and possibly also on the overall boundary conditions of the flow. While a log-normal distribution of $\varepsilon_r(x)$ can be associated with a multiplicative cascade~\cite{kolmogorov1962refinement}, a $\chi^2$ distribution of $\varepsilon_r(x)$ can be associated with an additive cascade (\emph{cf.}~\cite{beck2004superstatistics}), since it is defined as the sum of squared independent normally distributed variables. A log-Poisson fit was also tested but did not provide satisfactory agreement with the data. 
}

\newnew{For the remaining VTS h), which is excluded by the selection criteria, the two PDFs are also likewise highly self-similar. However, the PDF clearly consists of two superimposed distributions, thereby providing indirect support for the validity of the applied selection criteria.}

\newnew{Next, we focus on the shape of $p(u_r(x)|\varepsilon_r(x))$. In figure~\ref{figure_eight_examples_conditioned_PDF} in the appendix~\cite{SM}, we present $p(u_r(x)|\varepsilon_r(x))$ for three different values of $\varepsilon_r(x)$, respectively for the eight representative VTS. The corresponding intervals of $\varepsilon_r(x)$ are indicated as hatched rectangles in figure~\ref{figure_eight_examples_epsilon_r} and span the ranges from -1 to 0, 0 to 1 and 1 to 2 $\sigma_{r,\mathrm{ln}(\varepsilon_r/\overline{\varepsilon_r})}$, respectively. Figure~\ref{figure_eight_examples_conditioned_PDF} demonstrates that the assumption of a Gaussian shape for $p(u_r(x)|\varepsilon_r(x))$ is valid to first-order approximation, in particular as the PDFs follow Gaussian distributions in the range of the maxima $\pm 2 \sigma$ quite accurately.} 

\newnew{Finally, the variance of the conditional distributions $p(u_r(x)|\varepsilon_r(x))$ is examined. In figure~\ref{figure_eight_examples_castaing_assumption} in the appendix~\cite{SM}, we present $\sigma_{r,(u_r|{\varepsilon_r})}$ as a function of $\mathrm{ln}(\varepsilon_r/\overline{\varepsilon_r})$. The values of $\sigma_{r,(u_r|{\varepsilon_r})}$ are obtained from the PDFs shown in figure~\ref{figure_eight_examples_conditioned_PDF}, whereas the values of $\mathrm{ln}(\varepsilon_r/\overline{\varepsilon_r})$ are taken as the geometric mean of $\varepsilon_r$ within the corresponding hatched intervals shown in figure~\ref{figure_eight_examples_epsilon_r}. Although only three data points are available, a linear fit provides the best representation of the data. However, the actual slope and its dependence on the flow conditions remain an open question. Alternative fitting functions were also tested but did not provide better results of the observed relationship. Nevertheless, the primary objective here is not to identify the unique scaling law but rather to demonstrate the existence of a unique dependence between the two quantities. }

\newnew{Overall, this statistical analysis at a fixed scale $r$ demonstrates that the essential elements of the Kolmogorov--Castaing framework, originally developed for HIT, remain applicable within SST. In particular, the characteristic behavior of $\varepsilon_r(x)$ and the conditional distributions $p(u_r(x)|\varepsilon_r(x))$ is preserved to first-order approximation, allowing the framework to be extended within Beck's more general concept of superstatistics. The resulting description is therefore referred to as the Kolmogorov--Castaing--Beck (KCB) model. This terminology reflects the combination of Kolmogorov's scale-dependent description of coarse-grained dissipation, Castaing's superposition of conditional distributions, and Beck's explicit superstatistical interpretation. Further work is nevertheless required to determine the precise functional forms and scaling laws of the underlying distributions, to quantify the observed deviations from the idealized assumptions, and to establish the extent to which the KCB framework remains valid across different turbulent flows and ranges of scales.}

\FloatBarrier

\section{\new{Conclusion}}
\label{conclusion}

{\new{This work presents and discusses new empirical relations identified in inhomogeneous turbulent flows, analyzing more than 5000 VTS from various flow configurations. Since most turbulence quantities are defined for HIT, particular care is taken to examine and compare different estimation methods for each quantity. The usage of HIT methods for non-HIT flows is justified by the self-similar evolution of the coarse-grained dissipation rate PDFs $p(\varepsilon_r(x))$ across the scales $r$. Overall, based on the shown figures and tables, we can state that the selection criteria make striking sense.} 

\new{In particular, filtering out datasets with non-Gaussianity at large scales can be seen as the key to distinguish between a turbulent flow with one set of boundary conditions on one side and the superimposition of at least two such flows on the other side. Although we were not interested in the exact value of $\Lambda_0^2$ or $F_u$ besides surpassing a threshold, the value itself could play a major role in turbulent-non-turbulent and turbulent-turbulent interfaces. Regarding the exclusion of data with high values of $\Lambda_0^2$, it is important to mention that we are not excluding all VTS which exhibit coherent structures marked by small maxima in the large scale noisy part of the power spectra. In contrast, we exclude data where two different kinds of turbulence overlap, which leads then to an even higher value of $\mu$. This is exactly what was found by Neunaber \emph{et al.} which they called an intermittency ring that surrounds any wake~\cite{neunaber2020distinct}.} 

\new{Together with the criteria of solid inertial range scalings of the energy spectral density and the shape parameter, Gaussianity at the large scales defines a new class of turbulence, which we call SST, including and extending the known HIT. By focusing on SST, a lot more data can be represented in comparison to only HIT.}

\new{The most significant finding is the universal collapse of all used data onto a single curve in each representation (figures~\ref{figure_law_1} a), \ref{figure_law_2} a), and \ref{figure_law_3} a)), demonstrating that the turbulence parameter $C_\varepsilon$, $\mu$, $\gamma$, and $C_k$ are directly linked, such that one quantity is enough predict a good estimation of the other three, using the provided eqs.~(\ref{equation_law_1})–(\ref{equation_law_3}).}

\new{We extensively tested the robustness of the identified relations by ensuring that our analysis encompasses different tests, such as:}

\new{
\begin{itemize}
    \item only centerline data $vs.$ all data,
    \item each case separately,
    \item using different methods to compute quantities,
    \item estimating possible uncertainties,
    \item using different restrictions (from lowest to highest) and
    \item using different fitting functions.
\end{itemize}}

\new{The identified relations appear to be robust to leading-order approximation, exhibiting no general dependence on Reynolds number, turbulence intensity, estimation method, flow geometry, downstream position or any other quantity examined. Importantly, however, the most remarkable result is not the specific functional form of the individual relations itself, but rather the fact that the complete dataset collapses onto a single characteristic curve in each of the three representations shown in figures~\ref{figure_law_1} a), \ref{figure_law_2} a), and \ref{figure_law_3} a).}

\new{The Reynolds number occupies a unique role in the classical description of HIT. Within SST, however, the identified relations allow the relevant non-dimensional quantities to be reorganized from a different perspective. The set of parameters $C_\varepsilon$, $\mu$, $\gamma$, and $C_k$ jointly characterizes the state of turbulence, whereas $Re_\lambda$ and $TI$ remain related to this state in a case-dependent manner. Since the four state parameters are mutually related, any one of them may, in principle, be used to represent the state. Choosing for example $C_\varepsilon$ for this purpose, the present results suggest that, while the Reynolds number remains a key quantity in turbulence, $C_\varepsilon$ may provide an even more fundamental descriptor of the turbulence state within SST. This perspective also sheds new light on the relation between $C_\varepsilon$ and $Re_\lambda$ proposed by Vassilicos, which is found to remain applicable within SST.}

\new{Taken together, the identified relations reduce the non-dimensional description of SST to three principal parameters: $TI$, $Re_\lambda$, and $C_\varepsilon$ (here we take $\alpha$, $\beta$, $\theta$ and $\phi$ as universal fixed values). Figure~\ref{figure_pi_parameter} schematically illustrates their respective roles and mutual relations. The dissipation parameter $C_\varepsilon$ specifies the state of turbulence and thereby captures both the characteristic speed of the cascade process and the ratio between dissipation and large-scale energy input. This state is reflected, among others, in the slopes of the energy spectral density and the shape parameter within the inertial range. The Reynolds number $Re_\lambda$, in turn, characterizes the separation of scales, or equivalently the extent of the inertial range for a given turbulence state, since $L/\lambda \propto C_\varepsilon Re_\lambda$. Finally, $TI$ provides the connection between the turbulent fluctuations and the mean flow.}

\begin{figure}[htbp]
    \centering
    \begin{subfigure}[t]{0.6\textwidth}
        \centering
        \includegraphics[width=\linewidth]{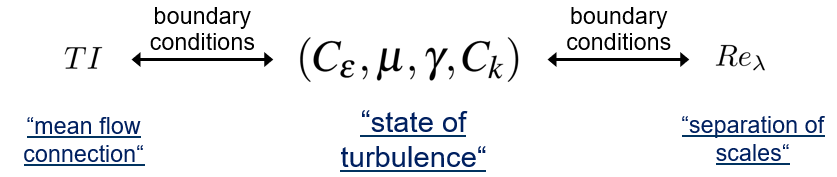}
    \end{subfigure}
    \caption{\new{For SST the tuple of $C_\varepsilon$, $\mu$, $\gamma$ and $C_k$ changes consistently and indicates the state of turbulence. Through the boundary conditions, the tuple is directly connected to the turbulence intensity $TI$ and the Reynolds number $Re_\lambda$. While $TI$ gives the relation between the turbulence and the mean flow, $Re_\lambda$ indicates the separation of scales.}}
    \label{figure_pi_parameter}
\end{figure}

\new{Based on the hypotheses formulated in section~\ref{assumptions}, the null hypothesis $H_0$ can therefore be rejected in favor of the alternative hypothesis $H_1$, which states that, for SST, a systematic relation between the values $C_\varepsilon$, $\mu$, $\gamma$ and $C_k$ exists. From this perspective, SST can be represented within a three-dimensional parameter space spanned by $TI$, $Re_\lambda$, and $C_\varepsilon$. Remarkably, these three quantities can be associated naturally with three fundamental elements of turbulence description: $TI$ with the Reynolds decomposition, $Re_\lambda$ with the balance between inertial and viscous effects embodied in the NSE~\cite{davidson2015turbulence}, and $C_\varepsilon$ with the corresponding energy-balance relations, such as the KHMH equation~\cite{hill2002exact,vassilicos2015dissipation}. Such a representation provides a broader framework for organizing, predicting, and interpreting the complex interactions encountered across different turbulent flows.}

At the same time, the present work can be interpreted as a unification of three levels of turbulence description, with the newly identified relations forming the highest level of the KCB framework, as illustrated in figure~\ref{all_schemes}. Level~3 specifies the central dimensionless quantities $C_\varepsilon$, $\mu$, $\gamma$, and $C_k$ and represents the most general classification of any VTS within SST. Each VTS can be located by a single value of $C_\varepsilon$ and, consequently, by a corresponding value of $\mu$ and $\gamma$, respectively, on the graph shown in figure~\ref{all_schemes} a). The dimensionless quantities $C_\varepsilon$, $\mu$, $\gamma$, and $C_k$ of a VTS determine the important statistical properties at level~2, such as the energy spectrum and the velocity increment distributions, as shown schematically in figure~\ref{all_schemes} b). The underlying KCB framework explains, at level~1 and for each scale $r$, the statistics of the velocity increments as a superposition of Gaussian probability distributions, or more generally distributions of the same functional form. In this way, the well-established small-scale intermittency of a VTS within SST is characterized, while an analytical expression for the statistics of small-scale velocity increments is provided. This, for example, enables the prediction of the occurrence probability of large velocity increments, provided that $L$ and $u^\prime$ are known, which may be relevant in situations in which such extreme fluctuations can have harmful consequences. Finally, we return to figure~\ref{data_selection} b) from the introduction, which illustrates the extended range of a non-ideal but realistic turbulent wake flow to which the concepts presented here can be applied.

\begin{figure}[htbp]
\centering
\begin{tabular}{c@{\hspace{1.2cm}}c@{\hspace{0.3cm}}c}
\raisebox{1.1\height}{\rotatebox{90}{\Large\textbf{level 3}}}&
\includegraphics[width=0.70\textwidth]{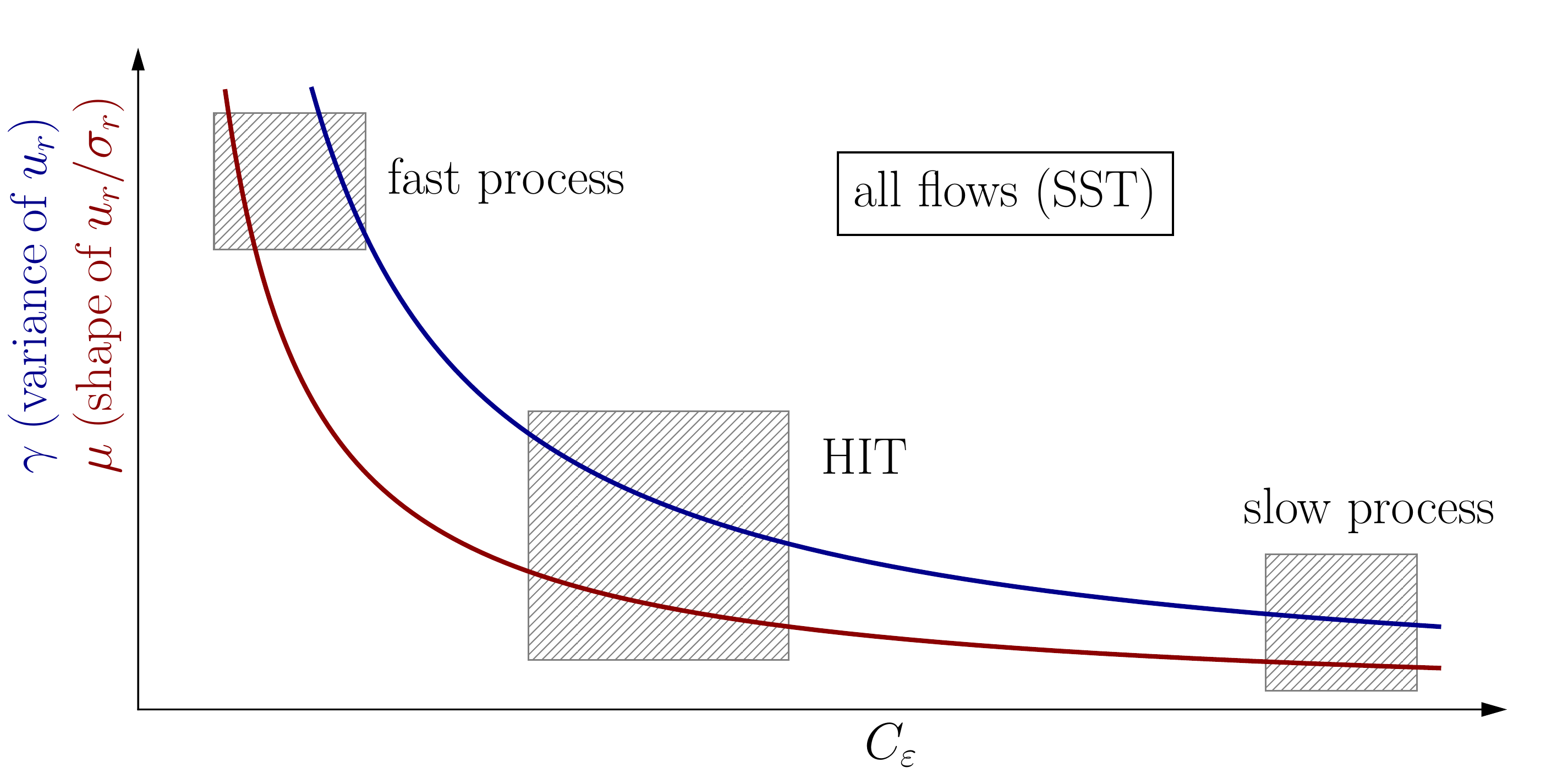}&
\raisebox{-0.45\height}{{(a)}}\\[1.3em]
\raisebox{1.2\height}{\rotatebox{90}{\Large\textbf{level 2}}}&
\includegraphics[width=0.55\textwidth]{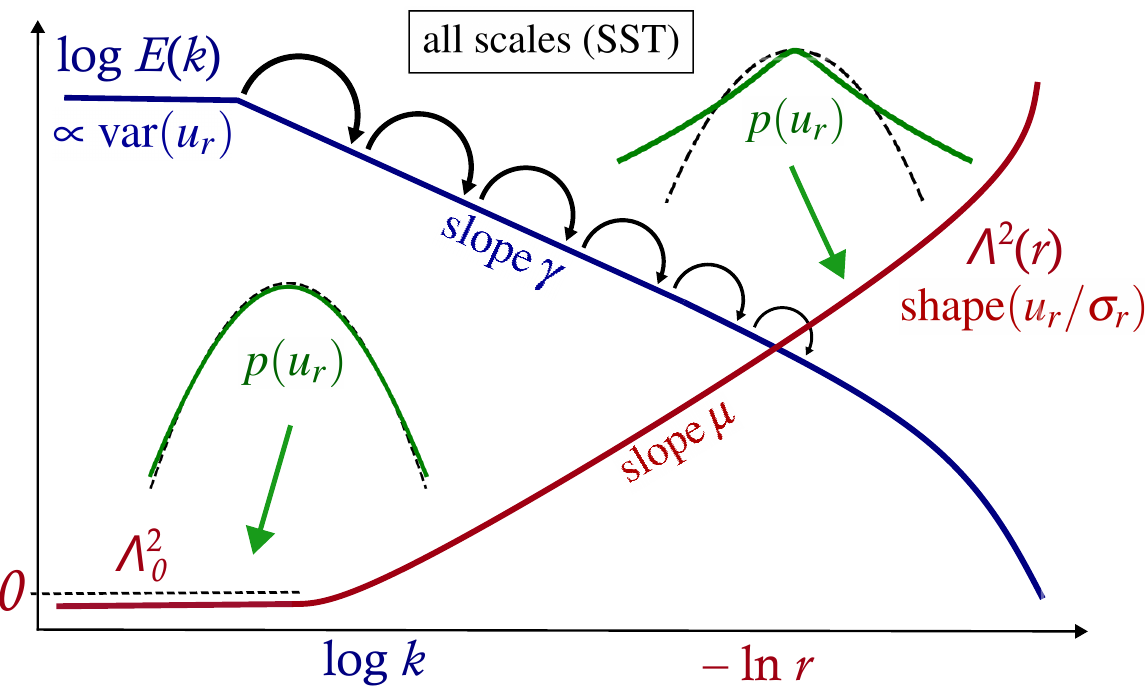}&
\raisebox{-0.45\height}{{(b)}}\\[1.3em]
\raisebox{2.0\height}{\rotatebox{90}{\Large\textbf{level 1}}}&
\includegraphics[width=0.55\textwidth]{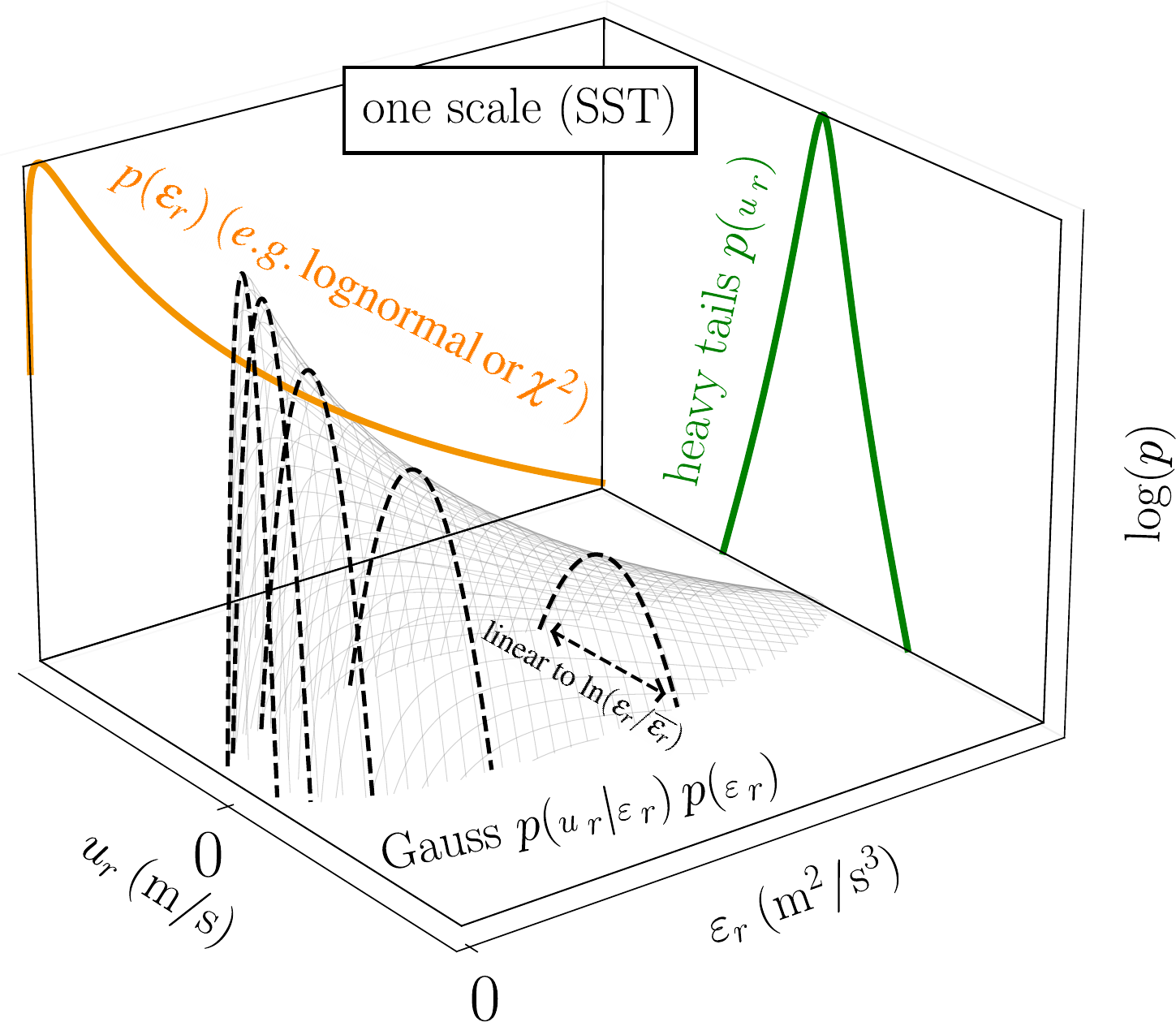}&
\raisebox{-0.45\height}{{(c)}}
\end{tabular}
\caption{\new{a) a schematic illustration of the relations between $C_\varepsilon$, $\mu$ and $\gamma$ for all flows within SST. b) a schematic illustration of all scales within SST showing the energy spectral density $E(k)$ and the shape parameter $\Lambda^2$ in $k$-space and $r$-space, respectively. The corresponding increment PDFs for two exemplary scales are shown in green. Note that the right green PDF corresponds to the green PDF shown in c). c) a schematic illustration of the Kolmogorov-Castaing-Beck model using the concept of superstatistics for one scale within SST. The orange curve shows the marginal distribution of $\varepsilon_r(x)$, while the black dashed curves indicate Gaussian increment PDFs conditioned on distinct values of $\varepsilon_r$. The green curve indicates the marginal distribution of the increments, which is generated by the integration of the Gaussian increment PDFs along all values of $\varepsilon_r(x)$.}}
\label{all_schemes}
\end{figure}

\new{As stated sufficiently within this paper, our analysis is only valid for stationary turbulence and the streamwise velocity component.  The upcoming challenge would be to shed light on the following questions:}

\new{
\begin{itemize}
    \item {How can the shape of $p(\varepsilon_r)$ be described across varying experimental conditions?}
    \item Do the identified relations still hold when analyzing all three velocity components?
    \item Are there other quantities that align with the identified relations, for example measures of isotropy or vorticity?
    \item How can the identified relations be observed in non-stationary turbulence?
    \item Are there other dissipative systems that exhibit similar relations, given that many systems show signs of intermittency, such as financial market data~\cite{ghashghaie1996turbulent}, seismic data~\cite{Manshour2009}, and physiological data~\cite{Ching2007}?
\end{itemize}}

\new{Overall, this lower-order analysis provides strong evidence that extending the analysis to non-homogeneous turbulence does not lead to a loss of the well-known structure of the cascade. Although quantities estimated in inhomogeneous flows using methods originally developed for HIT may no longer retain their strict physical interpretation, they remain statistically consistent. Rather than indicating a breakdown of the underlying framework, the results reveal a higher level of organization, suggesting that inhomogeneous turbulence follows the same fundamental statistical principles as HIT, but within a more general framework.}

\section*{Acknowledgements} 

\noindent We thank Christophe Penisson, Muriel Lagauzere, Stephane Pioz-Marchand and Sylvain Dauge for helping with the experiment,
\noindent Valentin Groß for proof-reading,
\noindent Olivier De Marchi and Sebastian Bergemann for the IT support,
\noindent Thomas Messmer, Lars Neuhaus and Daniela Moreno for validating parts of the used code and
\noindent Finn Köhne, Michael Hölling, Ingrid Neunaber, Jan Friedrich, and Clara Velte for fruitful discussion. 

\section*{Competing interests} 

The authors report no conflict of interest.



\section*{Declaration of funding}  

This work was supported by the LabEx Tec21 under Grant Investissements d’Avenir – grant agreement no. ANR-11-LABX-0030; and the Hanse-Wissenschaftskolleg (HWK Institute for Advanced Study, Delmenhorst, Germany) under a fellowship assigned to MO.

\section*{Appendix A}

\begin{table} [h!]
	\centering
	\begin{tabular}{lcccccccccc}
		\toprule
		   & $\textrm{L}_1$ & $\textrm{W}_1$ & $\textrm{W}_2$ & $\textrm{T}_2$& $\textrm{\cite{ferran2023characterising}}$ & $\textrm{\cite{mora2019energy}}$ & $\textrm{\cite{renner2001experimental}}$\\
		\midrule
		Cylinder, $ \o  =$ 10\,mm, $s=100\,$\%, $Re \approx 6\,$k & C1 \hspace{-0.3em}\raisebox{-0.2\height}{\includegraphics[height=1.2em]{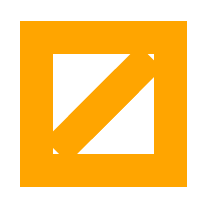}} & - & - & -& - & - & -\\
		Cylinder, $\o  =$ 20\,mm, $s=100\,$\%, $Re \approx 13\,$k & C2 \hspace{-0.3em}\raisebox{-0.2\height}{\includegraphics[height=1.2em]{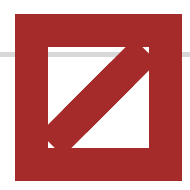}}  & C3 \hspace{-0.3em}\raisebox{-0.2\height}{\includegraphics[height=1.2em]{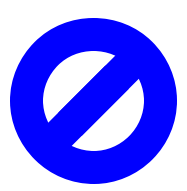}}  & C4 \hspace{-0.3em}\raisebox{-0.2\height}{\includegraphics[height=1.2em]{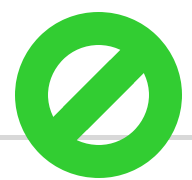}}  & -& - & - & -\\
		Cylinder, $\o  =$ 20\,mm, $s=58\,$\%, $Re \approx 13\,$k  & C5 \hspace{-0.3em}\raisebox{-0.2\height}{\includegraphics[height=1.2em]{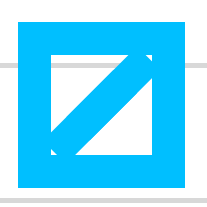}}  & - & C6  \hspace{-0.3em}\raisebox{-0.2\height}{\includegraphics[height=1.2em]{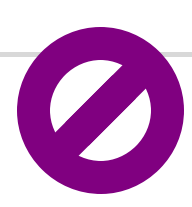}} & -& - & - & -\\
		Cylinder, $\o  =$ 20\,mm, $s=100\,$\%, $Re \approx 8\,$k  & - & - & C7 \hspace{-0.3em}\raisebox{-0.2\height}{\includegraphics[height=1.2em]{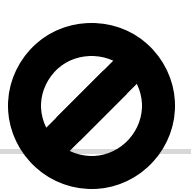}}  & C8 \hspace{-0.3em}\raisebox{-0.2\height}{\includegraphics[height=1.2em]{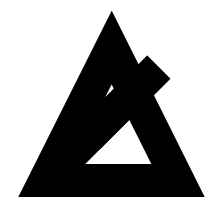}}  & - & - & - \\
		Cylinder, $\o  =$ 20\,mm, $s=100\,$\%, $Re \approx 5\,$k  & - & - & - & C9 \hspace{-0.3em}\raisebox{-0.2\height}{\includegraphics[height=1.2em]{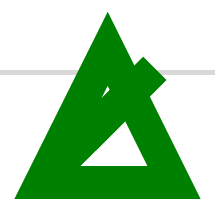}}  & - & - & -\\
		
		Disk, $\o  =$ 72\,mm, $s=34\,$\%, $Re \approx 47\,$k  & D10 \hspace{-0.3em}\raisebox{-0.2\height}{\includegraphics[height=1.2em]{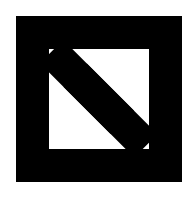}}  & - & D11 \hspace{-0.3em}\raisebox{-0.2\height}{\includegraphics[height=1.2em]{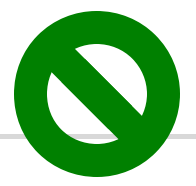}}  & -& - & - & -\\
		Disk, $\o  =$ 72\,mm, $s=48\,$\%, $Re \approx 47\,$k & D12 \hspace{-0.3em}\raisebox{-0.2\height}{\includegraphics[height=1.2em]{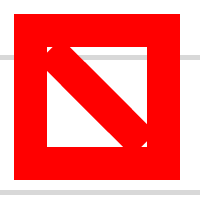}}  & - & D13 \hspace{-0.3em}\raisebox{-0.2\height}{\includegraphics[height=1.2em]{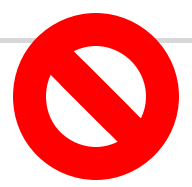}}  & -& - & - & -\\
		Disk, $\o  =$ 72\,mm, $s=34\,$\%, $Re \approx 28\,$k & D14 \hspace{-0.3em}\raisebox{-0.2\height}{\includegraphics[height=1.2em]{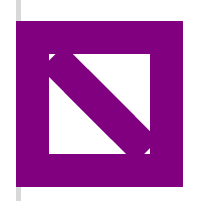}}  & -  & -  & -& - & - & -\\
		Disk, $\o =$  72\,mm, $s=48\,$\%, $Re \approx 28\,$k  & -  & - & -  & D15 \hspace{-0.3em}\raisebox{-0.2\height}{\includegraphics[height=1.2em]{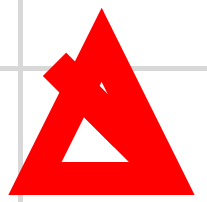}}  & - & - & -\\
		
		Grid, $\overline{u}_\infty=10\,$m/s  & - & G16 \hspace{-0.3em}\raisebox{-0.2\height}{\includegraphics[height=1.2em]{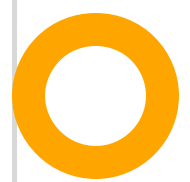}}  & G17 \hspace{-0.3em}\raisebox{-0.2\height}{\includegraphics[height=1.2em]{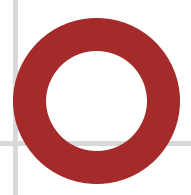}}  & -& - & - & -\\
		Grid, $\overline{u}_\infty=6\,$m/s  & - & - & G18 \hspace{-0.3em}\raisebox{-0.2\height}{\includegraphics[height=1.2em]{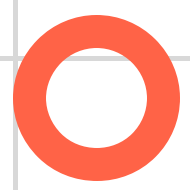}}  & G19 \hspace{-0.3em}\raisebox{-0.2\height}{\includegraphics[height=1.2em]{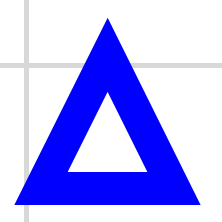}}  & - & - & -\\

            Grid, $\overline{u}_\infty=7\,$m/s  & - & - & -& -& G20 \hspace{-0.3em}\raisebox{-0.2\height}{\includegraphics[height=1.2em]{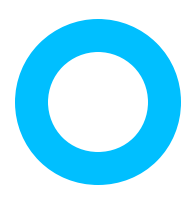}}  & - & -\\
		Grid, $\overline{u}_\infty=5\,$m/s  & - & - & -& -& G21 \hspace{-0.3em}\raisebox{-0.2\height}{\includegraphics[height=1.2em]{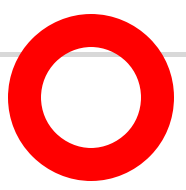}}  & - &\\
            Grid, $\overline{u}_\infty=4\,$m/s  & - & - & -& -& G22 \hspace{-0.3em}\raisebox{-0.2\height}{\includegraphics[height=1.2em]{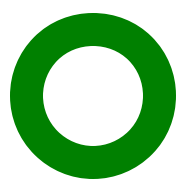}}  & - & -\\
		Grid, $\overline{u}_\infty=8.6\,$m/s  & - & - & -& -& - & G23  \hspace{-0.3em}\raisebox{-0.2\height}{\includegraphics[height=1.2em]{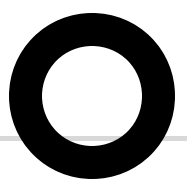}} & - \\
            Grid, $\overline{u}_\infty=17\,$m/s  & - & - & -& -& - & G24 \hspace{-0.3em}\raisebox{-0.2\height}{\includegraphics[height=1.2em]{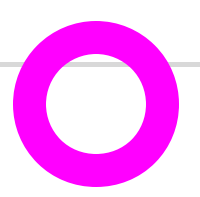}}  & - \\
            Jet, $\overline{u}_\infty=45.5\,$m/s  & - & - & -& -& - & - & J25 \hspace{-0.3em}\raisebox{-0.2\height}{\includegraphics[height=1.2em]{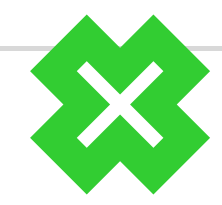}}  \\
		
		\bottomrule
	\end{tabular}
	\caption{\new{All measurement configurations used in this study. The first eleven lines correspond to the turbulent wake measurements. Within those, the first nine lines correspond directly to wake measurements and the two following lines correspond to the background flow measurements in the wind tunnel. The remaining lines of the table correspond to the studies based on other flows extracted from previous works. The column descriptions are sorted as follows: ``L'' means that there is no grid, ``W'' stands for a static grid which is made out of cylindrical wooden bars with a diameter of $20\,$mm, ``T'' stands for triple random mode of the active grid where all the bars are driven independently with independent rotational speed and in independent direction. The indices 1 and 2 are standing for the distance between the begin of the test section and the object ($1 = 630\,$mm, $2 = 1610\,$mm). The last three columns correspond to the studies from the literature~\cite{ferran2023characterising, mora2019energy, renner2001experimental}, respectively. For the actual wake measurements, $\o$ is the diameter of the wake-generating object, $s$ is the solidity of the object and $Re$ is the Reynolds number depending on the object diameter and the inflow velocity. For all other measurements, $\overline{u}_\infty$ is the mean inflow velocity.}}
	\label{tab:PhD measurements in LEGI 2023}
\end{table}

\FloatBarrier

\section*{Appendix B}

\new{Table~\ref{table_all_cases} summarizes, for each case, how the relevant turbulence quantities evolve along the streamwise direction. In this context, table~\ref{table_all_cases} presents the streamwise evolution of $Re_\lambda$, $TI$, $\overline{u}$, $u'$, $\varepsilon$, $L$, $\Lambda_0^2$, $\mu$, $C_\varepsilon$, and $\gamma$ for all cases in which retained data are available at two or more streamwise positions (case D14 is excluded because no data are retained, cases G16--G19 are omitted because measurements are available at only one streamwise location, and case J25 contains only a single data point). The streamwise evolution is presented as follow. A downward arrow (highlighted magenta) indicates a decrease, while an upward arrow (highlighted light-blue) indicates an increase. A largely stable behavior is represented by a rightward arrow (highlighted gray). In some cases, two different behaviors occur, which are shown as horizontally or vertically divided cells. A vertically divided cell indicates that the behavior of the retained data points changes along the streamwise direction (for example first decreasing and then increasing). A horizontally divided cell denotes distinct behavior of the inner and outer part of the flow, with the upper half of the cell representing the inner part. Note that the analysis that leads to this table was performed qualitatively to obtain an overview of atypical behavior. In some cases it is not entirely clear how to classify the available data, especially when the data originate from only a few streamwise positions. It should therefore be regarded as an initial qualitative guide. The last three columns show the proportions of the retained and discarded data points per case. ``use'' denotes the retained data relative to all measured VTS. ``out $\Lambda_0^2$'' denotes the proportion of data discarded because of their value of $\Lambda_0^2$. ``out (l.)'' denotes the fraction of data discarded because the VTS did not exhibit all signatures of turbulence. The criterion was that a VTS must permit estimation of one-point quantities such as the turbulence intensity, while the algorithm fails to compute all two-point statistics (in particular, successful estimation of $\Lambda_0^2$ was required). This was mostly the case when measurements were taken within the laminar background-flow region. The color of the last three columns gives an estimate of how much data per case are explained by the three preceding columns: if the sum exceeds $90\,\%$, the cells are highlighted in blue, if it exceeds $80\,\%$, the cells are highlighted in light-blue.} 

\begin{table}[h]
    \centering
    \setlength{\tabcolsep}{0pt}
    \begin{tabular}{ccccccccccccccc}
        \toprule
        case & $Re_\lambda$ & $TI$ & $\overline{u}$ & $u^\prime$ & $\varepsilon$ & $L$ & $\Lambda_0^2$ & $\mu$ & $C_\varepsilon$ & $\gamma$ & use $\;$& out($\Lambda_0^2$) $\;$& out(l.) $\;$\\
        \midrule

         C1 \hspace{1mm} & \splitV{magenta}{$\downarrow$}{gray}{$\rightarrow$} \hspace{0mm}&
        \singlecell{magenta}{$\downarrow$} \hspace{0mm} &
        \splitV{magenta}{$\downarrow$}{cyan}{$\uparrow$} \hspace{0mm}& \singlecell{magenta}{$\downarrow$} \hspace{0mm}&
        \singlecell{magenta}{$\downarrow$} \hspace{0mm} &
        \splitV{gray}{$\rightarrow$}{cyan}{$\uparrow$} \hspace{0mm} & \singlecell{gray}{$\rightarrow$} \hspace{0mm}&
        \splitV{magenta}{$\downarrow$}{gray}{$\rightarrow$} \hspace{0mm}&
        \splitV{cyan}{$\uparrow$}{gray}{$\rightarrow$} \hspace{0mm}&
        \splitV{magenta}{$\downarrow$}{gray}{$\rightarrow$} \hspace{0mm} & \cellcolor{blue!30}18$\,\%$ & \cellcolor{blue!30}45$\,\%$  & \cellcolor{blue!30}30$\,\%$  \\

         C2 \hspace{1mm} & \splitV{magenta}{$\downarrow$}{gray}{$\rightarrow$} \hspace{0mm}&
        \singlecell{magenta}{$\downarrow$} \hspace{0mm} &
        \splitV{magenta}{$\downarrow$}{cyan}{$\uparrow$} \hspace{0mm}& \singlecell{magenta}{$\downarrow$} \hspace{0mm}&
        \singlecell{magenta}{$\downarrow$} \hspace{0mm} &
        \singlecell{cyan}{$\uparrow$} \hspace{0mm} & \singlecell{gray}{$\rightarrow$} \hspace{0mm}&
        \splitV{magenta}{$\downarrow$}{gray}{$\rightarrow$} \hspace{0mm}&
        \splitV{cyan}{$\uparrow$}{gray}{$\rightarrow$} \hspace{0mm}&
        \splitV{magenta}{$\downarrow$}{gray}{$\rightarrow$} \hspace{0mm}& \cellcolor{blue!30}20$\,\%$ & \cellcolor{blue!30}41$\,\%$ & \cellcolor{blue!30}36$\,\%$ \\

        C3 \hspace{1mm} & \splitV{magenta}{$\downarrow$}{gray}{$\rightarrow$} \hspace{0mm}&
        \singlecell{magenta}{$\downarrow$} \hspace{0mm} &
        \singlecell{cyan}{$\uparrow$} \hspace{0mm}& \singlecell{magenta}{$\downarrow$} \hspace{0mm}&
        \singlecell{magenta}{$\downarrow$} \hspace{0mm} &
        \singlecell{cyan}{$\uparrow$} \hspace{0mm} & \singlecell{gray}{$\rightarrow$} \hspace{0mm}&
        \splitH{magenta}{$\downarrow$}{gray}{$\rightarrow$} \hspace{0mm}&
        \singlecell{cyan}{$\uparrow$} \hspace{0mm}& \singlecell{magenta}{$\downarrow$} \hspace{0mm}& \cellcolor{blue!30}59$\,\%$ & \cellcolor{blue!30}38$\,\%$ & \cellcolor{blue!30}0$\,\%$ \\

        C4 \hspace{1mm} & \singlecell{magenta}{$\downarrow$} \hspace{0mm}&
        \singlecell{magenta}{$\downarrow$} \hspace{0mm} &
        \splitV{magenta}{$\downarrow$}{cyan}{$\uparrow$} \hspace{0mm}& \singlecell{magenta}{$\downarrow$} \hspace{0mm}&
        \singlecell{magenta}{$\downarrow$} \hspace{0mm} &
        \singlecell{cyan}{$\uparrow$} \hspace{0mm} & \singlecell{gray}{$\rightarrow$} \hspace{0mm}&
        \splitH{magenta}{$\downarrow$}{gray}{$\rightarrow$} \hspace{0mm}&
        \splitH{cyan}{$\uparrow$}{gray}{$\rightarrow$} \hspace{0mm}& \splitH{magenta}{$\downarrow$}{gray}{$\rightarrow$}\hspace{0mm}& \cellcolor{blue!30}46$\,\%$ & \cellcolor{blue!30}46$\,\%$ & \cellcolor{blue!30}0$\,\%$ \\

        C5 \hspace{1mm} & \splitV{cyan}{$\uparrow$}{magenta}{$\downarrow$} \hspace{0mm}&
        \singlecell{magenta}{$\downarrow$} \hspace{0mm} &
        \singlecell{cyan}{$\uparrow$} \hspace{0mm}& \splitV{gray}{$\rightarrow$}{magenta}{$\downarrow$} \hspace{0mm}&
        \singlecell{magenta}{$\downarrow$} \hspace{0mm} &
        \singlecell{cyan}{$\uparrow$} \hspace{0mm} & \singlecell{gray}{$\rightarrow$} \hspace{0mm}&
        \splitV{cyan}{$\uparrow$}{magenta}{$\downarrow$} \hspace{0mm}&
        \splitV{magenta}{$\downarrow$}{cyan}{$\uparrow$} \hspace{0mm}&
        \splitV{cyan}{$\uparrow$}{magenta}{$\downarrow$} \hspace{0mm}& \cellcolor{blue!30}33$\,\%$ & \cellcolor{blue!30}39$\,\%$ & \cellcolor{blue!30}25$\,\%$  \\

        C6 \hspace{1mm} & \splitV{cyan}{$\uparrow$}{magenta}{$\downarrow$} \hspace{0mm}&
        \singlecell{magenta}{$\downarrow$} \hspace{0mm} &
        \singlecell{cyan}{$\uparrow$} \hspace{0mm}& \singlecell{magenta}{$\downarrow$} \hspace{0mm}&
        \singlecell{magenta}{$\downarrow$} \hspace{0mm} &
        \singlecell{cyan}{$\uparrow$} \hspace{0mm} & \singlecell{gray}{$\rightarrow$} \hspace{0mm}&
        \splitV{cyan}{$\uparrow$}{magenta}{$\downarrow$} \hspace{0mm}&
        \splitV{magenta}{$\downarrow$}{cyan}{$\uparrow$} \hspace{0mm}&
        \splitV{cyan}{$\uparrow$}{magenta}{$\downarrow$} \hspace{0mm}& \cellcolor{blue!30}63$\,\%$ & \cellcolor{blue!30}27$\,\%$ & \cellcolor{blue!30}0$\,\%$  \\
        
        C7 \hspace{1mm} & \singlecell{magenta}{$\downarrow$} \hspace{0mm}&
        \singlecell{magenta}{$\downarrow$} \hspace{0mm} &
        \singlecell{cyan}{$\uparrow$} \hspace{0mm}& \singlecell{magenta}{$\downarrow$} \hspace{0mm}&
        \singlecell{magenta}{$\downarrow$} \hspace{0mm} &
        \singlecell{cyan}{$\uparrow$} \hspace{0mm} & \singlecell{gray}{$\rightarrow$} \hspace{0mm}&
        \singlecell{magenta}{$\downarrow$} \hspace{0mm}&
        \singlecell{cyan}{$\uparrow$} \hspace{0mm}& \singlecell{magenta}{$\downarrow$} \hspace{0mm}& \cellcolor{blue!15}29$\,\%$ & \cellcolor{blue!15}44$\,\%$ & \cellcolor{blue!15}10$\,\%$ \\

        C8 \hspace{1mm} & \singlecell{gray}{$\rightarrow$} \hspace{0mm}&
        \splitV{cyan}{$\uparrow$}{magenta}{$\downarrow$} \hspace{0mm} &
        \splitV{magenta}{$\downarrow$}{cyan}{$\uparrow$} \hspace{0mm}& \splitV{cyan}{$\uparrow$}{magenta}{$\downarrow$} \hspace{0mm}&
        \splitV{cyan}{$\uparrow$}{magenta}{$\downarrow$} \hspace{0mm} &
        \singlecell{gray}{$\rightarrow$} \hspace{0mm} & \singlecell{gray}{$\rightarrow$} \hspace{0mm}&
        \singlecell{gray}{$\rightarrow$} \hspace{0mm}&
        \singlecell{gray}{$\rightarrow$} \hspace{0mm}&
        \singlecell{gray}{$\rightarrow$} \hspace{0mm}& 43$\,\%$ & 2$\,\%$ & 0$\,\%$ \\

        C9 \hspace{1mm} & \singlecell{gray}{$\rightarrow$} \hspace{0mm}&
        \singlecell{gray}{$\rightarrow$} \hspace{0mm} &
        \singlecell{gray}{$\rightarrow$} \hspace{0mm}& \singlecell{gray}{$\rightarrow$} \hspace{0mm}&
        \singlecell{gray}{$\rightarrow$} \hspace{0mm} &
        \singlecell{gray}{$\rightarrow$} \hspace{0mm} & \singlecell{gray}{$\rightarrow$} \hspace{0mm}&
        \singlecell{gray}{$\rightarrow$} \hspace{0mm}&
        \singlecell{gray}{$\rightarrow$} \hspace{0mm}&
        \singlecell{gray}{$\rightarrow$} \hspace{0mm}& 32$\,\%$ & 1$\,\%$ & 0$\,\%$  \\

        D10 \hspace{1mm} & \singlecell{magenta}{$\downarrow$} \hspace{0mm}&
        \singlecell{magenta}{$\downarrow$} \hspace{0mm} &
        \singlecell{cyan}{$\uparrow$} \hspace{0mm}& \singlecell{magenta}{$\downarrow$} \hspace{0mm}&
        \singlecell{magenta}{$\downarrow$} \hspace{0mm} &
        \singlecell{cyan}{$\uparrow$} \hspace{0mm} & \singlecell{gray}{$\rightarrow$} \hspace{0mm}&
        \singlecell{cyan}{$\uparrow$} \hspace{0mm}&
        \singlecell{cyan}{$\uparrow$} \hspace{0mm}&
         \singlecell{gray}{$\rightarrow$} \hspace{0mm}& \cellcolor{blue!30}1$\,\%$ & \cellcolor{blue!30}74$\,\%$ & \cellcolor{blue!30}19$\,\%$ \\

        D11 \hspace{1mm} & \singlecell{magenta}{$\downarrow$} \hspace{0mm}&
        \singlecell{magenta}{$\downarrow$} \hspace{0mm} &
        \splitH{cyan}{$\uparrow$}{gray}{$\rightarrow$} \hspace{0mm}& \singlecell{magenta}{$\downarrow$} \hspace{0mm}&
        \singlecell{magenta}{$\downarrow$} \hspace{0mm} &
        \singlecell{cyan}{$\uparrow$} \hspace{0mm} & \singlecell{gray}{$\rightarrow$} \hspace{0mm}&
        \singlecell{gray}{$\rightarrow$} \hspace{0mm}&
        \singlecell{cyan}{$\uparrow$} \hspace{0mm}&
        \splitH{magenta}{$\downarrow$}{gray}{$\rightarrow$} \hspace{0mm}& \cellcolor{blue!30}50$\,\%$ & \cellcolor{blue!30}44$\,\%$ & \cellcolor{blue!30}0$\,\%$ \\

        D12 \hspace{1mm} & \singlecell{magenta}{$\downarrow$} \hspace{0mm}&
        \singlecell{magenta}{$\downarrow$} \hspace{0mm} &
        \singlecell{cyan}{$\uparrow$} \hspace{0mm}& \singlecell{magenta}{$\downarrow$} \hspace{0mm}&
        \singlecell{magenta}{$\downarrow$} \hspace{0mm} &
        \singlecell{cyan}{$\uparrow$} \hspace{0mm} & \singlecell{gray}{$\rightarrow$} \hspace{0mm}&
        \singlecell{cyan}{$\uparrow$} \hspace{0mm}&
        \singlecell{gray}{$\rightarrow$} \hspace{0mm}&
        \singlecell{gray}{$\rightarrow$} \hspace{0mm}& \cellcolor{blue!30}3$\,\%$ & \cellcolor{blue!30}71$\,\%$ & \cellcolor{blue!30}25$\,\%$ \\

        D13 \hspace{1mm} & \singlecell{magenta}{$\downarrow$} \hspace{0mm}&
        \singlecell{magenta}{$\downarrow$} \hspace{0mm} &
        \splitH{cyan}{$\uparrow$}{gray}{$\rightarrow$} \hspace{0mm}& \singlecell{magenta}{$\downarrow$} \hspace{0mm}&
        \singlecell{magenta}{$\downarrow$} \hspace{0mm} &
        \singlecell{cyan}{$\uparrow$} \hspace{0mm} & \singlecell{gray}{$\rightarrow$} \hspace{0mm}&
        \singlecell{cyan}{$\uparrow$} \hspace{0mm}&
        \singlecell{gray}{$\rightarrow$} \hspace{0mm}&
        \singlecell{gray}{$\rightarrow$} \hspace{0mm}& \cellcolor{blue!15}32$\,\%$ & \cellcolor{blue!15}56$\,\%$ & \cellcolor{blue!15}1$\,\%$ \\

        D15 \hspace{1mm} & \singlecell{gray}{$\rightarrow$} \hspace{0mm}&
        \singlecell{magenta}{$\downarrow$} \hspace{0mm} &
        \singlecell{cyan}{$\uparrow$} \hspace{0mm}& \singlecell{magenta}{$\downarrow$} \hspace{0mm}&
        \singlecell{magenta}{$\downarrow$} \hspace{0mm} &
        \singlecell{cyan}{$\uparrow$} \hspace{0mm} & \singlecell{gray}{$\rightarrow$} \hspace{0mm}&
        \singlecell{gray}{$\rightarrow$} \hspace{0mm}&
        \singlecell{gray}{$\rightarrow$} \hspace{0mm}&
        \singlecell{gray}{$\rightarrow$} \hspace{0mm}& 59$\,\%$ & 3$\,\%$ & 0$\,\%$  \\

        G20 \hspace{1mm} & \singlecell{magenta}{$\downarrow$} \hspace{0mm}&
        \singlecell{magenta}{$\downarrow$} \hspace{0mm} &
        \splitV{magenta}{$\downarrow$}{cyan}{$\uparrow$} \hspace{0mm}& \singlecell{magenta}{$\downarrow$} \hspace{0mm}&
        \singlecell{magenta}{$\downarrow$} \hspace{0mm} &
        \singlecell{cyan}{$\uparrow$} \hspace{0mm} & \singlecell{gray}{$\rightarrow$} \hspace{0mm}&
        \singlecell{magenta}{$\downarrow$} \hspace{0mm}&
        \singlecell{cyan}{$\uparrow$} \hspace{0mm}& \singlecell{magenta}{$\downarrow$} \hspace{0mm}& 28$\,\%$ & 0$\,\%$ & 0$\,\%$\\

        G21 \hspace{1mm} & \singlecell{magenta}{$\downarrow$} \hspace{0mm}&
        \singlecell{magenta}{$\downarrow$} \hspace{0mm} &
        \splitV{magenta}{$\downarrow$}{cyan}{$\uparrow$} \hspace{0mm}& \singlecell{magenta}{$\downarrow$} \hspace{0mm}&
        \singlecell{magenta}{$\downarrow$} \hspace{0mm} &
        \singlecell{cyan}{$\uparrow$} \hspace{0mm} & \singlecell{gray}{$\rightarrow$} \hspace{0mm}&
        \singlecell{magenta}{$\downarrow$} \hspace{0mm}&
        \singlecell{cyan}{$\uparrow$} \hspace{0mm}& \singlecell{magenta}{$\downarrow$} \hspace{0mm}& 33$\,\%$ & 5$\,\%$ & 0$\,\%$\\

        G22 \hspace{1mm} & \singlecell{magenta}{$\downarrow$} \hspace{0mm}&
        \singlecell{magenta}{$\downarrow$} \hspace{0mm} &
        \splitV{magenta}{$\downarrow$}{cyan}{$\uparrow$} \hspace{0mm}& \singlecell{magenta}{$\downarrow$} \hspace{0mm}&
        \singlecell{magenta}{$\downarrow$} \hspace{0mm} &
        \singlecell{cyan}{$\uparrow$} \hspace{0mm} & \singlecell{gray}{$\rightarrow$} \hspace{0mm}&
        \singlecell{magenta}{$\downarrow$} \hspace{0mm}&
        \singlecell{cyan}{$\uparrow$} \hspace{0mm}& \singlecell{magenta}{$\downarrow$} \hspace{0mm}& 29$\,\%$ & 0$\,\%$ & 0$\,\%$ \\

        G23 \hspace{1mm} & \singlecell{magenta}{$\downarrow$} \hspace{0mm}&
        \singlecell{magenta}{$\downarrow$} \hspace{0mm} &
        \splitV{magenta}{$\downarrow$}{cyan}{$\uparrow$} \hspace{0mm}& \singlecell{magenta}{$\downarrow$} \hspace{0mm}&
        \singlecell{magenta}{$\downarrow$} \hspace{0mm} &
        \singlecell{cyan}{$\uparrow$} \hspace{0mm} & \singlecell{gray}{$\rightarrow$} \hspace{0mm}&
        \singlecell{magenta}{$\downarrow$} \hspace{0mm}&
        \singlecell{cyan}{$\uparrow$} \hspace{0mm}& \singlecell{magenta}{$\downarrow$} \hspace{0mm}& 55$\,\%$ & 14$\,\%$ & 0$\,\%$ \\

        G24 \hspace{1mm} & \singlecell{magenta}{$\downarrow$} \hspace{0mm}&
        \singlecell{magenta}{$\downarrow$} \hspace{0mm} &
        \singlecell{magenta}{$\downarrow$} \hspace{0mm}& \singlecell{magenta}{$\downarrow$} \hspace{0mm}&
        \singlecell{magenta}{$\downarrow$} \hspace{0mm} &
        \singlecell{cyan}{$\uparrow$} \hspace{0mm} & \singlecell{gray}{$\rightarrow$} \hspace{0mm}&
        \singlecell{magenta}{$\downarrow$} \hspace{0mm}&
        \singlecell{cyan}{$\uparrow$} \hspace{0mm}& \singlecell{magenta}{$\downarrow$} \hspace{0mm}& \cellcolor{blue!15}33$\,\%$ & \cellcolor{blue!15}50$\,\%$ & \cellcolor{blue!15}0$\,\%$ \\
        
        \bottomrule
        
        all \hspace{1mm} & &&& &&& &&& & \cellcolor{blue!15}35$\,\%$ & \cellcolor{blue!15}40$\,\%$ & \cellcolor{blue!15}14$\,\%$ \\
        
        \bottomrule
    \end{tabular}
    \caption{\new{For each case, the streamwise development of $Re_\lambda$, $TI$, $\overline{u}$, $u'$, $\varepsilon$, $L$, $\Lambda_0^2$, $\mu$, $C_\varepsilon$, and $\gamma$ is shown. An upward arrow ($\uparrow$) indicates an increase in the streamwise direction and is highlighted in green, while a downward arrow ($\downarrow$) indicates a decrease and is highlighted in red. A horizontal arrow ($\rightarrow$) denotes an approximately constant trend and is highlighted in gray. Cells divided into two vertical parts indicate that the quantity exhibits two distinct trends, listed in order from left to right. Cells divided into two horizontal parts indicate different streamwise trends at the centerline (upper part) and in the outer flow (lower part). The column ``use'' shows the percentage of used data per case, whereas the column ``out($\Lambda^2_0$)'' shows the portion of data discarded because of $\Lambda^2_0$. Column ``out(l.)'' shows the percentage discarded because the VTS shows almost laminar flow. In the last column, the color indicates the amount of data that can be described by the three columns before (dark blue $> 90\%$ and light blue  $> 80\%$). Cases D14, G16, G17, G18, G19 and J25 do not have any used data points with varying streamwise position. Additional details about each case are provided in table~I of the appendix~\cite{SM}.}}
    \label{table_all_cases}
\end{table}

\FloatBarrier

\section*{Appendix C}

\newnew{For the eight VTS from table~\ref{table_eight_examples}, figure~\ref{figure_eight_examples_conditioned_PDF} shows conditioned increment PDFs $p(u_r|\varepsilon_r)$. Additionally, figure~\ref{figure_eight_examples_castaing_assumption} presents the standard deviation of the conditioned PDFs $p(u_r|\varepsilon_r)$ as a function of ln$(\varepsilon_r/\overline{\varepsilon_r})$. The scales $r$ and the intervals $\Delta_1$, $\Delta_2$, and $\Delta_3$  correspond to figure~\ref{figure_eight_examples_epsilon_r}}. 

\begin{figure}[htbp]
    \centering
    \begin{subfigure}[t]{0.49\textwidth}
        \centering        \includegraphics[width=\linewidth]{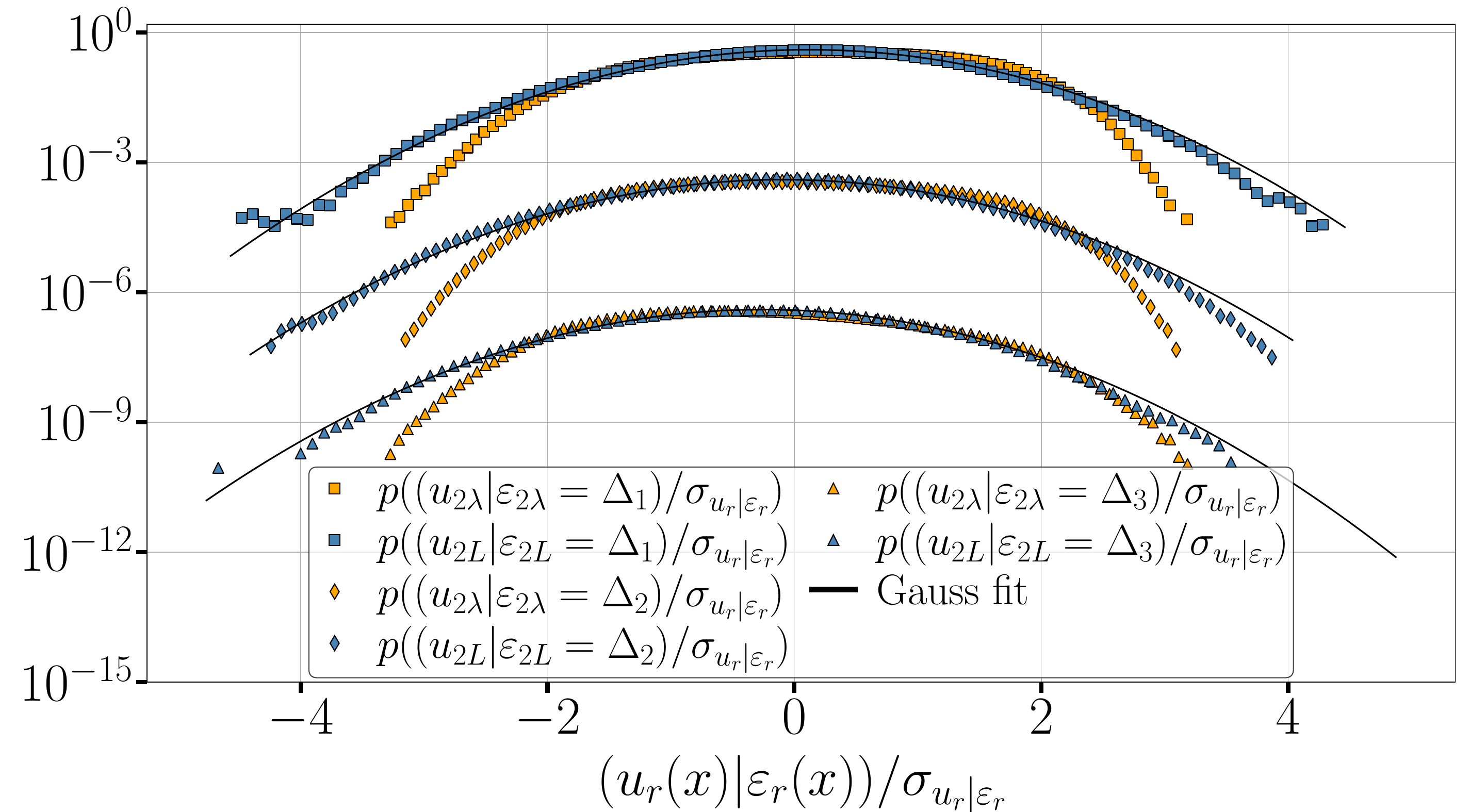}
        \caption{}
    \end{subfigure}
    \hfill
    \begin{subfigure}[t]{0.49\textwidth}
        \centering        \includegraphics[width=\linewidth]{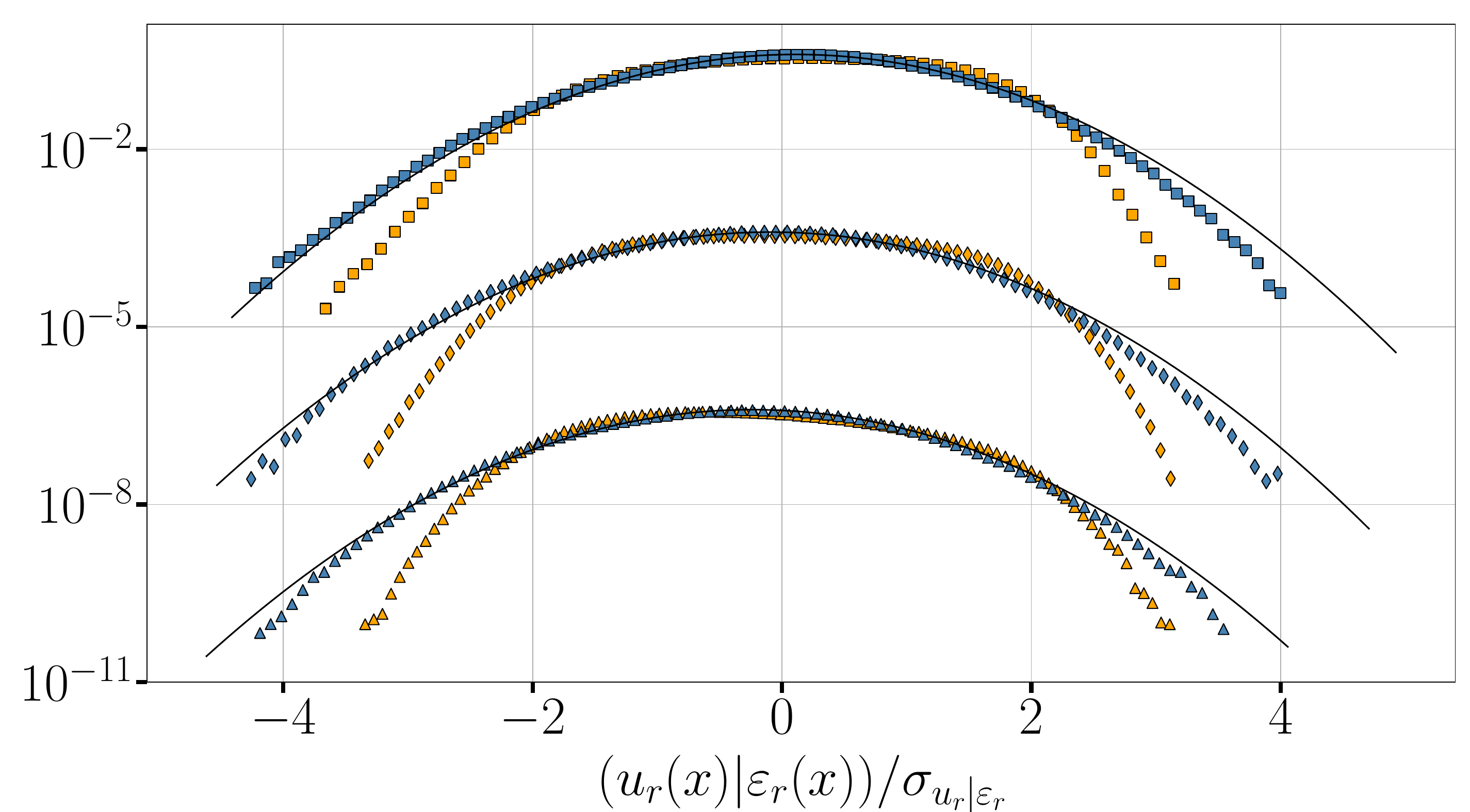}
        \caption{}
    \end{subfigure}
    \begin{subfigure}[t]{0.49\textwidth}
        \centering        \includegraphics[width=\linewidth]{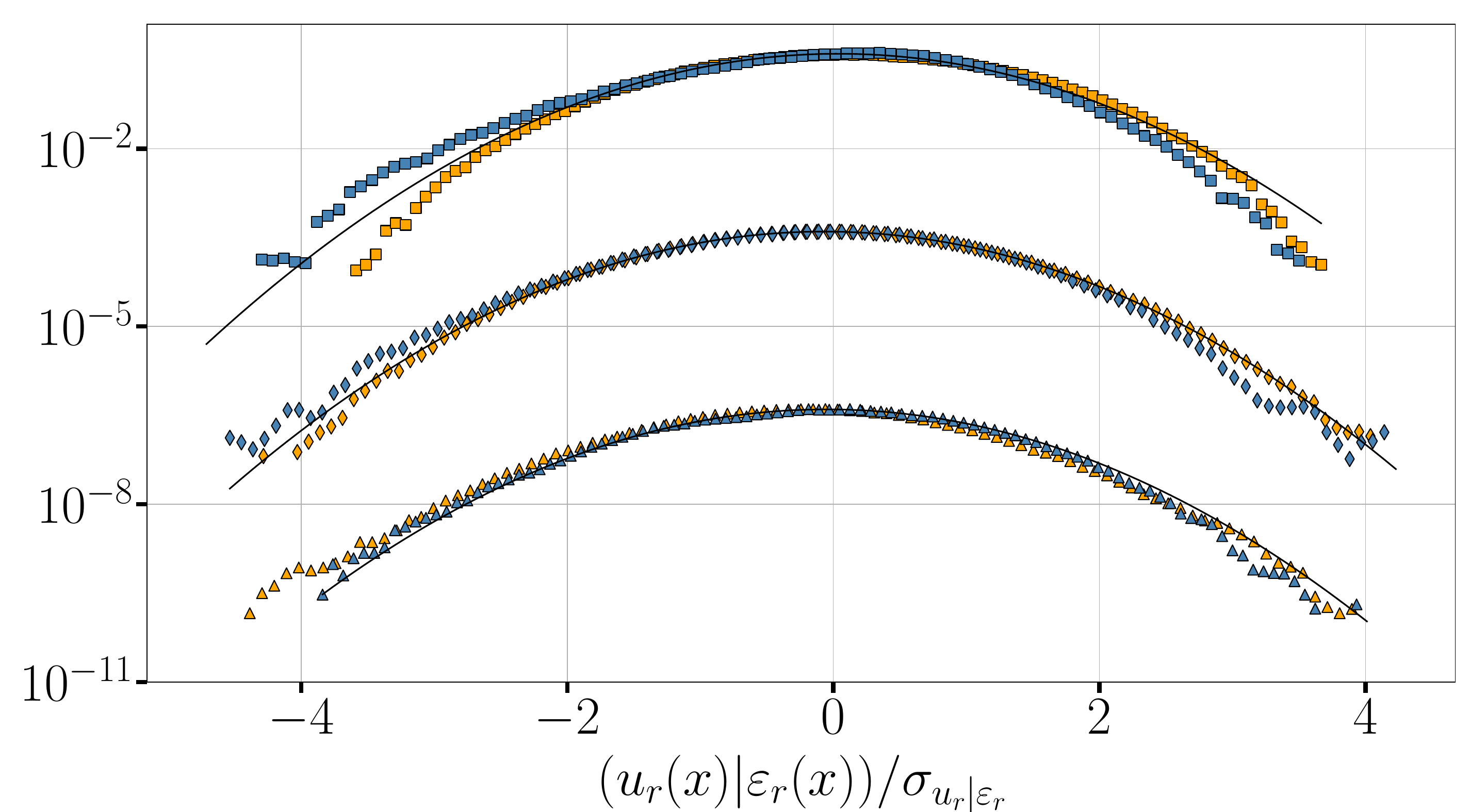}
        \caption{}
    \end{subfigure}
    \hfill
    \begin{subfigure}[t]{0.49\textwidth}
        \centering        \includegraphics[width=\linewidth]{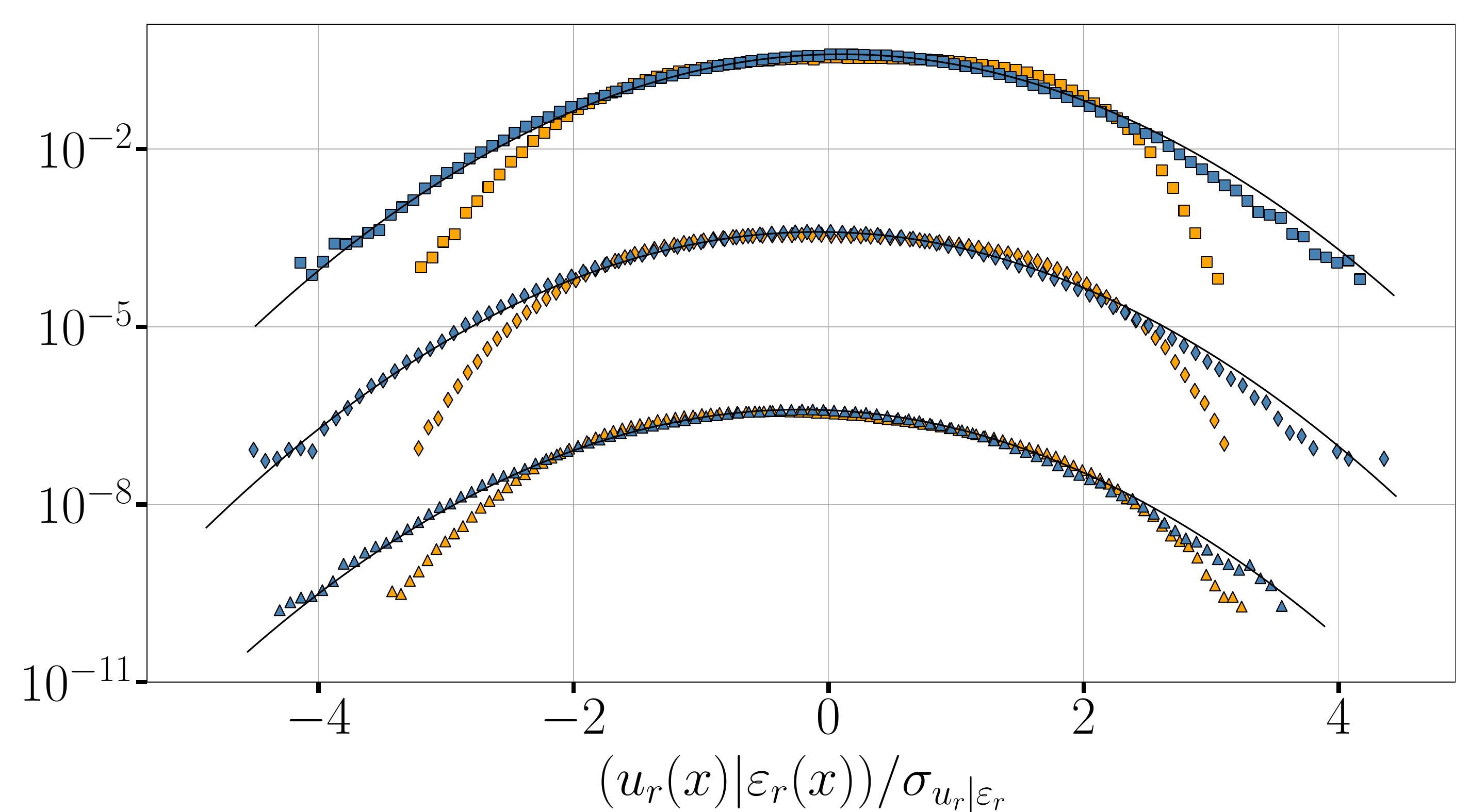}
        \caption{}
    \end{subfigure}
    \begin{subfigure}[t]{0.49\textwidth}
        \centering        \includegraphics[width=\linewidth]{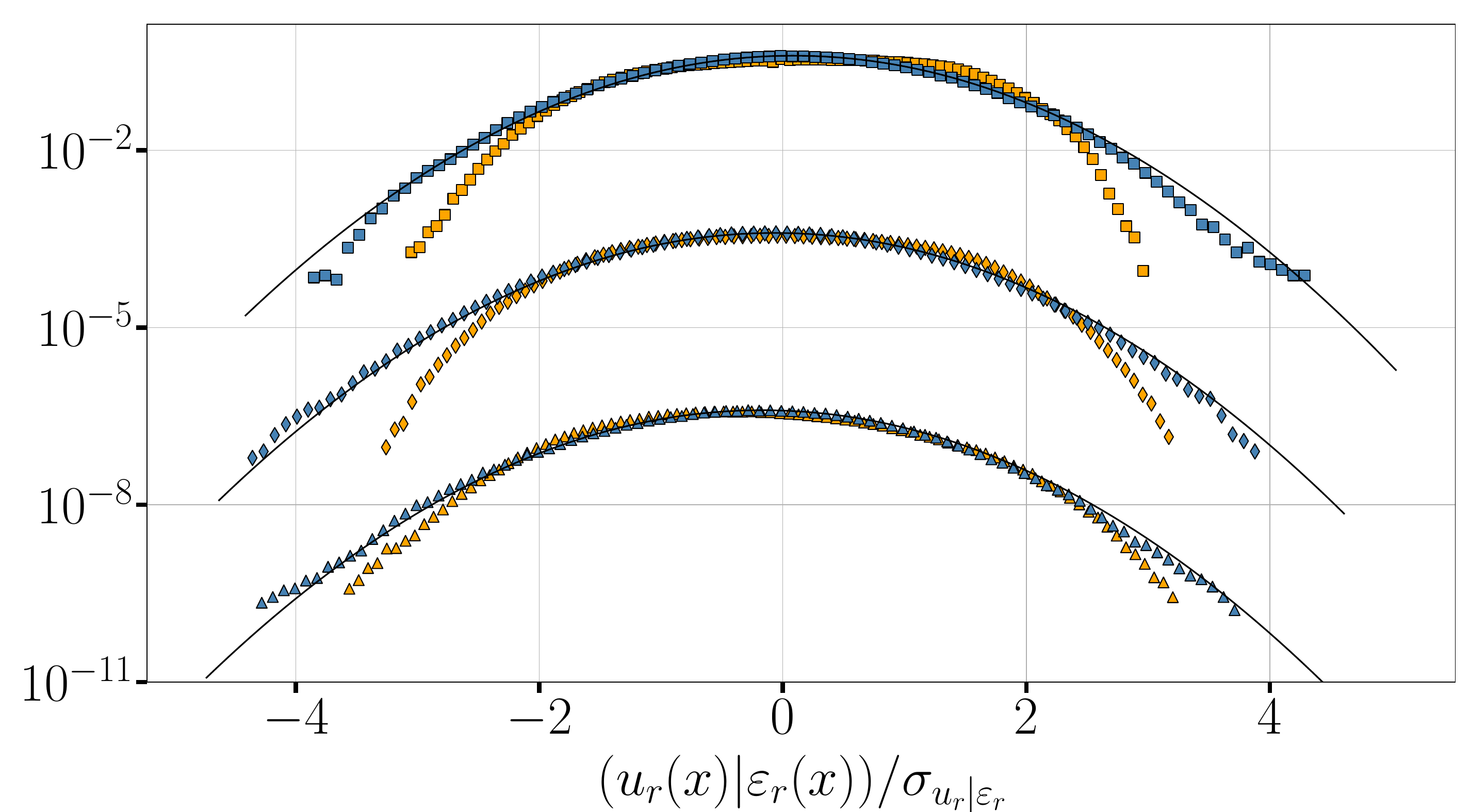}
        \caption{}
    \end{subfigure}
    \hfill
    \begin{subfigure}[t]{0.49\textwidth}
        \centering        \includegraphics[width=\linewidth]{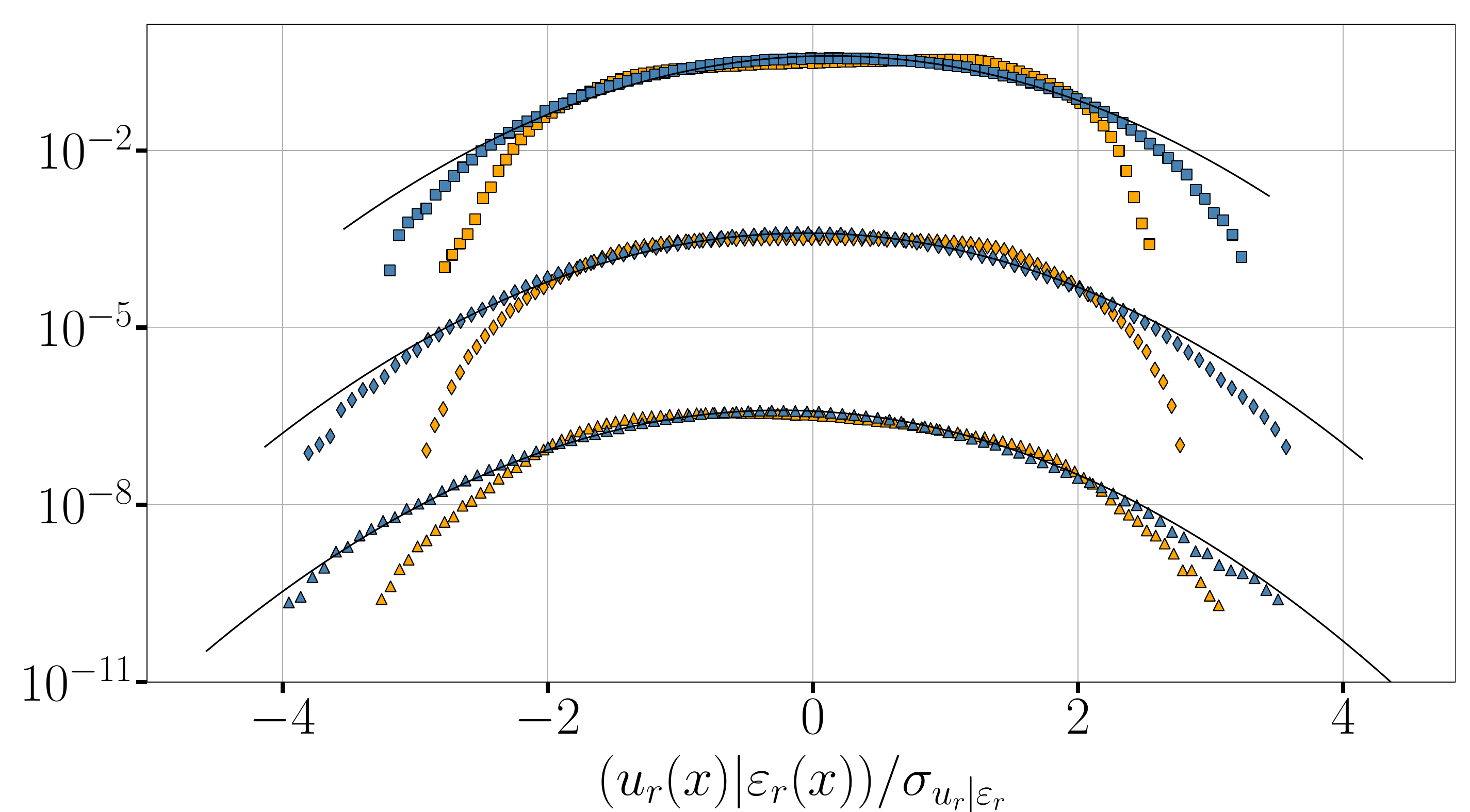}
        \caption{}
    \end{subfigure}
    \begin{subfigure}[t]{0.49\textwidth}
        \centering        \includegraphics[width=\linewidth]{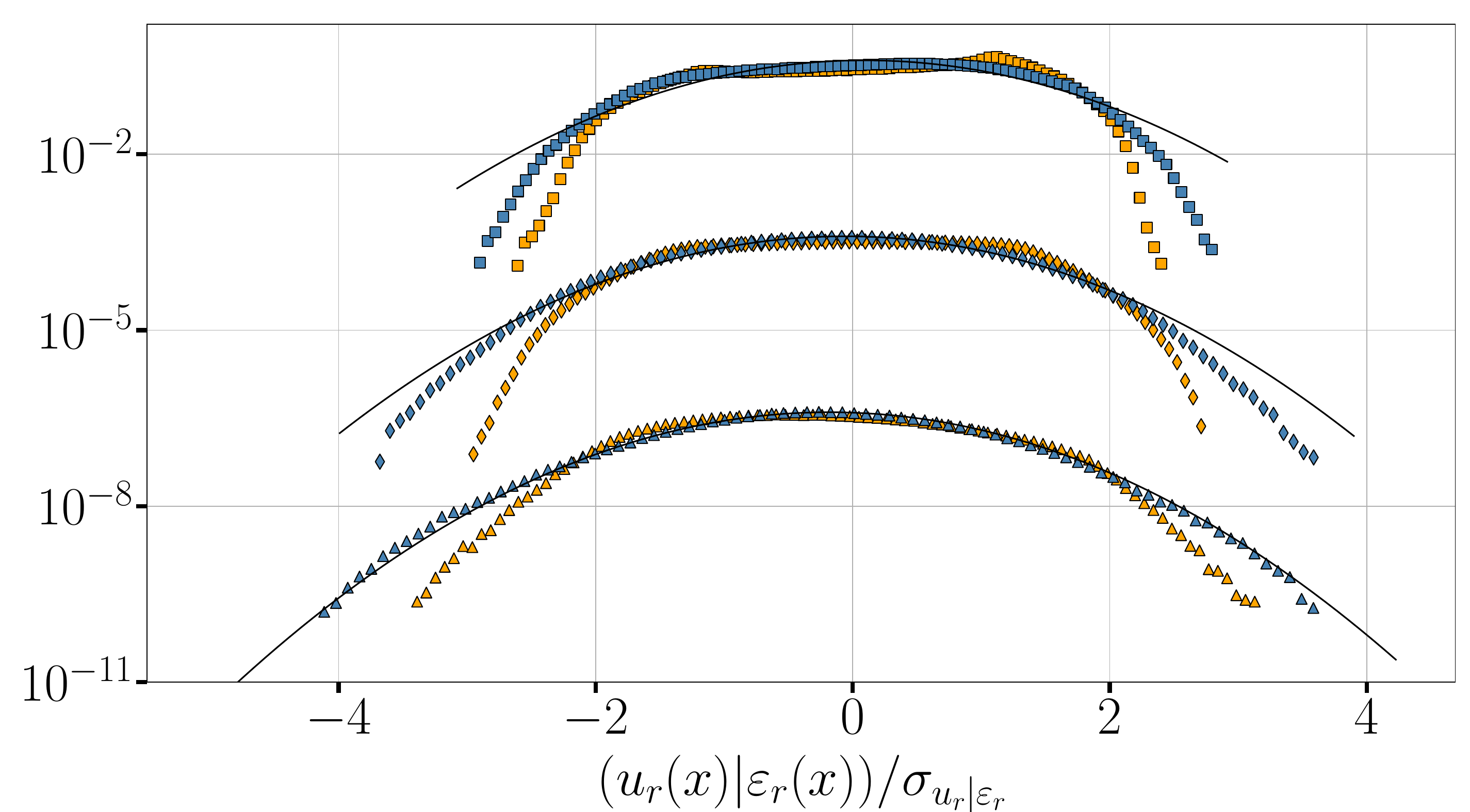}
        \caption{}
    \end{subfigure}
    \hfill
    \begin{subfigure}[t]{0.49\textwidth}
        \centering        \includegraphics[width=\linewidth]{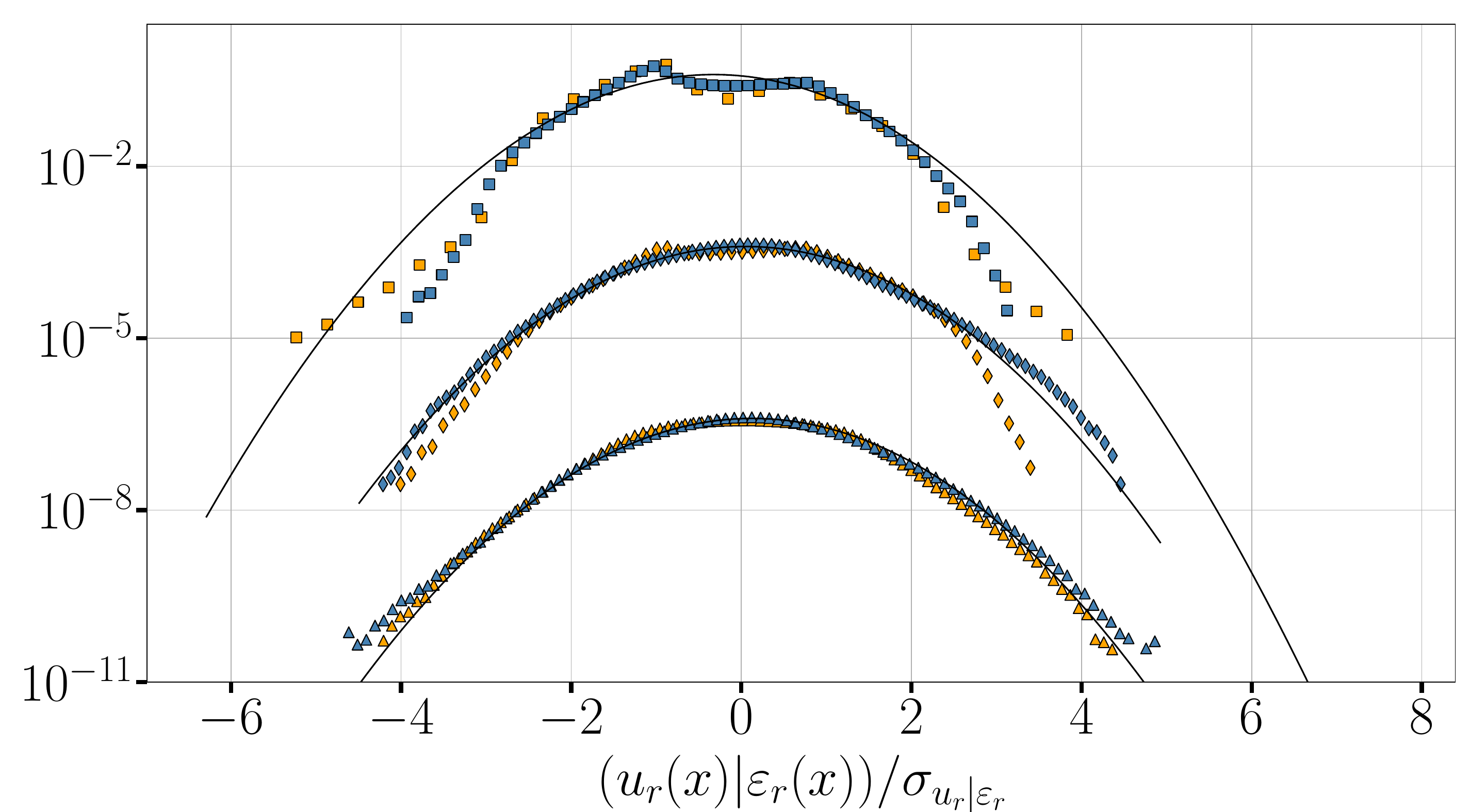}
        \caption{}
    \end{subfigure}
    \caption{\newnew{The conditioned increment PDFs $p(u_r|\varepsilon_r)$ are shown for the eight representative VTS. The scales $r$ and the intervals $\Delta_1$, $\Delta_2$, and $\Delta_3$  correspond to figure~\ref{figure_eight_examples_epsilon_r}. The black curves show Gaussian fits. Note that due to a better visibility, the PDFs are shifted vertically and that only bins are shown that have at least 10 values. a), b), c), e), f) and g) display VTS that satisfy the restriction criteria and taken together, span almost the full range of $\mu$-values. d) also fulfills the restriction criteria and exhibits the same value of $\mu$ as c) but with a $Re_\lambda$ more than three times smaller. h) however does not meet the restriction criteria since $\Lambda_0^2$ shows clear signatures of non-Gaussianity at large scales. The data stem from the cases G20, G24, C8, D11, C6, C1, C1, G23, respectively in order of appearance.} }
    \label{figure_eight_examples_conditioned_PDF}
\end{figure}

\begin{figure}[htbp]
    \centering
    \begin{subfigure}[t]{0.49\textwidth}
        \centering        \includegraphics[width=\linewidth]{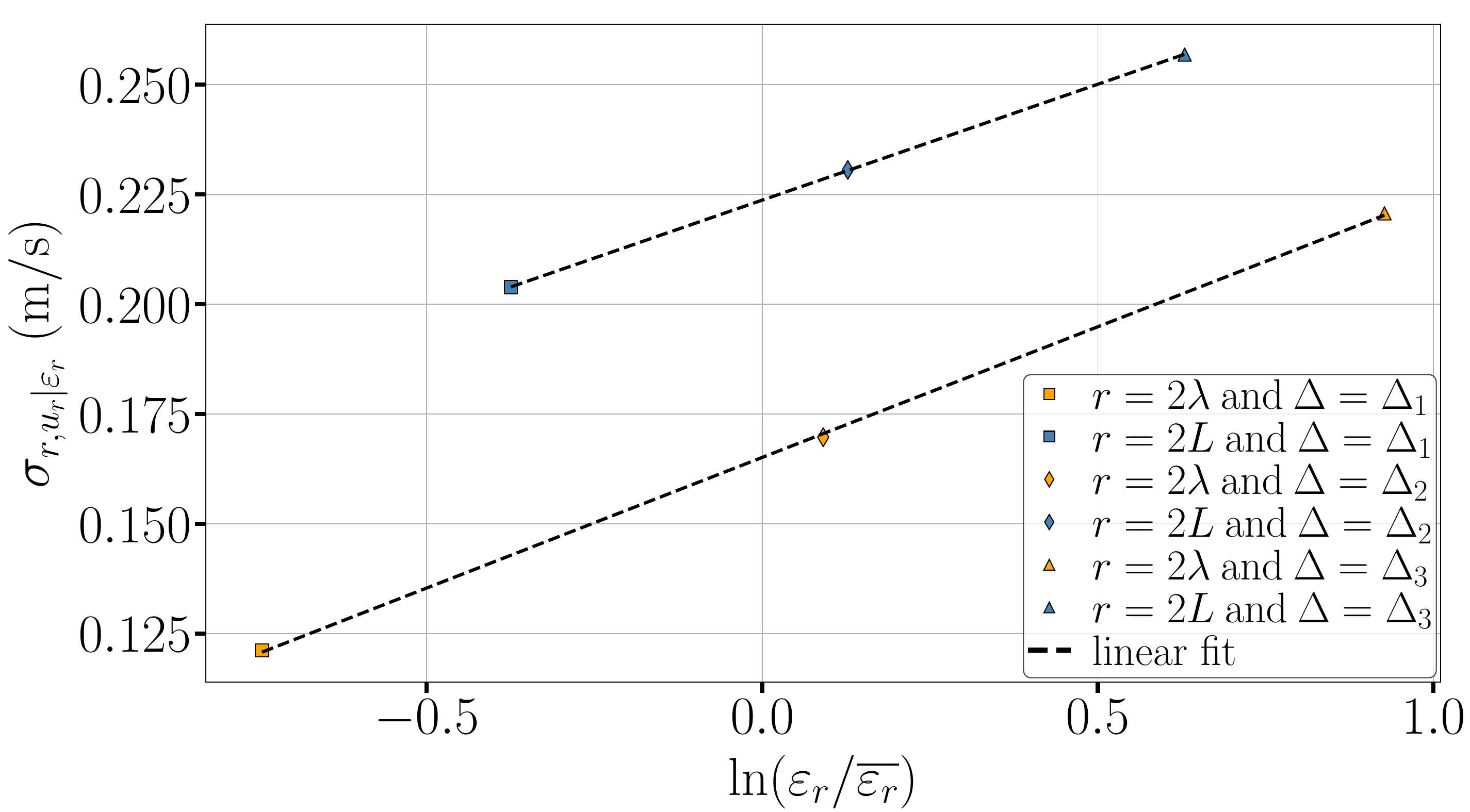}
        \caption{}
    \end{subfigure}
    \hfill
    \begin{subfigure}[t]{0.49\textwidth}
        \centering        \includegraphics[width=\linewidth]{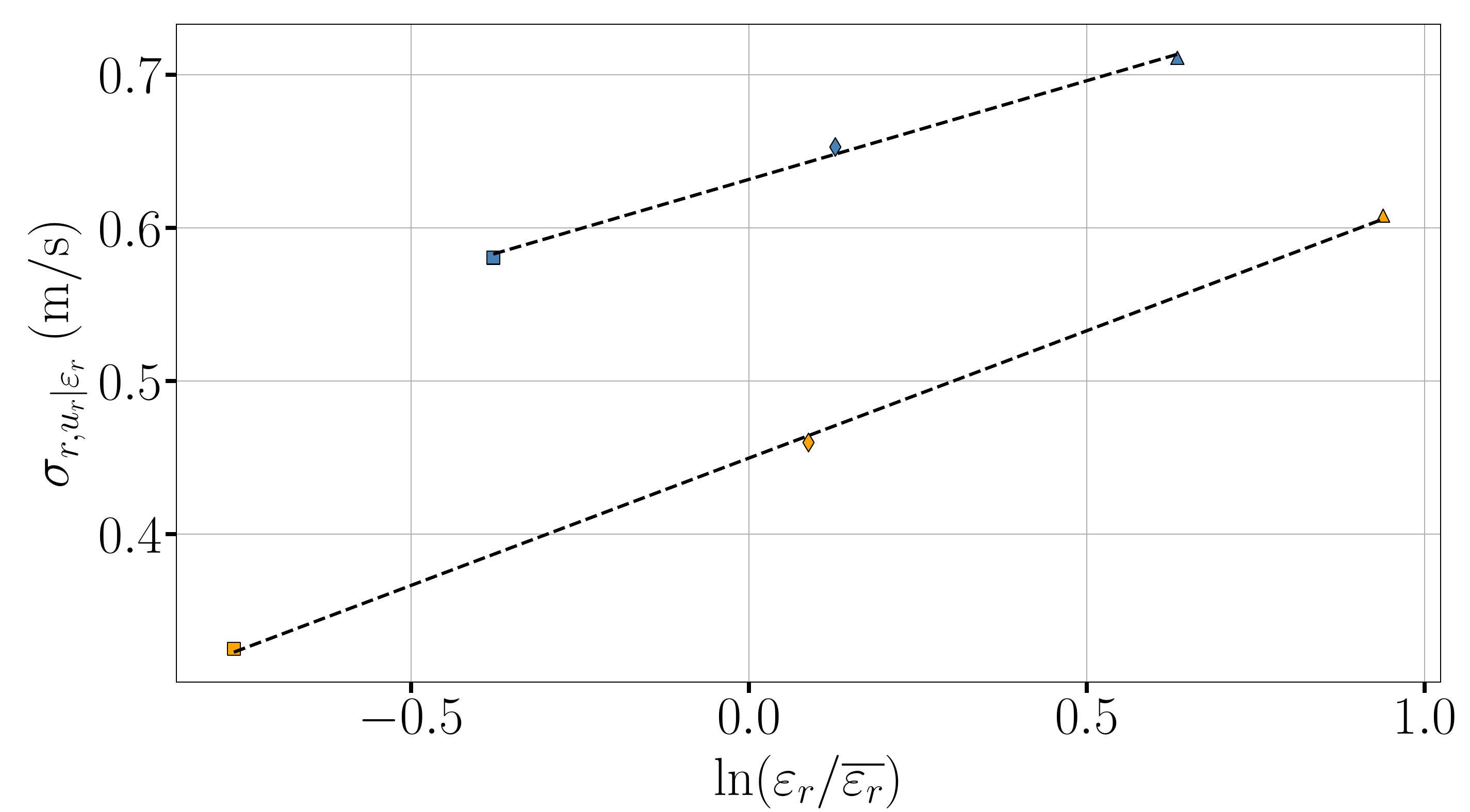}
        \caption{}
    \end{subfigure}
    \begin{subfigure}[t]{0.49\textwidth}
        \centering        \includegraphics[width=\linewidth]{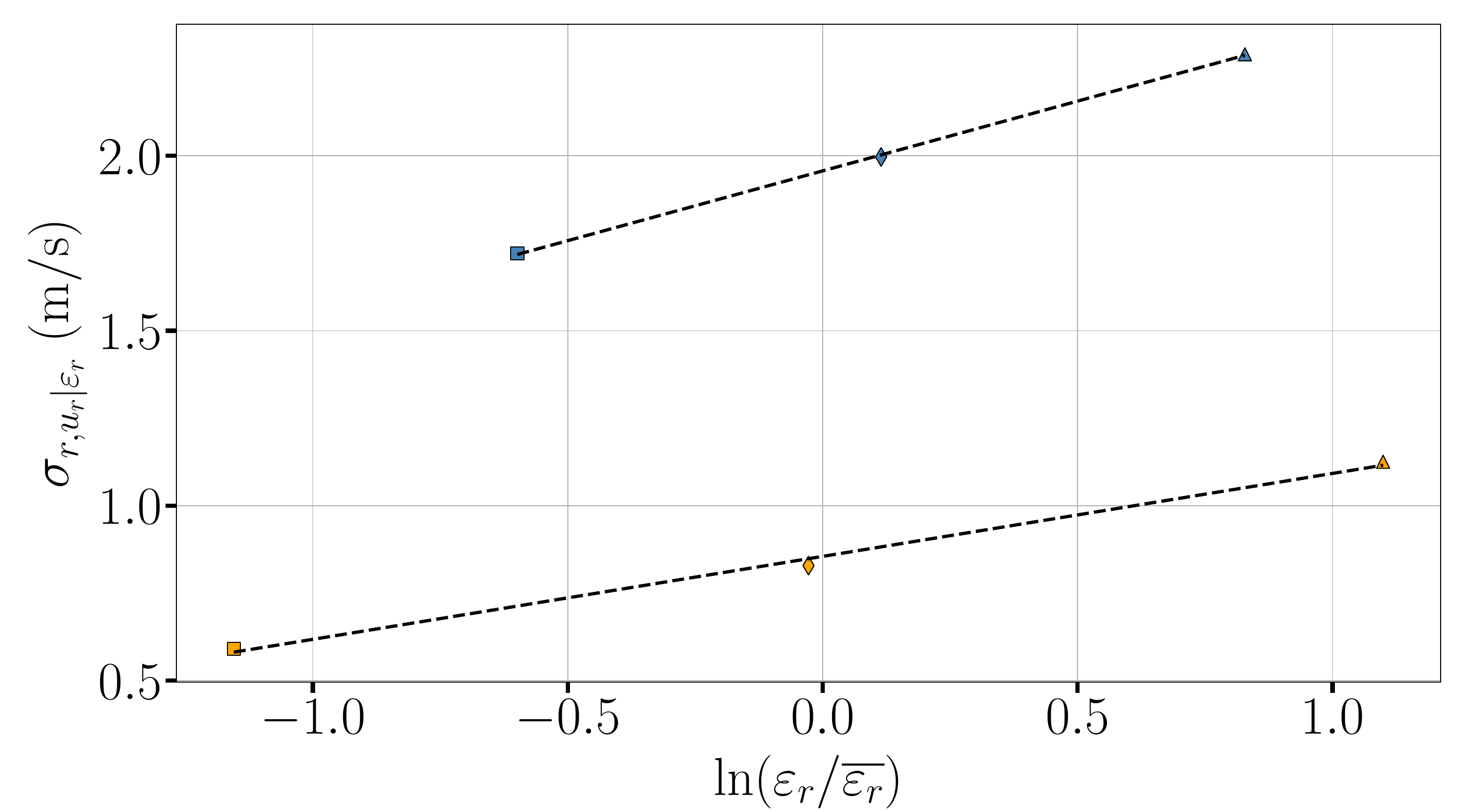}
        \caption{}
    \end{subfigure}
    \hfill
    \begin{subfigure}[t]{0.49\textwidth}
        \centering        \includegraphics[width=\linewidth]{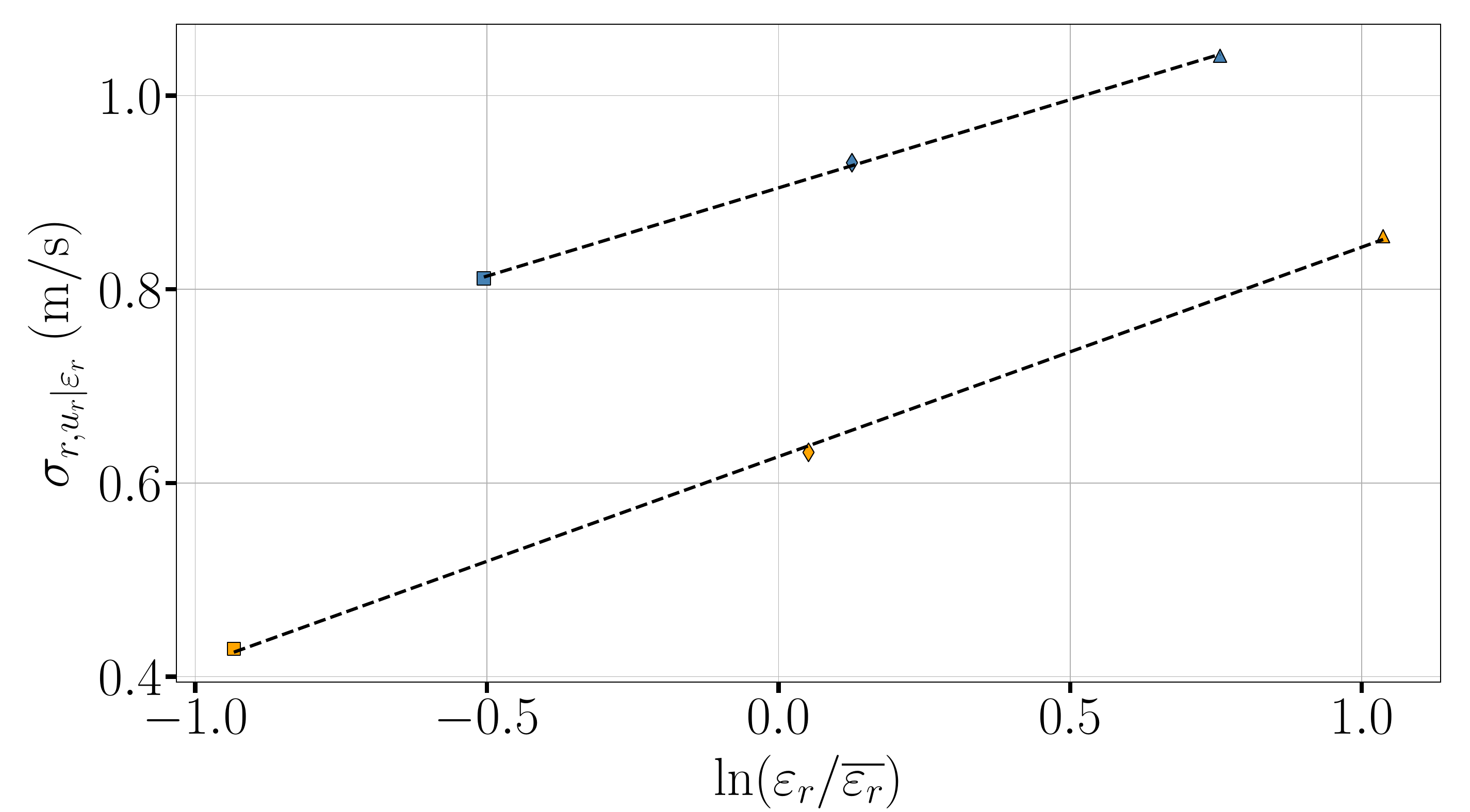}
        \caption{}
    \end{subfigure}
    \begin{subfigure}[t]{0.49\textwidth}
        \centering        \includegraphics[width=\linewidth]{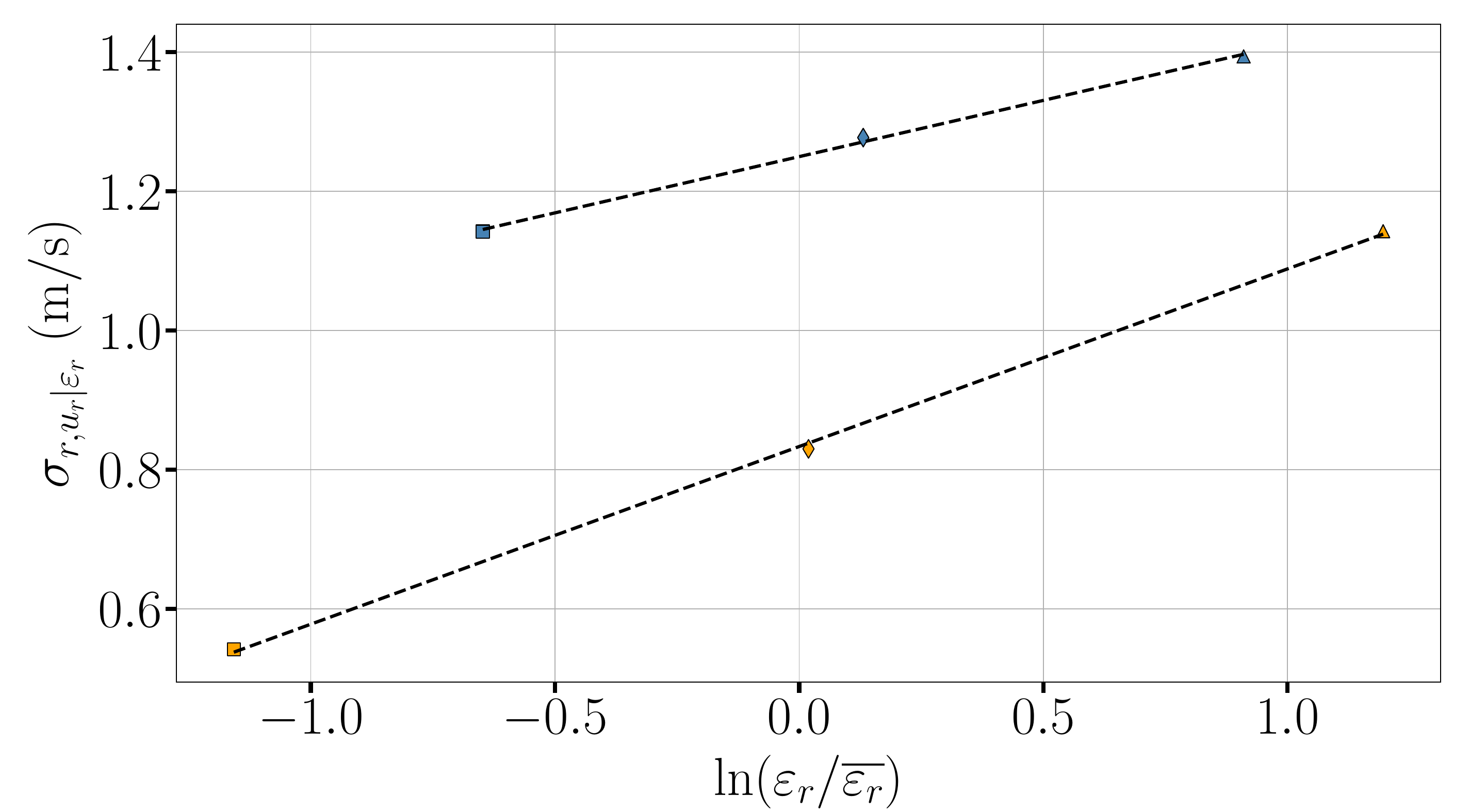}
        \caption{}
    \end{subfigure}
    \hfill
    \begin{subfigure}[t]{0.49\textwidth}
        \centering        \includegraphics[width=\linewidth]{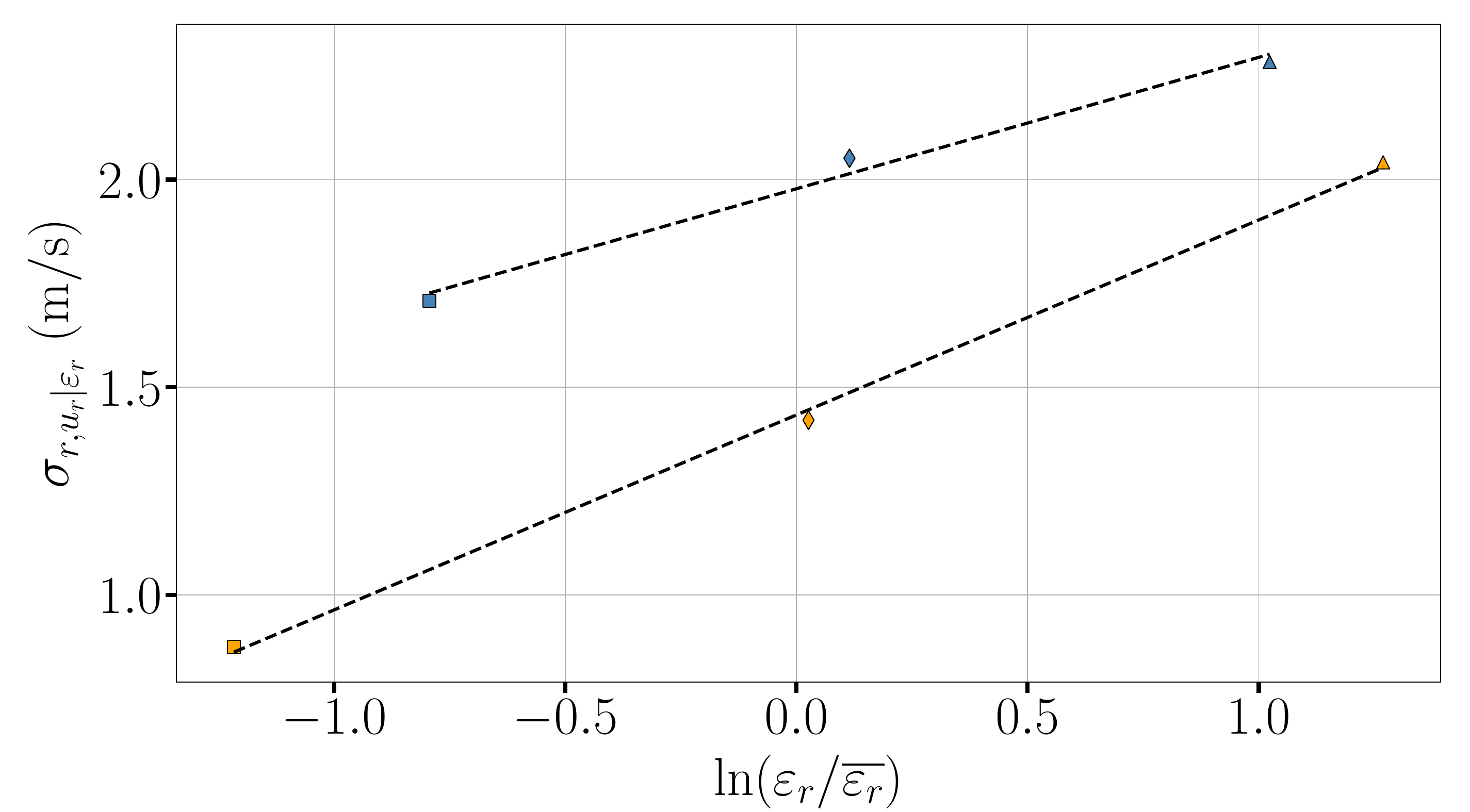}
        \caption{}
    \end{subfigure}
    \begin{subfigure}[t]{0.49\textwidth}
        \centering        \includegraphics[width=\linewidth]{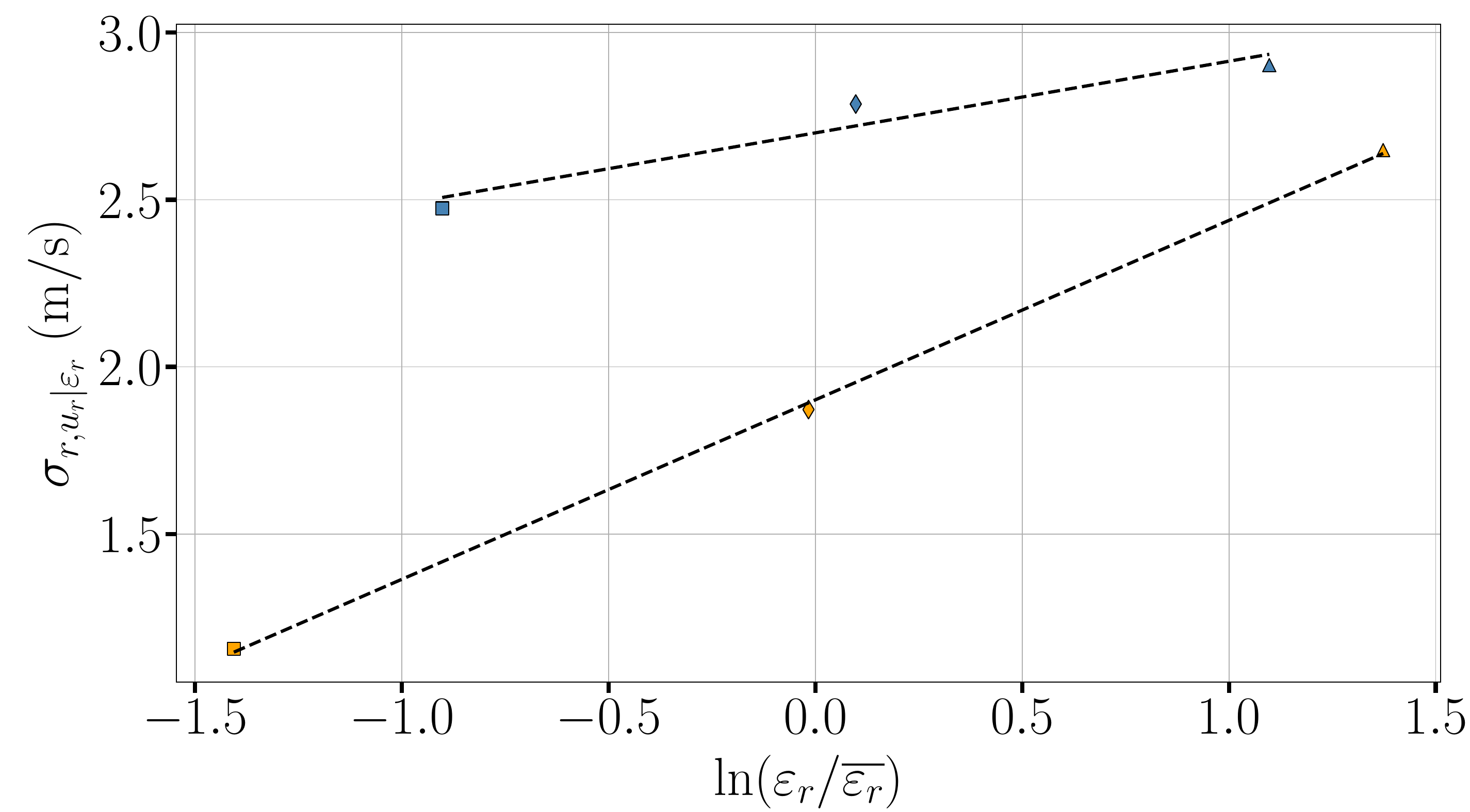}
        \caption{}
    \end{subfigure}
    \hfill
    \begin{subfigure}[t]{0.49\textwidth}
        \centering        \includegraphics[width=\linewidth]{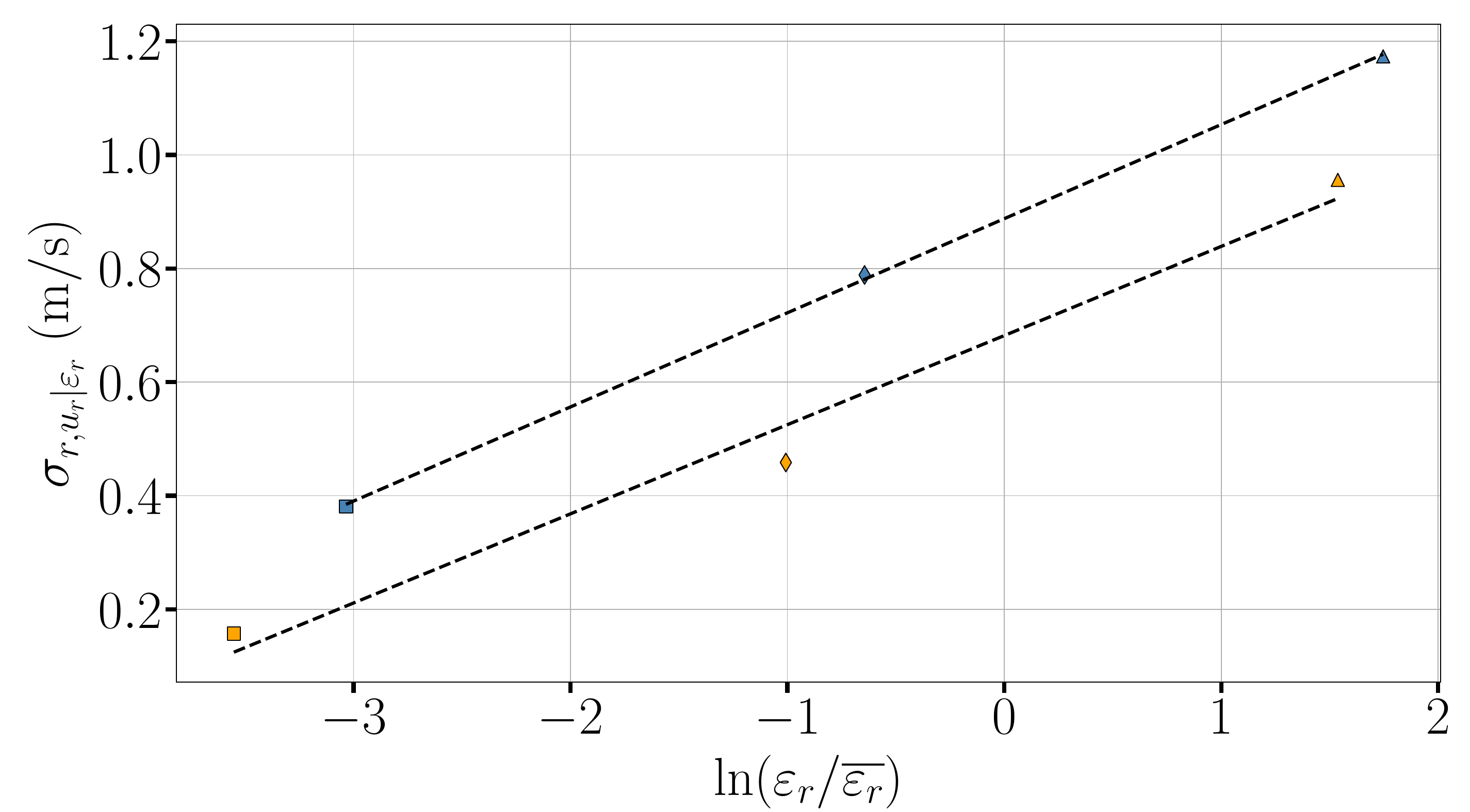}
        \caption{}
    \end{subfigure}
    \caption{\newnew{$\sigma_{r,(u_r|{\varepsilon_r})}$ as a function of $\mathrm{ln}(\varepsilon_r/\overline{\varepsilon_r})$ for the eight representative VTS. The scales $r$ and the intervals $\Delta_1$, $\Delta_2$, and $\Delta_3$ correspond to figure~\ref{figure_eight_examples_conditioned_PDF} and figure~\ref{figure_eight_examples_epsilon_r}. The black dashed lines show linear fits. a), b), c), e), f) and g) display VTS that satisfy the restriction criteria and taken together, span almost the full range of $\mu$-values. d) also fulfills the restriction criteria and exhibits the same value of $\mu$ as c) but with a $Re_\lambda$ more than three times smaller. h) however does not meet the restriction criteria since $\Lambda_0^2$ shows clear signatures of non-Gaussianity at large scales. The data stem from the cases G20, G24, C8, D11, C6, C1, C1, G23, respectively in order of appearance.}}
    \label{figure_eight_examples_castaing_assumption}
\end{figure}

\FloatBarrier

\bibliographystyle{spphys}
\bibliography{bib_tex_felix}

@article{rice1945mathematical,
	author = {Rice, Stephen O},
	journal = {Bell System Technical Journal},
	number = {1},
	pages = {46--156},
	publisher = {Wiley Online Library},
	title = {Mathematical analysis of random noise},
	volume = {24},
	year = {1945}}

@article{mcfadden1958axis,
	author = {McFadden, J},
	journal = {IRE Transactions on Information Theory},
	number = {1},
	pages = {14--24},
	publisher = {IEEE},
	title = {The axis-crossing intervals of random functions--II},
	volume = {4},
	year = {1958}}

@article{liepmann1953counting,
	author = {Liepmann, HW and Robinson, MS},
	title = {Counting methods and equipment for mean-value measurements in turbulence research},
	year = {1953}}

@article{mazellier2008turbulence,
	author = {Mazellier, Nicolas and Vassilicos, JC},
	journal = {Physics of Fluids},
	number = {1},
	pages = {015101},
	publisher = {AIP},
	title = {The turbulence dissipation constant is not universal because of its universal dependence on large-scale flow topology},
	volume = {20},
	year = {2008}}

@article{mora2019experimental,
  title={Experimental estimation of turbulence modification by inertial particles at moderate Re $\lambda$},
  author={Mora, DO and Cartellier, A and Obligado, Martin},
  journal={Physical Review Fluids},
  volume={4},
  number={7},
  pages={074309},
  year={2019},
  publisher={APS}
}

@article{sreenivasan1983zero,
	author = {Sreenivasan, KR and Prabhu, A and Narasimha, R},
	journal = {Journal of Fluid Mechanics},
	pages = {251--272},
	publisher = {Cambridge University Press},
	title = {Zero-crossings in turbulent signals},
	volume = {137},
	year = {1983}
}

@book{lesieur1987turbulence,
	author = {Lesieur, Marcel},
	publisher = {M. Nijhoff Boston},
	title = {Turbulence in fluids: stochastic and numerical modelling},
	year = {1987}}

@article{ferran2023characterising,
	author = {Ferran, Am{\'e}lie and Aliseda, Alberto and Obligado, Martin},
	journal = {Experiments in Fluids},
	number = {11},
	pages = {176},
	publisher = {Springer},
	title = {Characterising the energy cascade using the zero-crossings of the longitudinal velocity fluctuations},
	volume = {64},
	year = {2023}
}

@misc{SM,
	note = {For further information about the criteria used to select data and tests concerning the trends observed in different figues, see the appendix.}
}

@article{stevens2017flow,
	author = {Stevens, Richard JAM and Meneveau, Charles},
	journal = {Annual review of fluid mechanics},
	number = {1},
	pages = {311--339},
	publisher = {Annual Reviews},
	title = {Flow structure and turbulence in wind farms},
	volume = {49},
	year = {2017}}

@article{arneodo1996structure,
	author = {Arneodo, Alain and Baudet, Christian and Belin, F and Benzi, R and Castaing, B and Chabaud, B and Chavarria, R and Ciliberto, S and Camussi, R and Chilla, F and others},
	comment = {measuring an intermittency factor of 0.26},
	journal = {Europhysics Letters},
	number = {6},
	pages = {411},
	publisher = {IOP Publishing},
	title = {Structure functions in turbulence, in various flow configurations, at Reynolds number between 30 and 5000, using extended self-similarity},
	volume = {34},
	year = {1996}}

@article{vassilicos2015dissipation,
	author = {Vassilicos, J Christos},
	comment = {all about the dissipation constant},
	journal = {Annual review of fluid mechanics},
	pages = {95--114},
	publisher = {Annual Reviews},
	title = {Dissipation in turbulent flows},
	volume = {47},
	year = {2015}}

@article{castaing1990velocity,
	author = {Castaing, B and Gagne, Y and Hopfinger, EJ},
	comment = {introducing method of calculating mu},
	journal = {Physica D: Nonlinear Phenomena},
	number = {2},
	pages = {177--200},
	publisher = {Elsevier},
	title = {Velocity probability density functions of high Reynolds number turbulence},
	volume = {46},
	year = {1990}}

@article{sreenivasan1995universality,
	author = {Sreenivasan, Katepalli R},
	comment = {blablabl},
	journal = {Physics of Fluids},
	number = {11},
	pages = {2778--2784},
	publisher = {American Institute of Physics},
	title = {On the universality of the Kolmogorov constant},
	volume = {7},
	year = {1995}}

@article{yeung1997universality,
	author = {Yeung, PK and Zhou, Ye},
	comment = {blablabl},
	journal = {Physical Review E},
	number = {2},
	pages = {1746},
	publisher = {APS},
	title = {Universality of the Kolmogorov constant in numerical simulations of turbulence},
	volume = {56},
	year = {1997}}

@article{praskovsky1994measurements,
	author = {Praskovsky, Alexander and Oncley, Steven},
	comment = {measured Kolmogorov constant and intermittency factor},
	journal = {Physics of Fluids},
	number = {9},
	pages = {2886--2888},
	publisher = {American Institute of Physics},
	title = {Measurements of the Kolmogorov constant and intermittency exponent at very high Reynolds numbers},
	volume = {6},
	year = {1994}}

@article{praskovsky1997comprehensive,
	author = {Praskovsky, Alexander and Oncley, Steven},
	comment = {meta study of intermittency factor},
	journal = {Fluid dynamics research},
	number = {5},
	pages = {331--358},
	publisher = {Elsevier},
	title = {Comprehensive measurements of the intermittency exponent in high Reynolds number turbulent flows},
	volume = {21},
	year = {1997}}

@article{rodriguez2023not,
	author = {Rodriguez Imazio, Paola and Mininni, Pablo D and Godoy, Alejandro and Rivaben, Nicol{\'a}s and D{\"o}rnbrack, Andreas},
	comment = {Large deviations from 5/3 with a mean of 5/3},
	journal = {Journal of Geophysical Research: Atmospheres},
	number = {2},
	pages = {e2022JD037491},
	publisher = {Wiley Online Library},
	title = {Not All Clear Air Turbulence Is Kolmogorov---The Fine-Scale Nature of Atmospheric Turbulence},
	volume = {128},
	year = {2023}}

@article{kolmogorov1962refinement,
	author = {Kolmogorov, Andrey Nikolaevich},
	comment = {blablabl},
	journal = {Journal of Fluid Mechanics},
	number = {1},
	pages = {82--85},
	publisher = {Cambridge University Press},
	title = {A refinement of previous hypotheses concerning the local structure of turbulence in a viscous incompressible fluid at high Reynolds number},
	volume = {13},
	year = {1962}}

@article{kolmogorov1941local,
  title={The local structure of turbulence in incompressible viscous fluid for very large Reynolds},
  author={Kolmogorov, Andrey Nikolaevich},
  journal={Numbers. In Dokl. Akad. Nauk SSSR},
  volume={30},
  pages={301},
  year={1941}
}

@article{morales2012characterization,
	author = {Morales, A and W{\"a}chter, M and Peinke, J},
	comment = {blablabl},
	journal = {Wind Energy},
	number = {3},
	pages = {391--406},
	publisher = {Wiley Online Library},
	title = {Characterization of wind turbulence by higher-order statistics},
	volume = {15},
	year = {2012}}

@article{neunaber2020distinct,
	author = {Neunaber, Ingrid and H{\"o}lling, Michael and Stevens, Richard JAM and Schepers, Gerard and Peinke, Joachim},
	comment = {blablabl},
	journal = {Energies},
	number = {20},
	pages = {5392},
	publisher = {MDPI},
	title = {Distinct turbulent regions in the wake of a wind turbine and their inflow-dependent locations: the creation of a wake map},
	volume = {13},
	year = {2020}}

@book{frisch1995turbulence,
	author = {Frisch, Uriel},
	comment = {blablabl},
	publisher = {Cambridge university press},
	title = {Turbulence: the legacy of AN Kolmogorov},
	year = {1995}}

@article{krogstad2010grid,
  title={Is grid turbulence Saffman turbulence?},
  author={Krogstad, P-{\AA} and Davidson, PA},
  journal={Journal of Fluid Mechanics},
  volume={642},
  pages={373--394},
  year={2010},
  publisher={Cambridge University Press}
}

@article{lima2026superstatistics,
  title={Superstatistics approach to turbulent circulation fluctuations},
  author={Lima, Henrique S and Pereira, Rodrigo M and Moriconi, Luca and Sreenivasan, Katepalli R and Tsallis, Constantino},
  journal={Proceedings of the National Academy of Sciences},
  volume={123},
  number={27},
  pages={e2612658123},
  year={2026},
  publisher={National Academy of Sciences}
}

@article{melina2016vortex,
  title={Vortex shedding effects in grid-generated turbulence},
  author={Melina, G and Bruce, PJK and Vassilicos, JC},
  journal={Physical Review Fluids},
  volume={1},
  number={4},
  pages={044402},
  year={2016},
  publisher={APS}
}

@article{cafiero2020length,
  title={Length scales in turbulent free shear flows},
  author={Cafiero, Gioacchino and Obligado, Martin and Vassilicos, John Christos},
  journal={Journal of Turbulence},
  volume={21},
  number={4},
  pages={243--257},
  year={2020},
  publisher={Taylor \& Francis}
}

@article{chilla1996multiplicative,
	author = {Chilla, F and Peinke, J and Castaing, B},
	comment = {easy definition of lambda_square},
	journal = {Journal de Physique II},
	number = {4},
	pages = {455--460},
	publisher = {EDP Sciences},
	title = {Multiplicative process in turbulent velocity statistics: A simplified analysis},
	volume = {6},
	year = {1996}}

@article{reinke2018universal,
	author = {Reinke, Nico and Fuchs, Andr{\'e} and Nickelsen, Daniel and Peinke, Joachim},
	journal = {Journal of Fluid Mechanics},
	pages = {117--153},
	publisher = {Cambridge University Press},
	title = {On universal features of the turbulent cascade in terms of non-equilibrium thermodynamics},
	volume = {848},
	year = {2018}}

@article{renner2001experimental,
	author = {Renner, Christoph and Peinke, Joachim and Friedrich, Rudolf},
	journal = {Journal of Fluid Mechanics},
	pages = {383--409},
	publisher = {Cambridge University Press},
	title = {Experimental indications for Markov properties of small-scale turbulence},
	volume = {433},
	year = {2001}}

@article{mora2019energy,
	author = {Mora, DO and Mu{\~n}iz Pladellorens, E and Riera Turr{\'o}, P and Lagauzere, Muriel and Obligado, Martin},
	journal = {Physical Review Fluids},
	number = {10},
	pages = {104601},
	publisher = {APS},
	title = {Energy cascades in active-grid-generated turbulent flows},
	volume = {4},
	year = {2019}}

@article{buckingham1914physically,
	author = {Buckingham, Edgar},
	journal = {Physical review},
	number = {4},
	pages = {345},
	publisher = {APS},
	title = {On physically similar systems; illustrations of the use of dimensional equations},
	volume = {4},
	year = {1914}}

@inproceedings{vaschy1892lois,
	author = {Vaschy, Aim{\'e}},
	booktitle = {Annales t{\'e}l{\'e}graphiques},
	pages = {25--28},
	title = {Sur les lois de similitude en physique},
	volume = {19},
	year = {1892}}

@article{mydlarski1996onset,
	author = {Mydlarski, Laurent and Warhaft, Zellman},
	journal = {Journal of Fluid Mechanics},
	pages = {331--368},
	publisher = {Cambridge University Press},
	title = {On the onset of high-Reynolds-number grid-generated wind tunnel turbulence},
	volume = {320},
	year = {1996}}

@article{sreenivasan2025turbulence,
	author = {Sreenivasan, Katepalli R and Schumacher, J{\"o}rg},
	journal = {Annual Review of Condensed Matter Physics},
	number = {1},
	pages = {121--143},
	publisher = {Annual Reviews},
	title = {What is the turbulence problem, and when may we regard it as solved?},
	volume = {16},
	year = {2025}}

@article{nedic2017dissipation,
	author = {Nedi{\'c}, Jovan and Tavoularis, Stavros and Marusic, Ivan},
	journal = {Physical Review Fluids},
	number = {3},
	pages = {032601},
	publisher = {APS},
	title = {Dissipation scaling in constant-pressure turbulent boundary layers},
	volume = {2},
	year = {2017}}

@article{sinhuber2017dissipative,
	author = {Sinhuber, Michael and Bewley, Gregory P and Bodenschatz, Eberhard},
	journal = {Physical review letters},
	number = {13},
	pages = {134502},
	publisher = {APS},
	title = {Dissipative effects on inertial-range statistics at high Reynolds numbers},
	volume = {119},
	year = {2017}}

@article{schroder2024estimating,
	author = {Schr{\"o}der, Marcel and B{\"a}tge, Tobias and Bodenschatz, Eberhard and Wilczek, Michael and Bagheri, Gholamhossein},
	journal = {Atmospheric Measurement Techniques},
	number = {2},
	pages = {627--657},
	publisher = {Copernicus Publications G{\"o}ttingen, Germany},
	title = {Estimating the turbulent kinetic energy dissipation rate from one-dimensional velocity measurements in time},
	volume = {17},
	year = {2024}}

@article{schmitt2024universal,
	author = {Schmitt, F and Fuchs, A and Peinke, J and Obligado, M},
	journal = {arXiv preprint arXiv:2407.15953},
	title = {A Universal Relation Between Intermittency and Dissipation in Turbulence},
	year = {2024}}

@article{mora2020estimating,
	author = {Mora, Daniel Odens and Obligado, Martin},
	journal = {Experiments in fluids},
	number = {9},
	pages = {199},
	publisher = {Springer},
	title = {Estimating the integral length scale on turbulent flows from the zero crossings of the longitudinal velocity fluctuation},
	volume = {61},
	year = {2020}}

@article{welch1967use,
  title={The use of fast Fourier transform for the estimation of power spectra: A method based on time averaging over short, modified periodograms},
  author={Welch, Peter},
  journal={IEEE Transactions on audio and electroacoustics},
  volume={15},
  number={2},
  pages={70--73},
  year={1967},
  publisher={IEEE}
}

@article{vigneron2019wiener,
	author = {Vigneron, Francois},
	journal = {arXiv preprint arXiv:1909.06078},
	title = {On the Wiener-Khinchin transform of functions that behave as approximate power-laws. Applications to fluid turbulence},
	year = {2019}}

@article{tamburrino2024navier,
	author = {Tamburrino, Aldo},
	journal = {Fluids},
	number = {1},
	pages = {15},
	publisher = {MDPI},
	title = {From Navier to Stokes: Commemorating the bicentenary of Navier's equation on the lay of fluid motion},
	volume = {9},
	year = {2024}}

@article{fefferman2006existence,
	author = {Fefferman, Charles L},
	journal = {The millennium prize problems},
	number = {67},
	pages = {22},
	title = {Existence and smoothness of the Navier-Stokes equation},
	volume = {57},
	year = {2006}}

@article{jacobitz2024revisiting,
	author = {Jacobitz, Frank G and Schneider, Kai},
	journal = {Physical Review Fluids},
	number = {4},
	pages = {044602},
	publisher = {APS},
	title = {Revisiting Taylor's hypothesis in homogeneous turbulent shear flow},
	volume = {9},
	year = {2024}}

@article{roy2021deviations,
	author = {Roy, Sukesh and Miller, Joseph D and Gunaratne, Gemunu H},
	journal = {Communications Physics},
	number = {1},
	pages = {32},
	publisher = {Nature Publishing Group UK London},
	title = {Deviations from Taylor's frozen hypothesis and scaling laws in inhomogeneous jet flows},
	volume = {4},
	year = {2021}}

@article{hardle2003bootstrap,
	author = {H{\"a}rdle, Wolfgang and Horowitz, Joel and Kreiss, Jens-Peter},
	journal = {International Statistical Review},
	number = {2},
	pages = {435--459},
	publisher = {Wiley Online Library},
	title = {Bootstrap methods for time series},
	volume = {71},
	year = {2003}}

@book{lahiri2013resampling,
	author = {Lahiri, Soumendra Nath},
	publisher = {Springer Science \& Business Media},
	title = {Resampling methods for dependent data},
	year = {2013}}

@article{wiener1930generalized,
	author = {Wiener, Norbert},
	journal = {Acta mathematica},
	number = {1},
	pages = {117--258},
	publisher = {Springer},
	title = {Generalized harmonic analysis},
	volume = {55},
	year = {1930}}

@article{khintchine1934korrelationstheorie,
	author = {Khintchine, Alexander},
	journal = {Mathematische Annalen},
	number = {1},
	pages = {604--615},
	publisher = {Springer},
	title = {Korrelationstheorie der station{\"a}ren stochastischen Prozesse},
	volume = {109},
	year = {1934}}

@book{davidson2015turbulence,
	author = {Davidson, Peter},
	publisher = {Oxford university press},
	title = {Turbulence: an introduction for scientists and engineers},
	year = {2015}}

@inproceedings{velte2021dynamic,
	author = {Velte, Clara M and Buchhave, Preben},
	booktitle = {iTi Conference on Turbulence},
	organization = {Springer},
	pages = {3--12},
	title = {Dynamic triad interactions and non-equilibrium turbulence},
	year = {2021}}

@article{wyngaard1977taylor,
  title={Taylor's hypothesis and high--frequency turbulence spectra},
  author={Wyngaard, JC and Clifford, SF},
  journal={Journal of Atmospheric Sciences},
  volume={34},
  number={6},
  pages={922--929},
  year={1977}
}

@article{siefert2004different,
  title={Different cascade speeds for longitudinal and transverse velocity increments of small-scale turbulence},
  author={Siefert, M and Peinke, J},
  journal={Physical Review E—Statistical, Nonlinear, and Soft Matter Physics},
  volume={70},
  number={1},
  pages={015302},
  year={2004},
  publisher={APS}
}

@article{apostolidis2022scalings,
  title={Scalings of turbulence dissipation in space and time for turbulent channel flow},
  author={Apostolidis, Argyrios and Laval, Jean-Philippe and Vassilicos, JC},
  journal={Journal of Fluid Mechanics},
  volume={946},
  pages={A41},
  year={2022},
  publisher={Cambridge University Press}
}

@article{yin2024dynamics,
  title={Dynamics of turbulent energy and dissipation in channel flow},
  author={Yin, Le and Hwang, Yongyun and Vassilicos, John Christos},
  journal={Journal of Fluid Mechanics},
  volume={996},
  pages={A12},
  year={2024},
  publisher={Cambridge University Press}
}

@article{dubrulle2019beyond,
  title={Beyond kolmogorov cascades},
  author={Dubrulle, B{\'e}reng{\`e}re},
  journal={Journal of Fluid Mechanics},
  volume={867},
  pages={P1},
  year={2019},
  publisher={Cambridge University Press}
}

@article{chen2022scalings,
  title={Scalings of scale-by-scale turbulence energy in non-homogeneous turbulence},
  author={Chen, JG and Vassilicos, John Christos},
  journal={Journal of Fluid Mechanics},
  volume={938},
  pages={A7},
  year={2022},
  publisher={Cambridge University Press}
}

@article{hill2002exact,
  title={Exact second-order structure-function relationships},
  author={Hill, Reginald J},
  journal={Journal of Fluid Mechanics},
  volume={468},
  pages={317--326},
  year={2002},
  publisher={Cambridge University Press}
}

@article{ghashghaie1996turbulent,
    author = {Ghashghaie, Shoaleh and Breymann, Henriette and Peinke, Joachim and Talkner, Peter and Dodge, Yadolah},
    year = {1996},
    month = {06},
    pages = {767-770},
    title = {Turbulent Cascades in Foreign Exchange Markets},
    volume = {381},
    journal = {Nature},
}

@article{Manshour2009,
  title = {Turbulence-like Behavior of Seismic Time Series},
  author = {Manshour, P. and Saberi, S. and Sahimi, Muhammad and Peinke, J. and Pacheco, Amalio F. and Rahimi Tabar, M. Reza},
  journal = {Phys. Rev. Lett.},
  volume = {102},
  issue = {1},
  pages = {014101},
  numpages = {4},
  year = {2009},
  month = {Jan},
}

@article{Ching2007,
  title = {Multifractality and scale invariance in human heartbeat dynamics},
  author = {Ching, Emily S. C. and Lin, D. C. and Zhang, C.},
  journal = {Physical Review E},
  volume = {76},
  issue = {4},
  pages = {041910},
  numpages = {8},
  year = {2007},
  month = {Oct},
  publisher = {American Physical Society},
}

@article{pope2001turbulent,
  title={Turbulent flows},
  author={Pope, Stephen B},
  journal={Measurement Science and Technology},
  volume={12},
  number={11},
  pages={2020--2021},
  year={2001}
}

@inproceedings{schmitt2026small,
  title={On Small-Scale Intermittency for General Turbulence},
  author={Schmitt, FH and K{\"o}hne, F and Peinke, J and Obligado, M},
  booktitle={Journal of Physics: Conference Series},
  volume={3173},
  number={1},
  pages={012024},
  year={2026},
  organization={IOP Publishing}
}

@article{beck2004superstatistics,
  title={Superstatistics in hydrodynamic turbulence},
  author={Beck, Christian},
  journal={Physica D: Nonlinear Phenomena},
  volume={193},
  number={1-4},
  pages={195--207},
  year={2004},
  publisher={Elsevier}
}

@article{anselmet1984high,
  title={High-order velocity structure functions in turbulent shear flows},
  author={Anselmet, Fabien and Gagne, Yves and Hopfinger, Emil J and Antonia, Robert A},
  journal={Journal of Fluid Mechanics},
  volume={140},
  pages={63--89},
  year={1984},
  publisher={Cambridge University Press}
}

@article{saddoughi1994local,
  title={Local isotropy in turbulent boundary layers at high Reynolds number},
  author={Saddoughi, Seyed G and Veeravalli, Srinivas V},
  journal={Journal of Fluid Mechanics},
  volume={268},
  pages={333--372},
  year={1994},
  publisher={Cambridge University Press}
}

@article{gagne2004reynolds,
  title={Reynolds dependence of third-order velocity structure functions},
  author={Gagne, Yves and Castaing, Bernard and Baudet, Christophe and Mal{\'e}cot, Yann},
  journal={Physics of Fluids},
  volume={16},
  number={2},
  pages={482--485},
  year={2004},
  publisher={AIP Publishing}
}

@article{zhou2023appearance,
  title={Appearance of the- 5/3 scaling law in spatially intermittent flows with strong vortex shedding},
  author={Zhou, Yi and Nagata, Koji and Ito, Yasumasa and Sakai, Yasuhiko and Hattori, Yuji},
  journal={Physics of Fluids},
  volume={35},
  number={4},
  year={2023},
  publisher={AIP Publishing}}

@article{naert1998conditional,
  title={Conditional statistics of velocity fluctuations in turbulence},
  author={Naert, Antoon and Castaing, B and Chabaud, B and Hebral, B and Peinke, J},
  journal={Physica D: Nonlinear Phenomena},
  volume={113},
  number={1},
  pages={73--78},
  year={1998},
  publisher={Elsevier}
}

\end{document}